\documentclass[11pt,a4paper]{report}
\usepackage[dvips]{graphicx}
\usepackage[small]{caption}
\usepackage{epsfig,multicol,graphics}
\usepackage{verbatim}
\usepackage{axodraw}
\usepackage{rotating}
\usepackage{here}
\usepackage{anysize}

\def \igex {I{\sc gex}}
\def \bb {$\beta\beta$}
\def \nnbb {$2\nu\beta\beta$}
\def \onbb {$0\nu\beta\beta$}
\def \onECEC {$0\nu$2$EC$}

\def \qbb {$Q_{\beta\beta}$}

\def \Ra {$^{226}$\rm{Ra}}

\def \Rn {$^{222}$\rm{Rn}}
\def \U  {$^{238}$\rm{U}}
\def \Th {$^{232}$\rm{Th}}
\def \Co {$^{60}$\rm{Co}}
\def \Cs {$^{137}$\rm{Cs}}
\def \Bi {$^{214}$\rm{Bi}}
\def \Po {$^{214}$\rm{Po}}
\def \Tl {$^{208}$\rm{Tl}}

\def \Ge {$^{76}$\rm{Ge}}
\def \Se {$^{76}$\rm{Se}}
\def \As {$^{76}$\rm{As}}
\def \Ar {$^{36}$\rm{Ar}}
\def \Ba {$^{133}$\rm{Ba}}
\def \K {$^{40}$\rm{K}}

\def \bec {}

\def \znbb {$0\nu\beta\beta$ }
\def \tnbb {$2\nu\beta\beta$ }
\def \be {\begin{equation}}
\def \ee {\end{equation}}
\def \HdMo {H{\sc eidel\-berg}-M{\sc oscow}}
\def \GTF {G{\sc enius}-Test-Facility}
\def \genius {G{\sc enius}}
\def \HM {HdM}

\newcommand{\gerda}{G{\sc erda}}

\usepackage{amssymb}
\marginsize{4cm}{3cm}{3cm}{3cm}
\begin{document}
\flushbottom
 \pagenumbering{roman}
\thispagestyle{empty}

\baselineskip=0.75cm

\centerline{\LARGE\textbf{ Dissertation}}

\centerline{\LARGE\textbf{ submitted to the}}

\centerline{\LARGE\textbf{ Combined Faculty for Natural Sciences and
for Mathematics}}

\centerline{\LARGE\textbf{ of the Ruperto-Carola University of
Heidelberg, Germany}}

\centerline{ \LARGE\textbf{ for the degree of}}

\centerline{ \LARGE\textbf{ Doctor of Natural Sciences}}


\vskip 300 pt

\centerline{\Large presented by} \centerline{\Large Diplom
Engineer-Physicist \Large\textbf {Oleg Chkvorets}}
\centerline{\Large born in Horlivka, Ukraine}

\vskip 10 pt

\centerline{\Large Oral examination: 16.07.2008}


\newpage
\thispagestyle{empty} \vskip 80 pt

\baselineskip=0.75cm

\centerline{\LARGE\textbf{Search for Double Beta Decay with
HPGe Detectors}} \centerline{\LARGE\textbf{at the Gran Sasso
Underground Laboratory}}

\vskip 450 pt

\begin{tabular}{l l}
\Large Referees:     & \Large\bf Prof. Dr. Karl-Tasso Kn\"opfle \\
              & \\
              & \Large\bf Prof. Dr. Bogdan Povh \\
\end{tabular}


\newpage
\thispagestyle{plain}
 \centerline{\Large Zusammenfassung}
 \vskip 15 pt
\small { Der neutrinolose doppelte Betazerfall (\onbb) ist die
einzige Methode, die Majoranaeigenschaft des Neutrinos
nachzuweisen. Seine Zerfallsrate erlaubt es, die effektive
Neutrinomasse zu bestimmen. Experimente zum neutrinolosen
doppelten Betazerfall zeichnen sich durch lange Me{\ss}zeiten
in unterirdischen Labors aus. Sie erfordern eine starke
Reduktion der Umgebungsradioaktivit\"{a}t und eine hohe
Langzeitstabilit\"{a}t. Diese Probleme stehen im Mittelpunkt
der vorliegenden Arbeit, die im Zusammenhang mit den
Experimenten \HdMo, \GTF\ und \gerda\ entstanden ist. Die
Datennahme des \HdMo\ Experiments erstreckte sich \"{u}ber die
Jahre 1990 bis~2003. Im Rahmen dieser Arbeit wird eine
verbesserte Datenanalyse der \HdMo\ Daten pr\"{a}sentiert. Bei
\GTF\ handelt es sich um einen Testaufbau, in dem gepr\"{u}ft
wurde, ob nackte Germanium Detektoren in fl\"{u}ssigem
Stickstoff betrieben werden k\"{o}nnen. Die Daten des ersten
Jahres dieses Experiments werden diskutiert. Das \gerda\
Experiment wurde entwickelt, um die experimentelle
Empfindlichkeit weiter zu verbessern, indem nackte Germanium
Detektoren direkt in eine hochreine Kryofl\"{u}ssigkeit
eingebracht werden. Letztere dient sowohl als K\"{u}hlmedium,
als auch zur Abschirmung gegen radioaktiven Untergrund.
Hier\-zu wurde zun\"{a}chst die Untergrundradioktivit\"{a}t am
Ort des \gerda\ Experiments in der Halle A des Gran Sasso
Untergroundlabors gemessen. Zudem wurden die angereicherten
Detektoren der Experimente \HdMo\ und \igex\ im unterirdischen
Detektorlabor der \gerda\ Kollaboration charakterisiert und
die Langzeitstabilit\"{a}t eines nackten HPGe Detektors in
fl\"{u}ssigem Argon untersucht. Dabei wurde erstmals eine
untere Grenze f\"{u}r die Halbwertzeit des neutrinolosen
doppelten Elektroneneinfangs in \Ar\ ermittelt:
1.85$\cdot10^{18}$\,a bei 68\% statistischer Sicherheit.
 \vskip 25 pt
 \centerline{\Large Abstract}
 \vskip 15 pt
\small {Neutrinoless double-beta (\onbb) decay is practically
the only way to establish the Majorana nature of the neutrino
mass and its decay rate provides a probe of an effective
neutrino mass. Double beta experiments are long-running
underground experiments with specific challenges concerning
the background reduction and the long term stability. These
problems are addressed in this work for the \HdMo, \GTF\ and
\gerda\ experiments. The \HdMo\ (HdM) experiment collected
data with enriched \Ge\ detectors from 1990 to 2003. An
improved analysis of \HdMo\ data is presented, exploiting new
calibration and spectral shape measurements with the HdM
detectors.
\GTF\ was a test-facility that verified the feasibility of
using bare germanium detectors in liquid nitrogen. The first
year results of this experiment are discussed. The \gerda\
experiment has been designed to further increase the
sensitivity by operating bare germanium detectors in a high
purity cryogenic liquid, which simultaneously serves as a
shielding against background and as a cooling media. In the
preparatory stage of \gerda, an external background gamma flux
measurement was done at the experimental site in the Hall~A of
the Gran~Sasso laboratory. The characterization of the
enriched detectors from the \HdMo\ and \igex\ experiments was
performed in the underground detector laboratory for the
\gerda\ collaboration. Long term stability measurements of a
bare HPGe detector in liquid argon were carried out. Based on
these measurements, the first lower limit on the half-life of
neutrinoless double electron capture of \Ar\ was established
to be 1.85$\cdot10^{18}$ y (68\% C.L).}

\normalsize 
\newpage
\tableofcontents
\newpage \pagenumbering{arabic}
\chapter{Introduction} \label{ch:intro}
Existence of massive neutrinos and violation of the total
lepton number will require a new physics, because the present
Standard model of matter interaction assumes that neutrinos
are massless and lepton number is strictly conserved. However,
neutrino oscillation experiments confirm a non-vanishing
neutrino mass, but without providing any information on the
absolute mass scale. Neutrinoless double beta (\onbb) decay
may be the most sensitive way to look for lepton number
violation and to conclude the Dirac or Majorana nature of the
neutrino mass, while yielding the absolute scale of the
neutrino mass.\\
This thesis focuses on the experimental search for \onbb\
decay with germanium detectors in the framework of the \HdMo\
(\HM) and \gerda\ (GERmanium Detectors Array) experiments.
Both experiments are located in the Gran Sasso underground
laboratory (LNGS) of INFN in Italy. The \HM\ experiment had
searched for \onbb\ decay using five Ge diodes enriched in
\Ge. The \HM\ has collected data from 1990~to~2003. The \HM\
detectors were conventionally operated high purity germanium
(HPGe) detectors enclosed in vacuum copper cryostats and
cooled down via a cold finger with liquid nitrogen. A massive
shield was used to reduce external gamma radiation. One of the
main results presented in this thesis has lead to a further
increase in the sensitivity of the \HM\ experiment using an
improved data analysis based on investigation of detector
calibration \cite{KK-NewAn-NIM04}. This analysis has some
advantages with respect to the old procedure. The energy
resolution of the sum spectrum is improved by 20\%,
increasing the sensitivity of the \HM\ experiment by up to 10\%.\\
This thesis confirms previously obtained result
\cite{KK-Doer03,  Bi-KK03-NIM, Diss-Dipl-Dietz,
Diss-Dipl-Doer} that the \HM\ background is formed by the
radioactive sources located mainly in the detectors'
constructive materials: vacuum cryostats, detectors holders
and electrical contacts. A significant reduction of the
background is planned in the \gerda\ experiment \cite{Gerda},
which will use a new technique, proposed by G.\,Heusser
\cite{heu95}.
\begin{figure}
  \centering
  \includegraphics[width=12cm]{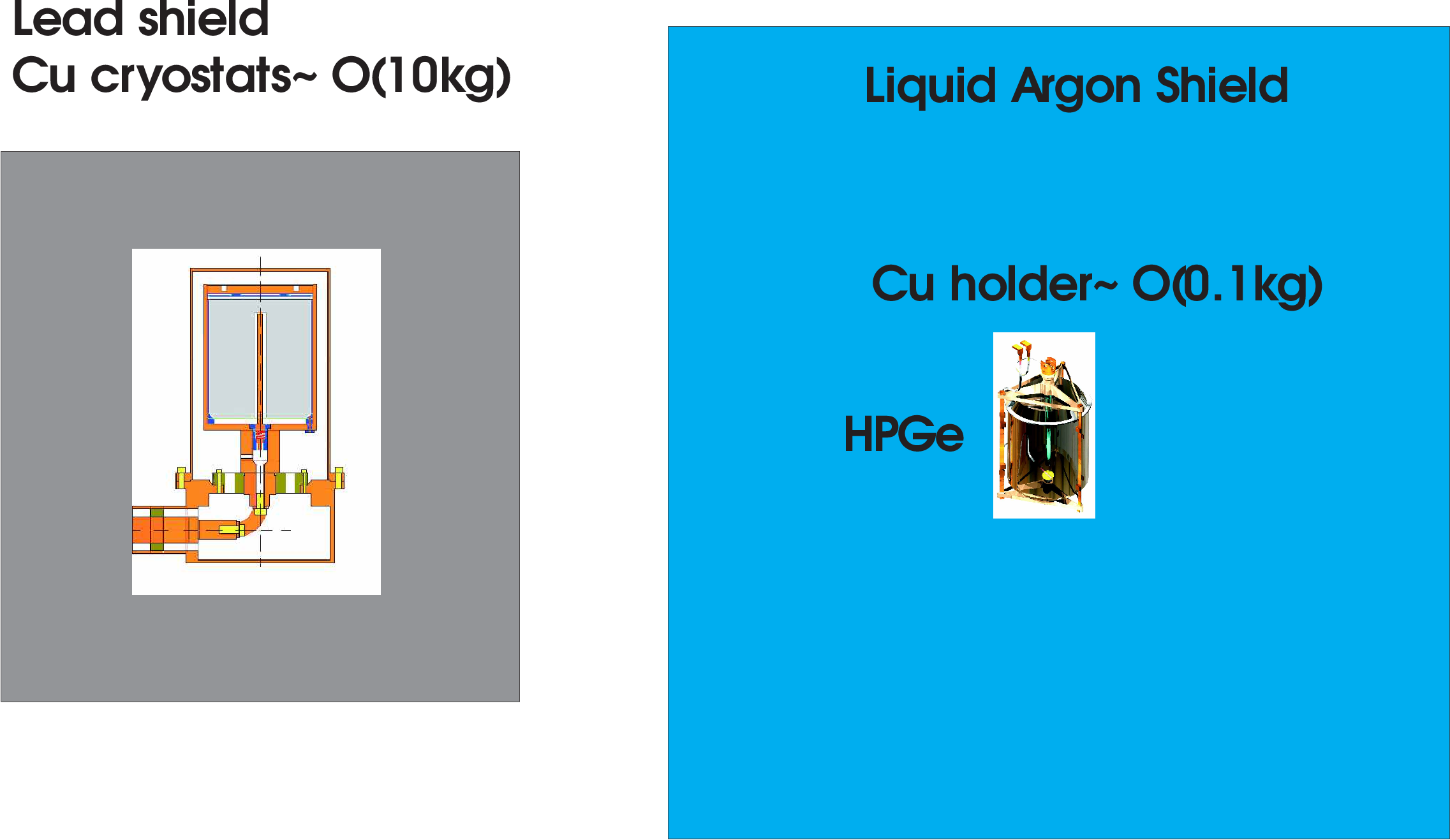}\\
  \caption{Old and new approaches for a \Ge\ double beta decay search.}\label{New concept}
\end{figure}
According to this novel technique, bare germanium detectors
will be operated inside high purity liquid nitrogen or argon,
which act both as a cooling medium and as shielding from
external radiation (Fig.~\ref{New concept}). Without the use
of vacuum cryostats, the amount of numerous pieces of solid
materials surrounding the detectors is significantly reduced.
This technique was considered in the \genius\ (GErmanium in
liquid NItrogen Underground Setup) and GEM proposals as well
\cite{HVKK-Prop97, GEM}. A \genius\ test facility
(\genius-TF)\cite{NIM02-TF} had been commissioned at LNGS in
May 2003 \cite{NIM-TF-03}. Four bare germanium detectors, with
a total mass of about 10 kg had been operated in liquid
nitrogen. During the first year of the \genius-TF operation,
detector parameters remained stable \cite{NIM-TF-04},
showing principal feasibility of the technique.\\
The following is an outline of this thesis. Chapter
\ref{ch:dbd} presents an introduction to \onbb\ decay and its
potential. 
Also, the principle of detecting rare events like the \onbb\
decay using HPGe detectors is discussed. Chapter \ref{ch:hdmo}
introduces the setup of the \HM\ experiment, its technical
parameters and the details of calibration and data analysis.
The improved analysis was used to evaluate the final \HM\ data
set. The recent claim about \onbb\ decay observation by
H.\,V.\,Klapdor-Kleingrothaus et al. \cite{KK-NewAn-NIM04,
KK-NewAn-PL04} is reviewed in Chapter \ref{ch:bkg}, exploiting
new spectral shape measurements with the \HM\ detectors. In
Chapter \ref{ch:genius}, the \genius-TF setup is described.
The operation of bare HPGe detectors during the first year of
measurements is summarized. In addition, the radon background
sources and methods to suppress it are discussed. The
measurements of the external gamma background on the \gerda\
experimental site in Hall A of LNGS are presented in Chapter
\ref{ch:halla}. Chapter \ref{ch:detchar} describes
characterization of the enriched detectors performed in the
underground \gerda\ Detector Laboratory (GDL) at LNGS after
the end of the HdM and IGEX experiments. The performance of
these detectors before refurbishment for \gerda\ is given.
Chapter \ref{ch:ar36} illustrates the operation of a bare HPGe
germanium detector in the GDL test setup. Long term stability
measurements of the HPGe detector with \Co\ were performed.
The background measured for 10 days was used to derive a
half-life limit for the radiative neutrinoless double electron
capture in \Ar\ isotope naturally occurring in liquid argon.
For the first time the limit on the half-life of radiative
neutrinoless double electron capture (\onECEC) decay of \Ar\
was obtained.\\
This work has been carried out by the author in a four year
period, during which the author was involved in the operation
and analysis of the \HdMo, HDMS and \genius-TF experiments,
and in preparatory experimental studies for the \gerda\
experiment for the Max-Planck-Institut f\"ur Kernphysik.

\newpage
\chapter{Neutrinoless Double Beta Decay}
\label{ch:dbd}

\section{Lepton number violation and the Majorana neutrino}
Lepton number violation, which is the case if neutrinos are
Majorana particles, creates a possibility to explain the
excess of the matter over antimatter, thus explaining the
overwhelming dominance of matter in the Universe. In a
standard model of particle physics, neutrinos are strictly
massless, the neutrinos and antineutrinos are different
particles and the lepton number is conserved
\cite{bbteoreview}. Experimental evidence states that the
neutrino has a non-zero mass, as deduced from the neutrino
flavor oscillations observed in atmospheric-SuperKamiokande,
reactor-KamLAND and solar neutrino-GALLEX/GNO-SAGE-SNO
experiments (For review see e.g.\cite{OscillReview}). Neutrino
oscillation experiments determine the mass squared differences
but not the absolute mass. The nuclear neutrinoless double
beta decay (\onbb) is a lepton number violating process (A, Z)
$\rightarrow$ (A, Z+2) + $2e^{-}$. It can only exists if the
neutrino is a massive Majorana ($\nu\equiv\bar\nu$) particle.
For the non-standard \onbb\ process to happen, the emitted
neutrino in the first neutron decay must equal to its
antineutrino and match the helicity of the neutrino absorbed
by the second neutron (Fig.~\ref{dbd_diagram}(a)).
\begin{center}
\begin{figure}[width=8cm]
  \begin{picture}(328,143) (-16,-11) 
    \SetWidth{0.5}
    \SetColor{Black}
    \Photon(118,34)(118,85){3}{6}
    \ArrowLine(60,85)(29,116)
    \ArrowLine(118,85)(150,116)
    \Photon(60,34)(60,85){3}{6}
    \ArrowLine(29,2)(61,34)
    \ArrowLine(149,2)(118,34)
    \ArrowLine(118,34)(150,65)
    \ArrowLine(60,34)(29,65)
    \ArrowLine(23,2)(23,65)
    \ArrowLine(17,2)(17,65)
    \ArrowLine(155,2)(155,65)
    \ArrowLine(161,2)(161,65)
    \Text(17,-11)[lb]{\Large{\Black{$n$}}}
    \Text(78,-20)[lb]{\Large{\Black{$a)$}}}
    \Text(149,-11)[lb]{\Large{\Black{$n$}}}
    \Text(35,116)[lb]{\Large{\Black{$e^-$}}}
    \Text(130,116)[lb]{\Large{\Black{$e^-$}}}
    \Text(67,52)[lb]{{\Black{$W^-$}}}
    \Text(98,52)[lb]{{\Black{$W^-$}}}
    \Text(17,72)[lb]{\Large{\Black{$p$}}}
    \Text(149,72)[lb]{\Large{\Black{$p$}}}
    \ArrowLine(60,85)(92,85)
    \ArrowLine(117,85)(92,85)
    \Vertex(92,85){2.83}
    \Text(86,95)[lb]{\Large{\Black{$\nu_{\sc M}$}}}
    \Photon(282,34)(282,85){3}{6}
    \ArrowLine(225,85)(194,116)
    \ArrowLine(282,85)(314,116)
    \Photon(225,34)(225,85){3}{6}
    \ArrowLine(194,2)(226,34)
    \ArrowLine(313,2)(282,34)
    \ArrowLine(282,34)(314,65)
    \ArrowLine(225,34)(194,65)
    \ArrowLine(187,2)(187,65)
    \ArrowLine(181,2)(181,65)
    \ArrowLine(320,2)(320,65)
    \ArrowLine(326,2)(326,65)
    \Text(181,-11)[lb]{\Large{\Black{$n$}}}
    \Text(248,-20)[lb]{\Large{\Black{$b)$}}}
    \Text(314,-11)[lb]{\Large{\Black{$n$}}}
    \Text(200,116)[lb]{\Large{\Black{$e^-$}}}
    \Text(295,116)[lb]{\Large{\Black{$e^-$}}}
    \Text(231,52)[lb]{{\Black{$W^-$}}}
    \Text(263,52)[lb]{{\Black{$W^-$}}}
    \Text(181,72)[lb]{\Large{\Black{$p$}}}
    \Text(314,72)[lb]{\Large{\Black{$p$}}}
    \Text(262,97)[lb]{\Large{\Black{$\bar\nu$}}}
    \ArrowLine(281,85)(269,117)
    \ArrowLine(225,85)(238,117)
    \Text(237,97)[lb]{\Large{\Black{$\bar\nu$}}}
\end{picture}
\label{dbd_diagram}\caption{Feynman diagrams of
neutrinoless~(a) and two-neutrino~(b) double beta
decay.}\label{dbd_diagram}
\end{figure}
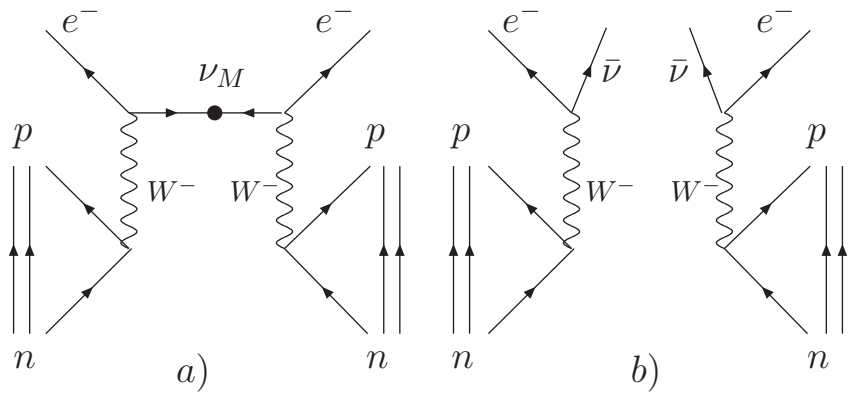
\end{center}
The Standard Model allowed the two neutrino double beta decay
(\tnbb), (A, Z) $\rightarrow$ (A, Z+2) + $2e^{-}$ + $2\bar\nu$
(Fig.~\ref{dbd_diagram}(b)) is a second order effect of weak
interaction in the nucleus and was observed for many nuclei.
The Schechter-Valle theorem \cite{blackbox} shows that in any
gauge theory, whatever mechanism is responsible for the
neutrinoless double beta decay, a massive Majorana neutrino is
required. The neutrinoless decay half-life (assuming the light
$\nu$ exchange mechanism) is expressed as \cite{Vogel}:
\begin{equation}\label{DBD_rate}
(T_{1/2}^{0\nu})^{-1}= G_{0\nu} \mid M^{0\nu} \mid ^2\langle
m_\nu\rangle^2
\end{equation}
where, $M^{0\nu}$ is the nuclear matrix-element,
$M^{0\nu}=M_{GT}^{0\nu}-(g_V/g_A)^2\ M_F^{0\nu}$, with
$M_{GT,F}^{0\nu}$ the corresponding Gamow-Teller and Fermi
contributions, $M^{0\nu}$ is in the range 3.3-5.7
\cite{Rodin08}, and $G_{0\nu}$ is an integrated kinematic
factor $G_{0\nu}=2.44\cdot10^{-26}[1/y]$ \cite{Vogel} for
\onbb\ decay of \Ge. The quantity $\langle m_\nu\rangle
=\Sigma_j\lambda_jm_jU_{ej}^2$ is the effective neutrino mass
parameter, where $U_{ej}$ is a unitary matrix describing the
mixing of neutrino mass eigenstates to electron neutrinos,
$\lambda_j$ a CP phase factor, and $m_j$ the neutrino mass
eigenvalue. The effective Majorana neutrino mass is than
expressed as a function of the half-life of the \Ge\ \onbb\
decay as:
\begin{equation}\label{Ge76NuMass}
    \langle m_\nu\rangle = \frac{6.4\cdot10^{12}}{\mid M^{0\nu} \mid \cdot
    \sqrt{T^{0\nu}_{1/2}[y]}}[eV].
\end{equation}
The discovery of a \onbb\ decay will tell that the Majorana
neutrino has a mass equal or larger than $\langle
m_\nu\rangle$. On the contrary, when only a lower limit of the
half-life is obtained, one gets only an upper bound on
$\langle m_\nu\rangle$, but not an upper bound on the mass of
any neutrino. In fact, $\langle m_\nu\rangle_{exp}$ can be
much smaller than the actual neutrino masses. The $\langle
m_\nu\rangle$ bounds crucially depend on the nuclear model
used to compute the \onbb\ matrix element.

\section{Search for double beta decay with Germanium detectors}

Double beta decay of \Ge,
\begin{equation}\label{G76-decay}
    ^{76}\rm{Ge} \rightarrow ^{76}\!\rm{Se} + 2e^- (+ 2\bar\nu_e),
\end{equation}
bypassing \As\ is possible because the pairing energy makes a
nucleus with an even number of neutrons and protons more
tightly bound than its odd-odd neighbor \As\
(Fig.~\ref{GeLevels}).
\begin{figure}[ht]
\centering
  \includegraphics[width=10 cm]{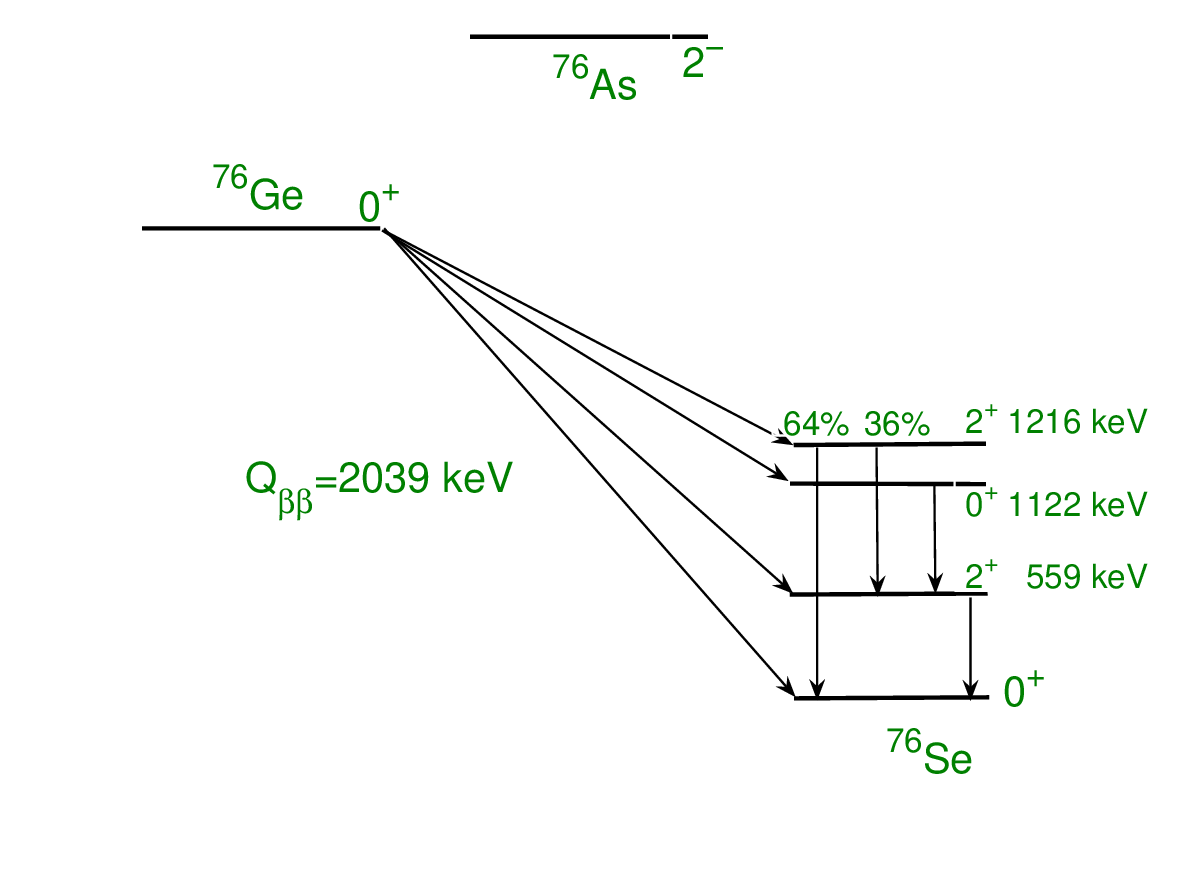}\\
  \caption{Lowest energy levels of isobar triplet A=76 with double beta decay of
  \Ge\ \cite{TOI99}.}\label{GeLevels}
\end{figure}
Because double beta decay is the most rare process known, its
experimental investigation requires a large amount of emitters
and low-background detectors with capability of selecting the
signal from the background reliably. Ge detectors provide
excellent energy resolution, so the peak at the \qbb\ value at
2039\,keV expected for \onbb\ decay could be seen with a full
width at half maximum (FWHM) of about 3 keV. This helps
considerably in reducing background counts in a region of
interest around \qbb. The material for HPGe detector
production is supposed to be of highest purity (impurity level
as low as $\sim$ 10$^9$\,atoms/cm$^3$). Germanium double beta
decay experiments start with the cleanest source of double
beta emitter material, which is also used as a detector at the
same time. The experimental signatures of the \onbb\ decay to
the ground state is a peak at the \qbb\ value in the
two-electron summed energy spectrum and a continuous \tnbb\
spectrum (Fig.~\ref{dbd_spectrum}).
\begin{figure}[!h]
\centering
  \includegraphics[width=10 cm]{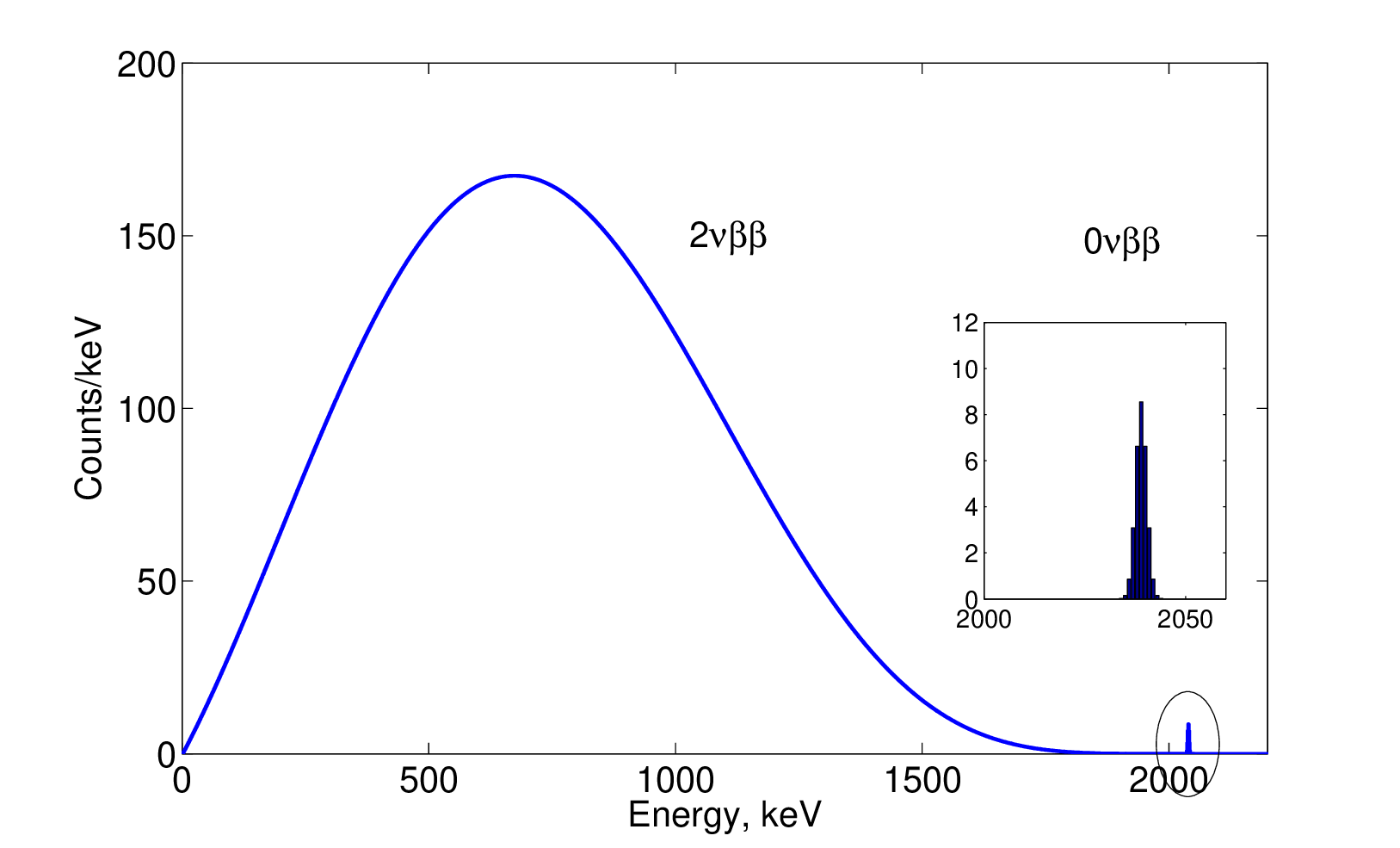}\\
  \caption{A calculated two-electron summed energy
spectrum of double beta decay of \Ge\ to ground state. Number
of counts in the continuous spectrum corresponds to \nnbb\
decay of \Ge\ with a half-life
$T^{2\nu}_{1/2}~=~1.7\cdot10^{21}~y$. The peak at Q-value
corresponds to \onbb\ decay with a half-life
$T^{0\nu}_{1/2}~=~1.2\cdot10^{25}~y$. The exposure is
72\,kg\,y. A \onbb\ peak in the inset has a typical resolution
for HPGe detectors (3\,keV FWHM at
2\,MeV).}\label{dbd_spectrum}
\end{figure}
The \onbb\ decay to the excited states of the \Se\ is
suppressed by the kinematic phase space factor, which depends
on the \qbb\ value as $\sim{Q^5}$. Despite such characteristic
signal, the rarity of the process makes their identification
very difficult. Probable signals have to be disentangled from
background due to natural radioactivity, cosmogenic-induced
activity and anthropogenic radioactivity, which deposit energy
in the \onbb\ region. The general approach followed to perform
a \onbb\ decay experiment is dictated by the expression of
half-life:
\begin{equation}\label{halflife}
T_{1/2} = ln2\cdot\varepsilon_E\cdot\frac{N\cdot t}{S}
\end{equation}
where $\varepsilon_E$ is the detector efficiency, $N$ is the
number of decaying nuclei and $S$ is the number of recorded
peak counts during time $t$ (or the upper limit of peak counts
consistent with the observed background). In the case when S
is the 1$\sigma$ background fluctuation and the detector is
made of the \bb\ source, the \onbb\ decay experiment
sensitivity at 68\%~C.L. can be derived from
Eq.~\ref{halflife} using:
\begin{eqnarray}\label{DBD_sens}
\nonumber N &=& N_A\cdot\frac{a\cdot M}{A},\\
\nonumber\\
\nonumber S &=& \sqrt{\Delta E\cdot B\cdot M\cdot t },\\
\nonumber\\
T_{1/2}^{0\nu}\,[y]&=&4.17\cdot10^{26}\times\frac{\varepsilon_E\cdot
a}{A} \sqrt{\frac{M\cdot t}{\Delta{E}\cdot B}},
\end{eqnarray}
where $B$\,[keV\,kg\,y]$^{-1}$ is the background index,
M\,[kg] is the active mass of \bb\ emitters, $N_A$ is the
Avogadro number, $\Delta E$\,[keV] is the energy window around
Q$_{\beta\beta}$ ($\Delta E = 3\sigma$), $t$\,[y] is the
live-time of the measurement, $a$ and $A$\,[g/mol] are
respectively the isotopic abundance and the atomic mass of the isotope.\\
The use of germanium as both the source and detector of double
beta decay has been suggested almost forty years ago by the
Milan group, which has also carried out the first experiments
\cite{Fiorini67}. Many experiments were performed since the
first attempt with a consistently increased sensitivity, as
summarized in Table \ref{tab:Ge76experiments}.
\begin{table}[ht]
 \small
  \centering
  \begin{tabular}{|l|c|c|c|c|c|c|}
    \hline
    &&&&&&\\
    Experiment &  Detector & \Ge\ &Exposure  & Background & Sensitivity & Ref. \\
    & Type& \%& kg\,y & {[keV\,kg\,y]$^{-1}$} & T$_{1/2}^{0\nu}$\,[y]\,(68\%\,C.L) &\\
    \hline
    &&&&&&\\
    Milan (1967) &  Ge(Li) & 7.8 & $0.0073$ &  $10^{3}$ &$3\cdot10^{20}$ & \cite{Fiorini67}\\
     &&&&&&\\
    S.\,Carolina-PNL &  HPGe & 7.8 & $ 0.3 $ &  $ 108 $ & $1.16\cdot10^{23}$ & \cite{Avignone83}\\
    (1983) &&&&&&\\
    Guelf-Queens- &  HPGe  & 7.8 & $ 0.2 $ &  39 & $3.2\cdot10^{22}$ & \cite{Simpson84}\\
    Aptec (1984)&&&&&&\\
     &&&&&&\\
    Milan (1984)&  Ge(Li) & 7.8 & 1.5 &  17 & $1.2\cdot10^{23}$ & \cite{Bellotti84}\\
     &&&&&&\\
    Osaka (1986)&  HPGe &  7.8 & 0.28 & 6.0 &$4.6\cdot10^{22}$ & \cite{Ejiri86}\\
     &&&&&&\\
    UCSB-LBL (1987)&  HPGe & 7.8 & 22.6 &  1.2 &$2.2\cdot10^{24}$ & \cite{Coldwell87}\\
     &&&&&&\\
    Caltech-PSI- &  HPGe &7.8 & 3.9 &  2.5 &$2.7\cdot10^{23}$ & \cite{CaltechPSI89}\\
    Neuchatel (1989)&&&&&&\\
    &&&&&&\\
    Zaragoza-INPN- &  HPGe & 7.8 &1.6 &  28 &$4.8\cdot10^{22}$ & \cite{ZaragBordoStr90}\\
    Bordeaux (1990)&&&&&&\\
    &&&&&&\\
    ITEP-YePI (1991) &  Ge(Li) &  85 & 1.6 &  2.5 &$1.8\cdot10^{24}$ & \cite{Kirpich91}\\
    &&&&&&\\
    IGEX (1999)&  HPGe & 86 & 10.1 &   0.16 & $1.6\cdot10^{25}$  & \cite{IGEX00}\\
    &&&(4.6 PSD$^*$)&0.06&$1.8\cdot10^{25}$&\\
    &&&&&&\\
    Heidelberg- &  HPGe & 86 & 71.7 &  0.17 & $4.6\cdot10^{25}$ & \cite{KK-NewAn-NIM04}\\
    Moscow (2003)&&&(51.4 PSD)&0.07&$6.1\cdot10^{25}$&\\
     &&&&&&\\
    \hline
  \end{tabular}
  \caption{Experiments for search of neutrinoless double beta
\onbb\ decay with germanium detectors and their sensitivities
for \onbb\ decay. ($^*$PSD-Pulse Shape
Discrimination).}\label{tab:Ge76experiments}
\end{table}
A major progress had been achieved in the ITEP-YePI experiment
\cite{Kirpich91}, which used germanium detectors enriched in
\Ge\ for the first time. Two experiments had been operated
recently to look for the \onbb\ decay of \Ge, the \igex\ and
the \HM\ experiments. The \igex\ Collaboration \cite{IGEX00}
operated a set of three detectors with a total mass of 6.3\,kg
in the Canfranc Underground Laboratory, Spain. The \HM\
Collaboration \cite{HDM97} used a set of five detectors with a
total mass 11.5\,kg at LNGS, Italy. Both experiments were
designed to get the highest possible sensitivity
(Eq.~\ref{DBD_sens}) to look for the \onbb\ decay of the \Ge\
using a large amount of isotope, good energy resolution
detectors and very low radioactive background conditions. In
the next Chapter the HdM experiment is described in detail.

\newpage
\chapter{Improved Analysis of the Data
from the Heidelberg-Moscow (HdM) Experiment} \label{ch:hdmo}

The \HM\ experiment was proposed in 1987 \cite{Prop87-HM-HVKK}
to look for neutrinoless double beta decay of \Ge\ at the Gran
Sasso Underground Laboratory at LNGS. The experiment collected
data from 1990 to 2003. In 2002 and 2004, evidence of
neutrinoless  double beta decay of \Ge\ was claimed by
Klapdor-Kleingrothaus et al. \cite{KK-NewAn-NIM04,
KK-NewAn-PL04}. This chapter gives an overview of the
experiment and describes the improved data analysis, which has
led to an enhanced sensitivity of the experiment and
subsequent publications \cite{KK-NewAn-NIM04, KK-NewAn-PL04}.
{\it A posteriori} analysis of
the background is given in the next chapter.\\

\section{\HM\ experiment overview}
\label{sec:hd-rev}
The experiment was carried out with five high-purity p-type
germanium detectors enriched in the isotope \Ge\ to $\sim$86\%
\cite{HDM97}. The total active mass was 10.96 kg, resulting in
a \Ge\ source strength of 125\,mol. The HdM detectors were the
first enriched high-purity Ge detectors ever produced. The
energy resolution of the Ge detectors was $\sim$3\,keV in the
region of interest around the \qbb\ value, which assures zero
background in the \onbb\ line from the \nnbb\ decay. The \qbb\
value (2039.006$\pm$0.050\,keV) has been determined recently
with high precision \cite{Ge76Qvalue}. The detection
efficiency of Ge detectors for the \onbb\ decay events was
calculated to be 95\%.  The background index was
0.17\,counts/kg\,y\,keV in the \onbb\ decay region for the
total exposure of 71.7\,kg\,y.

\section{HdM setup}
\label{sec:hd-setup}

    The technical parameters of all the HdM detectors
    are given in Table \ref{Techn_Param}.
\begin{table}[ht]
\centering
\begin{tabular}{c|c|c|c}
\hline
            &   Total       & Active          &   Enrichment\\

    Detector    &   Mass($^*$),        &    Mass, &   in \Ge,\\
                &   [g]       &   [g]        &   [\%] \\
\hline
    ANG1        &   968.7 & 920 &   85.9 $\pm$ 1.3\\
    ANG2        &   2878.3       &
    2758       &   86.6 $\pm$ 2.5    \\
    ANG3        &   2446.5       &
    2324       &   88.3 $\pm$ 2.6    \\
    ANG4        &   2401.2       &
    2295       &   86.3 $\pm$ 1.3    \\
    ANG5        &   2782.1       &
    2666       &   85.6 $\pm$ 1.3    \\
\hline
\end{tabular}
\caption{Masses and enrichments of the five enriched
$^{76}{Ge}$ detectors \cite{HDM97}. ($^*$) The actual masses
of the detectors were measured after dismounting from their
cryostats in 2006.}\label{Techn_Param}
\end{table}
The degree of enrichment (86-88\%) was verified by
investigation of pieces of Ge after crystal production using a
mass spectrometer at MPI-K \cite{Dipl-Echt}. The experimental
setup is shown in Figure \ref{HDM-Set-up}.
\begin{figure}[!ht]
\centering
\includegraphics[width=75 mm]{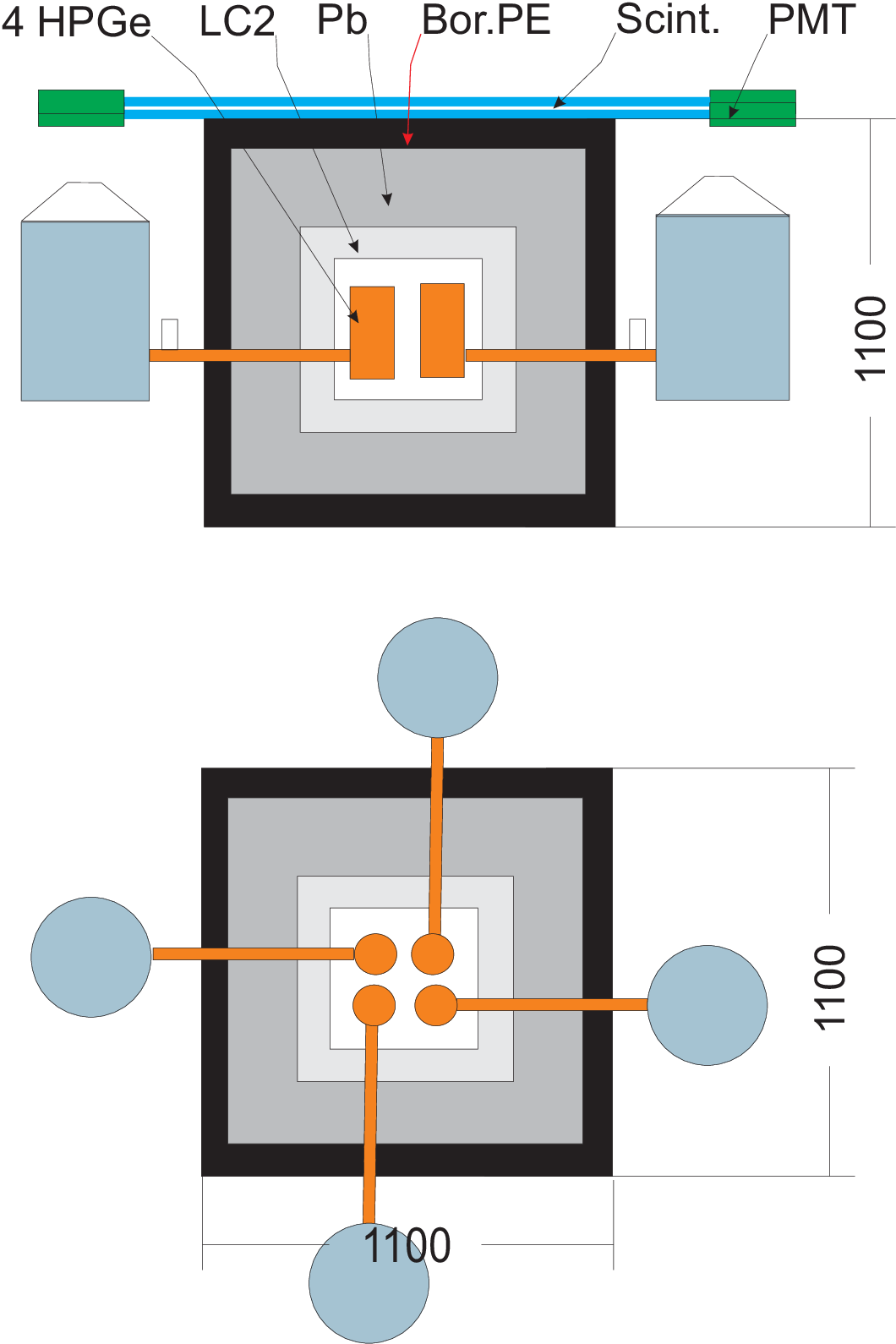}
\includegraphics[width=55 mm]{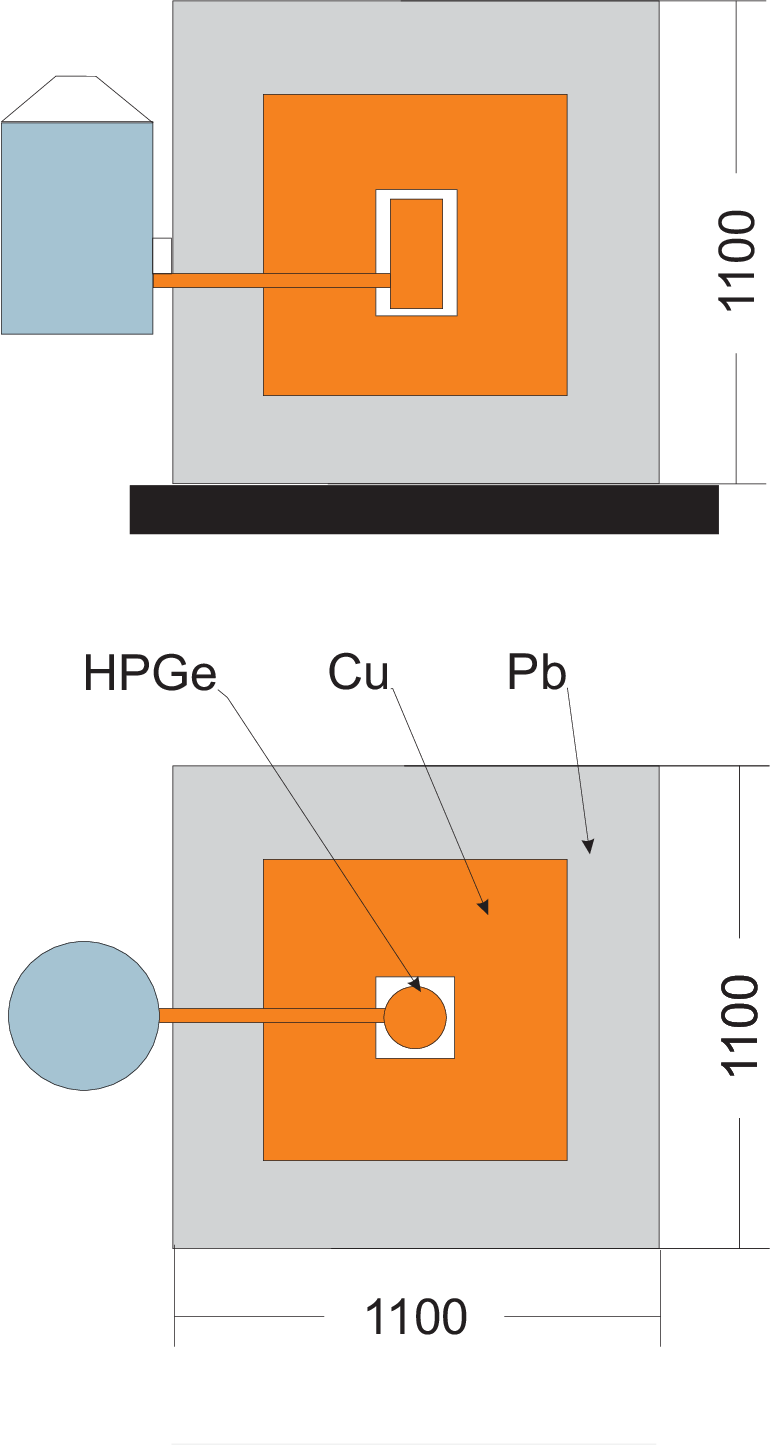}\\
\caption{
    Schematic
    drawings of the \HM\ experiment setup. Four enriched
    detectors, ANG1, ANG2, ANG3 and ANG5, are installed
    in a common shield (left panel).
    ANG4 detector is installed in a separate
    shield made of electrolytic copper and lead (right panel).}
\label{HDM-Set-up}
\end{figure}
All detectors, except ANG4, were operated in a common lead
shield, which consisted of an inner 10 cm thick layer of
low-activity LC2-grade lead followed by 20 cm of Boliden lead.
The lead shield was enclosed in a steel box, which was
surrounded by 10 cm of boron-loaded polyethylene shielding to
decrease the neutron flux from outside. Since 1995, an active
anticoincidence muon veto was placed on top  of the setup.
Detector ANG4 was installed in a separate setup, which had an
inner shielding of 27.5 cm electrolytic Cu and 20 cm of lead.
The shield was enclosed in a steel box with a plate of
boron-loaded polyethylene shielding below. No muon veto was
implemented for this detector. The two setups were flushed
with nitrogen in order to suppress the \Rn\ contamination. The
setup was kept hermetically closed since February 1995. Before
this time the shielding of the experiment was opened several
times to add new detectors. Since November 1995, the data was
acquired using a CAMAC system and a CETIA processor in event
by event mode. The energy spectra from the preamplifiers
(model 2002C) were recorded with 13\,bit ADCs developed at
MPI-K Heidelberg. The spectra from each detector were recorded
in parallel using two ADCs with 8192 channels each, one
ranging from the threshold up to about 3 MeV, and another one
up to about 8 MeV. The timing signals from the preamplifiers
of all detectors, except ANG1, were differentiated by ORTEC's
Timing Filter Amplifiers (models TFA 474 and 579) and the
pulse shapes were recorded with 250\,MHz flash ADCs of type
Analog Devices 9038 JE (in DL515 modules), with 8\,bit
resolution \cite{Diss-Dipl-Helmig}.
In the event by event mode, the event time, the high voltage
applied on the detectors, the temperature in the detector
cave, the computer room and the electronics crate, the
information from the muon shield and the status of the DAQ
were recorded. In addition, an operator was daily checking the
high voltage of the detectors, the temperature, the nitrogen
flow flushing the detector boxes, the muon anticoincidence
signal, the leakage current of the detectors and the trigger
rates.

\clearpage
\section{Development of the experiment}
\label{sec:hd-chrono}

For each detector, the exposure and the background index
during the full live time of the experiment, divided into two
periods, are presented in Table \ref{activities}.
\vspace{0.5cm}
\begin{table}[th]
\small \centering
\setlength\tabcolsep{1.pt}
\begin{tabular}{|c|p{1.5cm}|p{1.5cm}|p{1.5cm}|p{1.5cm}|c|c|c|}
\hline
    Detector      &
\multicolumn{4}{|c|}{Live Time and Exposure}     &
    Date        &  Background   &  Pulse shape \\
\cline{2-5}
          &
 \multicolumn{2}{|c|}{of all data}&\multicolumn{2}{|c|}{accepted for}      & &
{~~[counts/}& recording  \\
 &   \multicolumn{2}{|c|}{} &\multicolumn{2}{|c|}{analysis} & Start End & {~{[keV y
kg]}}&       \\
\cline{2-5}
   & ~~[days]  & {~{[kg d]}} & ~~[days]  &{~{[kg d]}} & &   2000 - 2100 & \\
    &           & &&&&    [keV] &       \\
\hline \hline
    ANG1 & 1237.0  & 1138.04& 930.9   &   856.43  &   8/90 - 8/95 &
    0.31    &   no  \\
    ANG2   &   1070.0      & 2951.06&   997.2   &   2750.28 & 9/91 - 8/95   &
    0.21    &   no  \\
    ANG3   &   834.7       & 1939.84&   753.1   &   1750.20 & 9/92 - 8/95   &
    0.20    &   no  \\
    ANG4   &   147.6       & 338.74&   61.0    &   139.99  & 1/95 - 8/95   &
    0.43    &   no  \\
    ANG5   &   48.0        & 127.97&   {\centerline -} &   {\centerline -} & 12/94 - 8/95  &
    0.23    &   no  \\
\hline \multicolumn{8}{|c|}
{ After summing of all 5 detectors over period 1990 - 1995:}\\
\multicolumn{8}{|c|}{Accepted exposure = {\bf 15.05\,kg\,y} }\\
\hline \hline
\multicolumn{8}{|c|}{ Full Setup, over period 1995 - 2003:}\\
\multicolumn{8}{|c|}{ Four detectors in common shielding,
ANG4 detector separate}\\
\hline
    ANG1   &   2123.90     & 1967.25
&   2090.61 &   1923.36 &   11/95 - 5/03    &
    0.20    &   no  \\
    ANG2   &   1953.65     & 5427.94
&   1894.11 &   5223.96 & 11/95 - 5/03  &
    0.11    &   yes \\
    ANG3   &   2120.22     & 4960.83
&   2079.46 &   4832.67 & 11/95 - 5/03  &
    0.17    &   yes \\
    ANG4   &   2123.90     & 4907.44
&   1384.69 &   3177.86 & 11/95 - 5/03  &
    0.21    &   yes \\
    ANG5   &   2110.66     & 5665.46
&   2076.34 &   5535.52 & 11/95 - 5/03  &
    0.17    &   yes \\
\hline \hline
\multicolumn{8}{|c|}{ After summing of all 5 detectors over period 1995 - 2003}\\
\multicolumn{8}{|c|}{Accepted exposure = {\bf 56.655\,kg\,y}}\\
\multicolumn{8}{|c|}{Total exposure = {\bf 71.71 kgy}}\\ 
\hline
\end{tabular}
\caption{\label{activities}Exposure and
    background index for the enriched detectors
    of the HEIDELBERG-MOSCOW experiment for the period
    from 1990 to 2003 \cite{KK-NewAn-NIM04}.
    The live time and exposure accepted for analysis are calculated
    according to Table \ref{conditions}.}
\end{table}
From 1990 to 1995 the setup was opened several times for
installation of new detectors. A muon veto was not implemented
and no individual event information was recorded.

\clearpage

\begin{figure}[ht]
\begin{center}
\includegraphics[height=4cm]{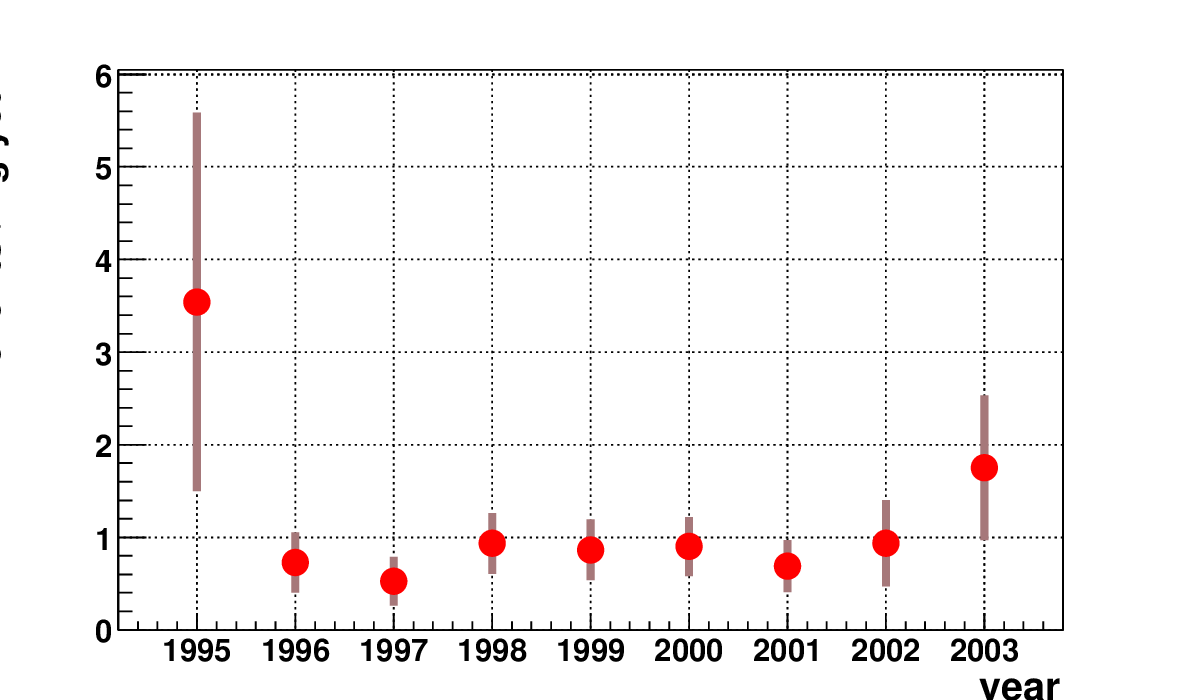}
\end{center}
\caption[]{The time distribution of all events in the energy
interval from 2036.5 to 2041.0\,keV, for the period from
November 1995 to May 2003 \cite{KK-NewAn-PL04}.}
\label{fig:RatesIndivDet}
\end{figure}

The time dependence of the measured rate of the energy
interval from 2036.5 to 2041.0\,keV is shown in Fig.
\ref{fig:RatesIndivDet}. The arrival-time distribution of the
events was analyzed. Figure \ref{fig:KSprob} shows the
arrival-time distribution of all events observed in the energy
interval (2035.5 - 2042.5)\,keV as a function of time. The
events distribution is consistent with a uniform one,
according to the Kolmogorov-Smirnov distribution test. It
demonstrates that the arrival-time distribution is not
affected by technical operations during the measurement, such
as the calibration procedure (introducing the thorium source
into the detector chamber through a thin tube), refilling of
liquid nitrogen, etc.
\begin{figure}[ht]
\centering
\includegraphics[height=7cm]{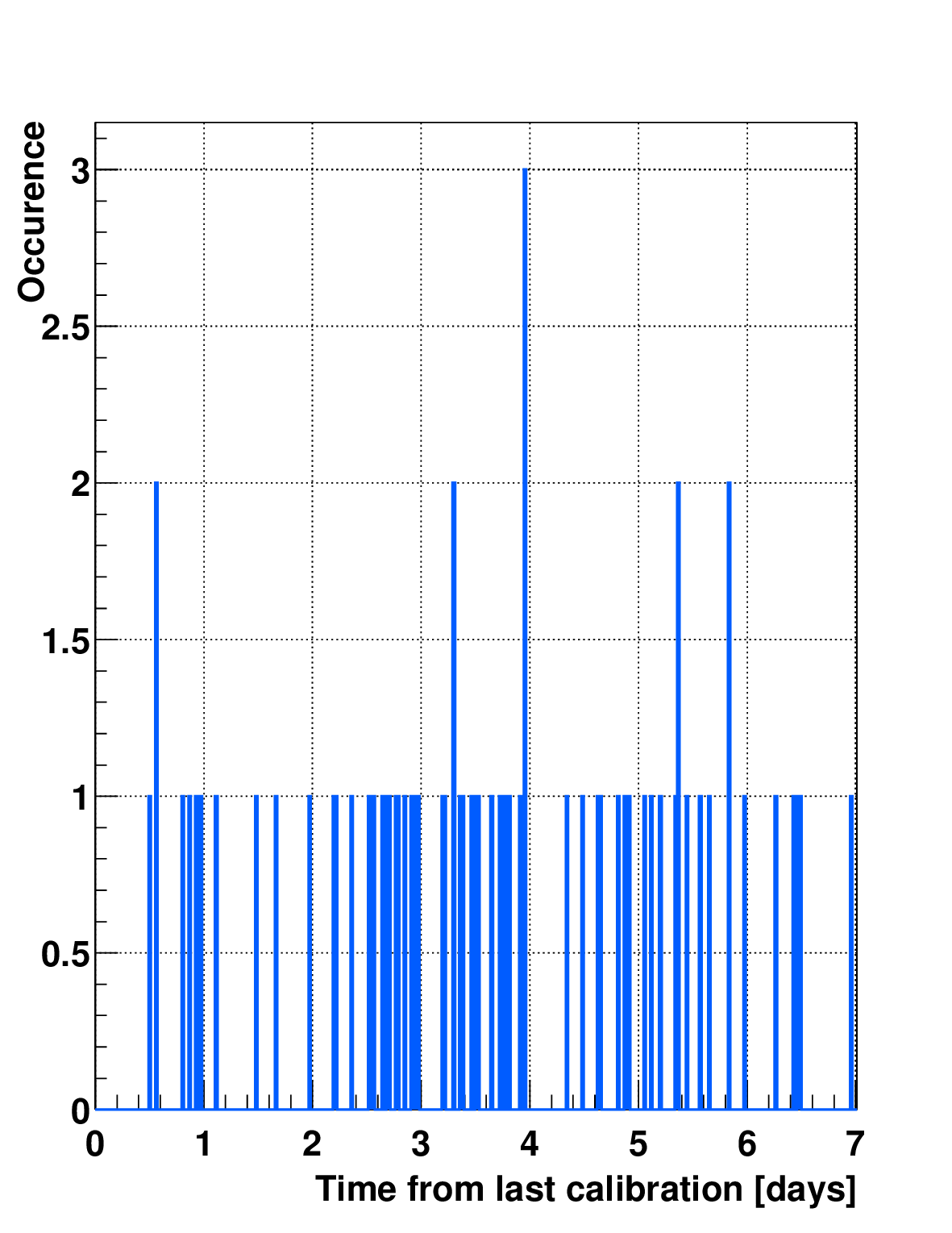}
\includegraphics[height=5cm]{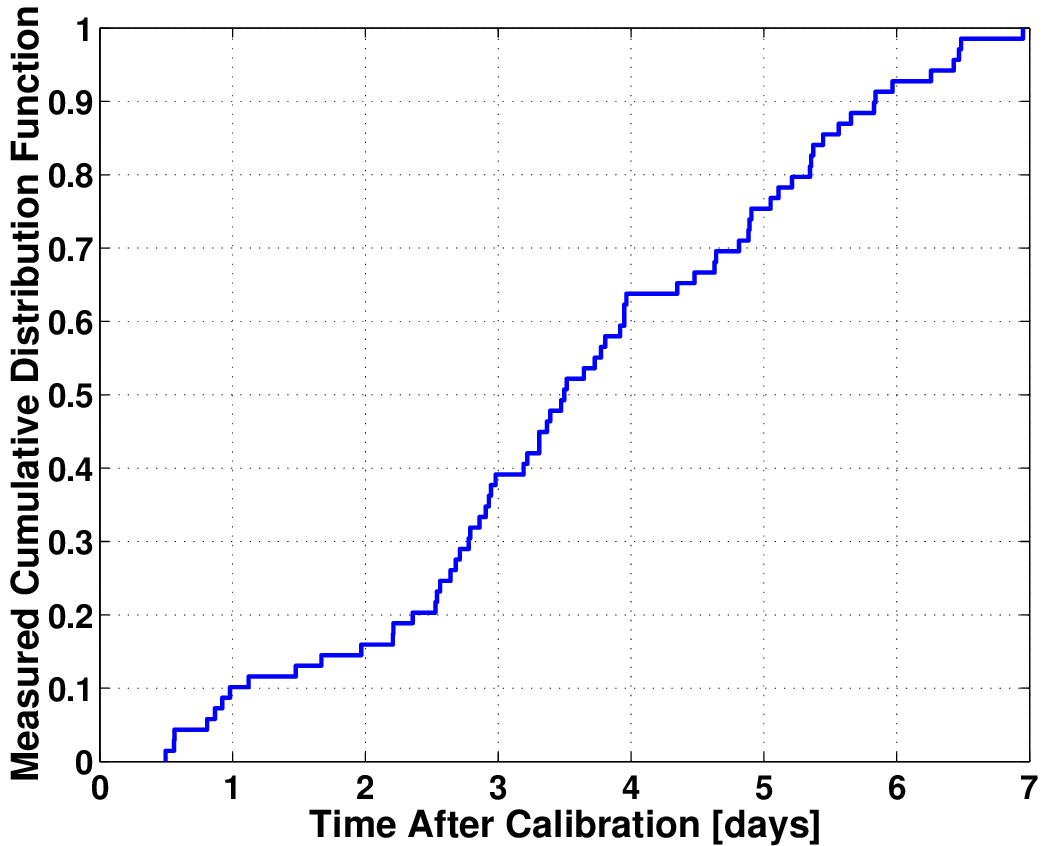}
\caption{Left: arrival time for all 70 events
    in the interval 2035.5 - 2042.5\,keV
    as function of time relative to the last time of calibration
    for the period 1995 - 2003. The acquisition was turned off
    during the first six hours after each calibration for a liquid
    nitrogen filling. The corresponding cumulative distribution
    (right) is consistent with a uniform one at 87\% C.L.,
    according to the Kolmogorov-Smirnov test \cite{KK-NewAn-NIM04}.}
    \label{fig:KSprob}
\end{figure}

\section{Energy calibration}
\label{sec:hd-calib}
A precise energy calibration of all detectors, before summing
the individual runs
and summing the spectra of all detectors,
is important in order to achieve a good final energy
resolution, and therefore an optimal sensitivity.

\subsection{Calibration procedure}

To control the energy scale stability of the experiment,
calibrations were performed weekly with a $^{228}$Th (110\,kBq
in 2001) source for the detectors ANG1,2,3,5 and with
$^{152}$Eu (10\,kBq) + $^{228}$Th (7\,kBq in 2001) sources for
ANG4. The sources were mounted on steel wires and were
inserted into the setup approximately 15\,cm away from the
detectors trough teflon tubes. The calibrations lasted 15
minutes for the four detectors setup and 45 minutes for the
ANG4 setup. Only energy spectra were collected during the
calibration, no individual event information was recorded.

\subsection{Method of energy calibration}

After completion of the experiment, the linearity of the
electronics of the HdM detectors was studied with a \Ra\
source. It was observed that the integral nonlinearity is not
negligible. Figure \ref{CalNonlin} presents the residual from
a linear energy calibration with the \Ra\ source as a function
of energy for each detector.
\begin{figure}[ht]
\begin{center}
\includegraphics[width=6.5cm]{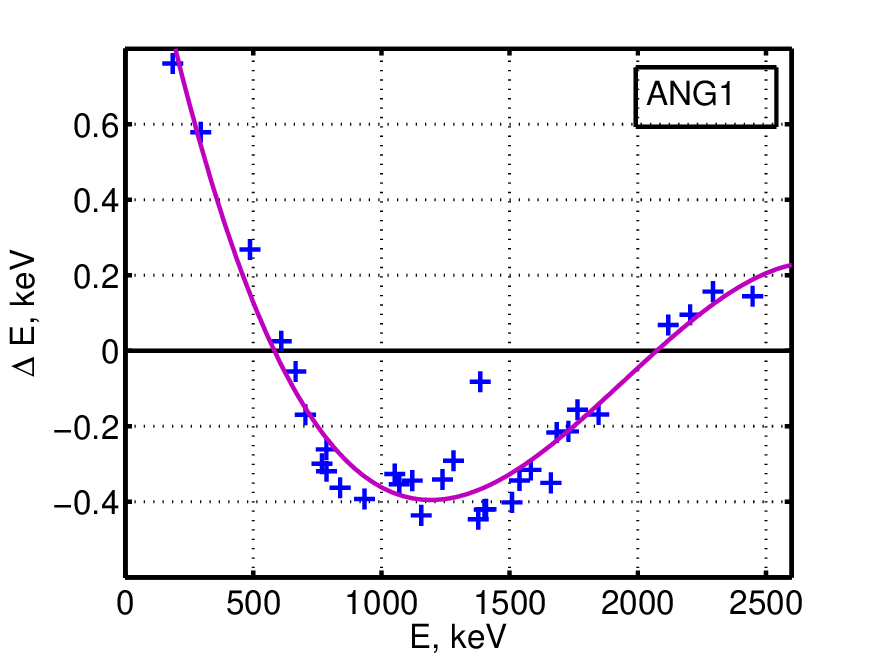}
\includegraphics[width=6.5cm]{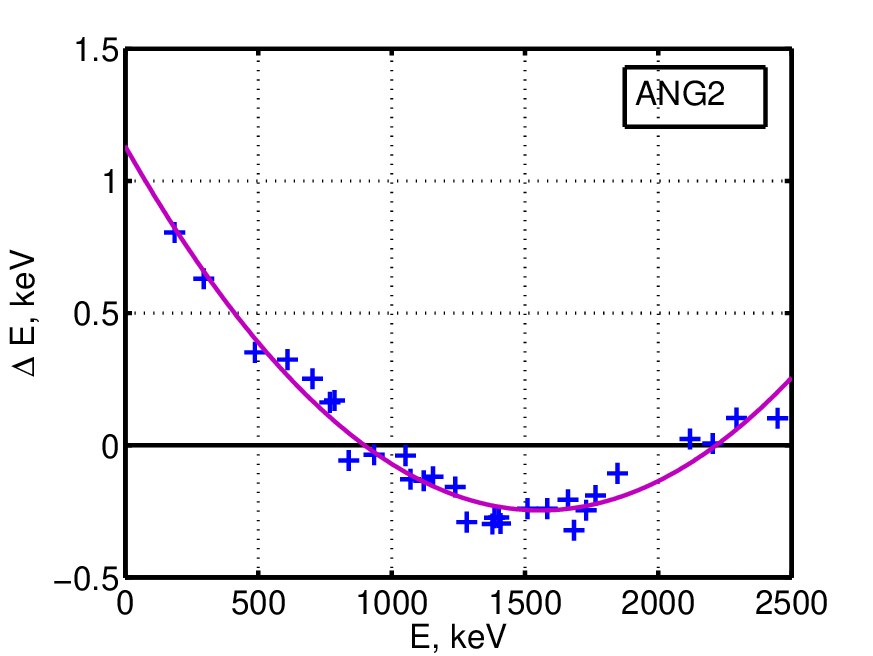}
\vskip 1 cm
\includegraphics[width=6.5cm]{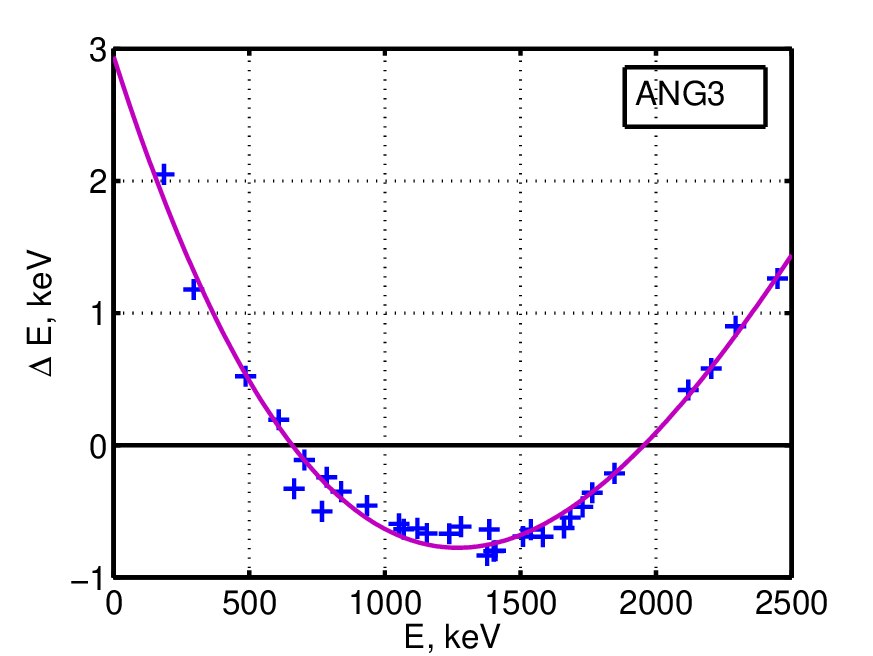}
\includegraphics[width=6.5cm]{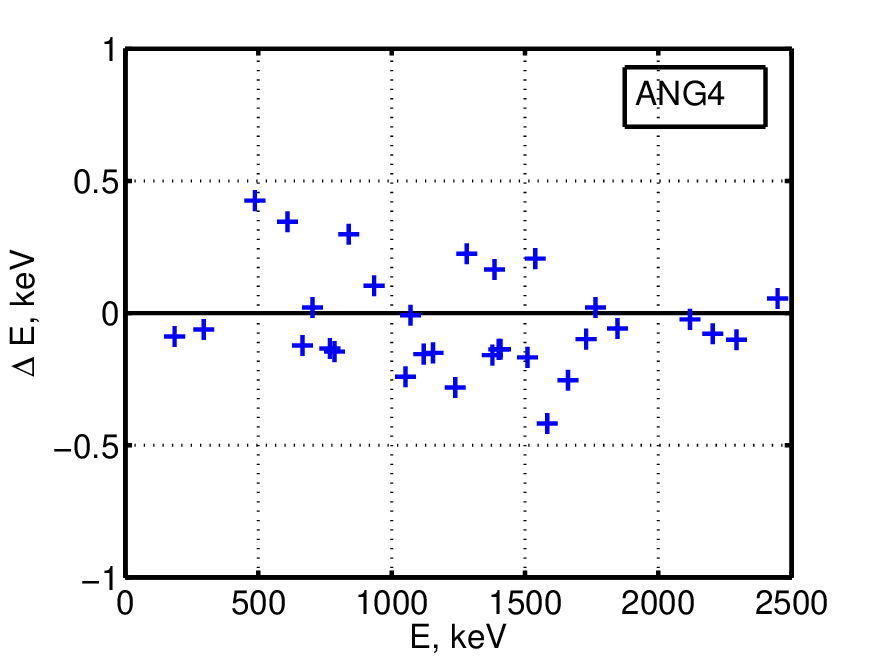}
\vskip 1 cm
\includegraphics[width=6.5cm]{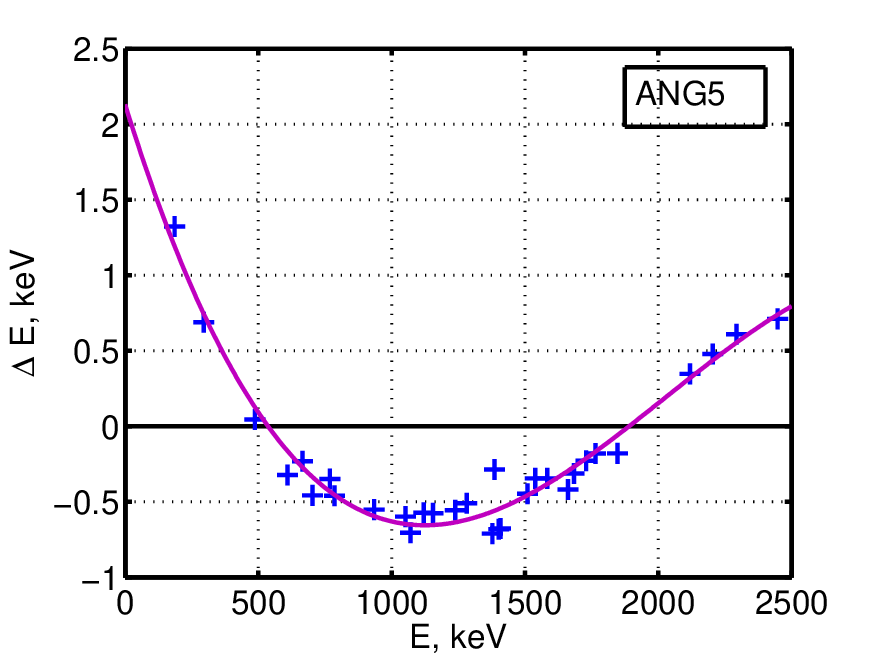}
\end{center}
\caption[]{Residuals $\Delta E$ of the linear calibration with
a \Ra\ source as a function of the energy for the five HdM
detectors. Measurements were performed in November 2003. The
cubic fits of the data are shown.} \label{CalNonlin}
\end{figure}
Figure \ref{fig:FWHM-RaTh} presents the energy resolution
(FWHM) as a function of energy of the $\gamma$-peaks measured
with a \Ra\ and a \Th\ source.
\begin{figure}[ht]
\centering
\includegraphics[width=12cm]{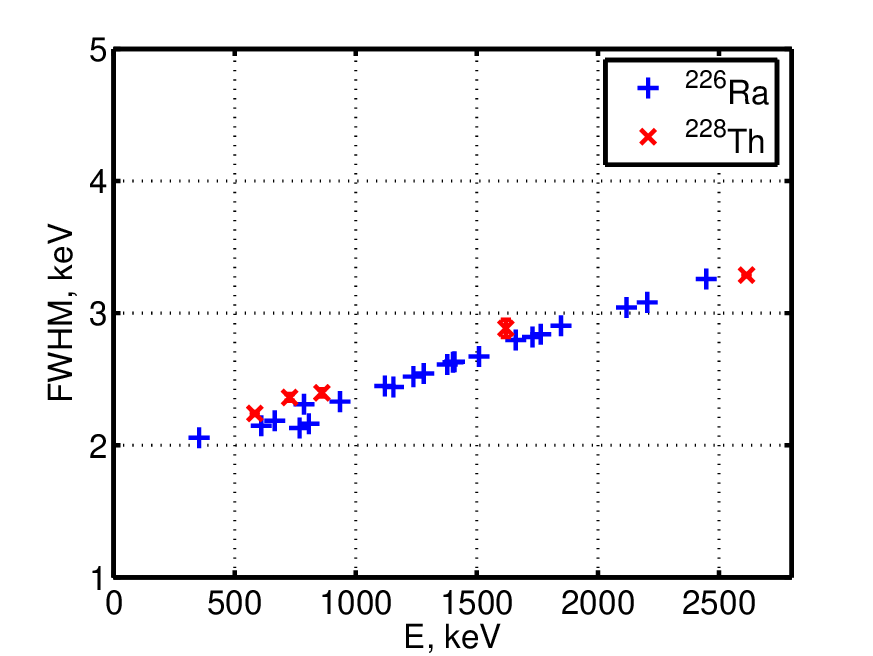}
\caption[]{Energy resolution (FWHM) as a function of energy of
the $\gamma$-peaks in the \Ra\ spectrum measured in November
2003 and in the $^{228}$Th calibration spectrum, measured in
1999 (maximum error of FWHM is 0.05\,keV). No resolution
degradation was observed in November 2003.}
\label{fig:FWHM-RaTh}
\end{figure}
The fact that $^{228}$Th does not have many spectral lines and
that the electronics are not linear limit the energy
calibration accuracy in the range of interest around the
Q$_{\beta\beta}$ value. There are only two peaks in the
$^{228}$Th spectrum above 2000 keV, the $\gamma$-line at
2614.5\,keV and its single escape peak at 2103.5\,keV. The
next line is at 1620 \,keV, which is too far to make proper
correction for linearity. In order to minimize the errors
introduced by the nonlinearity, the low and the high energy
parts of the spectra were calibrated separately. For the
energy above 1640\,keV, the $^{228}$Th lines at 2103.5 and
2614.5\,keV were used. For the low-energy part of the spectra,
the $^{228}$Th lines at 583 and 1620\,keV were chosen. The
combination of these linear calibrations successfully
reproduced the energies of other known lines in the spectrum.
In order to check this calibration method, the individual
calibration spectra were relocated to the same energy scale
before being summed. The relocation method is explained in
detail at the end of this section. The peak positions and
energy resolution of the 2103.5 and 2614.5\,keV calibration
lines are presented in Table \ref{tab:Calibr-Peaks}.

\clearpage

\begin{table}[ht]
\centering
\renewcommand{\arraystretch}{1.4}
\setlength\tabcolsep{3.5pt}
    \begin{tabular}{|c|c|p{1.5cm}c|p{1.5cm}c|}
\hline
 &  & \multicolumn{2}{|c}{Single escape (SE) $^{228}$Th} &   \multicolumn{2}{|c|}{$^{228}$Th}\\
Calibration   &  Detector & \multicolumn{2}{|c}{E$_0$=2103.5\,keV}& \multicolumn{2}{|c|}{E$_0$=2614.53\,keV} \\
\cline{3-6}
    period &   & E$_{peak}$    & FWHM & E$_{peak}$ & FWHM\\
&   & [keV] & [keV] & [keV] & [keV]\\
\hline One     &   ANG1   &   2103.36 &   3.65  &  2614.40 &   2.96    \\
of the &   ANG2   &   2103.58 &   3.85    &  2614.46 &   3.43    \\
weekly  &   ANG3   & 2103.35 &   3.65  &  2614.66 &   3.00    \\
spectra &   ANG4   &    2103.49 &   4.47 &  2614.61 &   3.48    \\
    &   ANG5   &    2103.59 &   3.82  &  2614.52 &   3.36    \\
\hline After   &   ANG1  & 2103.53 &   3.63    & 2614.52 &   3.00    \\
summing &   ANG2   &     2103.53 &   3.93    &   2614.53 &   3.40    \\
over the   &   ANG3   &     2103.53 &   3.69    &   2614.52 &   3.04    \\
period  &   ANG4   &     2103.52 &   4.55    &    2614.53 &   3.98    \\
1995-2003  &  ANG5      &    2103.54 &   3.95    &  2614.54 &   3.38    \\
\hline
\multicolumn{6}{|c|}{After summing all 5 detectors over the period 1995-2003}\\
\hline
  &   &  2103.53 &   3.86    & 2614.53 &   3.27    \\
\hline
\end{tabular}
\caption[]{Positions (error $\pm$0.05\,keV) and energy
resolutions (error $\pm$0.05\,keV) of two most important
calibration lines measured over the period from 1995 to 2003.
The summed spectrum shows that the calibration method does not
introduce broadening of the peaks \cite{KK-NewAn-NIM04}.}
    \label{tab:Calibr-Peaks}
\end{table}

One can see that the positions and the energy resolution of
the peaks are not deteriorated by this calibration method. The
single escape (SE) peak at 2103.5\,keV has generally a broader
resolution compared to the full-energy peak. This is a known
effect, arising because the positron and the atomic electron
are not at rest at the moment of annihilation. The resulting
net momentum of the particles increases the statistical
uncertainty of energy, leading to the broadening of the peak
(see, e.g. \cite{Knoll89}), but it does not affect the peak
position at 2103.5 keV. The slightly worse resolution of the
SE peak does not degrade the energy calibration, but instead
provides the advantage of having a calibration line close in
energy to the \znbb Q$_{\beta\beta}$ value.

\subsubsection{Method of spectra relocation}
Relocation of multichannel spectra is necessary in order to
compare or add different spectra. The relocation of a
multichannel spectrum into new energy bins E$_k$ is performed
as follows \cite{NIMA378-Binning}: First, the energies E$_j$
for all channels $j$ of the original spectrum are calculated
with a known calibration equation ($E_j=f_{cal}(j)$). Second,
if the energy interval [E$_j$,E$_{j+1}$] is fully contained in
a new bin [E$_k$,E$_{k+1}$], then the full content of channel
$j$ is added to bin $k$. Otherwise, the content is distributed
between bins proportionally to the fraction of energy in each
covered bin. As an example, a schematic diagram of the applied
method is shown in Figure \ref{fig:Binning}.
\begin{figure}[ht] \centering
\includegraphics[width=10cm]{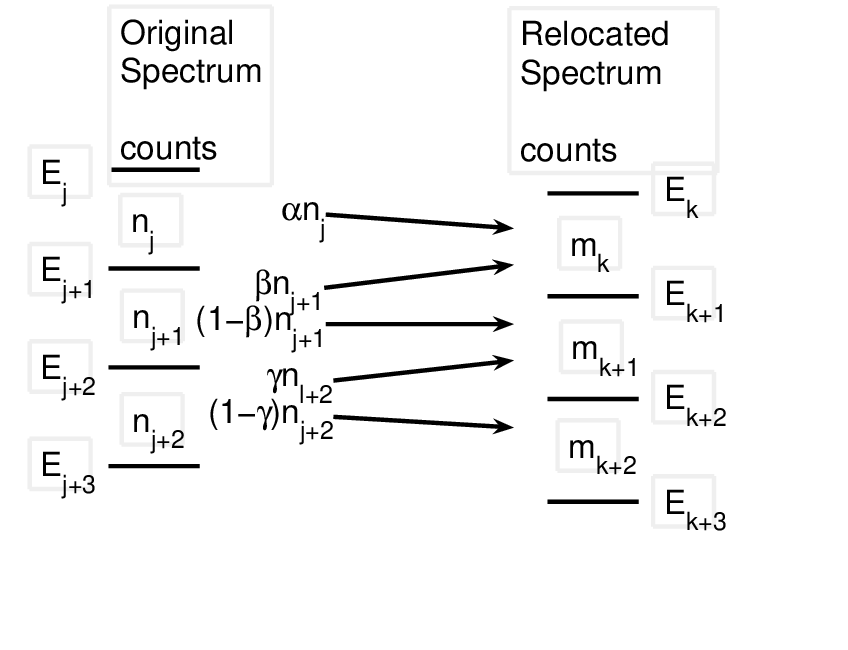}
 \caption[]{Scheme of the distribution of multichannel
  spectrum \cite{NIMA378-Binning}.} \label{fig:Binning}
\end{figure}
The fraction of counts from channel $j+1$ to be added to the
new bin $k$ is $\beta =
\frac{E_{k+1}-E_{j+1}}{E_{j+2}-E_{j+1}}$ and the fraction of
counts to be added to the following bin $k+1$ is $(1-\beta)$.
The content of the new bin $k$, $m_k=\alpha n_j+\beta
n_{j+1}$, is not necessarily an integer number. In general, it
does not obey a Poisson distribution because the variance
$\sigma^2_{m_k} =\alpha^2 n_j + \beta^2 n_{j+1}$ is not equal
to $m_k$.\\
The MATLAB implementation of the algorithm \cite{SAND0} was
used for re-binning of all HdM calibration spectra.

\clearpage
\subsection{Energy resolution and the accuracy of energy
calibration of the sum spectrum}
The energy resolution obtained with the combined spectrum over
the 13 years of data acquisition of the HdM experiment is
presented in Figure \ref{fig:FWHM-9003}.
\begin{figure}[ht]
\centering
\includegraphics[width=12cm]{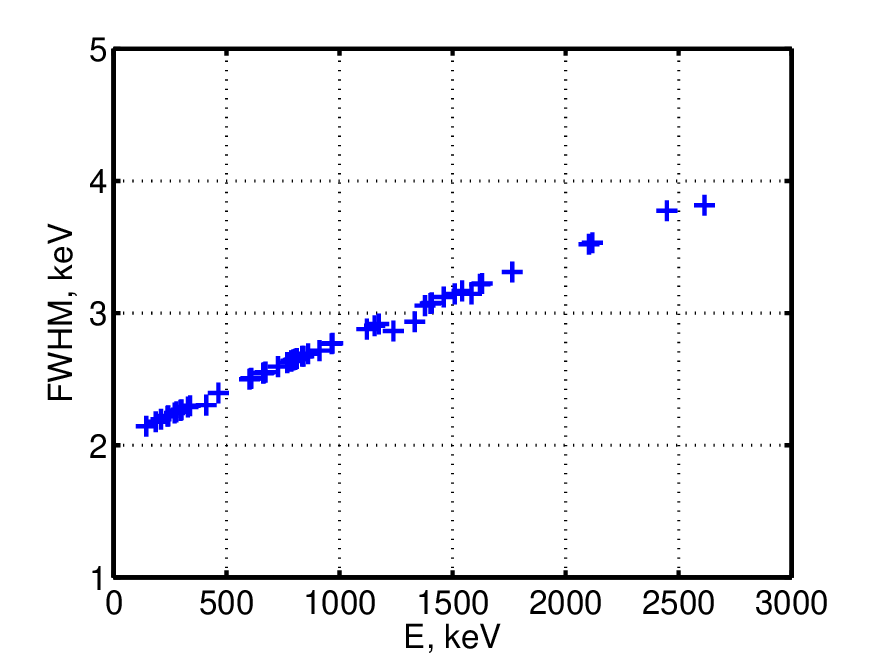}
\caption[]{Energy resolution (FWHM) as a function of energy of
the $\gamma$-peaks in the sum spectrum (1990 - 2003). Maximum
error of FWHM is 0.1\,keV} \label{fig:FWHM-9003}
\end{figure}
At the Q$_{\beta\beta}$ value, the interpolated energy
resolution is found to be 3.49$\pm$0.03\,keV, which is better
than 4.23$\pm$0.14\,keV presented in an earlier analysis
\cite{HDM01}. The 20\% improvement in the energy resolution of
the sum spectrum, leading to a 10\% increase in the
sensitivity of the experiment, is a consequence of the refined
summing procedure which was used for the individual 9570 data
sets. Figure \ref{fig:DeltaE-9003} shows the residuals
$\Delta$E of energy of the $\gamma$-peaks in the sum spectrum
from 1990 to 2003.
\begin{figure}[ht]
\centering
\includegraphics[width=12cm]{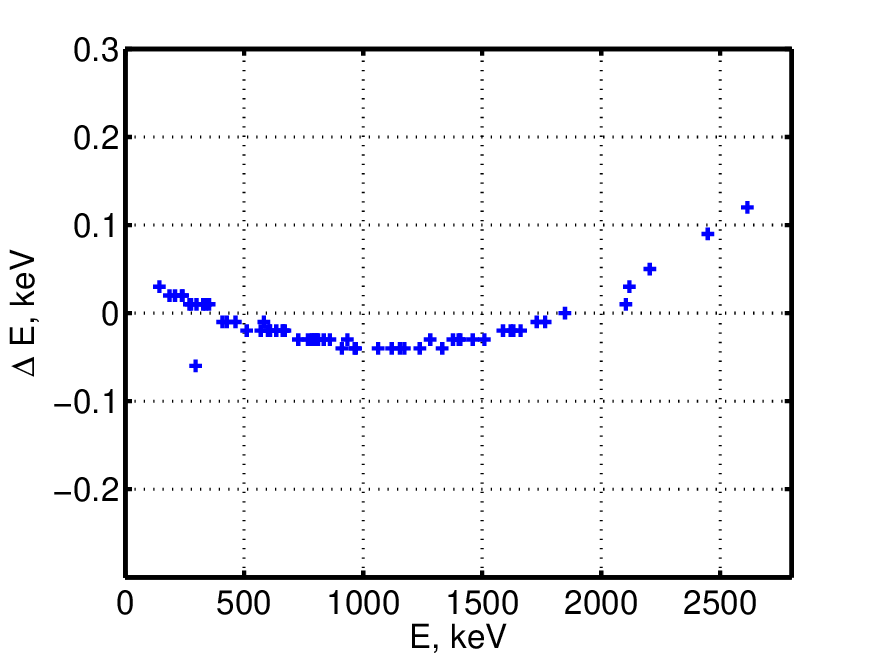}\\
\caption[]{$\Delta$E residuals from the reference values of
the $\gamma$-peaks positions in the HdM spectrum (1990 -
2003). The maximum error for each point is $\pm$0.1\,keV.}
\label{fig:DeltaE-9003}
\end{figure}
The energy calibration accuracy is better than 0.1\,keV in the
range 100-3000\,keV.

\subsubsection{An optimal bin width}
The histogram's bin width controls the tradeoff between
presenting a spectrum with too much or too little details.
Gilmore and Hemingway \cite{Gilmore} and Knoll \cite{Knoll89}
recommend four bins for FWHM of germanium spectra. In the case
of the HdM spectrum, FWHM is 3.46\,keV at 2039\,keV, thus
1\,keV bin width is optimal and convenient. Scott's reference
rule \cite{Scott79} for optimal bin width is
$w=3.49\cdot\sigma{/}\sqrt[3]{n}$, where $\sigma$ is an
estimate of the standard deviation and $n$ is the number of
counts in ROI. For $\sigma$=1.48\,keV and $n$=100 the bin
width is 1.1\,keV, which is consistent with 1 keV bins.

\section{Event selection in the summed spectrum}
\label{sec:hd-selection}

For the data collected in the first phase of the experiment
(1990 - 1995), no information about individual events was
recorded. Therefore, only a summed spectrum over this period
with an exposure of 15.05\,kg.y is used for the present
analysis. In total 2142 runs (10 513 data sets for the five
detectors) were taken since 1995. The duration of each run is
on average one day. The first 200\,days of operation of each
detector, corresponding to about three half-lives of the
$^{56}$Co cosmogenic contamination ($T_{1/2}$=77.27\,days),
are excluded. In addition, 792 data sets, $\sim$7.5\% of all
data sets, were found to be corrupted because of an electronic
problem, and have been rejected. To check the quality of each
event, the 'Energy over Integral' (EoI) value was calculated
as the ratio of the deposited energy measured by the ADC and
the integral of the current pulse measured by the Flash ADC in
the timing channel. Both values are proportional, unless the
ADC or Flash ADC fails. The raw data sets and events
considered for further analysis are only those which satisfy
the following conditions:
\begin{enumerate}
\item
    no coincidences with another Ge detector and/or muon veto;
\item
    the deviation from the average count rate of each detector
    is within $\pm$5$\sigma$;
\item
    the deviation from the average EoI value of each detector
    is within $\pm$3$\sigma$;
\end{enumerate}
Table \ref{conditions} summarizes the data sets and events
remaining after the applied cuts.
\begin{table}[h]
\centering
\renewcommand{\arraystretch}{1.2}
\setlength\tabcolsep{5.9pt}
\begin{tabular}{c|c|c}
\hline
    &   Data Sets   &   Events
\\
\hline
    Full measurement    &   10 513  &   951 044
\\
\hline
    Corrupted data sets &   792 &   92 553
\\
\hline
   Run Count Rate $>$ $\pm 5 \sigma$ &   151 &   32 922
\\
\hline
    Muon veto coincidence $^{*}$ &       &   3 672
\\
\hline
    Ge - Ge coincidence &     &   23 563
\\
\hline
    EoI selection $>$ $\pm 3 \sigma$  &       &   13 158
\\
\hline \hline
    Data accepted       &   9 570   &   786 941
\\
\hline
\end{tabular}
\caption{ Events and data sets accepted for the analysis of
the \HdMo\ experiment for the period from 1995 to 2003.($^*$)
Some muon veto events coincide with Ge - Ge coincidences
\cite{KK-NewAn-NIM04}.} \label{conditions}
\end{table}
Starting from 951 044 events, 786 941 events remain for
analysis.\\
In \cite{Kurch03}, the presence of anomalous peaks in the HdM
data in the low-energy range (500 - 700\,keV) was discussed.
The peaks at 550\,keV and 640\,keV were observed in the
spectrum of the ANG4 detector, but they had disappeared after
application of the EoI cut, as demonstrated in Figure
\ref{UnderThresh}(left).
\begin{figure}[ht]
\centering
\includegraphics[width=6.8cm]{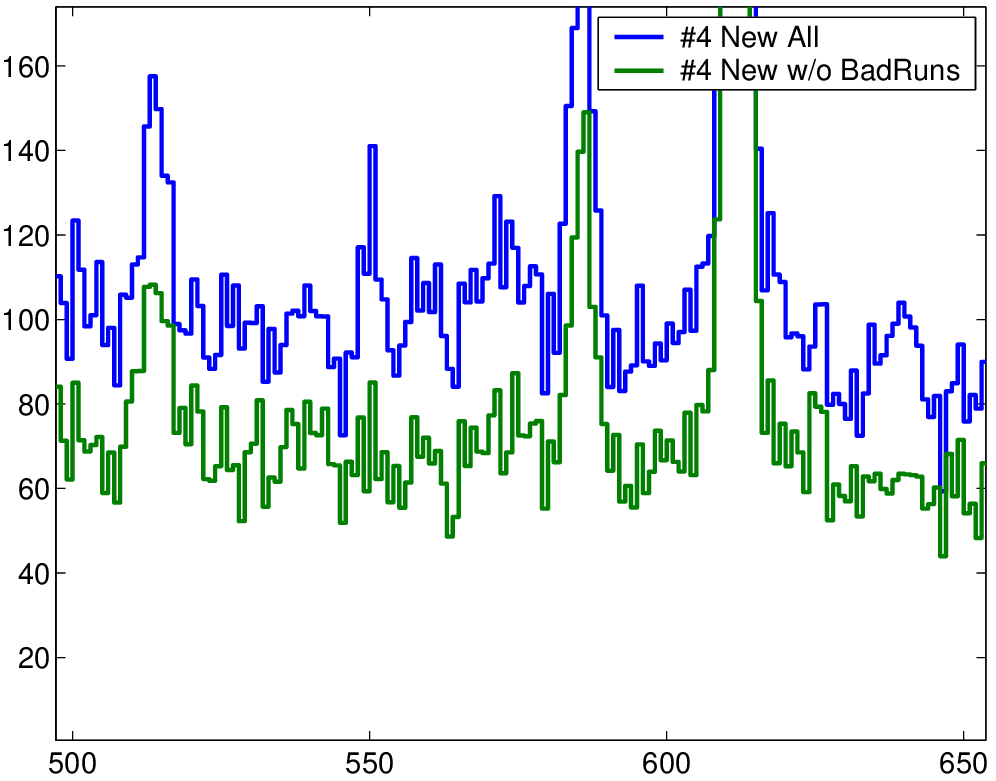}
\includegraphics[width=6.7cm]{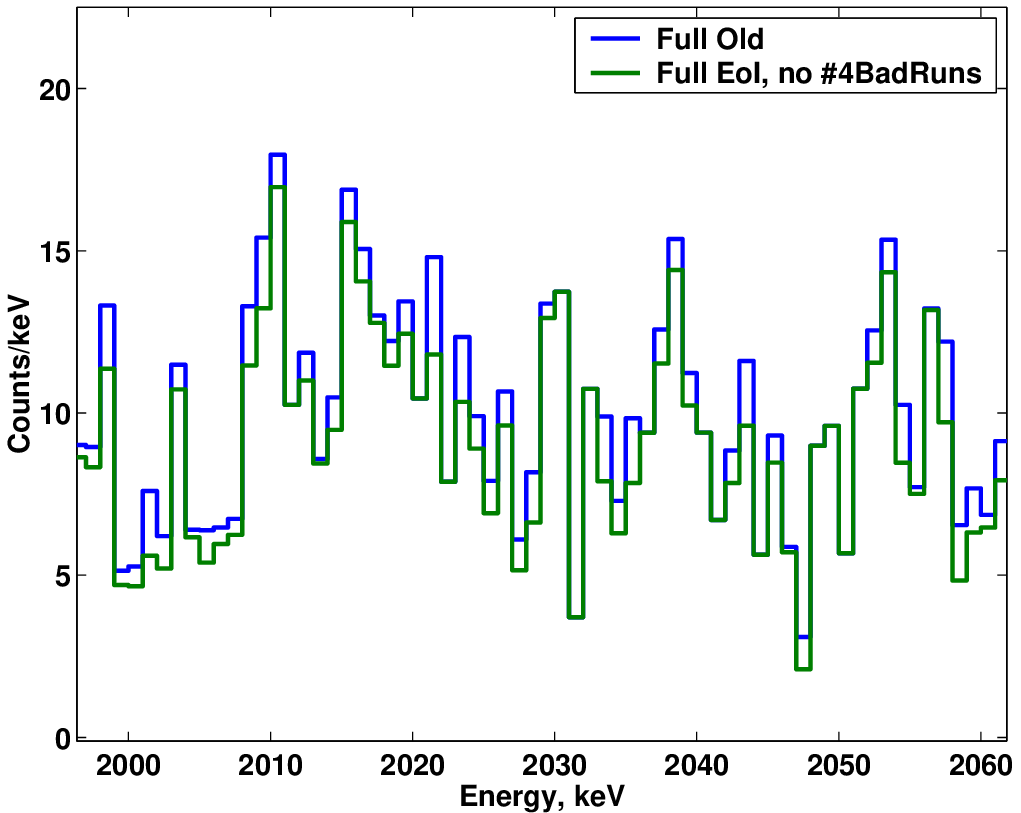}
\caption[]{Left: the peaks at 550\,keV and 640\,keV in the
low-energy range (500-700\,keV) of the ANG4 detector disappear
after applying the EoI cut and removing corrupted data sets.
Right: the effect of these cuts in the sum spectrum around the
\qbb\ value.} \label{UnderThresh}
\end{figure}
The EoI data selection rejects particulary the events which
have an energy below the ADC threshold. The effect of these
cuts in the region of interest around the \qbb\ value is also
shown in Figure \ref{UnderThresh}. One can see that, after
applying the EoI cut, the structure of this part of the
spectrum remains unchanged.

\clearpage

\section{Identification of peaks in the sum spectrum}
\label{sec:Identif}
The total sum spectrum measured over the full energy range
with all five detectors for the period from August 1990 to May
2003 is shown in Figure \ref{fig:LowAll90-03}.
\begin{figure}[ht]
\begin{center}
\includegraphics[height=10cm]{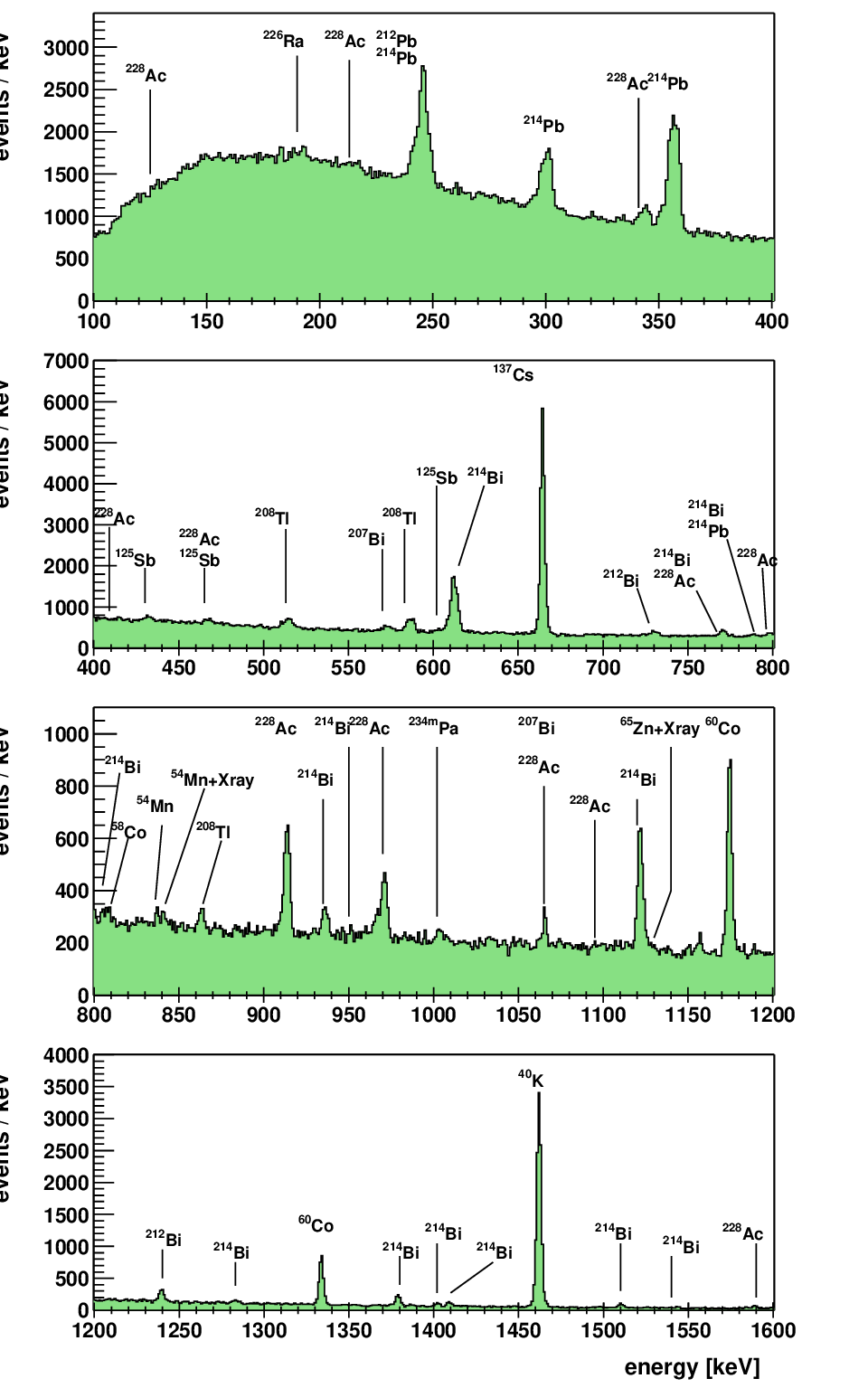}
\includegraphics[height=10cm]{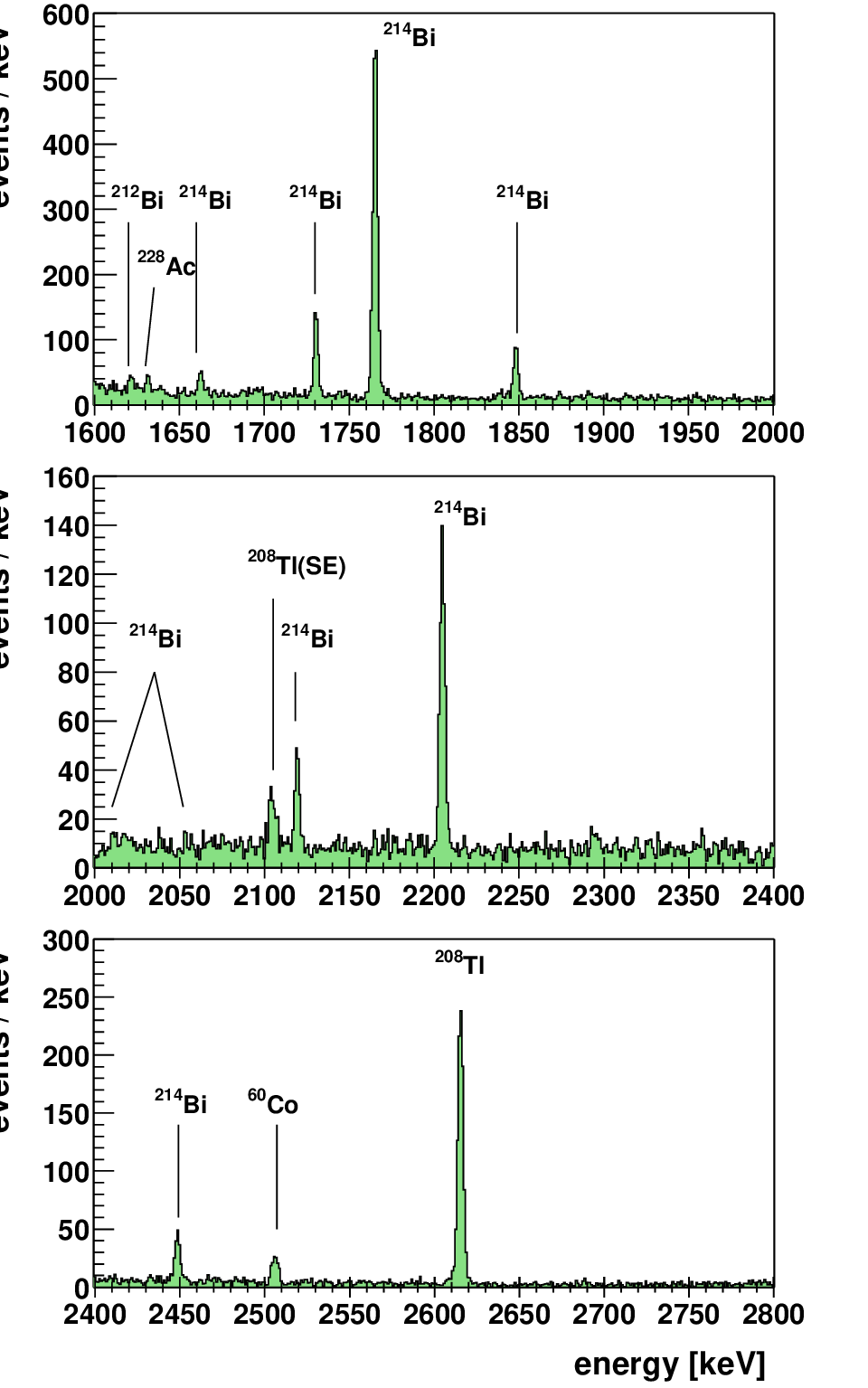}\\
\end{center}
\caption[]{The total sum spectrum measured over the full
energy range of all five detectors for the period from August
1990 to May 2003 \cite{KK-NewAn-NIM04}.}
\label{fig:LowAll90-03}
\end{figure}
The background identified by the measured $\gamma$ lines in
the spectrum consists of primordial activities ($^{238}$U,
$^{232}$Th and $^{40}$K), anthropogenic radioactivity
($^{137}$Cs, $^{125}$Sb, $^{207}$Bi) and cosmogenic isotopes
produced by cosmic ray activation ($^{54}$Mn, $^{58}$Co,
$^{60}$Co). Figure \ref{fig:LoHiEnergy} presents the
comparison of the spectra collected with the low-energy (0-3
MeV) and the wide-energy (0-8 MeV) ADCs. The two spectra
collected simultaneously coincide up to $\sim$3\,MeV, showing
good consistency.
\begin{figure}[h]
\centering
\includegraphics[width=8cm]{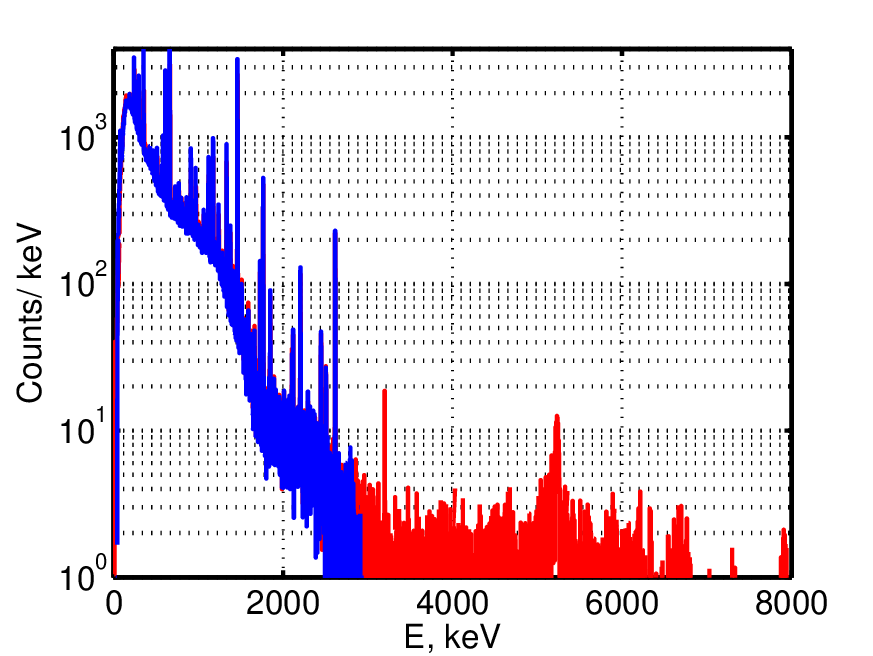}
\caption[]{Comparison of the summed spectra collected in two
energy ranges: 0 - 3\,MeV and 0 - 8\,MeV
\cite{KK-NewAn-NIM04}.} \label{fig:LoHiEnergy}
\end{figure}
The intensities of the background peaks are determined from
the spectrum. The background sources can be located using the
measured and simulated relative peak intensities of the
$\gamma$ lines. The external $\alpha$ and $\beta$ radiation
does not contribute to the background because it cannot
penetrate the 0.7\,mm external dead layer of the p-type
diodes. The detectors ANG1, ANG2 and ANG3 show negligible
$\alpha$-peaks but detectors ANG4 and ANG5 are contaminated
with $^{210}$Pb (see Fig.~\ref{fig:AlphaPeaks}). This
contamination, identified by $\alpha$-peaks at 5245\,keV from
the daughter of $^{210}$Pb ($^{210}$Po), is most likely
located on the surface of the inner contact of the detectors.
The energy of the \Po\ $\alpha$ particles (5304\,keV)
decreases by 64 keV as they pass through the 0.3\,$\mu$m boron
implantation contact in the boreholes of the diodes.\\
\begin{figure}
\centering
  \includegraphics[width=12cm]{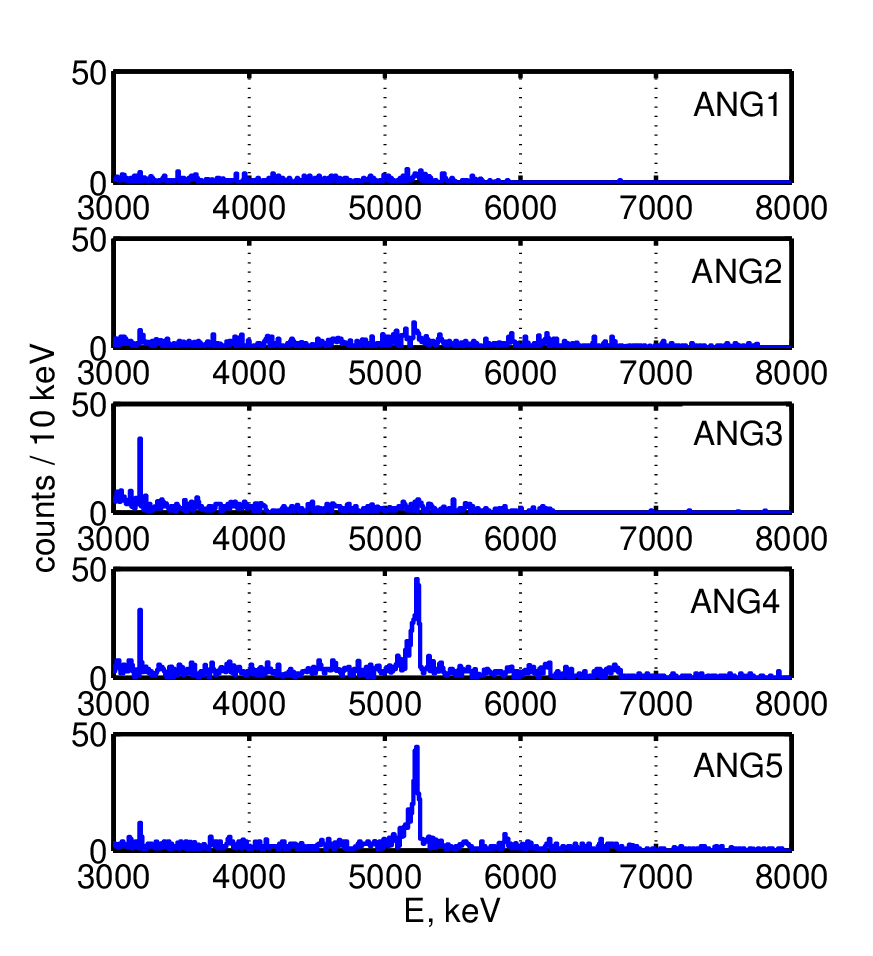}\\
  \caption{The high-energy part of the HdM detector spectra showing the
  $\alpha$-peaks. The peaks from $^{210}$Po, the daughter of $^{210}$Pb, are
  clearly seen in ANG4 and ANG5 \cite{KK-NewAn-NIM04, Diss-Dipl-Dietz}.}\label{fig:AlphaPeaks}
\end{figure}
The most significant background contribution in the region of
interest comes from the Compton continuum of the 2614\,keV
gamma line of \Tl. Other contributions to HdM background come
from \U\ and \Ra\ decay chain, mainly from \Bi, and cosmogenic
contribution from summation of \Co. Table
\ref{tab:Backgr-Peaks} lists all the spectral lines present in
the HdM background spectrum.
\begin{table}[ht]
\centering
\setlength\tabcolsep{3.5pt}
\begin{tabular}{|c|c|c|c|}
\hline
Nuclide& E$_\gamma$ [keV] & Peak Area & FWHM [keV]\\
\hline
RA-226&    186.0 &     1152 $\pm$      180 &     2.18 \\
AC-228&    209.4 &      451 $\pm$      180 &     2.20 \\
PB-212&    238.6 &     6419 $\pm$      197 &     2.22 \\
PB-214&    241.9 &     2831 $\pm$      215 &     2.22 \\
AC-228&    270.3 &      170 $\pm$       68 &     2.24 \\
TL-208&    277.4 &      169 $\pm$       86 &     2.25 \\
PB-214&    296.1 &     4910 $\pm$      184 &     2.26 \\
PB-212&    300.1 &      276 $\pm$      159 &     2.27 \\
AC-228&    328.1 &      315 $\pm$      113 &     2.29 \\
AC-228&    338.4 &     1302 $\pm$      144 &     2.30 \\
PB-214&    352.6 &     9948 $\pm$      343 &     2.62 \\
AC-228&    409.6 &      252 $\pm$      213 &     2.30 \\
SB-125&    428.1 &     1113 $\pm$      228 &     3.29 \\
AC-228+
SB-125&    463.5 &      850 $\pm$      132 &     2.40 \\
TL-208&    511.2 &     2143 $\pm$      236 &     3.55 \\
BI-207&    570.3 &      867 $\pm$      151 &     2.74 \\
TL-208&    583.1 &     2566 $\pm$      228 &     2.77 \\
SB-125&    600.8 &      450 $\pm$       87 &     2.50 \\
SB-125&    606.8 &      589 $\pm$      87  &     2.50 \\
BI-214&    609.3 &     7552 $\pm$       96 &     2.51 \\
SB-125&    635.2 &      389 $\pm$      179 &     1.98 \\
CS-137&    661.6 &    20201 $\pm$      164 &     2.55 \\
BI-214&    665.5 &      289 $\pm$      187 &     2.55 \\
SB-125&    671.7 &      135 $\pm$       82 &     2.55 \\
BI-212&    727.2 &      579 $\pm$       88 &     2.59 \\
BI-214&    768.4 &      769 $\pm$       86 &     2.63 \\
BI-212+
PB-214&    785.4 &      242 $\pm$       82 &     2.64 \\
PB-214&    786.0 &      176 $\pm$       82 &     2.64 \\
AC-228&    794.8 &      660 $\pm$      168 &     2.64 \\
      &    801.8 &      178 $\pm$       83 &     2.65 \\
BI-214&    806.2 &      275 $\pm$       77 &     2.65 \\
CO-58 &    810.8 &      171 $\pm$       75 &     2.66 \\
MN-54 &    834.8 &      420 $\pm$      106 &     2.67 \\
AC-228&    840.8 &      200 $\pm$       80 &     2.67 \\
TL-208&    860.5 &      449 $\pm$       78 &     2.69 \\
AC-228&    910.8 &     2135 $\pm$      115 &     2.72 \\
BI-214&    933.9 &      630 $\pm$      131 &     3.15 \\
\hline
\end{tabular}
\label{tab:Backgr-Peaks} \caption[]{Background peaks in the
measured sum spectrum for all five detectors in the energy
range 140-2800\,keV, for 71.7 kg\,y exposure (1990 - 2003),
their energies (error $\pm$0.1\,keV), intensities and widths
(error $\pm$0.05\,keV).}
\end{table}
\addtocounter{table}{-1}  
\begin{table}[ht]
\centering
\begin{tabular}{|c|c|c|c|}
\hline
Nuclide& E$_\gamma$ [keV]& Peak Area & FWHM [keV]\\
\hline
AC-228&    964.6 &      458 $\pm$       86 &     2.77 \\
AC-228&    968.9 &     1259 $\pm$       82 &     2.77 \\
PA-234m&   1001.3 &      250 $\pm$       78 &     3.77 \\

BI-207&   1063.3 &      634 $\pm$       98 &     3.39 \\
BI-214&   1120.3 &     1926 $\pm$       86 &     2.88 \\
BI-214&   1155.2 &      250 $\pm$       62 &     2.90 \\
CO-60 &   1173.2 &     3955 $\pm$       88 &     2.92 \\
BI-214&   1238.1 &      807 $\pm$       89 &     2.86 \\
BI-214&   1281.5 &      274 $\pm$       66 &     3.54 \\
CO-60 &   1332.3 &     3690 $\pm$       90 &     2.93 \\
BI-214&   1377.7 &      675 $\pm$       50 &     3.06 \\
BI-214&   1401.5 &      220 $\pm$       42 &     3.07 \\
BI-214&   1408.0 &      292 $\pm$       45 &     3.08 \\
K-40  &   1460.8 &    13010 $\pm$      134 &     3.12 \\
BI-214&   1509.2 &      261 $\pm$       38 &     3.15 \\
BI-214&   1542.2 &      105 $\pm$       32 &     3.17 \\
BI-214&   1583.4 &       50 $\pm$       37 &     3.15 \\
AC-228&   1587.6 &      157 $\pm$       42 &     3.71 \\
BI-212&   1620.6 &      127 $\pm$       26 &     3.22 \\
AC-228&   1630.4 &       75 $\pm$       27 &     3.22 \\
BI-214&   1661.8 &      163 $\pm$       34 &     3.88 \\
BI-214&   1729.9 &      491 $\pm$       40 &     3.05 \\
BI-214&   1764.5 &     2204 $\pm$       51 &     3.31 \\
BI-214&   1847.7 &      324 $\pm$       30 &     3.21 \\
SE(2614)&   2103.6 &      134 $\pm$      21 &     3.52 \\
BI-214&   2118.5 &      185 $\pm$       21 &     3.53 \\
BI-214&   2204.3 &      579 $\pm$       33 &     3.89 \\
BI-214&   2448.0 &      193 $\pm$       15 &     3.77 \\
$\Sigma$CO-60 &   2505.8 &      156 $\pm$       14 &     4.73 \\
TL-208&   2614.5 &     1184 $\pm$       36 &     3.82 \\
$\Sigma$TL-208\,$^*$&   3198.0 &     80 $\pm$       9 &     4.48 \\
PO-210 $\alpha\,^*$& 5245 & 454$\pm$22 & 53 \\
\hline
\end{tabular}
\caption[]{(Continued). The energy range 1900 - 2100\,keV is
analyzed separately in the Section \ref{sec:hd-resfull} and
Chapter \ref{ch:bkg}. ($^*$) The summation peak from \Tl\ and
the $^{210}$Po $\alpha$-peak are measured with exposure 56.6\,
kg\,y (1995-2003). (SE-single escape peak, $\Sigma$ -
summation peak.)}
\end{table}
All nuclei contributing to the background in the region of
interest have been commented. Gamma lines from $^{228}$Ac,
$^{212}$Pb, $^{212}$Bi and \Tl\ are present in the background.
The parent long-lived nucleus of this chain is \Th\ or
$^{228}$Ra. In the high energy part of the spectra of ANG3,
ANG4 and ANG5 (Fig.~\ref{fig:AlphaPeaks}), a peak at 3198 keV
is identified as a summation peak from the gamma-ray cascade
583.2+2614.5=3197.7 keV. The presence of this true coincidence
summation peak indicates that the sources are very close to
the detectors (within 1\,cm). It is believed that this \Th\
contamination is located on the detector holders.\\
No secular equilibrium of \U\ with its daughter is observed.
The 1001.9\, keV gamma line in the HdM spectrum can be
attributed to the $^{234m}$Pa decay. The intensity of this
gamma line is $250\pm 45$ counts and the main contribution
comes from the ANG3 and ANG5 detectors. Therefore, it shows a
contamination of these detectors with \U. The relative
intensities of the \Bi\ lines indicate that the \Ra\
contamination is located minimum 5\,cm away from the diodes,
probably in the copper cryostats or inner surface of the
shield.\\
The 1173.2 and 1332.5\,keV lines and their summation peak at
2505.7\,keV from \Co\ are visible in the spectra of all
detectors. The relative intensities of the summation peaks
show that the holders of ANG1, ANG2 and ANG5 and the cryostats
of ANG3 and ANG4 are contaminated with \Co. The method for
localization of sources using relative intensity of
$\gamma$-lines is described in Chapter \ref{ch:bkg}.

\clearpage

\section{Analysis of the spectrum around Q$_{\beta\beta}$}
\label{sec:hd-resfull}

\subsection{Method of fitting}
\label{ssec:hd-method}
The least squares method with the Levenberg-Marquardt
algorithm has been used to analyze the spectra. The
Levenberg-Marquardt method \cite{Marq63} is one of the most
developed and tested minimization algorithms, finding fits
most directly and efficiently \cite{Beving}. It is also
reasonably insensitive to the starting values of the
parameters. A nonlinear fit in the range 2000 - 2060\,keV was
performed. The following procedure was used: The histogram
$N_i$, where $i$ is the bin number, corresponds to a spectrum
with bin energy $E_i$. The spectrum is fitted using $n$
Gaussians $G(E_i,E_{\circ j},\sigma_j)$, where $n$ is the
number of lines to fit, $E_{\circ j}$ and $\sigma_j$ are the
estimated centroid and width of the Gaussian $j$:
\begin{equation}
G(E_i,E_{\circ j},\sigma_j)=
\frac{1}{\sigma_j\sqrt{2\pi}}\exp\Bigl[-\frac{(E_i-E_{\circ
j})^{2}}{2\sigma_j^{2}}\Bigr]. \label{Gaussian}
\end{equation}
Different background models $B(E_i)$ were used to fit the
data: simulated background (linear with fixed
slope)\cite{KK-Doer03}, linear and constant background. The
fitting function $M(E_i)$ is a sum of Gaussians and background
models:
\begin{equation}\label{Model}
M(E_i) = \sum^n_j{S_j\cdot{G(E_i,E_{\circ
j},\sigma_j)}}+b\cdot
    B(E_i),
\end{equation}
where $S_j$ are the estimated peak intensities and $b$ is the
background index. The peak intensities, the mean centroids,
the widths and the background index are found by minimizing
the norm of errors:
\begin{equation}
    S_j,E_{\circ j},\sigma_j, b =
\arg\min{\sum^m_i\left[M(E_i) - N(E_i)\right]^2}, \label{LSQ}
\end{equation}
where $m$ is the number of bins in the analyzed part of the
spectrum. Following the suggestion of the Particle Data Group
\cite{RPD00}, the number of events in the measured spectrum
was compared to the number of events given by the fit. Because
these numbers are identical on a percent level, no
normalization is required. This method has advantages over
others of providing an estimate of the full error matrix. The
MATLAB Statistical Toolbox \cite{MATLAB} provides a function
for calculating the confidence interval of the parameters. It
uses the residuals of the fit and the Jacobian matrix $J_{ij}$
of $M_{i}$ around the solution of Eq.~\ref{LSQ} to obtain the
error of the parameters. The confidence intervals calculated
by the MATLAB statistical function were tested with a MC
simulation. 100 000 spectra were simulated with a
Poisson-distributed background and a Gaussian-shaped peak of
given intensity. An analysis of the true number of counts
inside the calculated confidence interval was performed. Table
\ref{tab:Non-Lin-Opt} shows the results of this confidence
test. For example, peaks with 30\,counts on a background of
10\,events per bin were analyzed with 100 000 simulated
spectra. In 99 979 cases, the true value of the peaks was
within the 4$\sigma$ calculated confidence interval (compare
to the expected value of 99994 cases for 4$\sigma$). It means
that the calculated confidence interval is underestimated by
$\sim$ 0.3 $\sigma$.
\begin{table}[ht]
\small
\centering
\renewcommand{\arraystretch}{1.7}
\setlength\tabcolsep{2.4pt}
\begin{tabular}{|c|c|c|c|c|c|c|c|c|c|c|c|c|}
\hline B   & \multicolumn{4}{|c|}{S = 10} &
\multicolumn{4}{|c|}{S = 20}
& \multicolumn{4}{|c|}{S = 30}\\
\hline &   1$\sigma$   &   2$\sigma$   & 3$\sigma$   &
4$\sigma$   &
    1$\sigma$   &   2$\sigma$   &
3$\sigma$   &       4$\sigma$   & 1$\sigma$   &   2$\sigma$
&
3$\sigma$   &       4$\sigma$   \\
\cline{2-13}
    2   &
66505   &   93781   &   99183   &   99932   & 63967   &
93176   & 99219   &   99952   &
62094   &   92191   &   99121   &   99932   \\
    7   &
66879   &   91683   &   98476   &   99900   & 67618   &
94733   & 99498   &   99961   &
66332   &   94443   &   99539   &   99967   \\
    10  &
64962   &   90029   &   98380   &   99918   & 68210   &
94921   & 99455   &   99933   &
67352   &   94784   &   99554   &   99979   \\
\hline
\multicolumn{13}{|c|}{Expected:}\\
\hline
    &
68269   &   95449   &   99730   &   99994   & 68269   &
95449   & 99730   &   99994   &
68269   &   95449   &   99730   &   99994   \\
\hline
\end{tabular}
\caption[]{Results of the simulations of 100 000 spectra
giving the number of cases where the true number of counts in
the peak is found in the calculated confidence interval. The
spectra have a Poisson-distributed background and a
Gaussian-shaped peak of a given intensity. The simulations
were performed for different backgrounds (B), peak areas (S)
and confidence levels. The least squares method, using the
Levenberg-Marquardt algorithm implemented in the MATLAB
function \cite{MATLAB}, was used \cite{KK-NewAn-NIM04}.}
\label{tab:Non-Lin-Opt}
\end{table}

\subsection{Fitting results}

Figure \ref{fig:Sum90-03} compares the fit with the measured
spectrum in the range around the \qbb\ value (2000 -
2060\,keV).
\begin{figure}[ht]
\centering
\includegraphics[width=8cm]{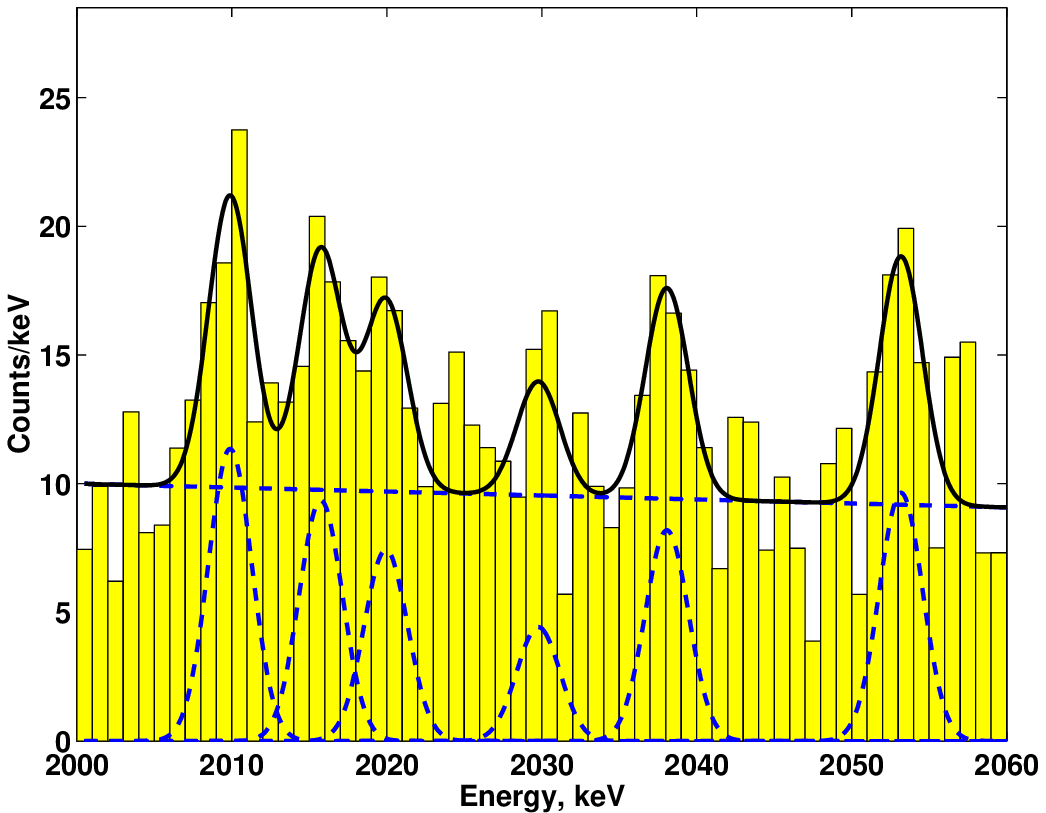}
\caption[]{ The total sum spectrum of all five detectors (in
total 10.96\,kg enriched in \Ge), from November 1990 to May
2003 (71.7\,kg\,y) in the 2000 - 2060\,keV range and its fit
(solid line) with fixed peak widths, but free centroids
\cite{KK-NewAn-NIM04}.} \label{fig:Sum90-03}
\end{figure}
The peak positions, their intensities and the background are
given simultaneously by the fit. The positions and the
intensities of the peaks are fitted but the peak widths are
fixed. Table \ref{tab:SpecAtQ} gives the intensities and the
positions of the spectrum peaks in the region of interest.
\begin{table}[ht]
\small
\renewcommand{\arraystretch}{1.7}
\centering
\begin{tabular}{|c|c|c|c|c|c|}
\hline
Line & \multicolumn{2}{|c|}{Detectors ANG1-5} & \multicolumn{3}{|c|}{Expected and Measured Intensity} \\
{[keV]} & \multicolumn{2}{|c|}{1990-2003, 71.7\,kg\,y, \cite{KK-NewAn-NIM04}} & \multicolumn{3}{|c|}{with a \Ra\ source}\\
\cline{2-6}
 &  Fit Energy,  &  Fit Intensity, &   Rel. Intensity, &  Expected & Measured \\
 &   [keV]   &   [counts]  &  [\%], TOI$^*$ \cite{TOI99}   &     TOI, [counts] & \Ra, [counts] \\

\hline

2010.7  &  2009.9$\pm$0.3  &  39.9$\pm$6.9  & 0.047 $\pm$ 0.003 & 5.6 $\pm$ 0.4 & 10.9$\pm$0.3 \\

\hline

2016.7 & 2015.7$\pm$0.4     & 32.8$\pm$7.1  &  0  &  0 & 13.4$\pm$0.3\\

\hline

2021.8  & 2019.9$\pm$0.5     & 26.1$\pm$7.1  &  0.020 $\pm$ 0.003&  2.4 $\pm$ 0.3 & 2.6$\pm$0.3\\

\hline

2030  & 2029.8$\pm$0.8     & 15.5$\pm$6.9  & --  & -- & -- \\

\hline

 2039.0  & 2038.1$\pm$0.4     & 28.75$\pm$6.9  & --  & -- & --\\

\hline

 2052.9  & 2053.2$\pm$0.4     & 33.9 $\pm$6.8  & 0.069$\pm$ 0.005  &  8.2$\pm$ 0.6 & 8.0$\pm$0.3\\

\hline \hline
 2204.2 & 2204.3$\pm$0.1              & 579 $\pm$33   & 5.08$\pm$0.4 & &\\
\hline
\end{tabular}
\caption[]{Peaks of the measured spectra (1990-2003) in the
2000-2060\,keV range, and their non-linear least squares fit
with unconstrained intensities and peak positions. The
energies and intensities from the fit, and their 1$\sigma$
errors are provided \cite{KK-NewAn-NIM04}. The expected
intensities of the weak \Bi\ lines are calculated using the
intensity of the 2204\,keV (5.08\%) line in the HdM spectrum
(579$\pm$33 counts). All peaks measured with the \Ra\ source
are normalized to the 2204\,keV peak area in the HdM spectrum.
The 2030\,keV line has not been identified. These measurements
are discussed in the next Chapter. ($^*$)TOI: Table of
Isotopes, TSC: true coincidence summing.}
    \label{tab:SpecAtQ}
\end{table}
The signal at Q$_{\beta\beta}$ in the full spectrum at $\sim$
2039\,keV reaches a 4.2 $\sigma$ confidence level for the
period 1990-2003. The intensities of the 2022 and 2053\,keV
\Bi\ lines are higher than expected relative to the intensity
of a strong \Bi\ line at 2204\,keV. The table of isotopes
\cite{TOI99} gives relative intensities of 5.08$\pm$0.04\% and
0.069$\pm$0.005\%, for the \Bi\ lines 2204.2 and 2052.9\,keV
respectively. The detection efficiency for photons in the
region 2000--2060\,keV and the detection efficiency for
2204\,keV photons are equal within a few percent. The
intensity of the 2204\,keV line, 579$\pm$33 counts
(Tab.~\ref{tab:Backgr-Peaks}), gives an expected count of
8.2$\pm$0.6 in the 2053\,keV peak. It is four times smaller
than the number of counts obtained from the presented fit
(33.9$\pm$6.8\,counts). In order to understand the discrepancy
in the line intensities from the spectrum, fits were performed
using wider energy intervals. Figure \ref{fig:FitEnergyWindow}
shows the same spectrum as in Figure \ref{fig:Sum90-03} but
the data are fitted over energy windows of 60, 80 and 100\,keV
width.
\begin{figure}[ht]
\centering
\includegraphics[width=14cm]{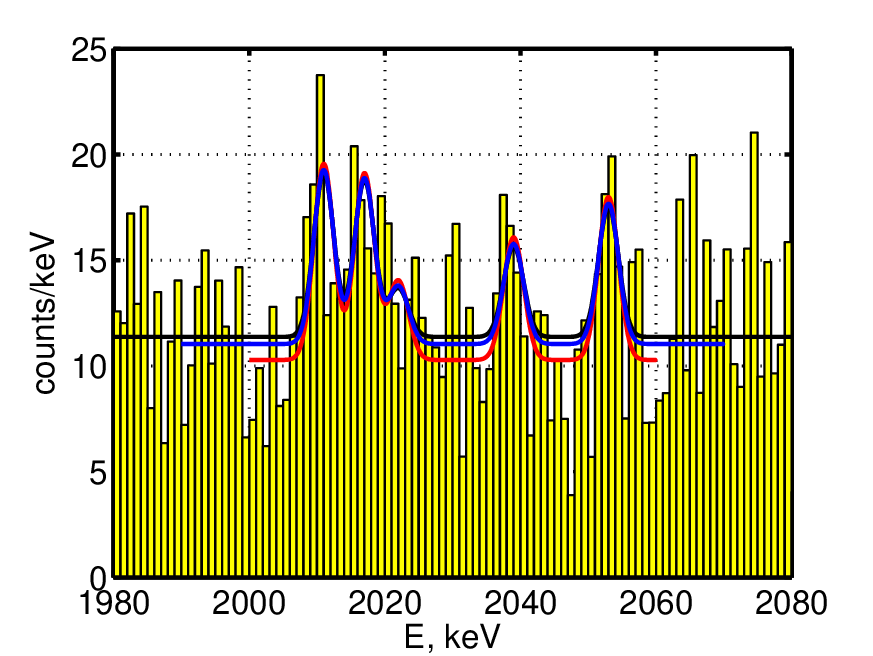}
\caption[]{ Fits of the HdM spectrum for three energy windows:
2000-2060\,keV, 1990-2070\,keV and 1980-2080\,keV. The
spectrum is fitted with fixed peak positions \cite{TOI99} and
fixed peak widths (3.48\,keV FWHM) defined by the energy
calibration. The fitted background depends on the energy
interval.} \label{fig:FitEnergyWindow}
\end{figure}
\begin{table}
  \centering
\begin{tabular}{|c|c|c|c|}
  \hline
  Line         & \multicolumn{3}{|c|}{Energy window}\\
  \cline{2-4}
  E, [keV] & 2000-2060\,keV & 1990-2070\,keV & 1980-2080\,keV\\
  \hline
  2010     & 34.5$\pm$   8.2 &    31.5 $\pm$   8.4 &    29.4 $\pm$   8.7\\
  2017     & 32.9 $\pm$  8.2 &    30.0 $\pm$   8.4 &    28.0 $\pm$   8.6\\
  2022     & 14.0 $\pm$  8.2 &    11.0 $\pm$   8.4 &     9.0 $\pm$   8.7\\
  2039     & 21.6 $\pm$  8.3 &    18.5 $\pm$   8.4&    16.4 $\pm$   8.7\\
  2053     & 28.7 $\pm$  8.3 &    25.6 $\pm$   8.4 &    23.4 $\pm$   8.7\\
  \hline
   B\,[counts/keV] &  10.3 $\pm$  0.6 & 10.9 $\pm$   0.5 &  11.3 $\pm$   0.4 \\
  \hline
\end{tabular}
\caption{Intensities of the peaks in the region of interest
around \qbb\ obtained from fit using different energy
windows.}\label{tab:FitEnergyWindow}
\end{table}

The peak intensities and the background level depend on the
energy windows used for fitting. An empirical model was
created based on measurements with sources and simulated
spectra to investigate the shape of the spectrum and to
optimize the fitted energy interval. This empirical model is
presented in the next Chapter.

\clearpage

\section{Conclusions}
The improved analysis of the HdM raw data collected during
1990--2003 was performed, based on the investigation of the
calibration curves of the HdM experiment. The better quality
of the energy calibration and the energy resolution of the
summed spectrum allows to fit the spectrum with accurate peak
positions and widths. Because the energy resolution determines
the sensitivity of the experiment, a 20\% improvement in the
energy resolution leads to a 10\% increase in sensitivity.
This has lead to publications \cite{KK-NewAn-NIM04, KK-NewAn-PL04}.\\
The sum spectrum was analyzed and the $\gamma$ and $\alpha$
peaks were identified and their intensities are presented. The
fits of the region around the \qbb\ value show that the
resulting peak intensities depend on the width of the analyzed
energy window. For an optimization of the fitted energy
interval a model of the HdM background was proposed.

\newpage
\chapter{{\sl A posteriori} Background Evaluation for the~HdM Experiment}
\label{ch:bkg}
The HdM background around the \qbb\ value consists of
contributions from the Compton scattering of the \Tl, \Bi\ and
\Co\ $\gamma$-rays as well as the continuum from neutrons and
muons. An interpretation of the HdM background was done, based
on measurements with sources using the HdM detectors after
their handover to the \gerda\ collaboration in 2004. \Th, \Ra\
and \Co\ sources were placed at various locations, with
different thicknesses of lead absorber between the source and
the detector. The goal was to reproduce the proper line shapes
and the relative intensities observed in the HdM background.
The resulting spectra together with the contributions of \Cs,
\K, $^{210}$\rm{Pb} and the calculated shape of \tnbb\ at
lower energies were normalized and used to fit the HdM
background spectrum.

\section{Using peak ratios for the localization of background sources}

The shape of the spectrum is highly dependent on the position
of the source, and the amount of absorber between the source
and the detector. Different peak intensity ratios result from
different combinations of distance and absorber thickness. The
true coincidence summing (TCS) effect, which happens when more
than one photon is emitted in a nuclear de-excitation, can
help to resolve these combinations.\\
There is a certain probability that both photons
simultaneously deposit their full energy in the same germanium
crystal. Because detection efficiency is inversely
proportional to squared distance, TCS probability depends
strongly on the distance between the source and the detector,
and on the thickness of absorbing material.\\
The intensity ratio between a TCS summation peak and a single
$\gamma$-peak depends on the distance between the source and
the detector. The decay of \Bi\ (decay daughter of \Ra) emits
different combinations of photons (Fig.~\ref{decayBiTh}(Top)).
The 1994\,keV, 2010.7\,keV and 2016.7\,keV peaks in the
spectrum can originate from TCS with the 609.3\,keV photon
(emission probability = 46.1\%) with the 1385.3\,keV photon
(probability = 0.757\%), the 1401.5\,keV photon (probability =
1.27\%) and the 1408.0\,keV photon (probability = 2.15\%),
respectively \cite{TOI99}. The ratio of intensities between
the summation peak at 2016.7\,keV and a major peak at
2204.2\,keV (probability = 5.08\%) decreases
with distance.\\
Single $\gamma$-peak intensity ratios do not depend on
distance, but on the thickness of absorber. The ratio of the
intensity of the 609.3\,keV and 2204.2\,keV \Bi\ lines
decreases with increasing absorber thickness, and therefore
can be used to determine the amount of material between the
source and the detector. This value leads to the distance
between the source and the detector
using TCS.\\
The same method was used to localize the \Tl\ and \Co\
contamination. For \Tl, the summation peak at 3197\,keV and
$\gamma$-lines at 2614.5\,keV (probability=99\%) and
583.4\,keV (probability= 85.5\%) were used. The summation peak
at 2505.7\,keV from 1173.2 and 1332.5\,keV lines was used to
localize \Co\ sources. The localization of \Co\ was performed
by using MC simulations, while the \Th\ and \Ra\
contaminations were localized by comparison with source
measurements.\\
\begin{figure}[!htp]
\begin{center}
\includegraphics[width=8cm]{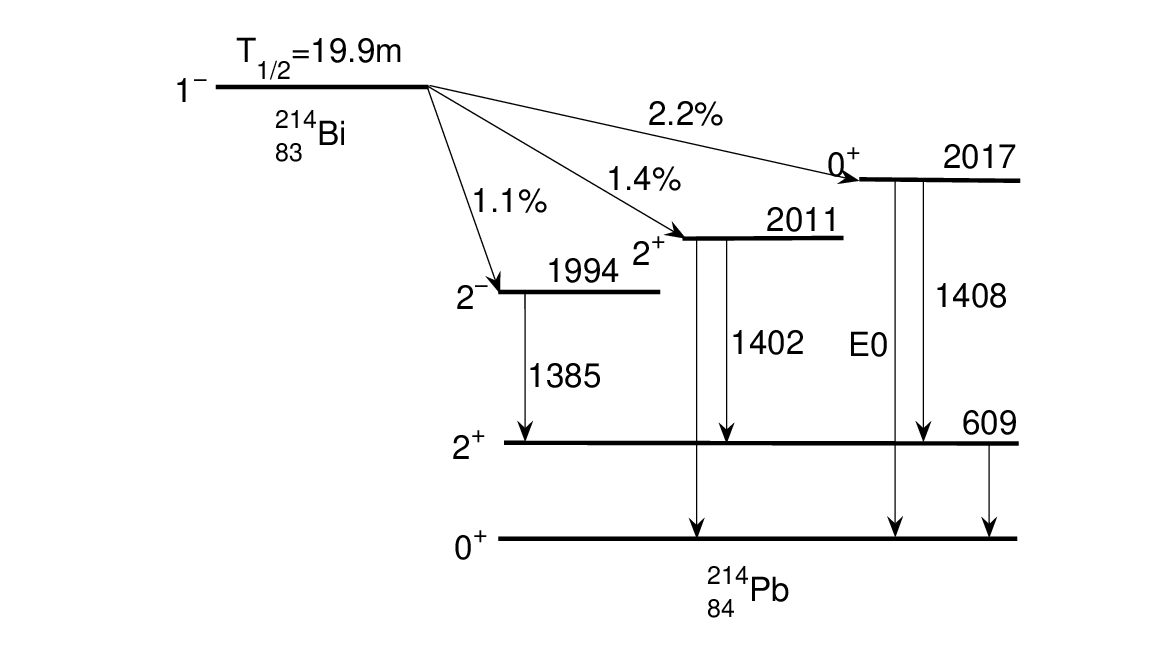}
\includegraphics[width=8cm]{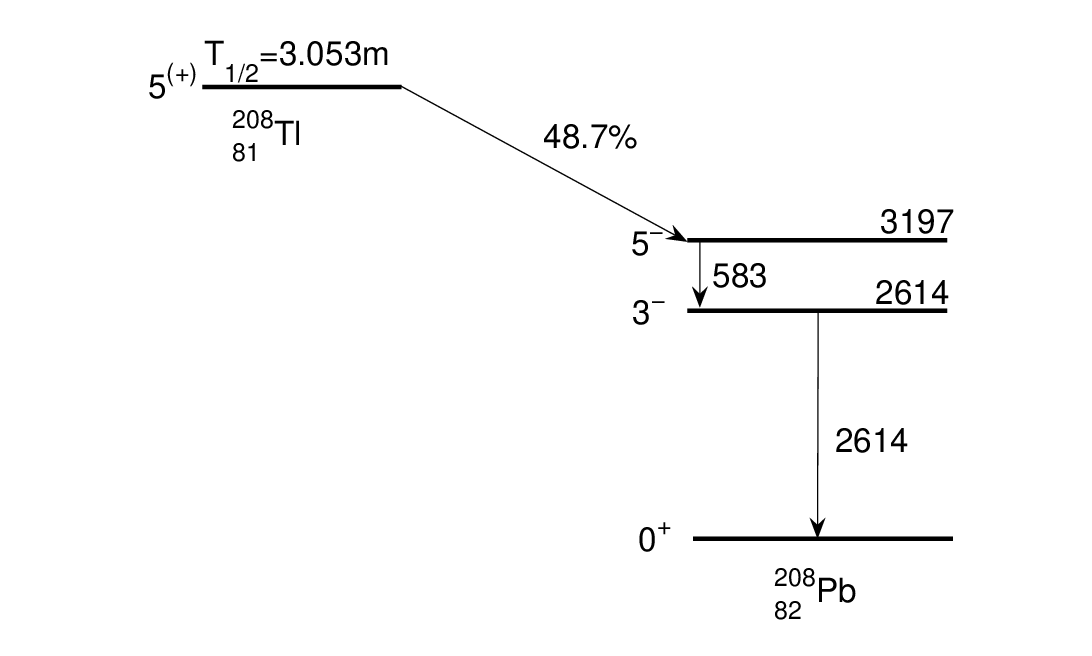}\\
\end{center}
\caption[]{Reduced decay scheme of \Bi\ and \Tl\ showing the
relevant levels for the true coincidence summing effect in
germanium detectors \cite{TOI99}.} \label{decayBiTh}
\end{figure}

\section{{\sl A posteriori} spectral shape measurements with sources}
Spectra were measured in 2005-2006 for different positions of
the \Ra\ and \Th\ sources using the ANG3 enriched detector
(2.44 kg). It has an external diameter of 7.85 cm and a length
of 9.35 cm. The distance between the top of crystal and the
copper cap is 2.85 cm. The detector setup with the lead
absorber and the position of the source are shown in
Fig.~\ref{ang3-detsetup}.
\begin{figure}[!htp]
\begin{center}
\includegraphics[width=6cm]{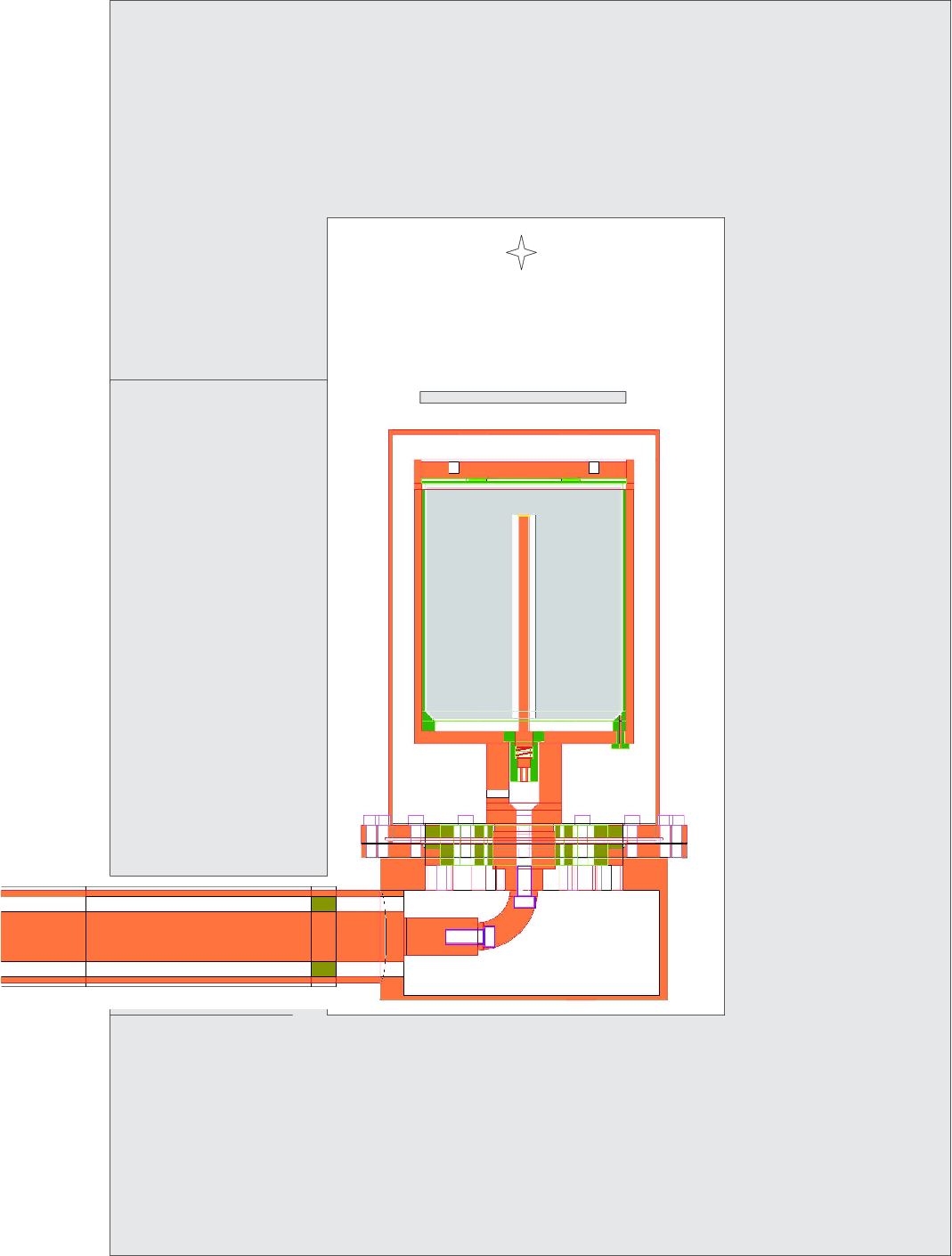}
\end{center}
\caption[]{Schematic view of the ANG3 detector used in the
spectral shape measurements, the source and the lead absorber.
The setup is shielded with 10\,cm of lead (gray area).}
\label{ang3-detsetup}
\end{figure}
The spectra were measured with a standard ORTEC spectroscopy
amplifier and a multichannel analyzer. The energy resolution
was $\sim$3keV at 1332\,keV. To reduce the external
background, the detector was surrounded by 10 cm of lead
shield. Consequently, the background contribution during the
measurements was not higher than 1\% of the total count rate
measured with the sources. To avoid random coincidences, the
count rates were not higher than 1000\,counts per second. Two
\Ra\ sources, both enclosed in stainless steel containers with
0.5\,mm walls, were used. One of the sources, with an activity
of 936\,Bq, was used for the measurements close to the
detector (0-5\,cm). The second source, with an activity of
95.2\,kBq, was used for measurements far from the detector cap
(10-20\,cm) and with a thick lead absorber. The activity of
the \Th\ source was 17\,kBq. The sources were measured at 0,
1, 2, 3, 5, 10, 15, and 20\,cm from the detector cap. At each
position a 3, 6, and 9\,mm thick lead absorber was used.\\
The TCS effect for the \Bi\ lines in the region of interest is
shown in Fig.~\ref{SpecSeq}. The \Ra\ spectra are normalized
to the area of the 2204\,keV peak. The peaks at 2022 and
2053\,keV are less affected by TCS than the peaks at 1994,
2010 and 2017\,keV. The intensities of the measured 2022 and
2053\,keV peaks are in agreement with TOI\cite{TOI99}
(Tab.~\ref{tab:SpecAtQ}).
\begin{figure}[!htp]
\begin{center}
\includegraphics[width=9cm]{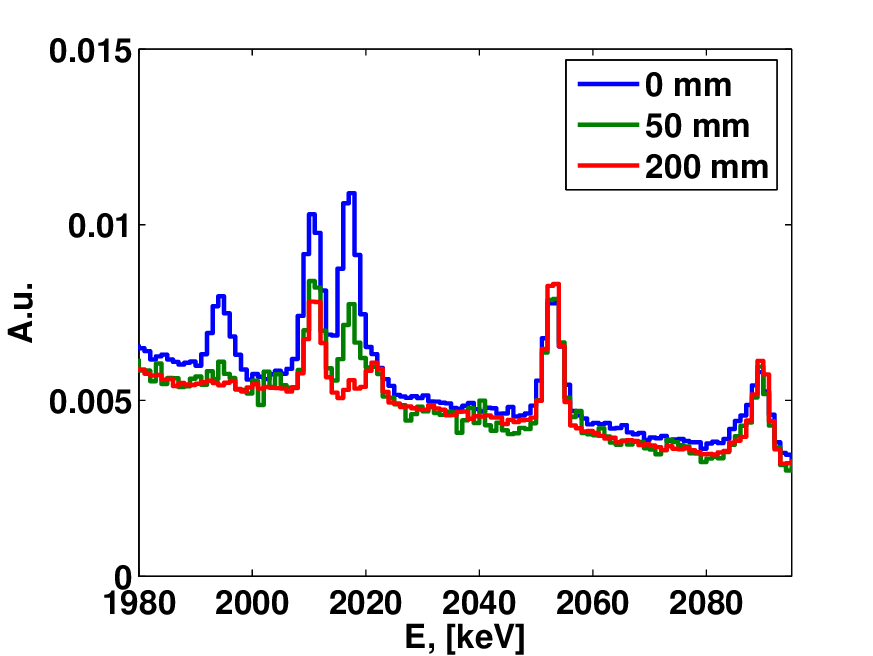}\\
\end{center}
\caption[]{Spectrum shapes for three positions of the \Ra\
source (0\,mm, 50\,mm and 200\,mm from the top of the detector
copper cap). The spectra are normalized to the area of the
2204\,keV peak. The peaks at 1994\,keV and 2017\,keV are not
present in the spectrum at 200\,mm, but become stronger when
the source is closer to the detector. The peak at 2053\,keV is
practically not affected by TCS.}\label{SpecSeq}
\end{figure}
Without any absorber, the summing line intensities depend only
on the inverse squared distance between the source and the
detector. In reality, the source is distributed in the
detector construction materials. To find the effective
absorbtion material thickness, the ratios of the major gamma
lines of \Ra\ and \Tl\ as a function of the absorber thickness
were measured. These ratios are approximately constant (within
5\%) for all measured distances from the detector, but depend
on the absorber thickness (Fig.~\ref{ra609-2016-2204} and
\ref{th583-3197-2614}).
\begin{figure}[!htp]
\begin{center}
\includegraphics[height=5cm]{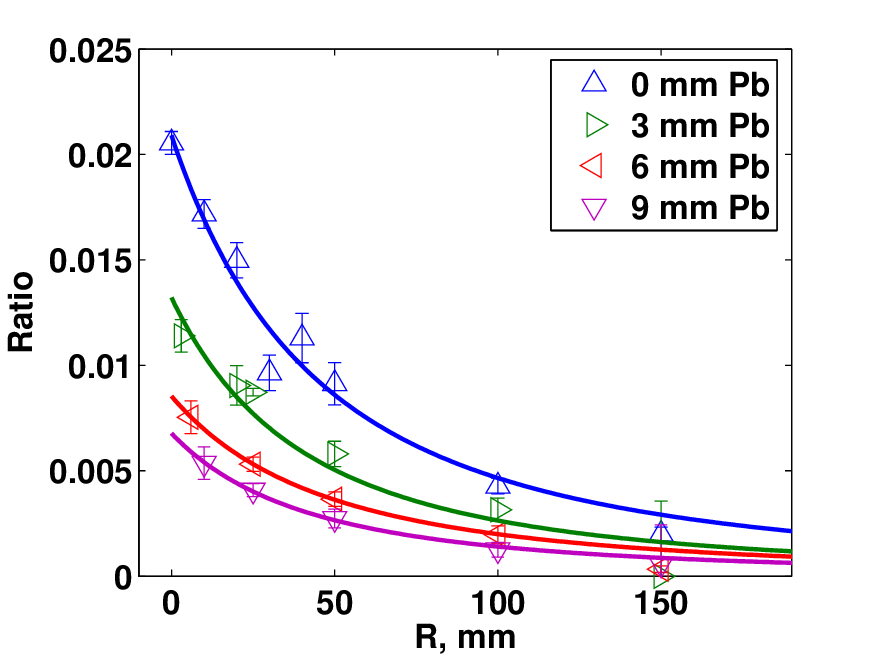}
\includegraphics[height=5cm]{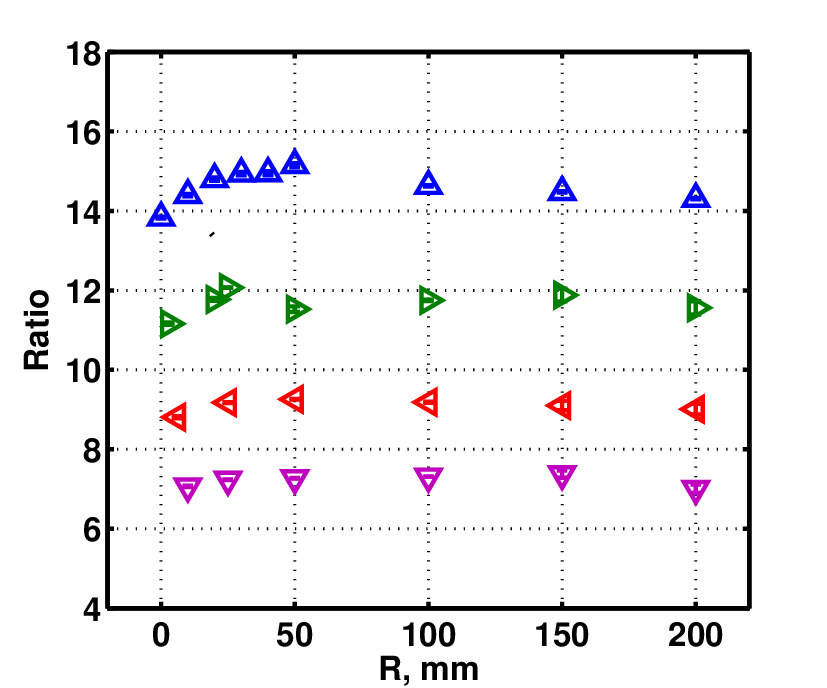}\\
\end{center}
\caption{Left: Ratios of the summation peak at 2016\,keV to
the $\gamma$-line at 2204\,keV vs. distance of the source to
the detector and thickness of the lead absorber. Right: Ratios
of 609\,keV line to 2204\,keV line vs. distance of the source
to detector and thickness of the lead absorber.}
\label{ra609-2016-2204}
\end{figure}
\begin{figure}[!htp]
\begin{center}
\includegraphics[height=5cm]{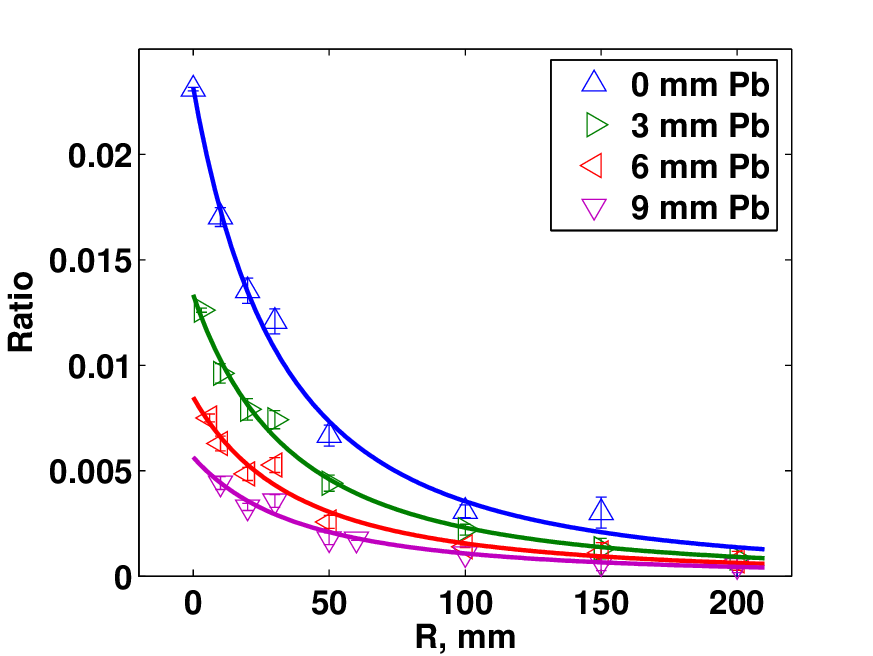}
\includegraphics[height=5cm]{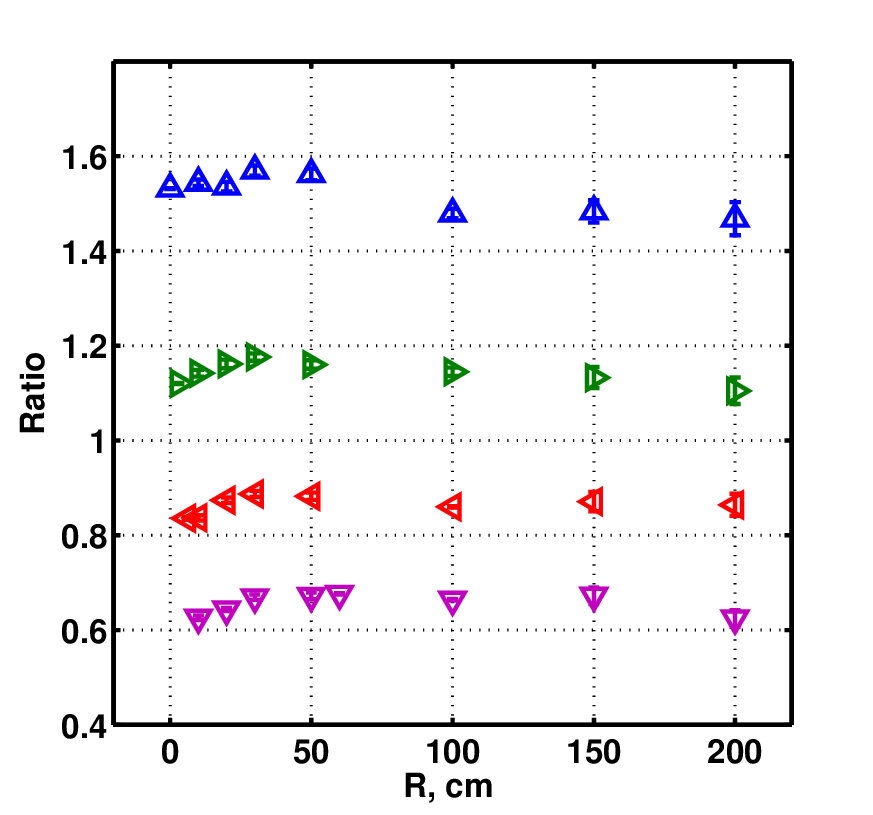}\\
\end{center}
\caption{Left: Ratios of the summation peak at 3197\,keV to
the $\gamma$-line at 2614\,keV vs. distance of the source to
the detector and thickness of the lead absorber. Right: Ratios
of 583\,keV line to 2614\,keV line vs. distance of the source
to detector and thickness of the lead absorber.}
\label{th583-3197-2614}
\end{figure}
From the measured ratios of the 609 and 2204\,keV peaks of the
\Bi\ decay and the 583 and 2614\,keV peaks of the \Tl\ decay,
the absorber thickness between the source and the cryostat was
estimated. Using these additional absorber thicknesses and the
relative intensities of the summation peaks
(Tab.~\ref{tab:HdMsummation}), the source locations were
found, as was shown in section \ref{sec:Identif}.
\begin{table}
  \centering
  \begin{tabular}{|c|c|c|c|c|c|c|}
\hline
    Isotope & E, & ANG1 & ANG2 & ANG3 & ANG4 & ANG5 \\
    & [keV] & [counts]&[counts] &[counts]&[counts]&[counts]\\
\hline
    \Th\ & 583 & 220$\pm$18 & 260$\pm$23 & 549$\pm$28 & 303$\pm$24 & 350$\pm$27 \\
     & 2614 & 86$\pm$9 & 154$\pm$13 & 308$\pm$18 & 154$\pm$13 & 242$\pm$16 \\
     & 3197 & 3$\pm$2 & 4$\pm$2 & 32$\pm$5 & 28$\pm$5 & 10$\pm$3 \\
\hline
    \Bi\ & 609 & 809$\pm$30 & 1450$\pm$41 & 1561$\pm$43 & 1290$\pm$40 & 1643$\pm$45 \\
     & 2204 & 39$\pm$6 & 118$\pm$11 & 105$\pm$11 & 81$\pm$9 & 148$\pm$13 \\
     & 2016 & 1$\pm$3 & 9$\pm$3 & 4$\pm$3 & 2$\pm$3 & 9$\pm$4 \\
\hline
    \Co\ & 1332 & 233$\pm$16 & 713$\pm$28 & 454$\pm$23 & 398$\pm$22 & 653$\pm$27 \\
       & 2505 & 5$\pm$2 & 42$\pm$7 & 17$\pm$5 & 14$\pm$4 & 35$\pm$6 \\
    \hline
  \end{tabular}
  \caption{The HdM background peak intensities used for source localization.
   The exposure is 56.6\,kg\,y, during the period
of 1995 - 2003.}
  \label{tab:HdMsummation}
\end{table}

\clearpage

\section{HdM background model}

Eight spectral shapes were included in the model matrix $M$:
\begin{itemize}
  \item \Ra\ spectral shape, measured at 5\,cm from the
  detector;
  \item \Th\ spectral shape, measured at 0\,cm from the
  detector;
  \item \Co\ spectral shape, simulated in the holders of the
  detectors;
  \item $^{210}$\rm{Pb} spectral shape, simulated in the HdM
  lead  shield \cite{Diss-Dipl-Doer};
  \item \K\ spectral shape, simulated in the HdM setup;
  \item \Cs\ spectral shape, simulated in the ANG5 cryostat;
  \item muons' and neutrons' contribution to the continuum, calculated
  from the fit of the simulation \cite{Diss-Dipl-Doer} and
  \cite{Mei-Hime};
  \item calculated shape of the \nnbb.
\end{itemize}
The contribution $C_j$ of each of the $N$ spectral shapes to
the measured spectrum $S_i$ is an approximate solution of a
system of $n$ linear equations:
\begin{equation}\label{LinEquation}
    \sum_{j=1}^N{M_{ij}\cdot C_j} \simeq S^{meas}_i,{}(i=1..n),
\end{equation}
where $n$ is the number of channels. It was solved using a
least squares method in the energy interval from 250\,keV to
2800\,keV. The sum of all components $S^{model}$ is expressed
as:
\begin{equation}\label{ModelSum}
    S^{model}_i = \sum_{j=1}^N{M_{ij}\cdot C_j},
\end{equation}
and is shown in Figure \ref{fig:HdModel} together with each
contribution.
\begin{figure}
  \centering
  \includegraphics[width=8cm,angle=270]{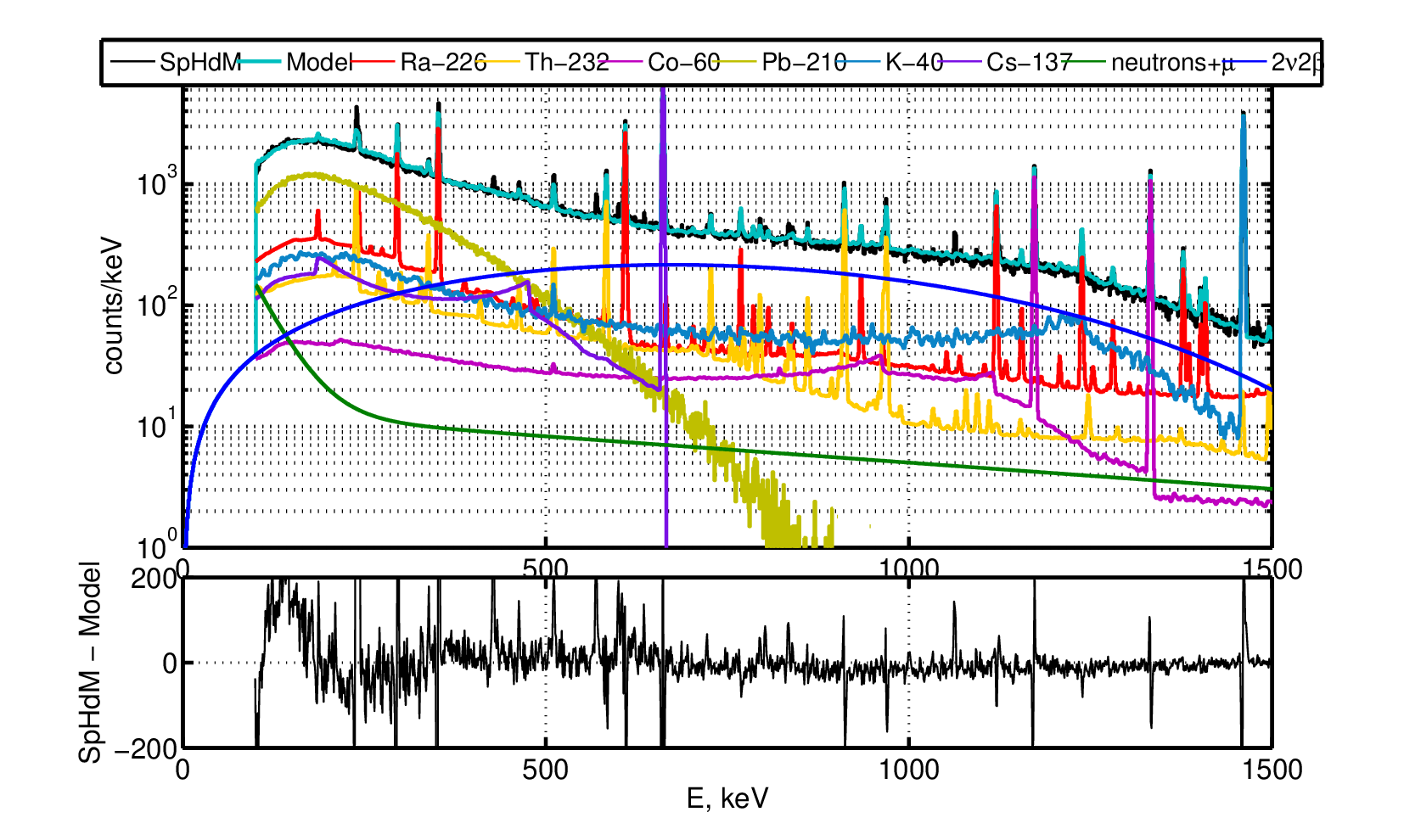}
  \includegraphics[width=8.5cm, angle=270]{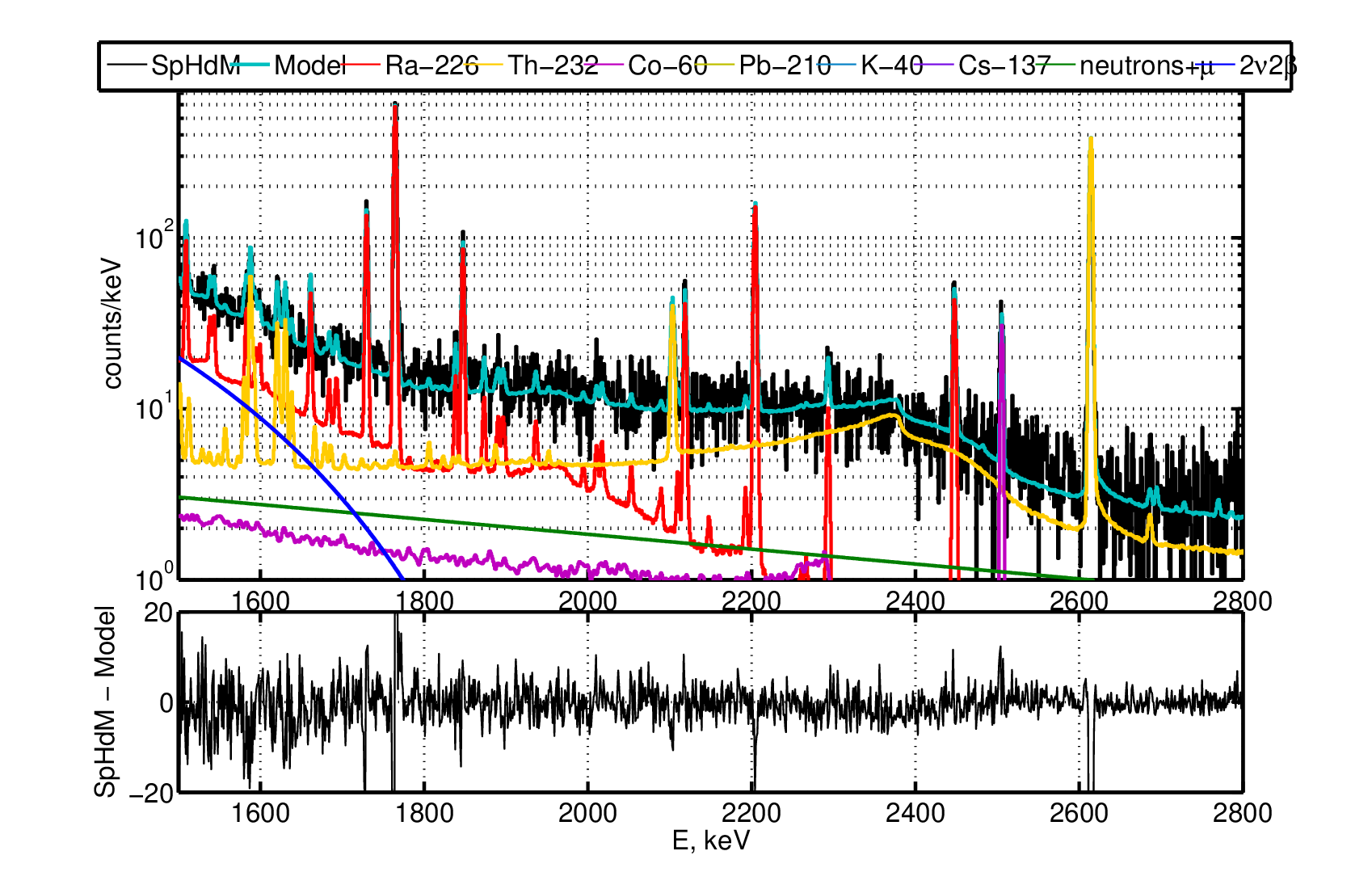}\\
  \caption{The HdM spectrum, its model and its components.}\label{fig:HdModel}
\end{figure}
The total counts from each contribution in the background
model are summarized in Table \ref{tab:HdM-contrib}.
\begin{table}
  \centering
  \begin{tabular}{|c|c|c|c|}
    \hline
    Source &  Counts \\
    \hline
    \Ra &  164060 $\pm$2378 \\
    \Th &  90390 $\pm$12959 \\
    \Co &  43960 $\pm$ 949 \\
    $^{210}$Pb & 268910$\pm$2832 \\
    \K  &  132930 $\pm$1578 \\
    \Cs & 83820$\pm$676 \\
    Neutrons+muons & 16660$\pm$2123\\
    \nnbb & 193250 $\pm$ 2822 \\
    \hline
    $\Sigma$ Model & 993980 \\
    $\Sigma$ HdM & 982580 \\
    \hline
  \end{tabular}
  \caption{Major contributions to the HdM spectrum calculated with
   a linear least squares method (Eq.~\ref{LinEquation}).}\label{tab:HdM-contrib}
\end{table}
The intensity of the major peaks in the HdM and model spectra
are compared in Table \ref{tab:HdModelPeaks}. The calculated
\Bi, \Co, \K\ and \Cs\ peak intensities are in agreement with
the HdM spectrum. The intensity of the 2614\,keV \Tl\ line is
higher in the model spectrum. According to Figure
\ref{th583-3197-2614}, the ratio of the measured intensities
of the \Tl\ lines (Tab. \ref{tab:HdMsummation}), suggests that
the location of the \Th\ contamination is closer to the
diodes, probably on the holder.
\begin{table}
  \centering
  \begin{tabular}{|c|c|c|c|}
    \hline
    Isotope & E\,[keV] & HdM spectrum & Model spectrum \\
    \hline
    \Cs\ & 662 & 20201$\pm$164 & 21220$\pm$150 \\
    \K\ & 1461 & 13010$\pm$134 & 12900$\pm$120 \\
    \Co\ &1332 & 3690$\pm$90 & 3782$\pm$66 \\
    $\Sigma$\Co\ & 2505 & 156$\pm$14 & 111$\pm$12 \\
    \Bi  & 1764 & 2204$\pm$51 & 2220$\pm$48 \\
    \Bi\ & 2204 & 579$\pm$33 & 600$\pm$25 \\
    \Bi\ & 2447 & 193$\pm$15 & 187$\pm$15 \\
    \Tl\ & 583 & 2566$\pm$58 & 2438$\pm$68 \\
    \Tl\ & 2614& 1184$\pm$36 & 1590$\pm$40 \\
    \hline
  \end{tabular}
  \caption{Comparison of the major peaks in the measured HdM spectrum
   and in the model spectrum (Eq.~\ref{ModelSum}).}\label{tab:HdModelPeaks}
\end{table}

\clearpage

The histograms of the normalized residuals
$\frac{S^{meas}-S^{model}}{\sqrt{S^{model}}}$, between the HdM
spectrum and the model spectrum, were fitted with Gaussian for
three energy ranges (Fig.~\ref{fig:NormResidDist}). The
distributions' standard deviations are equal to one (within
error) and their mean values are slightly below zero, in
agreement with the $\sim$1\% excess of total counts in the
model spectrum over HdM, as seen in Table
\ref{tab:HdM-contrib}. The deviation from the Gaussian at
tails is related to the slightly different peak widths in the
model and in the HdM spectrum.
\begin{figure}[ht]
\centering
\includegraphics[width=8cm]{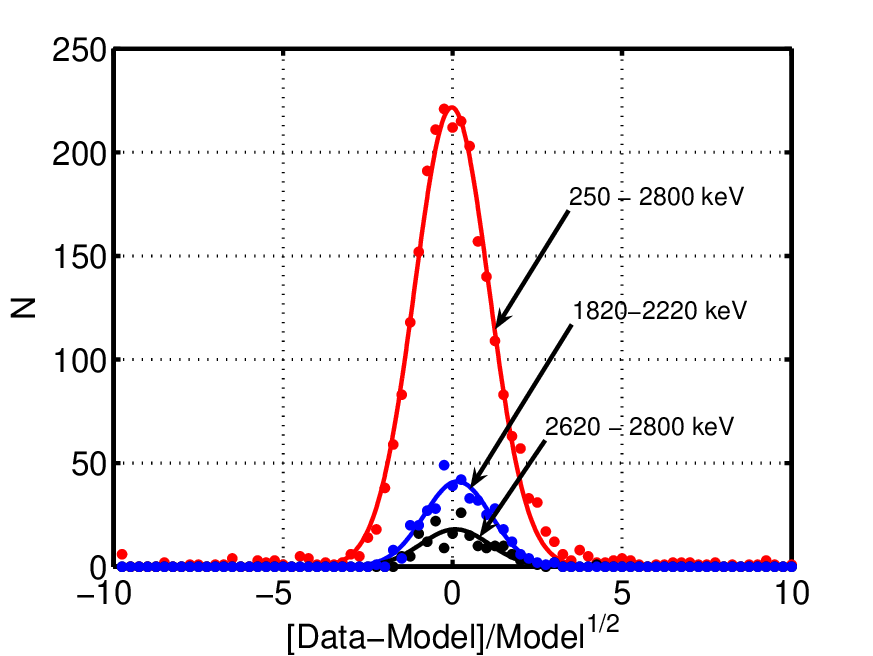}\\
\caption[]{Distributions of the normalized residuals between
the HdM spectrum and the model spectrum, fitted with Gaussian
for three energy ranges.} \label{fig:NormResidDist}
\end{figure}

The least squares fit, obtained for the full spectrum in the
range 250 - 2800\,keV, was examined at smaller energy
intervals around the \qbb\ value to check probable local
deviations.
The region of interest around \qbb\ of the HdM spectrum was
tested using several energy intervals $\Delta$E. The $\chi^2$
value was calculated according to Eq.~\ref{Chi2back}:
\begin{equation}\label{Chi2back}
    \chi^2(\alpha \cdot S^{model}_Q) =
    \sum_{i\in\Delta E}\frac{(S^{meas}_i-\alpha \cdot S^{model}_i)^2}{S^{meas}_i},
\end{equation}
where $\alpha$ is a scale parameter. All the five tested
energy intervals, including 2000 - 2060\,keV, give the same
value of the mean background at the \qbb\ value:
11.9$\pm$0.5\,counts/keV (68\% c.l.) as shown in Figure
\ref{fig:ChiFitEnergyWindow}.
\begin{figure}[ht]
\centering
\includegraphics[width=8cm]{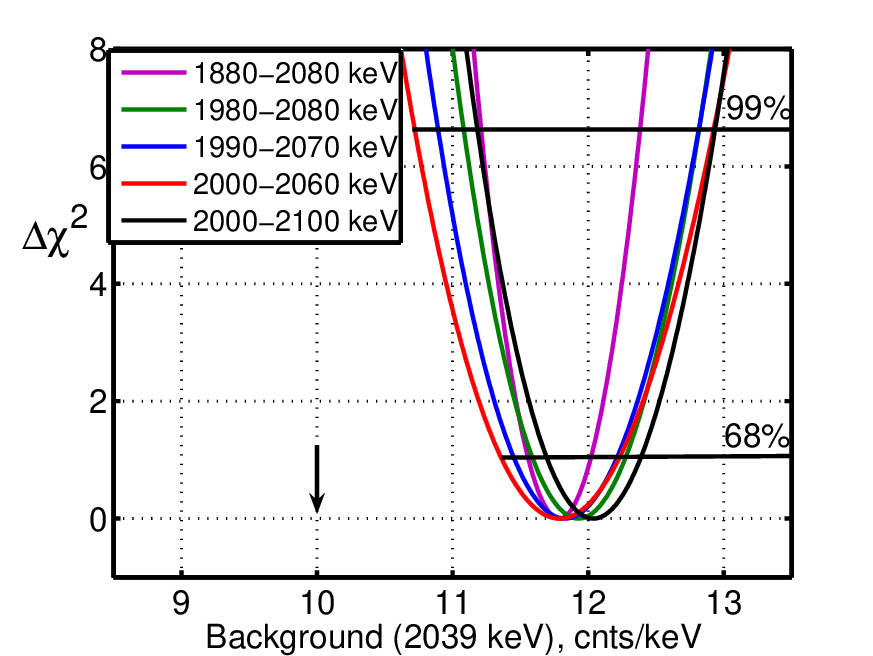}\\
\caption[]{$\chi^2$ as a function of the scaled model
background for the five energy windows (Eq.~\ref{Chi2back}).
The arrow shows the background value obtained with an
unconstrained fit (Fig.~\ref{fig:Sum90-03}) in the
2000-2060\,keV energy interval.}
\label{fig:ChiFitEnergyWindow}
\end{figure}
\\
The components of the model background in the region near
2039\,keV are shown in Table \ref{tab:ComposQ}.
\begin{table}
  \centering
  \begin{tabular}{|c|c|c|c|c|}
    \hline
    $\Sigma$ Bkg. & \Bi\ & \Tl\ & \Co\ & Neutrons+muons \\
      {[cts/keV]}  & [cts/keV] & [cts/keV] & [cts/keV] & [cts/keV] \\
      \hline
    11.9 & 3.0 & 5.2 & 1.2 & 2.5 \\
    \hline
    100\% & 25\% & 44\% & 10\% & 21\% \\
    \hline
  \end{tabular}
  \caption{Composition of the background model at 2039\,keV.}\label{tab:ComposQ}
\end{table}
The neutron and muon contributions to the background are 21\%
of the total background at the \qbb\ value, 57\% of this
contribution was from the ANG4 detector,
which operated without a muon veto.\\

\section{HdM fit results in the region of \qbb}

Given the accurate calibration of energy and energy
resolution, the fit of the HdM spectrum was performed using
the presented background model. An estimation of lower and
upper limits on the signal in each bin was performed using the
ADMA program \cite{ADMA}, according to Feldman-Cousins
\cite{Feldm98}. Table \ref{tab:SpFC} presents the HdM spectrum
in the energy interval 2000-2060\,keV (71.7\,kg\,y). The
fourth and fifth columns show the 68.27\% C.L. intervals for a
Poisson signal, assuming the mean background of the presented
model (11.8$\pm$0.5\,Cts/keV at \qbb).
\begin{table}
\renewcommand{\arraystretch}{1.2}
\setlength\tabcolsep{2.2pt}
\begin{minipage}[t]{0.48\linewidth}
\tiny
  \centering
\begin{tabular}{|c|c|c|c|c|}
  \hline
   E & HdM spectrum & Model spectrum  &LL &  UL \\
      {[keV]}    &  [cts/keV] & [cts/keV] &   &     \\
   \hline
    1980 &     12.6 &     13.0 &      0.0 &      4.3 \\
    1981 &     12.0 &     12.8 &      0.0 &      3.5 \\
    1982 &     17.2 &     12.7 &      1.1 &      9.1 \\
    1983 &     12.9 &     12.8 &      0.0 &      4.5 \\
    1984 &     17.5 &     12.8 &      1.6 &     10.0 \\
    1985 &      8.0 &     12.7 &      0.0 &      6.4 \\
    1986 &     13.5 &     12.7 &      0.0 &      5.6 \\
    1987 &      6.3 &     12.7 &      0.0 &      0.3 \\
    1988 &     11.2 &     12.8 &      0.0 &      2.3 \\
    1989 &     14.0 &     12.8 &      0.0 &      5.5 \\
    1990 &      7.2 &     12.7 &      0.0 &      0.5 \\
    1991 &     10.0 &     12.9 &      0.0 &      1.5 \\
    1992 &     13.7 &     13.3 &      0.0 &      5.0 \\
    1993 &     15.5 &     13.8 &      0.1 &      7.0 \\
    1994 &     10.1 &     13.9 &      0.0 &      1.0 \\
    1995 &     14.0 &     13.6 &      0.0 &      4.7 \\
    1996 &     11.9 &     13.2 &      0.0 &      3.2 \\
    1997 &     11.0 &     12.9 &      0.0 &      2.1 \\
    1998 &     14.7 &     12.6 &      0.3 &      6.7 \\
    1999 &      6.6 &     12.3 &      0.0 &      0.5 \\
    2000 &      7.4 &     12.5 &      0.0 &      0.9 \\
    2001 &      9.9 &     12.4 &      0.0 &      2.2 \\
    2002 &      6.2 &     12.2 &      0.0 &      4.2 \\
    2003 &     12.8 &     12.2 &      0.0 &      5.4 \\
    2004 &      8.1 &     12.2 &      0.0 &      9.9 \\
    2005 &      8.4 &     12.3 &      0.0 &      9.9 \\
    2006 &     11.4 &     12.5 &      0.0 &      3.0 \\
    2007 &     13.2 &     12.7 &      0.0 &      5.4 \\
    2008 &     17.0 &     13.5 &      1.5 &      9.9 \\
    2009 &     18.6 &     14.5 &      3.0 &     11.9 \\
    2010 &     23.8 &     15.1 &      7.4 &     17.4 \\
    2011 &     12.4 &     14.8 &      0.0 &      4.4 \\
    2012 &     13.9 &     13.9 &      0.1 &      6.4 \\
    2013 &     13.2 &     13.3 &      0.0 &      5.4 \\
    2014 &     14.6 &     13.1 &      0.6 &      7.4 \\
    2015 &     20.4 &     14.4 &      3.9 &     18.4 \\
    2016 &     17.8 &     15.5 &      2.2 &     10.9 \\
    2017 &     15.6 &     15.7 &      1.0 &      8.9 \\
    2018 &     14.4 &     14.6 &      0.1 &      6.4 \\
    2019 &     18.0 &     13.5 &      2.2 &     10.9 \\
    2020 &     16.7 &     13.0 &      1.5 &      9.9 \\
    2021 &     12.9 &     12.9 &      0.0 &      5.4 \\
    2022 &      9.9 &     12.6 &      0.0 &      2.2 \\
    2023 &     13.1 &     12.3 &      0.0 &      5.4 \\
    2024 &     15.1 &     12.2 &      0.0 &      5.4 \\
    2025 &     12.3 &     12.0 &      0.0 &      4.4 \\
    2026 &     11.4 &     12.0 &      0.0 &      3.0 \\
    2027 &     10.9 &     12.0 &      0.0 &      3.0 \\
    2028 &      9.5 &     12.1 &      0.0 &      2.2 \\
    2029 &     15.2 &     12.0 &      0.6 &      7.4 \\
  \hline
\end{tabular}
\end{minipage}
\begin{minipage}[t]{0.48\linewidth}
\tiny
  \centering
\begin{tabular}{|c|c|c|c|c|}
  \hline
   E & HdM spectrum & Model spectrum  &LL &  UL \\
      {[keV]}    &  [cts/keV] & [cts/keV] &   &     \\
   \hline
    2030 &     16.7 &     12.1 &      1.5 &      9.9 \\
    2031 &      5.7 &     11.8 &      0.0 &      0.4 \\
    2032 &     12.7 &     11.9 &      0.0 &      5.4 \\
    2033 &      9.9 &     11.8 &      0.0 &      2.2 \\
    2034 &      8.3 &     11.8 &      0.0 &      9.9 \\
    2035 &      9.8 &     11.8 &      0.0 &      2.2 \\
    2036 &     13.4 &     11.9 &      0.0 &      5.4 \\
    2037 &     18.1 &     11.9 &      2.2 &     10.9 \\
    2038 &     16.6 &     11.9 &      1.5 &      9.9 \\
    2039 &     14.4 &     11.8 &      0.1 &      6.4 \\
    2040 &     11.4 &     11.7 &      0.0 &      3.0 \\
    2041 &      6.7 &     11.8 &      0.0 &      0.6 \\
    2042 &     12.6 &     11.8 &      0.0 &      5.4 \\
    2043 &     12.4 &     11.8 &      0.0 &      4.4 \\
    2044 &      7.4 &     11.9 &      0.0 &      0.6 \\
    2045 &     10.3 &     11.9 &      0.0 &      2.2 \\
    2046 &      7.5 &     11.7 &      0.0 &      0.6 \\
    2047 &      3.9 &     11.7 &      0.0 &      0.2 \\
    2048 &     10.8 &     11.7 &      0.0 &      3.0 \\
    2049 &     12.2 &     11.8 &      0.0 &      4.4 \\
    2050 &      5.7 &     12.3 &      0.0 &      0.4 \\
    2051 &     14.3 &     13.0 &      0.1 &      6.4 \\
    2052 &     18.1 &     13.6 &      2.2 &     10.9 \\
    2053 &     19.9 &     13.6 &      3.9 &     13.4 \\
    2054 &     14.7 &     12.8 &      0.6 &      7.4 \\
    2055 &      7.5 &     12.2 &      0.0 &      1.0 \\
    2056 &     14.9 &     11.8 &      0.6 &      7.4 \\
    2057 &     15.5 &     11.7 &      1.0 &      8.9 \\
    2058 &      7.3 &     11.5 &      0.0 &      0.6 \\
    2059 &      7.3 &     11.4 &      0.0 &      0.6 \\
    2060 &      8.4 &     11.4 &      0.0 &      1.0 \\
    2061 &      8.7 &     11.4 &      0.0 &      1.8 \\
    2062 &     11.3 &     11.4 &      0.0 &      3.4 \\
    2063 &     17.9 &     11.3 &      2.6 &     11.5 \\
    2064 &      9.8 &     11.3 &      0.0 &      2.6 \\
    2065 &     20.0 &     11.6 &      4.2 &     13.7 \\
    2066 &      8.7 &     11.3 &      0.0 &      1.9 \\
    2067 &     15.9 &     11.3 &      1.3 &      9.5 \\
    2068 &     11.8 &     11.3 &      0.0 &      5.0 \\
    2069 &     13.1 &     11.2 &      0.0 &      6.1 \\
    2070 &     15.5 &     11.2 &      0.9 &      8.1 \\
    2071 &     10.1 &     11.2 &      0.0 &      2.7 \\
    2072 &      9.0 &     11.3 &      0.0 &      1.9 \\
    2073 &     15.5 &     11.2 &      1.3 &      9.6 \\
    2074 &     21.0 &     11.3 &      5.0 &      15.0 \\
    2075 &      9.5 &     11.3 &      0.0 &      2.6 \\
    2076 &     14.9 &     11.1 &      0.9 &      8.2 \\
    2077 &      9.6 &     11.2 &      0.0 &      2.7 \\
    2078 &     11.0 &     11.2 &      0.0 &      3.6 \\
    2079 &     15.9 &     11.0 &      1.5 &      9.8 \\
  \hline
\end{tabular}
\end{minipage}
 \caption{The HdM spectrum (71.7\,kg\,y) in
the energy interval 1980-2080\,keV. The last two columns show
the 68.27\% C.L. interval for a Poisson signal, with an
assumed model background (11.8$\pm$0.5\,cts/keV at \qbb),
which is shown in the third column. Confidence intervals are
calculated according to Feldman-Cousins \cite{Feldm98, ADMA}.
The starting energy of each 1-keV bin is
given.}\label{tab:SpFC}
\end{table}
\clearpage

The total number of counts in the range 2000--2060\,keV of the
HdM and the model spectra are 759 and 766 counts respectively.
The intensity of the peaks, found in the 2000 -- 2060\,keV
region are presented in Table \ref{tab:NewHdM-results1}.
\begin{table}[h]
\centering
\begin{tabular}{|c|c|c|c|}
  \hline
  E\,[keV] & $S_{FC}$\,[counts] & $S_{LSQ}$\,[counts] & $S_{model}$\,[counts]\\
  \hline
  2010 & 26$\pm$14 & 29$\pm$9 & 10.9$\pm$0.3\\
  2016 & 22$\pm$15 & 28$\pm$9 & 13.4$\pm$0.3\\
  2022 & 12$\pm$9 & 9$\pm$9 & 2.6$\pm$0.3\\
  2030 & 9$\pm$8  &  8$\pm$9 & 0\\
  2038 & 15$\pm$12 & 16$\pm$9 & 0\\
  2053 & 22$\pm$16 & 23$\pm$9 & 8.0$\pm$0.3\\
  \hline
\end{tabular}
  \captionof{table}{The peak intensities in the region of
  interest around \qbb\ for the weak \Bi\ lines and the line at 2039\,keV.
  The second column is the estimation by the Feldman-Cousins method and
   the third column is the least squares fit in the 1980--2080\,keV energy range.}
  \label{tab:NewHdM-results1}
\end{table}
\begin{figure}[h!]
\centering
\includegraphics[width=8cm]{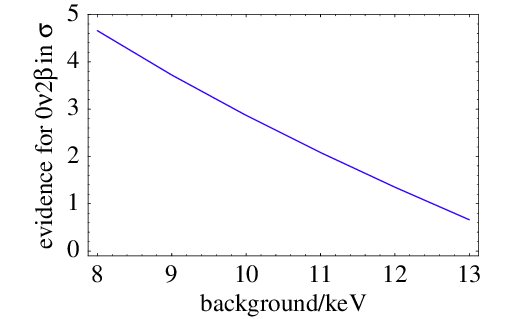}
  \caption{The confidence level of the \onbb\ peak as a function
   of the background at \qbb\ \cite{VissaniKlapdor}.}
  \label{fig:Vissani}
\end{figure}
The calculated intensities of the weak \Bi\ lines at 2022 and
2053\,keV (Tab.~\ref{tab:SpecAtQ}) are consistent with the
measured values at 68\% C.L.. The significance of the \onbb\
peak is 1.3$\sigma$. This value is in agreement with the
independent estimation of the expected statistical
significance of the \onbb\ signal at the assumed HdM
background, made by Strumia and Vissani \cite{VissaniKlapdor}
shown in Figure \ref{fig:Vissani}.

\clearpage

\section{Limits on the half-life of \onbb\ decay of $^{76}$Ge and the effective neutrino mass}
\label{sec:hd-halflife}

As shown in the previous section, the $\sim1\sigma$ signal
found around the \qbb\ value can be classified as \onbb\ with
quite low significance. The half-life of the \Ge\ \onbb\ decay
can be determined using Eq.~\ref{halflife} and the 2039\,keV
peak intensity S$_{FC}$ from Table \ref{tab:NewHdM-results2}.
\begin{table}[h]
\centering
\begin{tabular}{|c|c|c|}
  \hline
  Method & E$_{peak}$\,[keV] & Peak area\,[counts]\\
  \hline
    S$_{Feldman-Cousins}$ & 2039.0 & 15$\pm$12 \\
    S$_{Fit}$ & 2038.1$\pm$1.3 &13.0$\pm$8.5\\
  \hline
\end{tabular}
  \captionof{table}{The 2039\,keV peak intensity determined with Feldman-Cousins and LSQ
  methods using background 11.9$\pm$0.5\,counts/keV.}
  \label{tab:NewHdM-results2}
\end{table}
The half-life of the neutrinoless double beta decay and the
effective neutrino mass are derived using Eqs.\ref{halflife}
and \ref{Ge76NuMass} and presented in Table \ref{Results1}.
The sensitivity of the HdM experiment for 71.7\,kg\,y,
calculated with Eq.~\ref{DBD_sens} is 4.6$\cdot10^{25}$\,y
(68\%C.L.).
\begin{table}[h]
\centering
\begin{tabular}{|c|c|c|c|}
\hline
\multicolumn{4}{|c|}{Period 1990 $\div$ 2003}\\
\hline
Exposure    & Background,    & ${\rm T}_{1/2}^{0\nu}~~[y]$ & $\langle m_{\beta\beta} \rangle $ [eV] \\
{[kg\,y]} & [cts/(keV\,kg\,y)] &    (68\% C.L. interval) & (68\% C.L. interval)\\
 \hline
 &&&\\
    71.7    &  0.17 &   $2.2\cdot 10^{25}$ &  0.32  \\
 &&$(0.4 - 4.0)\cdot 10^{25}$ & (0.19 - 0.45) \\
\hline
\end{tabular}
\caption[]{
    Half-life of the \Ge\ neutrinoless double beta decay
    and the effective neutrino mass, calculated with
    the nuclear matrix element (M=4.2) of \cite{Sta90}.}
    \label{Results1}
\end{table}

\section{Conclusions}

The background model of the HdM spectrum was developed using
measurements with sources and MC simulations, which accounts
for contributions from \Ra, \Th, \Co\ and also neutrons with
muons. The model deviation from the HdM spectrum is 1\% within
the 250-2800\,keV energy interval. The background continuum in
the region of interest around the \qbb\ value was determined
using the model. The obtained value of the background is
(11.8$\pm$0.5)\,counts/keV, which is higher than the
background used in publication \cite{KK-NewAn-NIM04}:
(10.0$\pm$0.3)\,counts/keV. The model is still not accounting
for all data. However, the fit of the HdM spectrum with the
model background gives a better agreement of the \Bi\ line
intensities. The intensity of the $\sim$2039\,keV peak,
(15$\pm$12)\,counts, is less significant than the published
value \cite{KK-NewAn-NIM04} obtained before the model was
developed.\\
The corresponding half-life of the \onbb\ decay is $2.2\cdot
10^{25}$\,y and the 68\% C.L. interval is $(0.4 - 4.0)\cdot
10^{25}$\,y. The sensitivity of the HdM experiment for \onbb\
decay is $4.6\cdot 10^{25}$ y (68\%, C.L.). The effective
neutrino mass is 0.32\,eV within the interval (0.19-0.45)\,eV
at 68\% C.L.,
calculated using the nuclear matrix element of \cite{Sta90}.\\

\newpage
\chapter{The \genius-TF Setup -- Installation of Four HPGe
Detectors and Background Measurements}\label{ch:genius}

A novel technique to use germanium detectors without a
conventional cryostat was proposed by G.\,Heuser \cite{heu95}
to achieve extreme background reduction. The idea was to
operate bare germanium diodes in high purity liquid nitrogen,
which serves as cooling medium and clean shielding
simultaneously. \genius-TF, a test-facility for the \genius\
project \cite{HVKK-Prop97}, was based on this idea. The first
spectroscopy measurements with a germanium detector immersed
directly in liquid nitrogen were performed in 1997 at MPI-K.
According to the \genius-TF proposal \cite{NIM02-TF}, 16 bare
HPGe detectors with a total mass of $\sim$40 kg would be
operated in liquid nitrogen. Finally, the \genius-TF setup has
run a total of six HPGe detectors during three years
(2003--2006)
\cite{NIM-TF-03, NIM-TF-04, NIM-TF-06, PhSc-TF-06}.\\
In this chapter the experimental work concerning the
installation of the detector setup and the first year results
of \genius-TF measurements are presented. An analysis of the
\genius-TF background after the completion of the external
shielding and careful investigation of its origin will
also be given.\\

\section{The \genius-TF setup}

The first four \genius-TF detectors were produced in 2002 by
ORTEC (USA) and then shipped to Germany. To minimize
cosmogenic activation they were transported to LNGS to be
stored underground until their installation. The diodes were
equipped by ORTEC with the ultra-low mass high voltage and
signal contacts ($\sim$3 g each) made from stainless steel
wire, gold and teflon (Fig.~\ref{fig:Genius-Foto-4det}(left)).
The depth of the core of the detectors was reduced to provide
lower detector capacity and, consequently, lower noise.
Therefore, a very low threshold, estimated by ORTEC to be
around 0.5-0.7\,keV, could be achieved. Some characteristics
of the 4 detectors, as stated
by the manufacturer, are listed in Table~\ref{tab:TFdet}.\\
\begin{table}[!htp]
\centering
\begin{tabular}{|l|c|c|c|c|}
\hline
Detector & GTF1 & GTF2 & GTF3 & GTF4 \\
 \hline
Serial No. & P41045A&P41044A &P41032A &P41112A\\
Mass, g & 2580 & 2447 & 2367 & 3128 \\
Diameter, mm &87.4&85.3&89.4&85.0 \\
Length, mm &77.2&86.5&71.5& 104.5\\
Core depth, mm & 44.1&45.5 &38.7 &64.6 \\
Operating voltage, V & 3000 & 2600 & 3200 & 2500 \\
Depletion voltage, V & 1000 & 700  & 3100 & 2000 \\
Leakage current, pA  & 30   & 76   & 85   & 244 \\
FWHM @ 1.33 MeV, keV & 2.0 & 3.1 & 1.8 & 2.0 \\
FWHM @ 122 keV, keV & 0.72 & 0.94 & 0.80 & 0.76 \\
\hline
\end{tabular}
\caption[]{Parameters provided by ORTEC for the four
\genius-TF HPGe-detectors \cite{GTF_AMETEC}. The energy
resolution and the leakage current were measured with the
diodes installed in a vacuum cryostat.}\label{tab:TFdet}
\end{table}
\noindent On May 5, 2003 the four bare Ge detectors were
installed under dust reduced atmosphere into the \genius-TF
setup. The detectors were positioned in a holder made from
high-purity teflon (PA5), as shown in
Fig.~\ref{fig:Genius-Foto-4det} (right). The detector holder
is surrounded by a shield made of zone refined
poly-crystalline germanium bricks (40 bricks with total mass
of 212\,kg). The thickness of the germanium shield is 10\,cm
on all sides and 5\,cm on the top and the bottom of the
detectors holder.
\begin{figure}[!htp]
\centering
\includegraphics[width=5.8 cm]{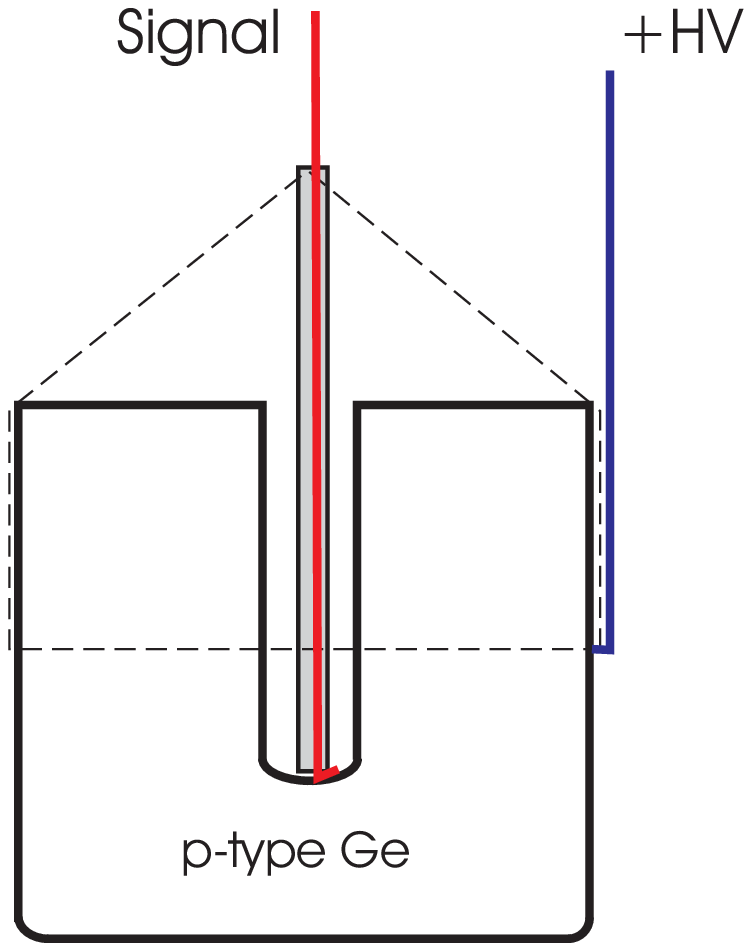}
\includegraphics[width=8 cm]{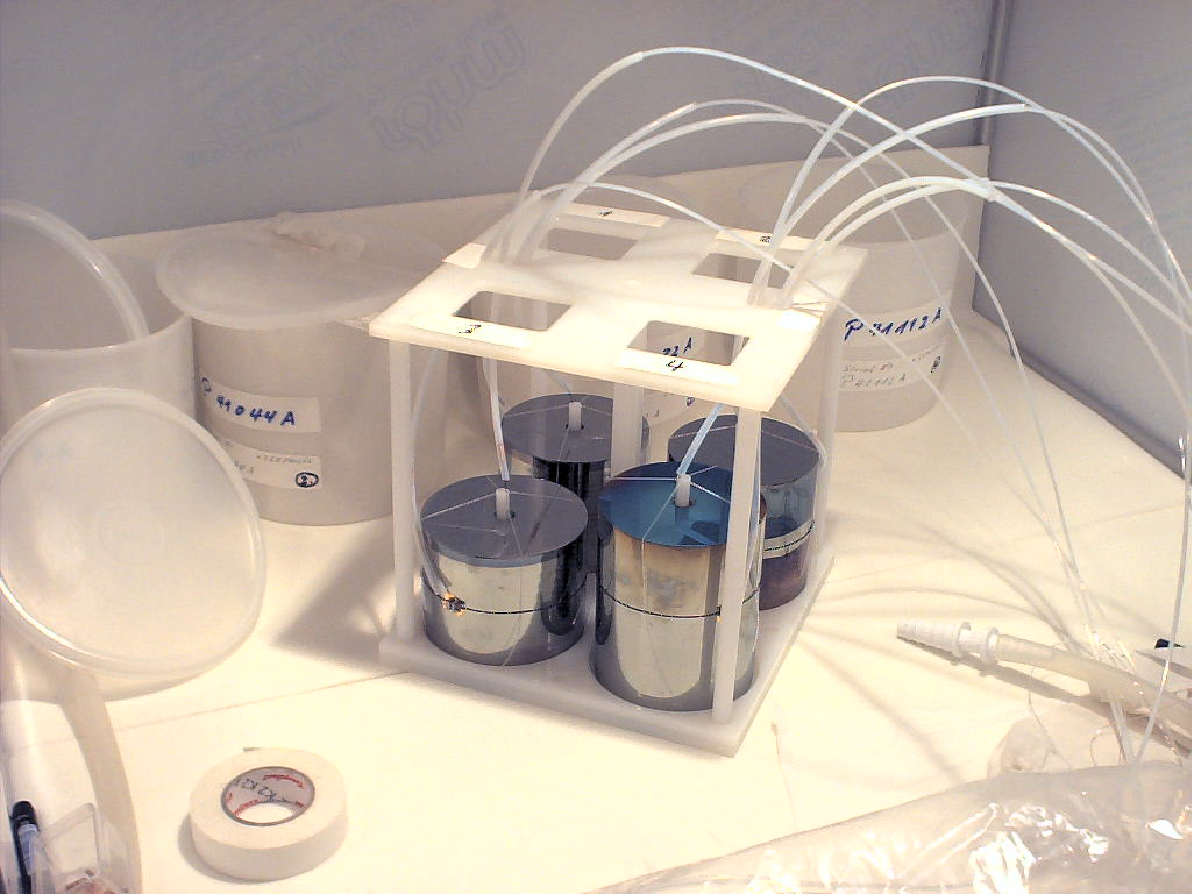}\\
\caption{The HV and signal cables are connected on the side
and central contacts of the diodes, respectively (left). The
four first \genius-TF bare diodes mounted in the PTFE detector
holder (right) \cite{NIM-TF-03}.} \label{fig:Genius-Foto-4det}
\end{figure}
\begin{figure}[!htp]
\centering
\includegraphics[width=7.5 cm]{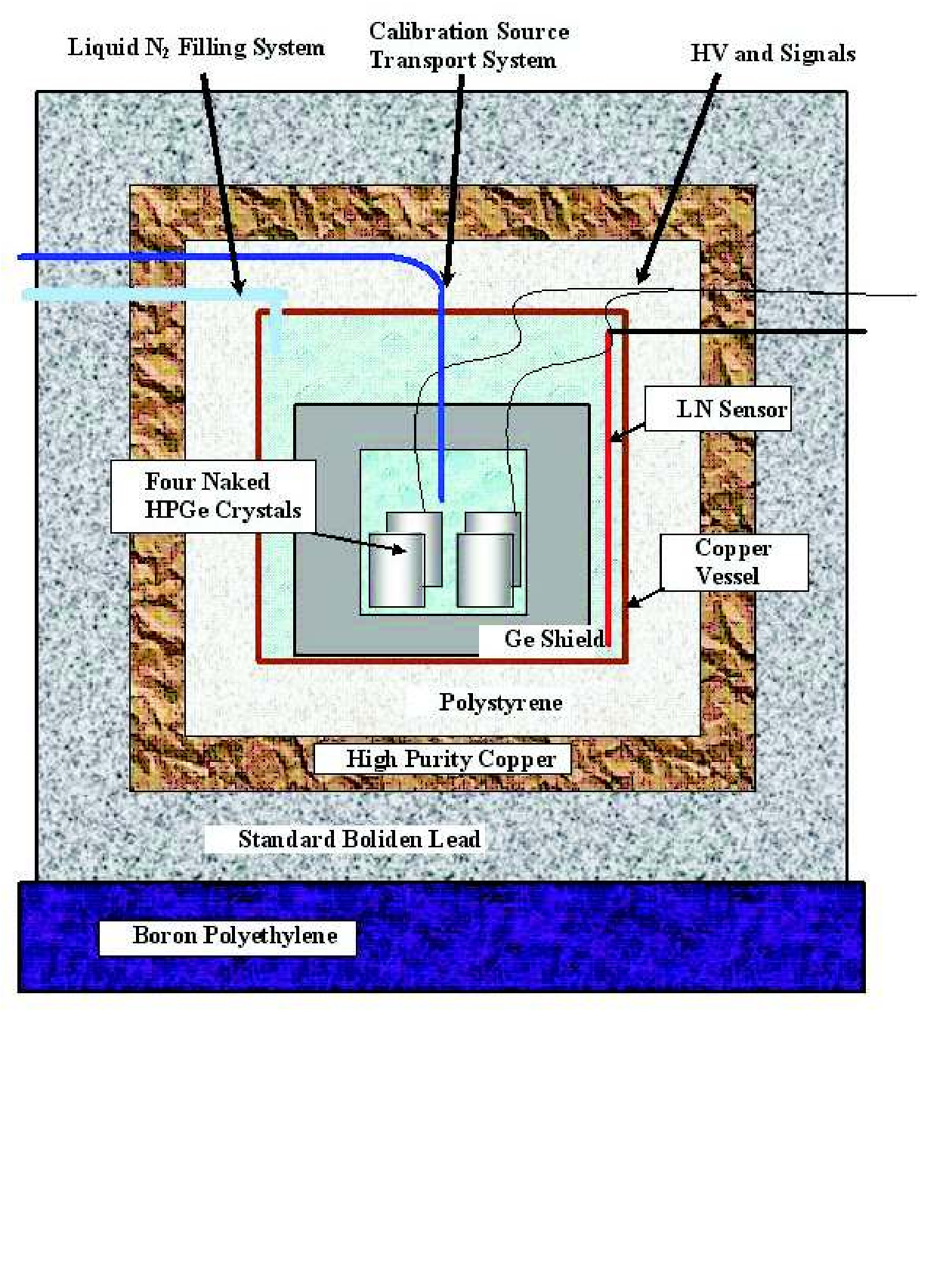}\\
\caption{Schematic drawing of the \genius-TF setup showing the
the external and the internal shield as well as the bare
germanium detectors in liquid nitrogen \cite{NIM-TF-03}.}
\label{fig:Geom-G-TF}
\end{figure}
The germanium shield is kept in a thin wall (1\,mm) box of
high-purity electrolytic copper of the size of $50 \times 50
\times 50$\,cm$^3$, which is filled with $\sim$70\,l of liquid
nitrogen. The copper box is thermally shielded by 20\,cm of
special low-radioactivity styrofoam, followed by a shield made
out of electrolytic copper (10\,cm, 15\,tons) and
low-radioactivity (Boliden) lead (20\,cm, 35\,tons). The
schematic drawing of the setup is shown in
Fig.~\ref{fig:Geom-G-TF}. It was partially shielded against
neutrons with 20\,cm Boron-polyethylene plates on the bottom.
The upper part of the
shield was completed in November 2003.\\
The liquid nitrogen, which is in direct contact with the
detectors must have a very high purity. It was produced by the
BOREXINO nitrogen purification plant - Low Temperature
Adsorber (LTA) \cite{Freudiger1}. Liquid nitrogen of technical
quality (99.99$\%$ purity) is directly purified in the liquid
phase by an adsorber column system filled with about 2 kg of
activated carbon. The pure liquid nitrogen was transported
from the production plant to the \genius-TF building by two
200\,l dewars. The filling of the copper container with
nitrogen was performed using a filling system made of
thermally isolated teflon tubes. The tube leading to the
copper dewar was equipped with a valve which was kept closed
between the fillings. Due to the evaporation of the liquid
nitrogen inside the copper box, \genius-TF had to be refilled
every two days. The liquid nitrogen level in the detector
chamber was measured by a capacitive sensor. It consists of
two 40-cm long isolated coaxial tubes made out of pure copper.
The electrical capacity of the sensor depends on the liquid
nitrogen filling level. For the capacity measurements the
electronics were installed and set according to calibration
tests. However, the operation of the capacitive sensor
introduces significant noise to the detector electronics, so
it was switched off during spectrometry measurements.\\
The infrared (IR) radiation coming from the upper unshielded
part of the setup pene\-trates through the styrofoam and is
partially absorbed by the upper lid of the copper box and the
germanium shield. The IR radiation which reaches the diodes
creates a significant leakage current when a reverse bias is
applied. To reduce the IR radiation in the setup, a copper
sheet was temporarily placed over the styrofoam. Then it was
possible to apply the operational voltage to the detectors.
The copper sheet was removed only after the completion of the
setup shield a few month later.\\
The read out system consisted of a linear amplifier, a Struck
Flash ADC (model SIS3301) with a sampling rate of up to
100\,MHz and 12-bit resolution and a VME computer
\cite{NIM-TF-DAQ}. A standard ORTEC MCA was also used for test
measurements. The detector calibrations were performed with a
$^{133}{\rm{Ba}}$ $\gamma$-source with an activity of
401\,kBq. The source was fixed on a steel wire inside a teflon
tube closed at both ends. The source was introduced into the
center of the detector array using a cylinder magnet.
Figure~\ref{fig:FirstSpect-G-TF} (top) shows the first
spectrum measured with a $^{60}{\rm{Co}}$ source positioned
outside the setup, on top of the styrofoam, and with the
$^{133}{\rm{Ba}}$ source inside the setup.
\begin{figure}[!htp]
\centering
\includegraphics[width=10cm]{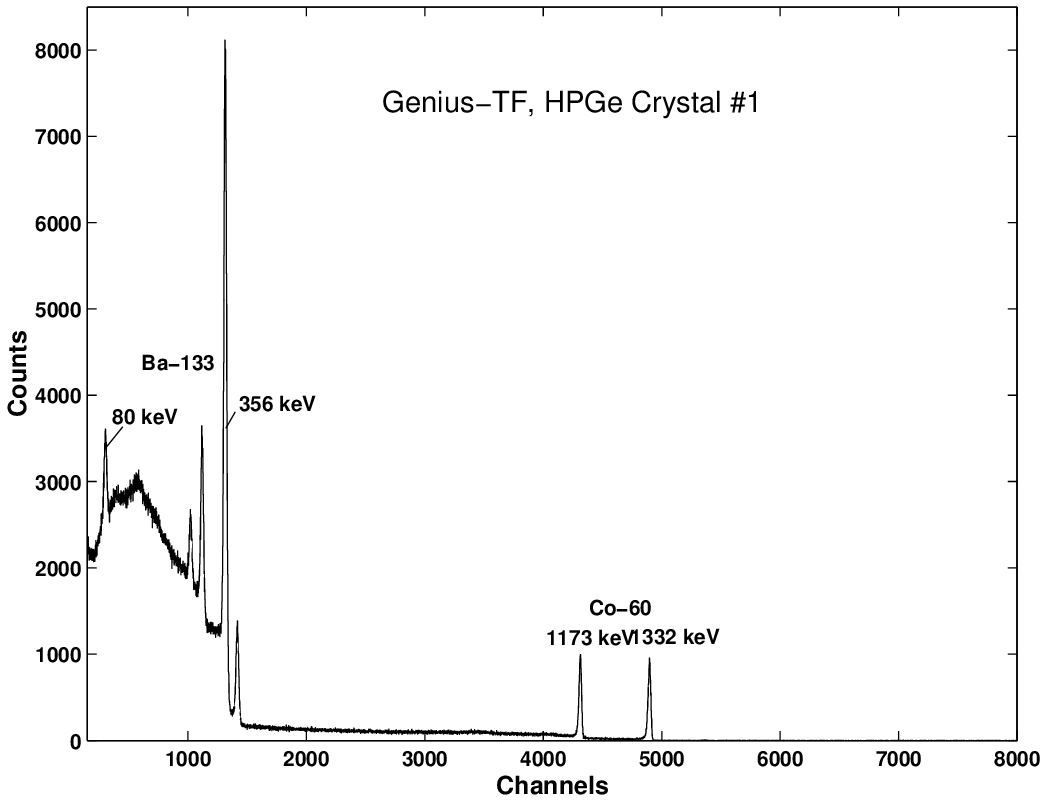}
\includegraphics[width=10cm]{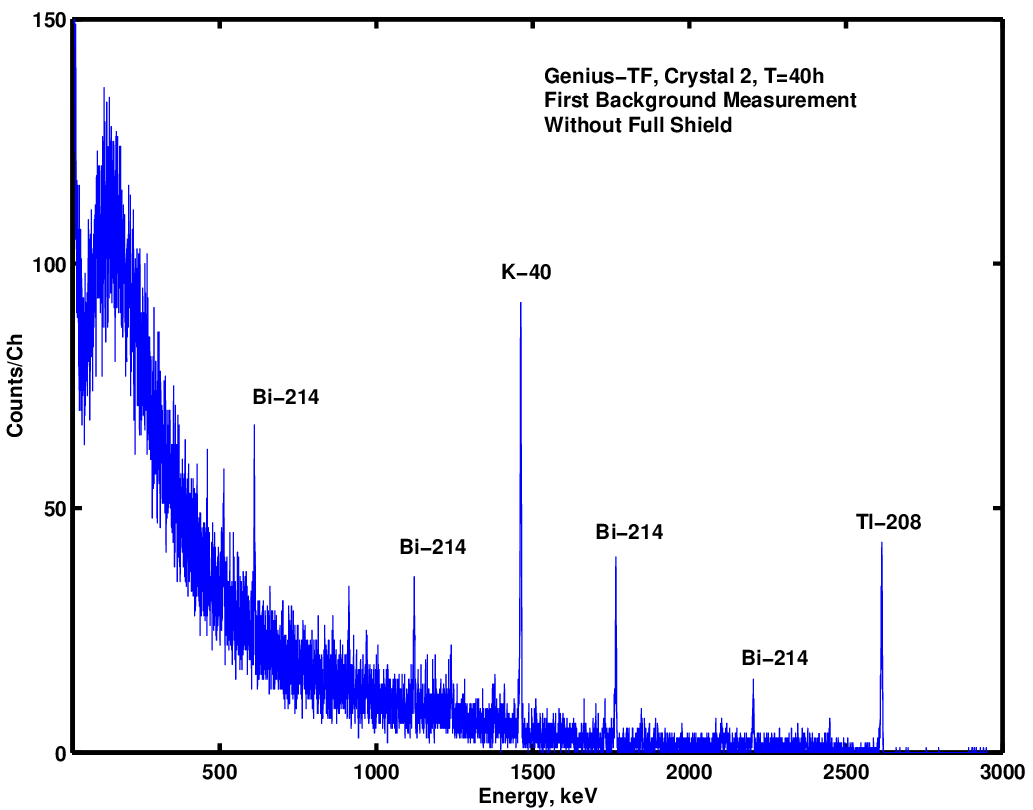}\\
\caption[]{Top: first spectrum measured with detector GTF1.
The $^{60}{\rm{Co}}$ source is outside, and the
$^{133}{\rm{Ba}}$ source is inside the setup. Bottom: first
background spectrum measured with detector GTF2 over 40\,hours
without the full shielding of the setup \cite{NIM-TF-03}.}
\label{fig:FirstSpect-G-TF}
\end{figure}
An increase of the signal noise was noticed when the source
was inserted inside the setup, probably because of
electro-magnetic pick-ups on the long source wire. The
measured energy resolution at 1.33 MeV of \Co\ was 7.4 keV
with the \Ba\ source inserted in the setup, and it improved to
5.1 keV when the \Ba\ source was removed. To reduce the noise,
the wire was cut and the source was attached to a thin teflon
tube. After this modification, the source had to be inserted
into the setup manually, because the teflon tube became rigid
in liquid nitrogen and the magnet force was not strong enough
to pull the source out of the setup. The teflon tube was kept
closed between the calibrations but radon contamination of the
setup could occurred during calibration measurements.

\clearpage
\section{$^{222}$Rn contamination}
In November 2003 the external shielding of \genius-TF was
completed using the lead shield from the decommissioned
Heidelberg-Moscow experiment. The external radiation was
suppressed by $\sim10^5$ times, while the detector count rates
decreased by only a factor of 100. The background was
dominated by the internal radioactive impurities in the setup.
\begin{figure}[!htp]
\begin{center}
\includegraphics[width=6.8cm]{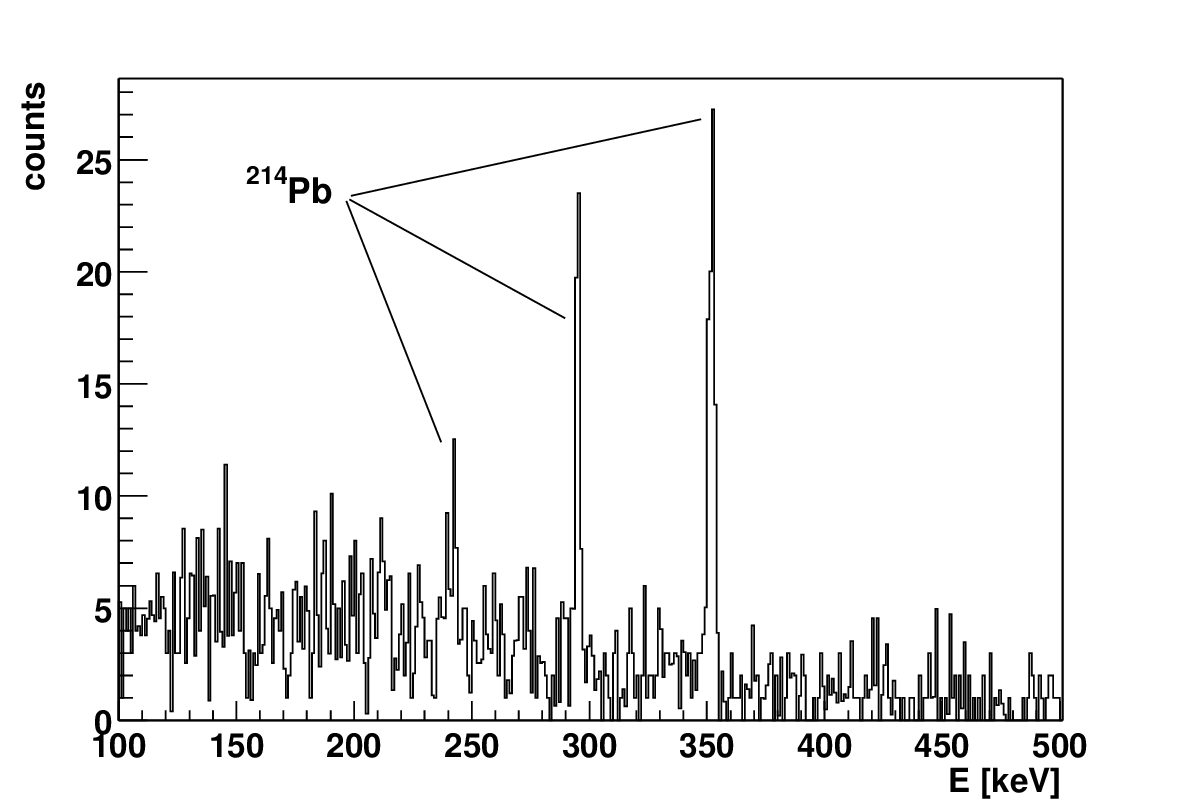}
\includegraphics[width=6.8cm]{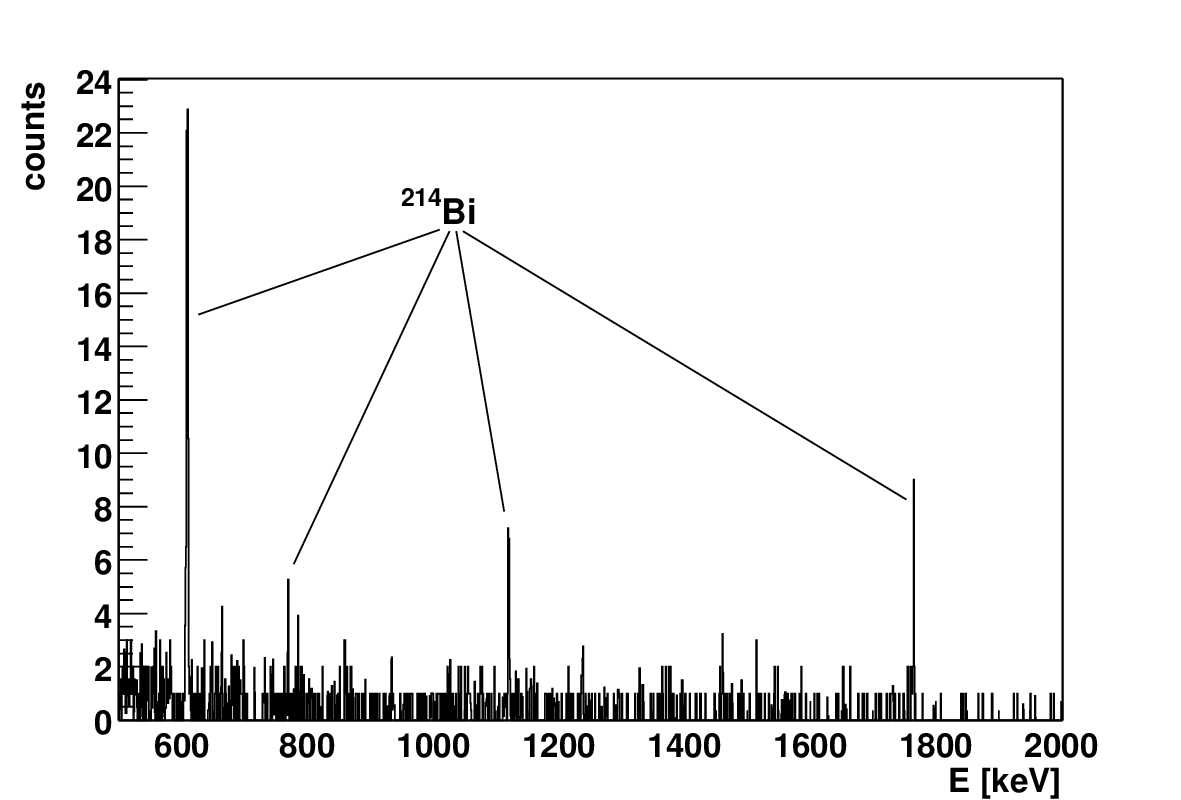}
\includegraphics[width=6.8cm]{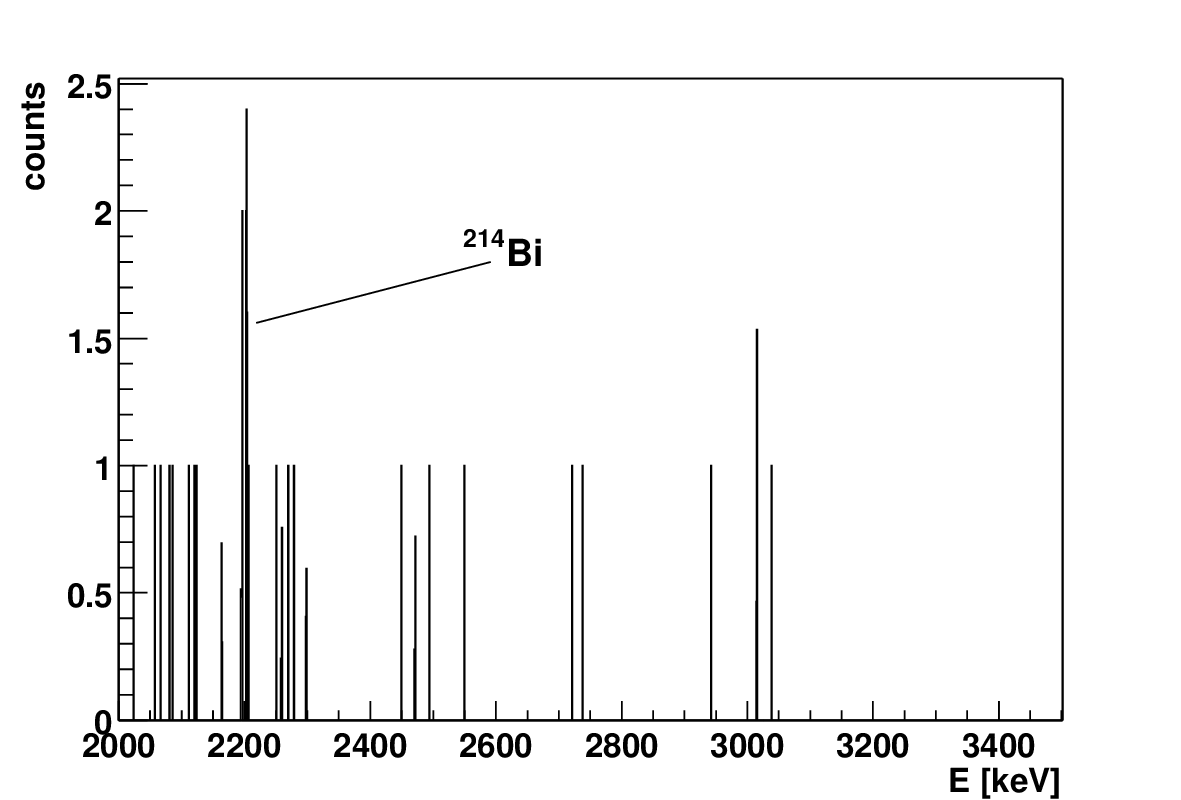}
\end{center}
\caption[]{Background spectrum measured with GTF2 detector a
few days after the completion of the external shielding. Only
peaks from $^{214}\rm{Pb}$ and $^{214}\rm{Bi}$ isotopes
(daughters of $^{222}\rm{Rn}$) are visible \cite{NIM-TF-04}.}
\label{det2_spectrum}
\end{figure}
In Fig.~\ref{det2_spectrum} a spectrum of the background
(live-time = 46 hours) measured with the GTF2 detector a few
days after the completion of the external shielding is shown.
The peaks from the gamma-lines of two isotopes,
$^{214}\rm{Pb}$ and $^{214}\rm{Bi}$, are presented. It
suggests that the peaks originate from the decay of
$^{222}\rm{Rn}$, introduced inside the \genius-TF setup with
liquid nitrogen fillings. \Rn\ is the only gaseous component
of the \U\ decay chain (Tab. \ref{tab:U238chain}) and, as a
noble gas, it has a high diffusion ability. Radon emanates
from the rock and the concrete inside the Gran Sasso
Laboratory. A high concentration of radon in the Gran Sasso
air (20-100 Bq/m$^3$,~\cite{GS_radon}) was observed. The
liquid nitrogen produced by the BOREXINO-purification plant is
especially treated to reduce the radon content by about two
orders of magnitude. The radon-content of unpurified nitrogen
has been measured to be around 0.1 - 0.3\,mBq/m$^3$ of gas at
STP. After the purification, the concentration is reduced to
$< 1 \mu$Bq/m$^3$ at STP~\cite{Freudiger1}. The background
produced by a concentration of radon in liquid nitrogen
equivalent to 0.5 $\mu$Bq/m$^3$ at STP for the four \genius-TF
detectors setup was simulated and is shown in
Fig.~\ref{sim_spectrum}.
\begin{figure}[!htp]
\begin{center}
\includegraphics[width=6.85cm]{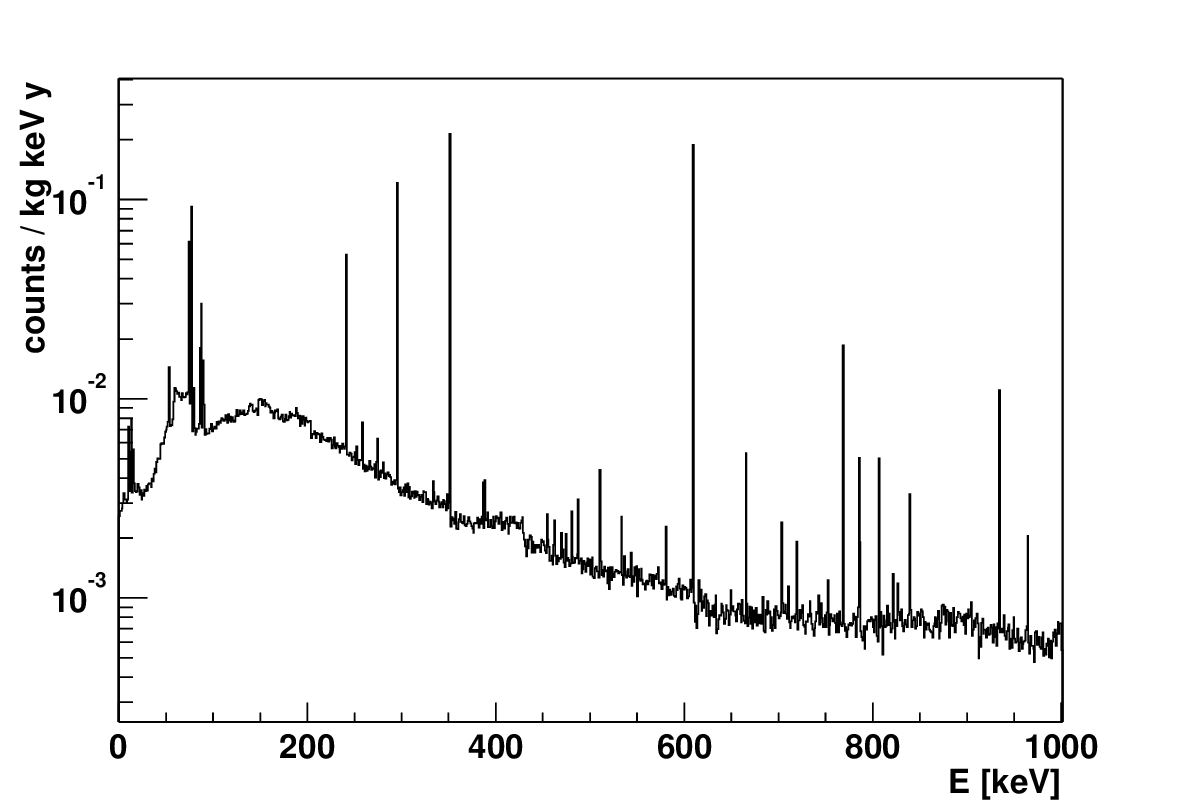}
\includegraphics[width=6.85cm]{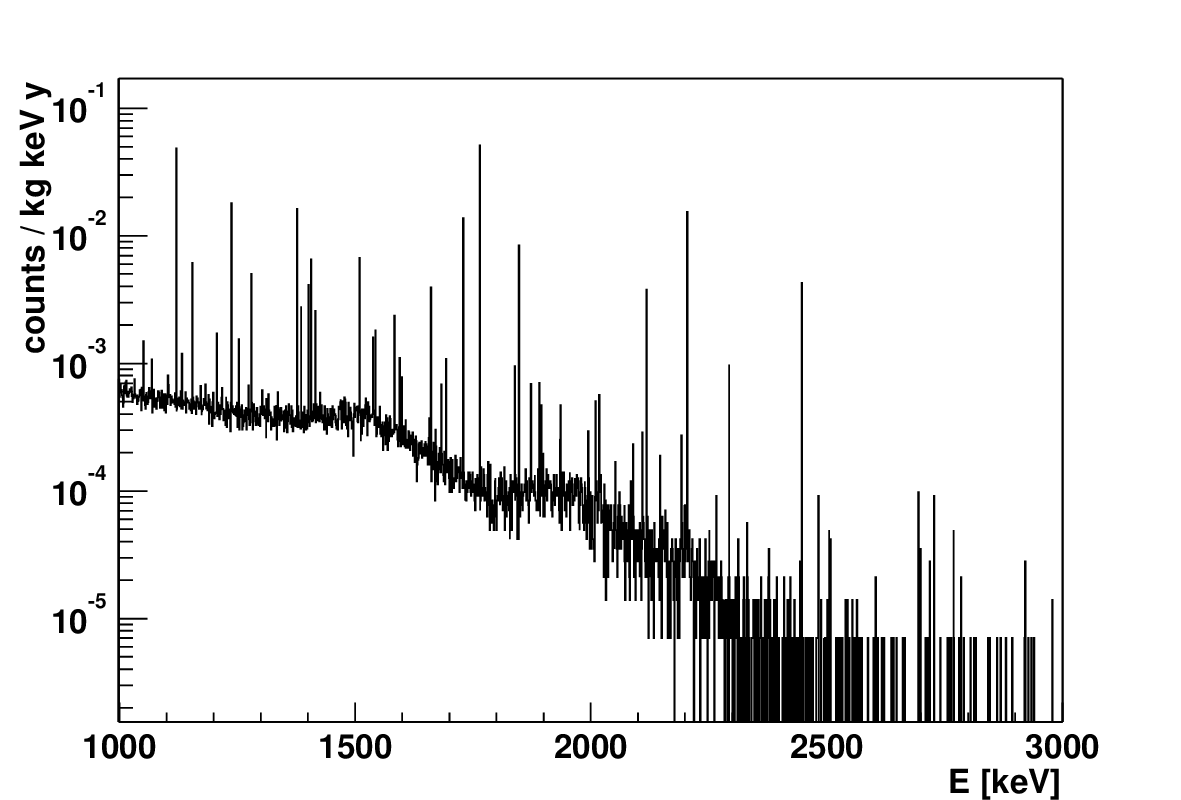}\\
\end{center}
\caption[]{GEANT4 simulation of the background in \genius-TF
setup produced by a radon concentration of 0.5\,$\mu$Bq/m$^3$
in nitrogen at STP \cite{Tomei-PhD}.} \label{sim_spectrum}
\end{figure}
The number of counts in the strongest peak from
$^{214}\rm{Pb}$ at 351.9 keV is calculated to be 0.16
counts/(kg\,y). In the measured spectrum of
Fig.~\ref{det2_spectrum}, the count-rate in the 351.9 keV peak
is 42 $\pm$ 5 counts/day or $\sim$6$\cdot10^{3}$
counts/(kg\,y). This count rate corresponds to a radon
concentration of 20\,mBq/m$^3$. To understand the origin of
this contamination, the intensity of the radon background with
time was studied. In Fig.~\ref{graph_rn}, the number of counts
per day observed in the 351.9 keV peak is plotted as a
function of time.
\begin{figure}[!htp]
\begin{center}
\includegraphics[width = 8cm]{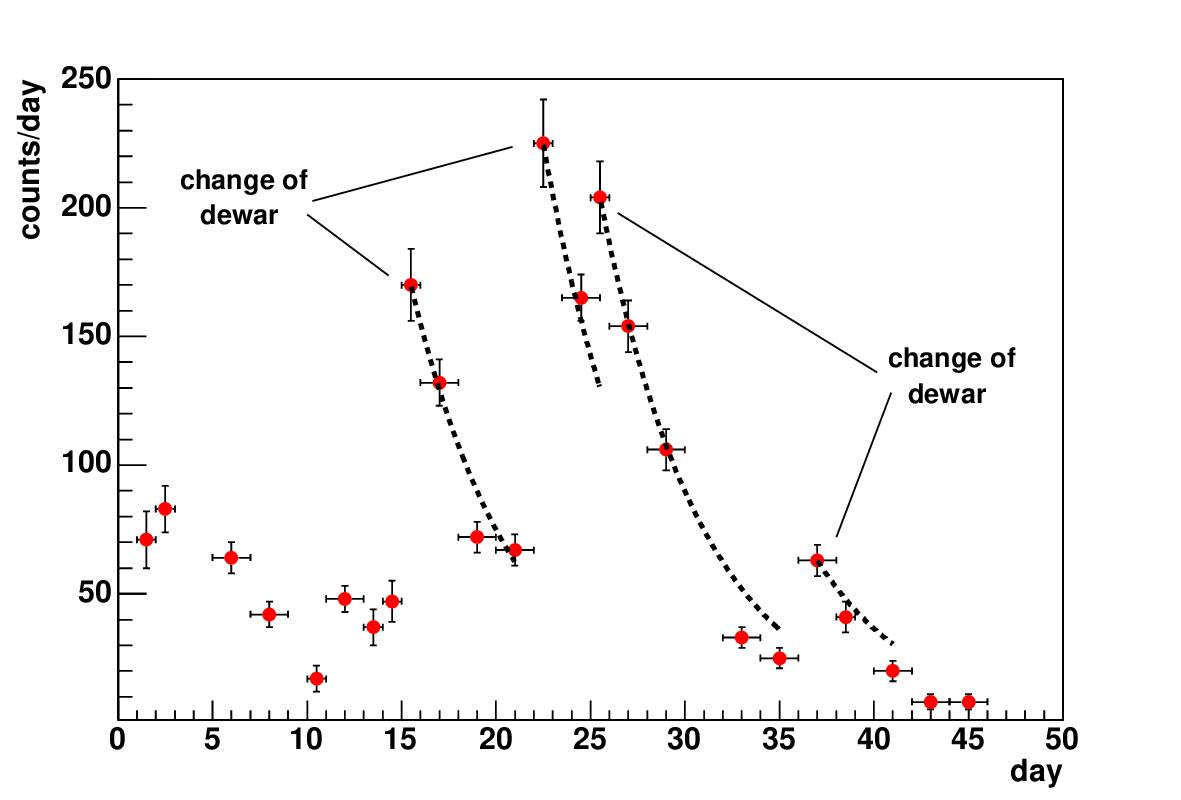}
\end{center}
\caption[]{Number of counts per day observed in the 351.9 keV
peak from $^{214}\rm{Pb}$ as a function of time. The dotted
lines are fits of the data with the exponents $\exp(-t/\tau$),
where $\tau=5.5$\,d is the $^{222}\rm{Rn}$ mean life-time. The
measurements were performed with GTF2 detector
\cite{NIM-TF-04}.} \label{graph_rn}
\end{figure}
\noindent It was noted that the number of counts from the
radon daughters was not constant but it increased strongly
when the refilling dewar was changed.
\begin{table}
  \centering
  \begin{tabular}{|l|l|c|c|l|}
    \hline
Nuclide     & Decay mode        &   Half life   & Q\,[MeV]     &   Decay products\\
\hline
238U        & $\alpha$          &   4.468·109 y & 4.270     &   234Th\\
234Th       & $\beta$-          &       24.10 d &   0.273   &   234Pa\\
234Pa       & $\beta$-          &       6.70 h  &   2.197   &   234U\\
234U        & $\alpha$          &   245500 y    &   4.859   &   230Th\\
230Th       & $\alpha$          &       75380 y &   4.770   &   226Ra\\
226Ra       & $\alpha$          &       1602 y  &   4.871   &   222Rn\\
222Rn       & $\alpha$          &   3.8235 d    &   5.590   &   218Po\\
218Po       & $\alpha$  99.98\% &   3.10 min    &   6.115   &   214Pb\\
            & $\beta$- 0.02\%   &               & 0.265     &   218At\\
218At       & $\alpha$  99.90\% &   1.5 s       &   6.874   &   214Bi\\
            & $\beta$- 0.10\%   &               &  2.883    &   218Rn\\
218Rn       & $\alpha$          &       35 ms   &   7.263   &   214Po\\
214Pb       & $\beta$-          &   26.8 min    & 1.024     &   214Bi\\
214Bi       & $\beta$- 99.98\%  &   19.9 min    &   3.272   &   214Po\\
            & $\alpha$   0.02\% &               &   5.617   &   210Tl\\
214Po       & $\alpha$          &   0.1643 ms   &   7.883   &   210Pb\\
210Tl       & $\beta$-          &   1.30 min    &   5.484   &   210Pb\\
210Pb       & $\beta$-          &   22.3 y      &   0.064   &   210Bi\\
210Bi       & $\beta$- 99.99987\% & 5.013 d     &   1.426   &   210Po\\
            & $\alpha$  0.00013\% &             &   5.982   &   206Tl\\
210Po       & $\alpha$          &   138.376 d   &   5.407   &   206Pb\\
206Tl       & $\beta$-          &   4.199 min   &   1.533   &   206Pb\\
\cline{2-5}
206Pb       & \multicolumn{4}{|c|}{stable}\\
    \hline
  \end{tabular}
  \caption{Uranium-238 decay chain \cite{TOI99}.}\label{tab:U238chain}
\end{table}
The \genius-TF setup was refilled every two days with liquid
nitrogen. One dewar of liquid nitrogen was enough for 3-4
fillings. When the current dewar was empty, it had to be
disconnected from the filling tubes and replaced. Following
the change of the refilling dewar, the number of counts in the
radon peaks decreased exponentially as a function of
$\sim\,e^{-t/\tau}$, where $t$ is the number of days after the
dewar change and $\tau = 5.5$\,d is the mean $^{222}\rm{Rn}$
lifetime (Fig.~\ref{graph_rn}). No jumps in the count rates
were observed after the fillings without change of the dewar.
Therefore, the sharp increases in the count-rate could be
explained by the radon entering the tubes during the change of
the dewar. The air has a high content of $^{222}\rm{Rn}$ and
this radon could be introduced in the inner copper dewar when
the valve was open. To avoid such contamination the tubes were
flushed with high-purity gaseous nitrogen before connecting
the new dewar. This procedure allowed to reduce the radon
content by an order of magnitude.
\begin{figure}[!htp]
\begin{center}
\includegraphics[width=6.8cm]{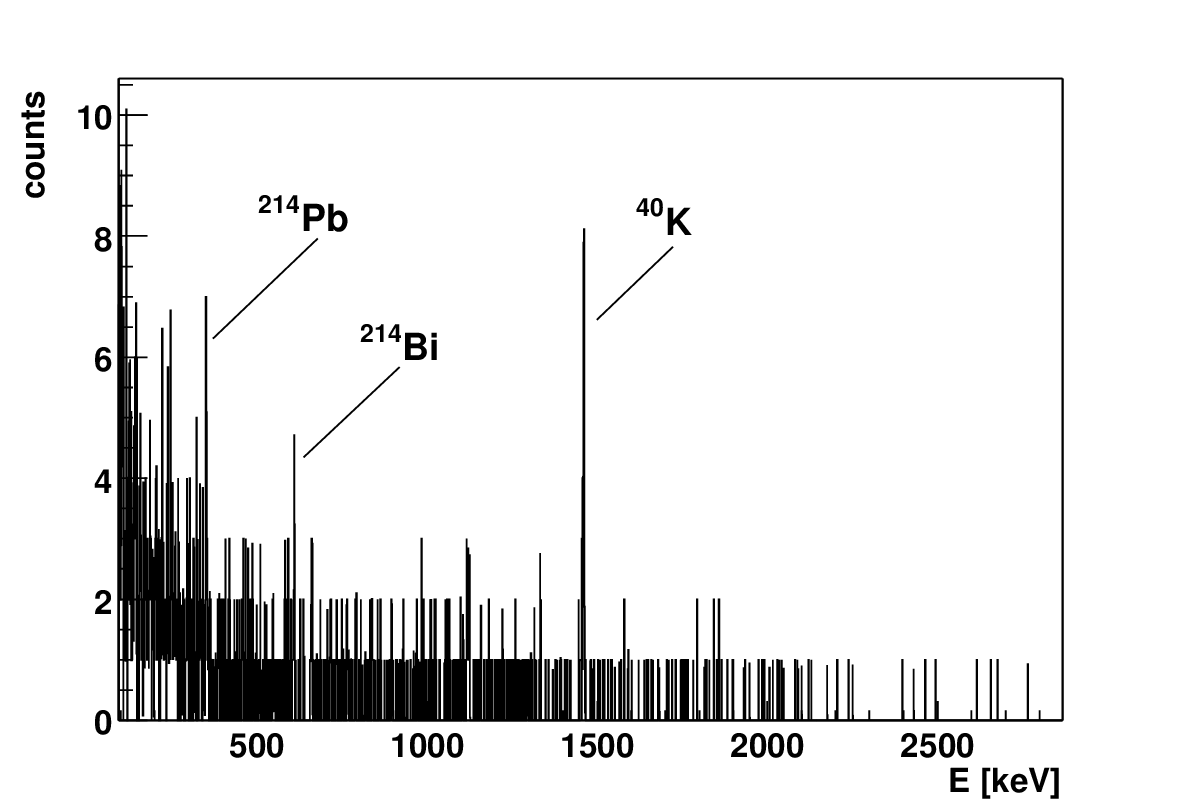}
\includegraphics[width=6.8cm]{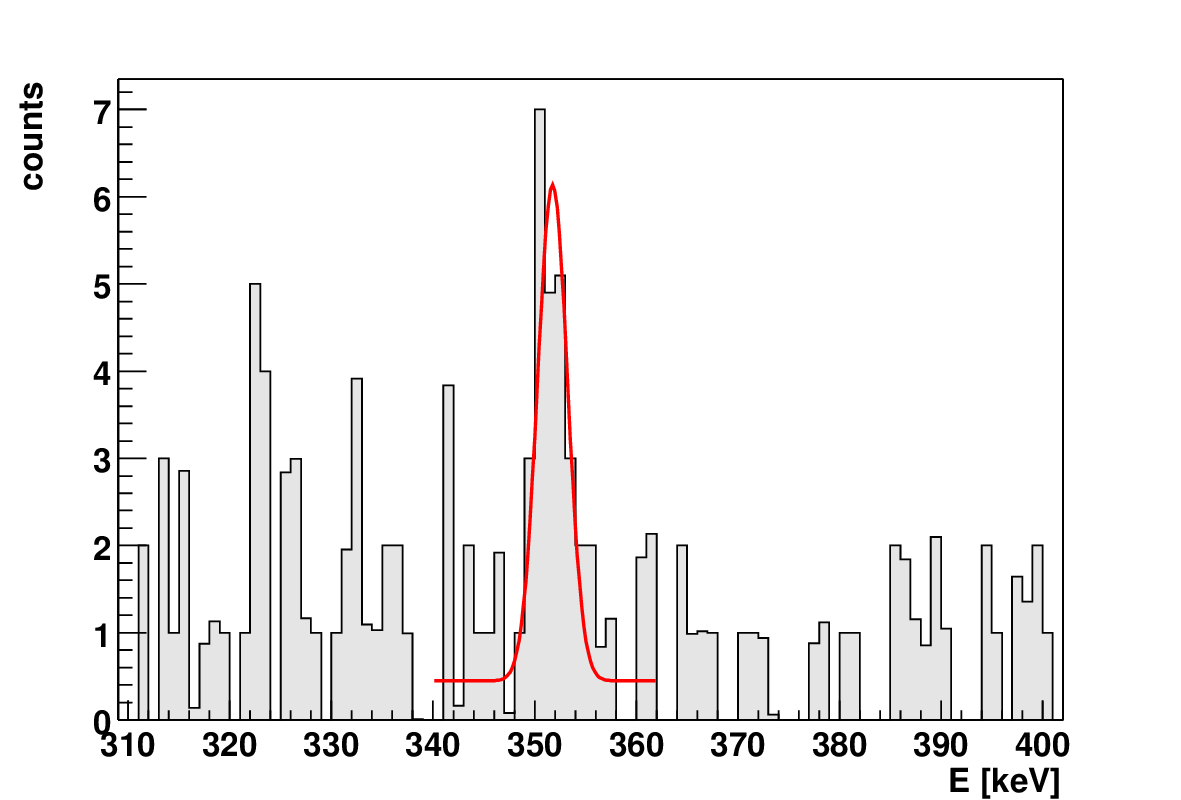}
\end{center}
\caption[]{Left: sum spectrum of the detector GTF2 from six
\genius-TF runs. The corresponding lifetime is 8.5 days.
Right: fit of the 351.9 keV line from $^{214}\rm{Pb}$ in the
same sum spectrum \cite{NIM-TF-04}.} \label{TF_sumspec}
\end{figure}
\noindent A fit of the 351.9\,keV peak from $^{214}\rm{Pb}$ in
the sum spectrum of Fig.~\ref{TF_sumspec} (right) gives a
count-rate of 2.7 $\pm$ 0.6\,counts/d, averaged over 8.5 days.
This count-rate correspond to an initial $^{222}\rm{Rn}$
concentration of $\simeq$2.4\,mBq/m$^3$ of gas at STP. A
reduction of this rather high level of specific activity can
be obtained with an appropriate isolation of the setup from
the external air with a radon tight cover, a constant flushing
with gaseous nitrogen, an improved filling procedure and an
improved calibration system.\\
\clearpage

\section{Stability of the \genius-TF detectors parameters}
The background and calibration measurements presented here
were performed during the first year of \genius-TF operation.
The energy resolutions were determined from the measured
spectra collected from May 2003 to May 2004. They were
obtained from the strongest lines in the background and the
calibration spectra ($^{214}\rm{Pb}$-352\,keV, \K-1461\,keV,
\Ba-356\,keV, \Co-1332\,keV). The long-term stability of the
detector energy resolution for these energies is shown in
Table \ref{tab:GeniusResol}. The bias voltage of the detectors
was set to the one recommended by ORTEC, unless the leakage
current had increased above 1\,nA. In that case, HV was set at
100--200\,V below the nominal value. The GTF3 detector bias
was initially set to 200\,V below depletion voltage, but in
April 2004 the detector worked at nominal bias of 3200\,V. The
resolution of the current measurements of the HV unit ampere
meter was 1\,nA. Detector GTF1 was temporary disconnected from
the DAQ because of high microphonic noise caused probably by
the vibrations of its cables in liquid nitrogen. This noise
created high trigger rate in the Struck Flash ADC, increasing
the DAQ dead time for the other detectors. The microphonic
noise is mostly attributed to the detector contacts and the
electronic wiring. The variation of the energy resolution of
the detectors during the first year of operation were caused
mostly by microphonic noise. The first year of \genius-TF had
demonstrated the possibility of long term operation of bare
HPGe detectors in liquid nitrogen without any irreversible
deterioration in energy resolution.
\begin{table} \small
  \centering
  \begin{tabular}{|c|c|c|c|c|r|}
    \hline
    & GTF1 & GTF2 & GTF3 & GTF4 & E,(keV) \\
\hline
Operating bias\,[V]&3000&2600&3200&2500&\\
Depletion bias\,[V]&1000&700&3100&2000&\\
FWHM\,[keV] at ORTEC & 2.0 & 3.1 & 1.8 &2.0 & \\
\hline
Date &   \multicolumn{4}{|c|}{Actual diodes bias\,[V]}&\\
\hline
   05/07/03& 2800&2601&3000&2401&\\
\hline
   05/07/03& 7.2&  Off & Off & Off &  356\\
   05/07/03& 7.4&  Off & Off & Off & 1332\\
   05/09/03& 5.1&  4.6&  4.9 & 4.5&  1332\\
   05/13/03& 7.5&  4.2&  3.8&  5.3&  1460\\
   07/26/03& 3.8&  5.3&  6.1&  5.4&  356\\
   07/26/03& 6.0&  9.2& 11.7& 11.8& 1332\\
   07/28/03& 3.1&  5.0&  5.2&  4.6&  356\\
   07/28/03& 5.3&  8.3&  9.7&  7.9& 1332\\
   07/30/03& 5.2&  8.2&  5.6&  7.6& 1460\\
   11/27/03& 3.0&  2.3&  5.5&  2.4&  356\\
\hline
   &\multicolumn{4}{|c|}{Actual diodes bias\,[V]}&\\
\hline
   12/10/03& 2404&2603&2879&2301&\\
\hline
   12/10/03& Off& 2.4& 2.5& 2.3& 356\\
   12/21/03& Off& 2.4& 2.7& 2.4& 352\\
   12/31/03& Off& 2.5& 2.7& 2.4& 352\\
   01/02/04& Off& 2.6& 2.8& 2.4& 352\\
   01/11/04& Off& 3.5& 2.3& 2.1& 356\\
   02/10/04& Off& 3.2& Off& 3.0& 356\\
   02/25/04& Off& 2.3& Off& Off& 356\\
   03/07/04&3.2& 2.0& 2.0& 2.0& 356\\
   03/11/04&5.2& 2.8& 2.6& 2.3& 352\\
   04/04/04&2.8& 4.0& 4.7& 2.4& 352\\
\hline
   &\multicolumn{4}{|c|}{Actual diodes bias\,[V]}&\\
\hline
   04/06/04& 2600&2220&2879&2301&\\
\hline
   04/15/04&4.1& 4.0& 5.0& 2.6& 352\\
   04/15/04&4.7& 4.7& 5.7& 3.8& 1332\\
   04/26/04&5.5& 3.0& 4.0& 2.3& 352\\
\hline
   &\multicolumn{4}{|c|}{Actual diodes bias\,[V]}&\\
\hline
   05/04/04& 2600&2220&3200&2500&\\
\hline
   05/19/04&4.4& 4.0& 3.3& 2.7& 352\\
   \hline
\end{tabular}
\caption{Energy resolution (FWHM\,[keV]) of the four
\genius-TF detectors as a function of time during the first
year of \genius-TF operation.}\label{tab:GeniusResol}
\end{table}

\clearpage

\section{Summary and outlook for GERDA}

The \genius-TF setup with four HPGe detectors immersed in
liquid nitrogen has been presented. The presented work covers
the first year of operation from the detectors assembly on May
5 2003, to June 2004 at the Gran Sasso underground laboratory.
First measurements showed high sensitivity of the diodes and
of the electronics to the environmental interferences:
vibrations in liquid nitrogen, electromagnetic pickups and
infrared radiation. Special measures were performed to reduce
these interferences and an energy resolution of $\sim$4\,keV
was achieved. The results and analysis of the \genius-TF
background after the complete assembly of the external
shielding were presented. The contribution to the background
coming from the $^{222}$\rm{Rn} decay chain was identified.
The high background from $^{222}$\rm{Rn} is caused by radon
coming from the laboratory air inside the setup during the
filling procedure. To eliminate radon influx additional
flushing lines, a radon tight copper box and high
quality valves were proposed to be installed.\\
The four first \genius-TF detectors were biased at their
initial HV during the first year of the \genius-TF operation.
Their energy resolution did not irreversibly deteriorate, but
was highly affected by the microphonic noise. These
measurements show principal feasibility of using bare
germanium detectors in cryogenic liquids serving as a shield
from external radiation. An irreversible deterioration of the
\genius-TF detectors operating in liquid nitrogen over three
years was observed in \cite{NIM-TF-06, PhSc-TF-06}. This could
be attributed to the probable improper handling during the
warm-up temperature cycles.

\newpage
\chapter{Measurements of the $\gamma$ Flux on the GERDA Site at LNGS}
\label{ch:halla}

\section{Introduction}
The necessary GERDA shielding depends on the intensity of the
external radiation. Therefore, the $\gamma$-ray flux at the
experimental site had to be determined. The GERDA experiment
is located underground in Hall A of LNGS, under 1400 m of
rock. The rock coverage results in a cosmic ray flux reduction
on the order of one million but the uranium, thorium and
potassium content in the mountain rock also produce gamma
radiation. The concentrations of \U,\,\Th\ and \K\ in the rock
and the concrete at LNGS have been measured by the Milan group
in 1985 \cite{Bellotti1985}. The authors concluded that the
normal rock has very low activity, but there are infiltrations
of black marnatic rock which contains much larger
radioactivity. These infiltrations are now hidden under a
concrete layer of 5-30 cm
thickness covering the laboratory walls.\\
To find a possible anisotropy or local anomalies at the
\gerda\ site, the angular distribution of the gamma flux was
first measured in 2004 using an HPGe detector with a
collimator. During the 2005-2006 reconstruction of LNGS, the
Hall A floor was covered with a new concrete of 30 cm
thickness and a layer of waterproof resin. Measurements showed
that the activity of the new concrete is higher than the
normal LNGS rock by one order of magnitude. Therefore, new
$\gamma$-ray flux measurements at the site of the GERDA
experiment were performed again in 2007. The following
sections describe the experimental setups, the calibration
procedure and the results of the $\gamma$-ray flux
measurements.

\section{Methods of flux determination}

The flux of photons $\Phi$ is the number of photons per unit
time passing through a unit surface area perpendicular to the
trajectory of the photons. The full-energy peak count rate $N$
of the detector irradiated with a photon flux
$\Phi(E,\theta,\,\varphi)$ at a given energy $E$ (Fig.
\ref{HallA:FluxDef}) is given by:
\begin{equation}\label{flux1}
\mathit N(E) = \int\limits_{\Omega}
\varepsilon(E,\theta,\,\varphi)\,
A_\perp(\theta,\,\varphi)\,\frac{d\Phi(E,\theta,\,\varphi)}
{d\Omega}\,d\Omega\\
\end{equation}\\
where $\varepsilon(E,\theta,\,\varphi)$, the intrinsic
efficiency, is the ratio of the counts in the full energy peak
to the number of photons hitting the detector surface,
$A_\perp$ is the projected area of the detector perpendicular
to the ($\theta,\,\varphi$) direction, and $\Omega$ is the
solid angle. The detector response ($S$) is the full energy
peak count rate per unit flux at a given photon energy:
$S(E,\theta,\,\varphi)\equiv\varepsilon(E,\theta,\,\varphi)
\,A_\perp(\theta,\,\varphi)$. In general, the flux $\Phi$
cannot be obtained by solving the integral equation
\ref{flux1}. However, if the detector has an isotropic
response $S(E,\theta,\,\varphi)=S(E)$, then the integral is
simplified to the product of the detector response and the
integral flux, which then equals to:
\begin{equation}
 \Phi(E) = \frac{N(E)}{S(E)}. \label{CountFlux}
\end{equation}
The HPGe detector used for the present $\gamma$-flux
measurements possesses a nearly isotropic response over a wide
range of energies, validating the application of the equation
\ref{CountFlux}, as will be shown in the section
\ref{HallA:Section:Response}. The isotropic response of the
detector allows a significant simplification of the total flux
determination.\\
\begin{figure}[h]
  \centering
  \includegraphics[height=60 mm]{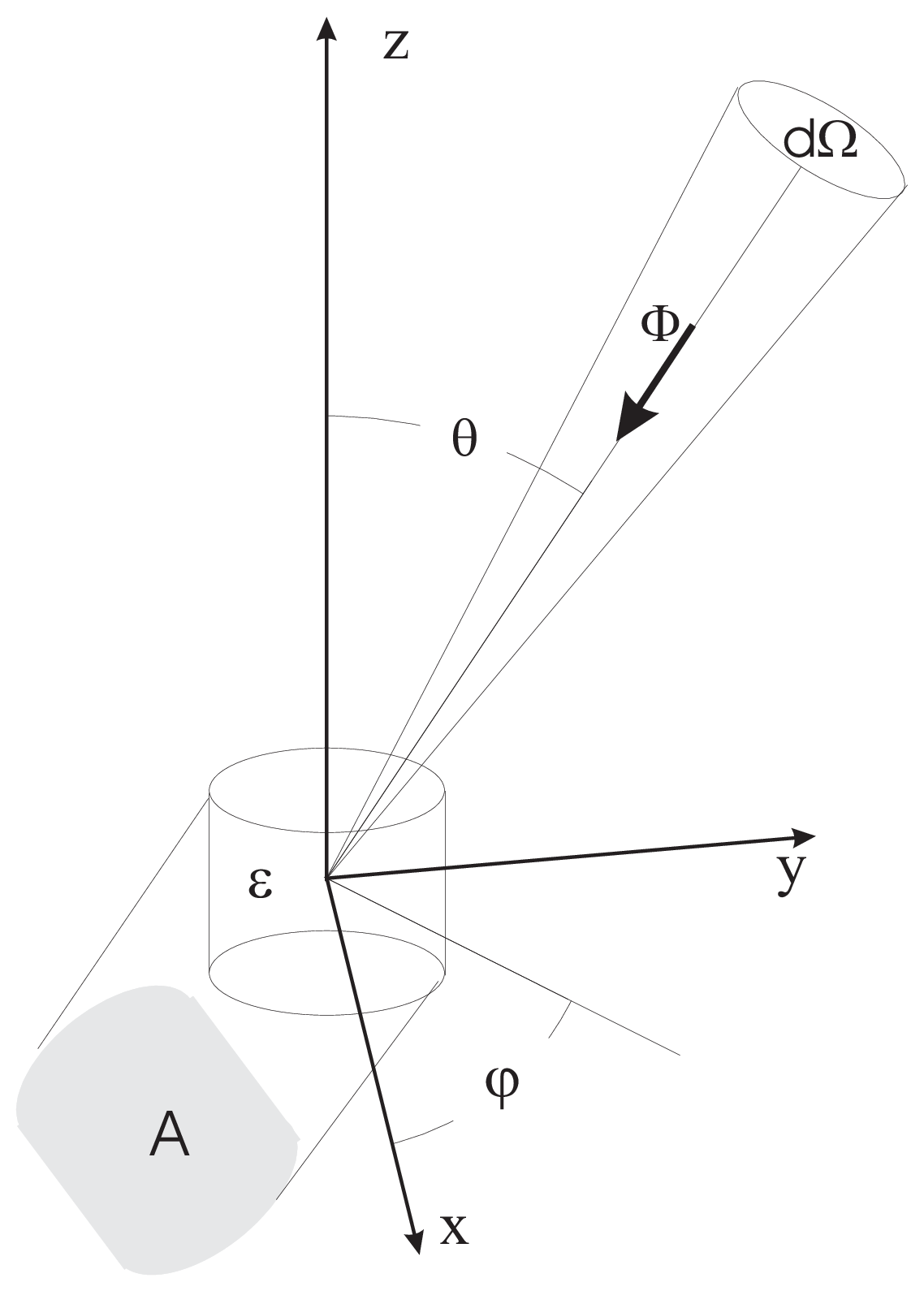}\\
\caption{Diagram showing the projected area of the detector
perpendicular to the flux $\Phi$ in the ($\theta$,$\varphi$)
direction.}\label{HallA:FluxDef}
\end{figure}
The flux can also be calculated using a source distribution.
Assuming a uniform volume distribution of the $\gamma$-source
with intensity $n_\gamma$ [$\gamma/s m^3$] in the material
surrounding a convex cavity (Fig. \ref{fig:FluxTheor}) and
neglecting absorption inside the cavity, a simple analytical
expression of the flux can be derived. The unscattered photon
flux $d\Phi$ inside the cavity, at a distance $R$ from the
volume element $dV=R^2\,dR\,d\,\Omega$ emitting $n_\gamma\,dV$
photons, can be expressed as:
\begin{equation}
  \mathit d\Phi = \frac{n_\gamma\,dV}{4\pi\,R^2}e^{-\mu_E\,(R-R_0)},\,R>R_0.\\
\label{FluxFormula1}
\end{equation}
\begin{figure}[h]
  \centering
  \includegraphics[width=8 cm]{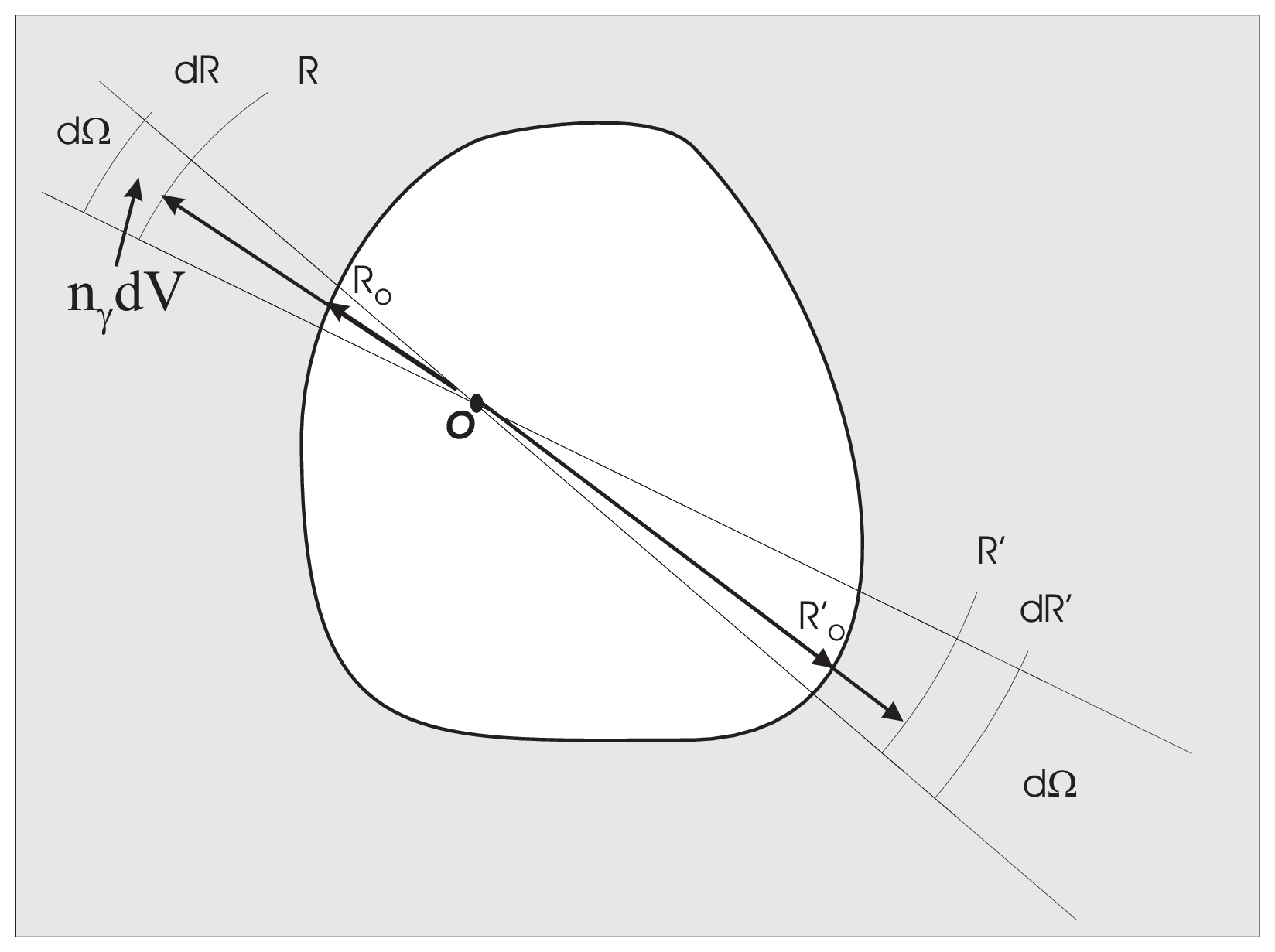}\\
  \caption{Geometry used for the calculation of the flux inside a cavity in
   a uniform infinite medium with a specific $\gamma$-activity
   ($n_\gamma$)
   and an attenuation coefficient $\mu_E$ (equation \ref{FluxFormula1})}\label{fig:FluxTheor}
\end{figure}
The absorption and scattering of the photons are accounted for
by the exponential term $e^{-\mu_E\,(R-R_0)}$, where $\mu_E$
is the attenuation coefficient for photons with energy $E$ and
$R_0$ is the distance from the point $O$ to the surface in the
$d\Omega$ solid angle direction. The total flux with energy
$E$ is given by the integration of the equation
\ref{FluxFormula1} over $R>R_0$ and a $4\pi$ solid angle
 as follows:
\begin{eqnarray}
\nonumber \Phi &=&
\int\limits_{\Omega}\int\limits_{R_0}^\infty
\frac{n_\gamma}{4\pi\,R^2}e^{-\mu_E\,(R-R_0)}\,R^2\,dR\,d\Omega\\
  &=& \int\limits_{\Omega} \frac{n_\gamma\,d\Omega}{4\pi}
\int\limits_{R_0}^\infty e^{-\mu_E(R-R_0)}\,dR =
           \frac{n_\gamma}{\mu_E}.\label{FluxFormula2}
\end{eqnarray}
In this idealized model the flux is defined by the emitted
photon density and the absorption in the surrounding media.
The flux is isotropic and uniform inside the cavity. Most of
the unscattered photons originate in the surface layer of
material with a thickness in the order of
$\mu_E^{-1}$($\mu_E$=10\,1/m for E$_\gamma$=2.6\,MeV in LNGS
rock). For example, the thorium activity in the first 10 cm of
rock walls produces 63\% of the 2.6 MeV photon flux.\\
The use of equation \ref{FluxFormula2} is justified if the
flux is isotropic. Flux isotropy measurement at the GERDA site
was performed in 2004 using an HPGe
detector with a collimator, described in the next section.\\

\clearpage

\section{Detector system for {\it in-situ} $\gamma$-flux
measurements}

For both measurements (2004/2007) a 114\,$cm^3$ closed-end
coaxial intrinsic germanium detector with a resolution of 2.0
keV FWHM at 1332 keV was used. The diode is housed in an
aluminum cap of 7.0 cm outside diameter, which also contains
the necessary signal processing electronics. The detector
capsule is attached to a electro-mechanical cryogenic cooler.
The cryogenic cooler allows measurements without interruptions
caused by liquid nitrogen refilling for one month of
measurements. For the first measurement in 2004 a collimator
was used. The copper collimator was made of inner parts of a
shielding from a low-background spectrometer (Fig.
\ref{HallA:GammaScope}). The parameters of the detector and
the collimator are presented in Table \ref{HallA:GammaScopeParam}.\\
\begin{table}[h]
\centering
\begin{tabular}{|l|c|}
  \hline
Detector Type & ORTEC HPGe 'PopTop'  with cryo-cooling\\
Relative Efficiency & 30\%\\
Energy Resolution & 2.0 keV FWHM at 1332 keV\\
Crystal Diameter and Length & 50 mm; 58 mm \\
\hline
Collimator Material & Copper \\
Collimator Dimensions & $20\times20\times30$ $cm^3$ \\
Collimator Well diameter & 8 cm \\
Wall thickness &min. 6 cm, max. 10 cm\\
Subtended solid angle  & 1.2 sr \\
\hline
\end{tabular}
\caption{The HPGe detector and collimator
parameters.}\label{HallA:GammaScopeParam}
\end{table}
\begin{figure}[h]
\centering
  \includegraphics[width=10 cm]{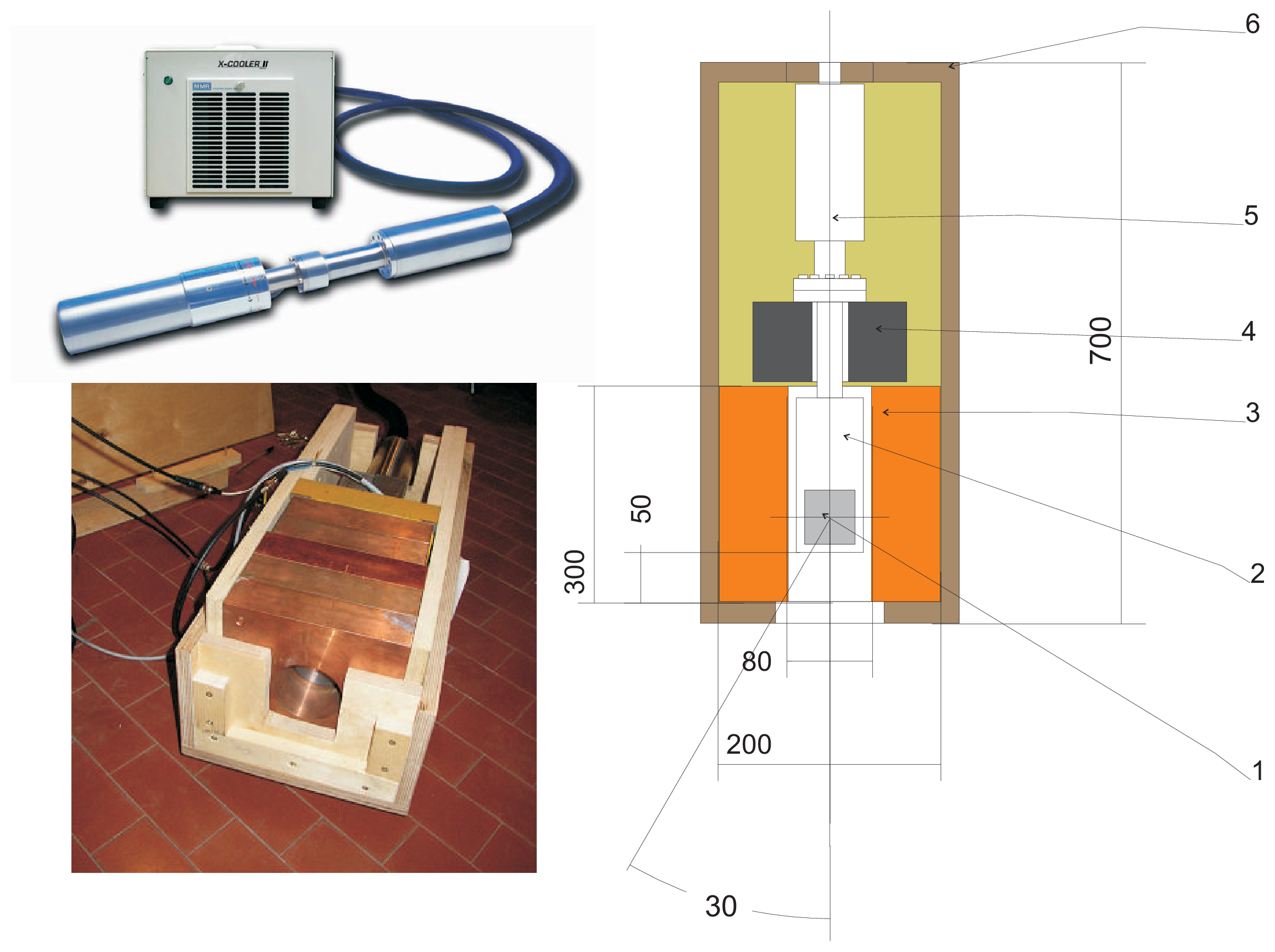}\\
\caption{The photographs show the HPGe detector connected to
the cryogenic cooler (left top), and the Cu collimator
assembly (left bottom). Right: diagram of the setup for
$\gamma$-flux anisotropy measurements with a germanium
detector using a collimator:
  1) HPGe diode,
  2) ORTEC PopTop capsule,
  3) copper collimator with a 30$^\circ$ half opening,
  4) lead shield,
  5) heat exchanger with tubes connected to external portable cryo-cooling system,
  6) case.}\label{HallA:GammaScope}
\end{figure}

\clearpage

\section{The detector system response to $\gamma$-radiation}
\label{HallA:Section:Response}

The response of the detector was calculated with Monte-Carlo
(MC) simulations. The use of simulations was validated by
performing measurements with point-like sources (\Co\ and
\Ra). The simulations and measurements agree to within 3\%, as
shown in Fig.~\ref{HallA:EfficMeasMC}. The MC simulation of
the detector response to the parallel photon flux with
energies ranging from 350 to 2614 keV was performed using
EGSnrc code. The angular response was determined for a
detector without (Fig.~\ref{HallA:AngularSens}) and with
(Fig.~\ref{HallA:AngularSensCollim}) a collimator.
\begin{figure}[h] \centering
  \includegraphics[width=75 mm]{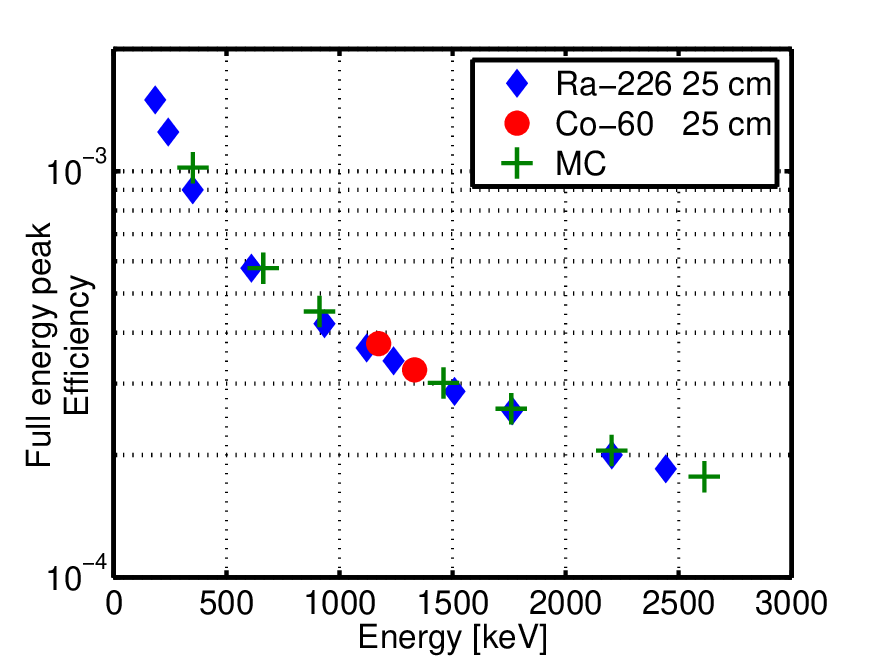}\\
\caption{Measured and simulated full energy peak efficiencies
of the HPGe detector as a function of energy.  The \Ra\ and
\Co\ sources were located 25\,cm away from the cap of the
detector.}\label{HallA:EfficMeasMC}
\end{figure}
\begin{figure}[!h]
\centering
  \includegraphics[width=65 mm]{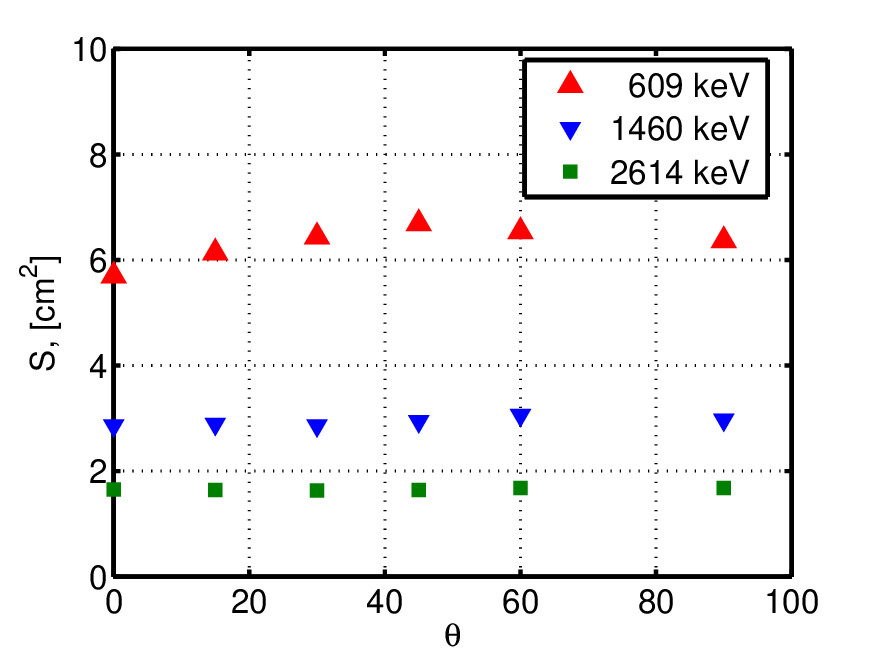}
  \includegraphics[width=65 mm]{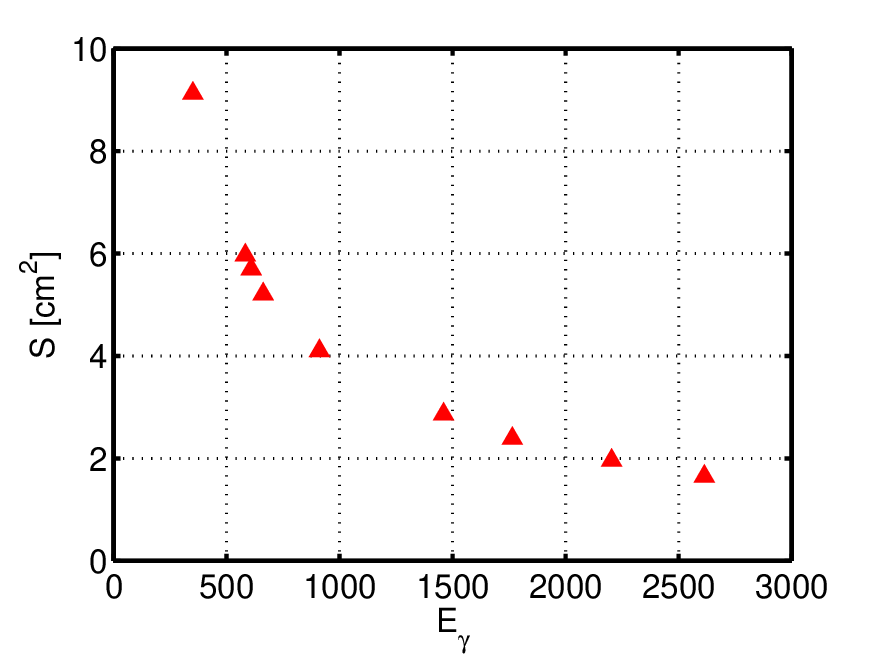}\\
\caption{Left: Angular response of the HPGe detector without a
collimator for a parallel flux of photons determined by MC
simulations. Right: Response of the HPGe detector without a
collimator as a function of photon
energy.}\label{HallA:AngularSens}
\end{figure}
\begin{figure}[h]
\centering
  \includegraphics[width=65 mm]{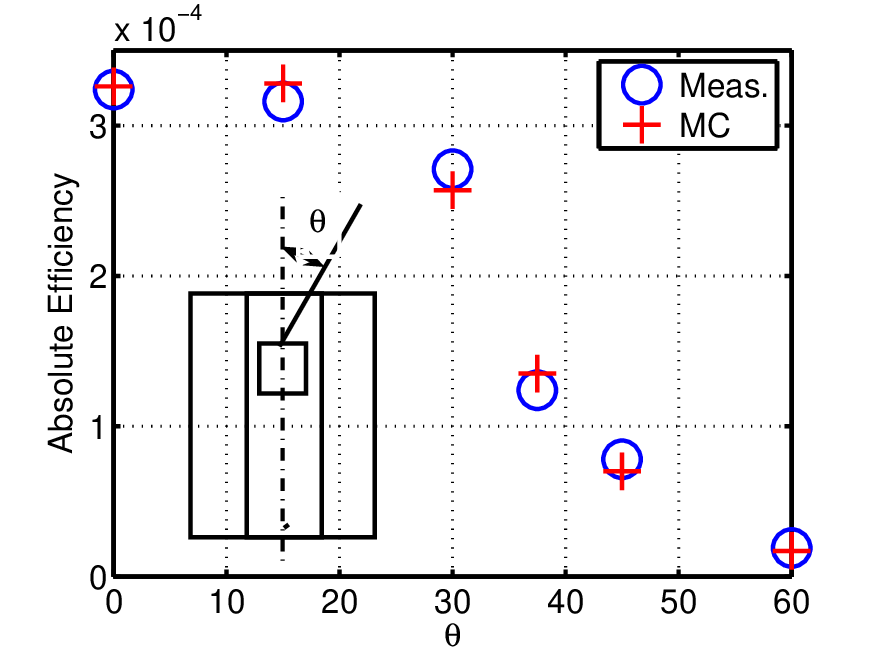}
  \includegraphics[width=65 mm]{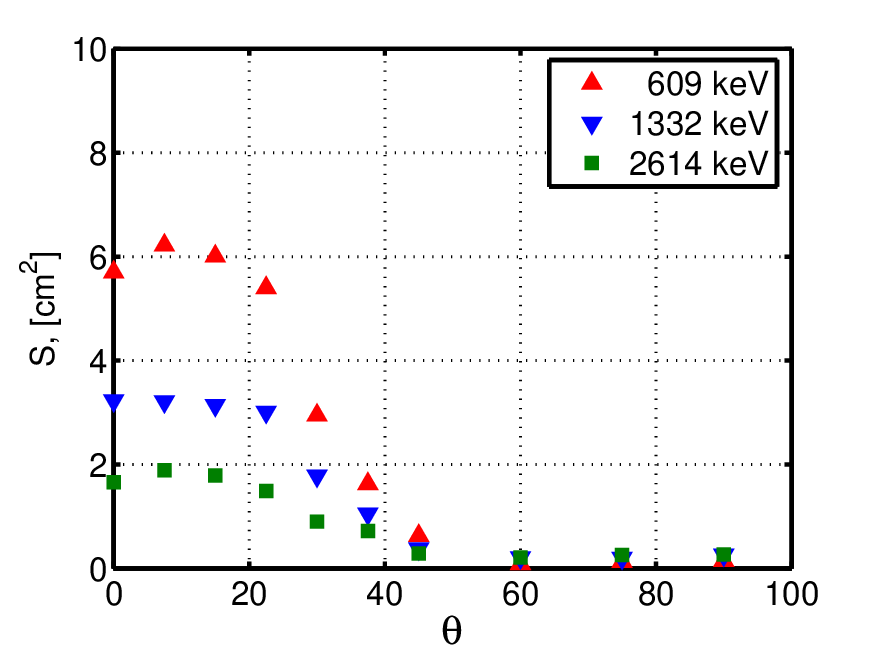}\\
\caption{Left: Angular dependence of the HPGe detector
efficiency with a collimator for 1332 keV photons from \Co\ at
25 cm from the cap. Right: Angular response of the HPGe
detector with the copper collimator for parallel flux of
photons from MC simulations.}\label{HallA:AngularSensCollim}
\end{figure}
As Figure~\ref{HallA:AngularSens} shows, for the HPGe detector
used in these measurements the response practically does not
depend on the direction of the incoming flux. The response
varies by 2\% for 2614 keV photons coming from all directions.
For lower energies (below 500 keV), the variations are mainly
defined by the geometrical factor $A_\perp$, which ranges from
20\,cm$^2$ at the front of the detector to 35\,cm$^2$ at the
side, while the intrinsic efficiency remains the same for low
energy photons.  It is seen in
Fig.~\ref{HallA:AngularSensCollim} that the simulation with
the collimator is also in good agreement with the angular
measurements using the \Co\ source.\\


\section{Measurements in Hall A and results}

First {\it in-situ} measurements of the gamma flux in Hall A
at LNGS were done in 2004. The general underground plan of
LNGS is presented in Fig. \ref{LNGS_plan} \cite{BellottiLNGS}.
Directional measurements in Hall A were performed by pointing
the collimated HPGe detector to the walls, the vaulting and
the floor. Each direction was measured for one day. The
directional measurements with a collimator were followed by
measurements of the background to account for the flux from
outside of the collimator opening. The side contribution to
the flux was measured by closing the collimator opening with a
10 cm thick copper cylinder. The background was measured for
each direction of the collimator. The measured intensities of
the background peaks were subtracted from the peak intensities
measured with an open collimator. Then, a 15 days long
measurement was done without a collimator to determine the
total flux. The spectrum is shown in Fig. \ref{FluxSpectrum}.
The floor of Hall A in 2004 was partly covered with the
massive magnet parts of the OPERA experiment, which was under
construction (Fig. \ref{HallA_setup}). In 2007 the flux
measurements were repeated without a collimator, using the
same detector as in 2004. In both cases the detector was
placed on a 2.5 m tall
scaffold as shown in Fig. \ref{HallA_setup}.\\
The peak intensities and the calculated fluxes
(Eq.~\ref{CountFlux}) are presented in Table \ref{TabFlux}.
The 2614 keV photon total flux is
(362$\pm$12)\,m$^{-2}$s$^{-1}$. As shown in Table
\ref{TabFlux}, the flux of the photons from the ceiling is two
times higher than the flux from the walls and the floor. The
$\gamma$-flux from the direction of the Large Volume Detector
(LVD) experiment was greatly suppressed. In 2007, the
intensity of the $\gamma$-flux from thorium was 20\% higher
than it was in 2004. However, this difference could be
explained by the absorption in the massive iron parts of the
OPERA experiment stored there in 2004. A 662 keV photon flux
from anthropogenic \Cs\ was measured in 2004 but this flux was
undetectable in 2007, probably because
of the new layer of concrete on the floor.\\
\begin{figure}
 \centering
  \includegraphics[width=9 cm]{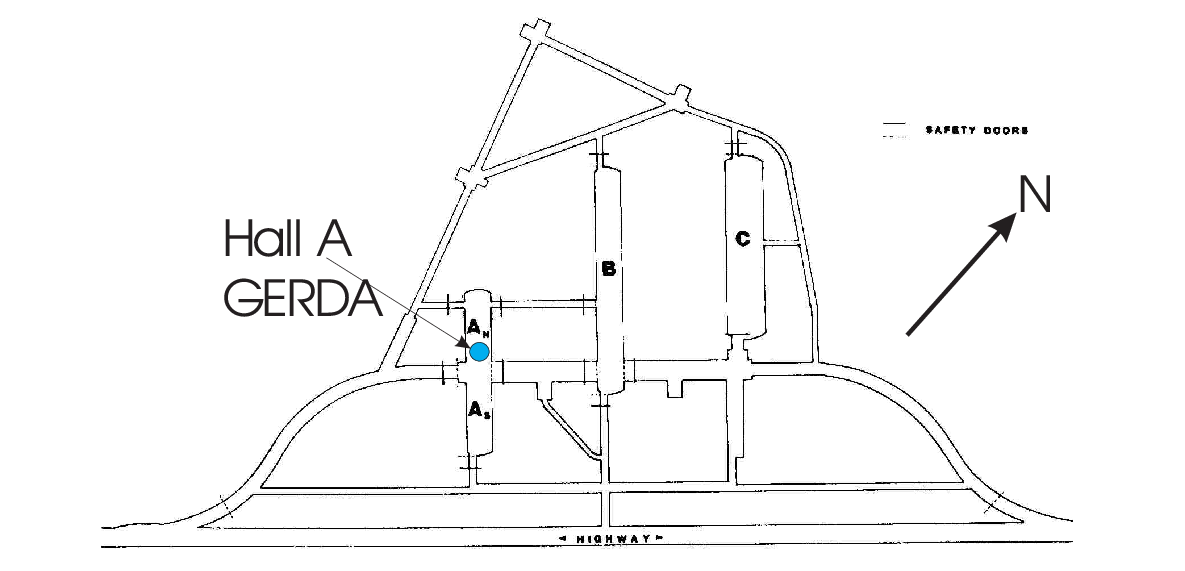}
  \includegraphics[width=4 cm]{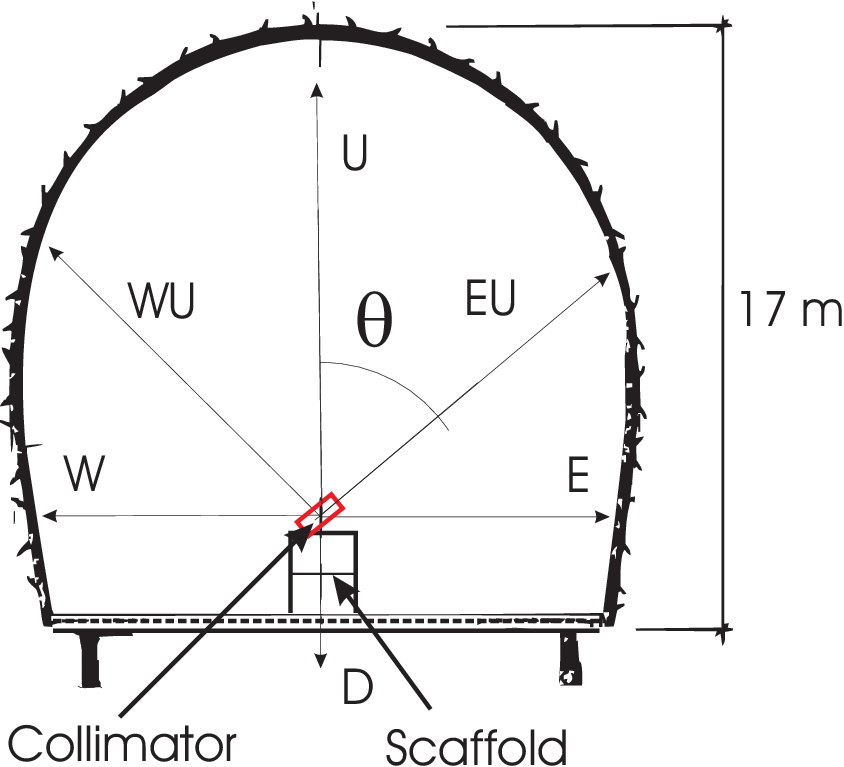}\\
  \caption{Left: floorplan of the LNGS underground laboratory. The
location of \gerda\ in Hall A indicated. Right: the detector
for the $\gamma$-flux measurements was positioned on a
scaffold $\sim$2.5 m above the ground. The angular
distribution of the $\gamma$-flux was measured with the
collimator pointing into the indicated directions, and also to
the north and to the south (direction assignments are: W-west,
WU-West-Up, etc., $\theta$=0 in zenith).}\label{LNGS_plan}
\end{figure}

\begin{figure}[h]
\centering
  \includegraphics[width=14 cm]{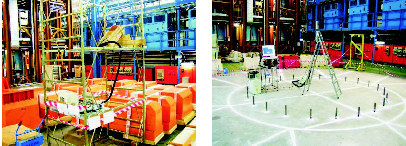}\\
  \caption{Setups for the flux measurements in Hall A in 2004 (left) and 2007 (right).
   In 2004 the massive parts of the OPERA experiment were stored at the same location as \gerda.
   In 2007 the flux measurements were performed when the bottom plate
   of the GERDA water tank was installed.}\label{HallA_setup}
\end{figure}

\begin{figure}
  \centering
  \includegraphics[width=20 cm, angle=90]{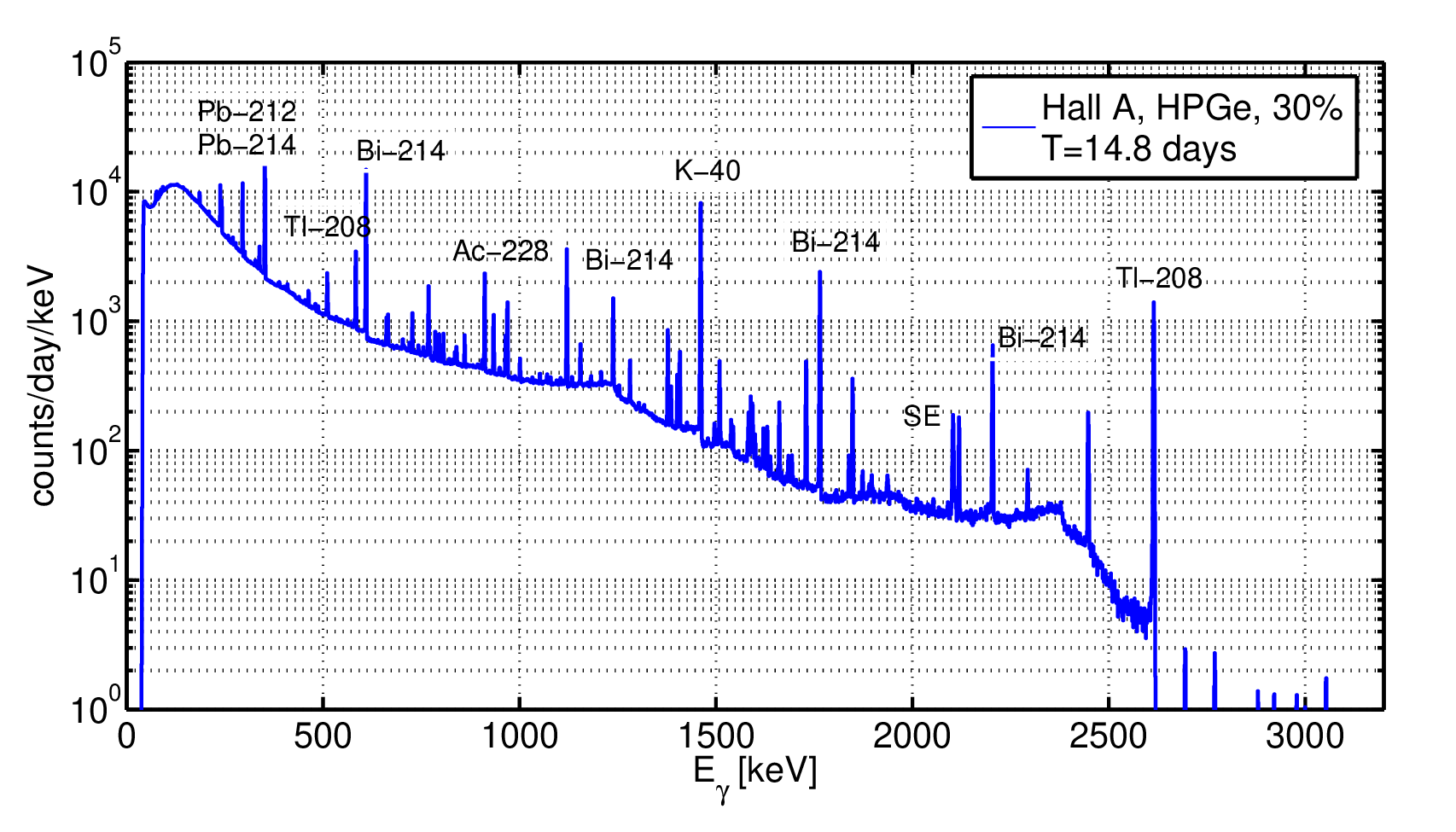}\\
  \caption{HPGe energy spectrum measured without
   collimation in Hall A of LNGS in 2004.}\label{FluxSpectrum}
\end{figure}

\clearpage
\begin{sidewaystable}[!h]

\begin{tabular}{|l|c|c c|c|c c c|c|c|c|}

\hline
Nuclide & &\multicolumn{2}{|c|}{Tl-208}  &    Ac228    &   \multicolumn{3}{|c|}{Bi-214} &   Pb-214 & K-40& Cs-137\\

$E_\gamma$,\,keV  &  & 2614 &   583 &   911  &  2204 &   1764
& 609 & 351 &  1460 & 662\\

\hline

Det. response, $S$, [cm$^2$] & & 1.65$\pm$0.05 & 6.0$\pm$0.3 &
4.1$\pm$0.2 & 2.0$\pm$0.1 & 2.4$\pm$0.1  & 5.7$\pm$0.3
& 9.1$\pm$0.5 & 2.9$\pm$0.1 & 5.2$\pm$0.3\\

\hline
Count rate,\,[10$^{-3}s^{-1}$] &&&&&&&&&&\\
August, 2004 & N & 46.5$\pm$0.3& 57.5$\pm$0.5&  46$\pm$1& 20.1$\pm$0.4 & 70$\pm$1 & 301$\pm$1& 319$\pm$2& 234$\pm$1& 8.6$\pm$0.6\\

&&&&&&&&&&\\

Flux, [$\gamma$/m$^2$\,s]& $\Phi$ &  284 $\pm$10& 96$\pm$7&
112$\pm$6& 101$\pm$7& 292$\pm$20&  528$\pm$37& 351$\pm$16&
807$\pm$45 & 17$\pm$2\\
\hline

Count rate,\,[10$^{-3}s^{-1}$] &&&&&&&&&&\\
November, 2007  & N & 59.8$\pm$0.9& 64.7$\pm$1.2&  56$\pm$1& 23.9$\pm$0.6 & 83$\pm$1 & 311$\pm$2& 318$\pm$3& 267$\pm$1& $<$0.4\\

&&&&&&&&&&\\

Flux [$\gamma$/m$^2$\,s] & $\Phi$ &  362 $\pm$12& 108$\pm$8&
137$\pm$7& 120$\pm$8&  346$\pm$24&  546$\pm$40& 349$\pm$16&
921$\pm$50 & $<$1\\
\hline\hline


South,\,CRESST  & N & 3.0$\pm$0.4 &  4.3$\pm$0.9& 1.4$\pm$0.8&
1.7$\pm$0.3& 6.3$\pm$0.5& 22$\pm$3&  40$\pm$1& 13$\pm$1 &
6.4$\pm$0.6\\
$\theta$=90, $\Omega=1.2$ sr  & $\Phi$  & 18$\pm$3& 7$\pm$2&
3$\pm$2&
9$\pm$2& 26$\pm$2& 39$\pm$5& 44$\pm$1&   45$\pm$3 & 12$\pm$1\\
\cline{1-11}

North,\, LVD & N &  0.6$\pm$0.8&   1.4$\pm$1.4& 1.1$\pm$1.5&
0.2$\pm$0.4& 2.0$\pm$0.7& 11$\pm$2&  17$\pm$4&
2.7$\pm$1.0 &4.2$\pm$0.8\\
$\theta$=90,
$\Omega=1.2$ sr & $\Phi$ &  4$\pm$5&   2$\pm$2&
3$\pm$4& 1$\pm$2& 9$\pm$3& 19$\pm$4&
19$\pm$5&   9$\pm$4& 8$\pm$2\\
\cline{1-11}

East & N &  2.5$\pm$0.4&   3.6$\pm$1.1& 2.4$\pm$1.1&
2.4$\pm$0.4& 8.5$\pm$0.6&   46$\pm$1&  47$\pm$2&
13$\pm$1 & $<$0.6\\
$\theta$=90, $\Omega=1.2$ sr & $\Phi$ &  15$\pm$3&  6$\pm$2&
6$\pm$2& 12$\pm$2&
36$\pm$3&  81$\pm$2&  52$\pm$2&   46$\pm$4 & $<$2\\
\cline{1-11}

West & N &  3.5$\pm$0.5&   6.1$\pm$1.3&   3.6$\pm$1.0 &
2.5$\pm$0.4&
10.0$\pm$0.7&  49$\pm$2&  70$\pm$2&  19$\pm$1 & 4.6$\pm$0.7\\
$\theta$=90, $\Omega=1.2$ sr & $\Phi$ &  21$\pm$3&  10$\pm$2&
9$\pm$2& 12$\pm$2& 42$\pm$3&
86$\pm$4&  77$\pm$2 & 66$\pm$4 & 9$\pm$1\\
\cline{1-11}

East-Up, & N &  3.6$\pm$0.6& 4.2$\pm$1.8&  2.7$\pm$1.3&
2.8$\pm$0.7& 5.9$\pm$0.8& 30$\pm$2&  34$\pm$3& 24$\pm$1 & 3$\pm$1\\
$\theta$=45, $\Omega=1.2$ sr & $\Phi$ &  22$\pm$4& 7$\pm$3&
7$\pm$3& 14$\pm$4& 25$\pm$4&
53$\pm$4&  37$\pm$4&  84$\pm$5 & 7$\pm$2\\
\cline{1-11}

West-Up & N &  6.0$\pm$0.5&   7.6$\pm$2.0&   2.3$\pm$1.5 &
2.9$\pm$0.5&
9.0$\pm$0.6&   30$\pm$3&  55$\pm$1&  27$\pm$1 & $<$0.6\\
$\theta$=45, $\Omega=1.2$ sr & $\Phi$ & 36$\pm$4&  13$\pm$3&
6$\pm$4& 15$\pm$3&
38$\pm$3&  53$\pm$4&  60$\pm$1 &  94$\pm$5 & $<$2\\
\cline{1-11}

Up, & N &  8.1$\pm$0.6&   13.0$\pm$1.6&  9.3$\pm$1.4 &
2.3$\pm$0.6&
          9.0$\pm$0.9&   38$\pm$3&  45$\pm$3&  30$\pm$2 & $<$0.7\\
$\theta$=0, $\Omega=1.2$ sr & $\Phi$ & 49$\pm$4&  22$\pm$3&
23$\pm$4& 12$\pm$3&  38$\pm$4&
 67$\pm$4&  49$\pm$3&  105$\pm$8 & $<$2\\
\cline{1-11}

Down,  & N &  4.5$\pm$0.5&   9.4$\pm$1.5&   4.3$\pm$1.2&
1.7$\pm$0.5  & 5.7$\pm$0.7&   26$\pm$2&  38$\pm$3&  26$\pm$1 & $<$0.6\\
$\theta$=180, $\Omega=1.2$ sr & $\Phi$ &  27$\pm$3&  16$\pm$3&
11$\pm$3& 9$\pm$2&   24$\pm$3&
46$\pm$4&  42$\pm$3&  91$\pm$4 & $<$2\\
\cline{1-11}

\hline \hline

\hline

\end{tabular}

\caption{The full energy peak count rates $N$,\,[10$^{-3}
s^{-1}$] of the major gamma rays and their fluxes
$\Phi,\,[m^2\,s]^{-1}$ in Hall A as measured in 2004 and 2007.
The directional measurements were performed in 2004. Intensity
and flux are given with 1$\sigma$ error.}\label{TabFlux}

\end{sidewaystable}

\clearpage

\section{Calculation of the flux from the natural radioactivity in Hall A}

The concentrations of \U,\,\Th\ and \K\ in the Hall A rock
samples have been measured by the Milano group
\cite{Bellotti1985} in 1985. The authors concluded that the
normal rock has very low activity (an order of magnitude lower
than granite rock from the Alps). Infiltrations of black
marnatic rocks have much higher activity, as shown in Table
\ref{Tab:Bellotti}. The high activity areas constitute,
however, only one percent of the total surface of the Hall A
walls. The contamination of concrete at the \gerda\
site was measured before the beginning of \gerda\ construction.\\
\begin{table}[h]
  \centering
\begin{tabular}{|l|c|c|c|}
  \hline
       & \Th\,[Bq/kg] & \U\,[Bq/kg] & \K\,[Bq/kg] \\
  \hline
  Hall A normal rock & $\lesssim$0.5 & $\lesssim$8 & $\lesssim$5 \\
  Hall A infiltration rock (average) & 8.7$\pm$0.4  & 91$\pm$23 & 248$\pm$16\\
  Hall A concrete & 10$\pm$1 & 8$\pm$1 & 90$\pm$10\\
  Hall B rock &  0.25$\pm$0.08&  5.2$\pm$1.3 & 5.1$\pm$1.3 \\
  Hall C rock & 0.27$\pm$0.10 & 8.2$\pm$1.7 & 2.9$\pm$1.4  \\
  \hline
\end{tabular}
  \caption{Activity of rock and concrete in the Gran Sasso
   tunnel \cite{Bellotti1985, Hampel07}.}\label{Tab:Bellotti}
\end{table}
The flux can now be determined independently using equation
\ref{FluxFormula2} for an idealized case of a uniform source
distribution using the measured contamination concentration.
The flux $\Phi$ is expressed as:
\begin{equation}\label{FluxEq}
        \Phi = \frac{n_\gamma}{\mu_E} = \frac{\eta\,A/\rho}{\mu_E}
\end{equation}
where  A\,[Bq\,g$^{-1}]$ is the specific activity, $\eta$ is
the number of photons per decay and $\mu_E$\,[cm$^{-1}$] is
the attenuation coefficient for photons with energy E
($\mu_E$=0.1\,cm$^{-1}$ for E$_\gamma$=2.6\,MeV in LNGS rock).
The density $\rho$ of the LNGS rock is 2.7\,g\,cm$^{-3}$ and
the density of the concrete is 2.3--2.5\,g\,cm$^{-3}$
\cite{NeutronsLNGS2004}. The average activity of the
surrounding material was calculated as a mean of the normal
rock and the new concrete activity. The resulting flux is
presented in Table \ref{FluxFromRock}.
\begin{table}[h]
  \centering
\begin{tabular}{|l|c c|c c c |c|}

\hline

Parent Nuclide & \multicolumn{2}{|c|}{Th-232}  &   \multicolumn{3}{|c|}{U-238} &  {K-40}\\
Average Activity & \multicolumn{2}{|c|}{5\,Bq/kg}  &   \multicolumn{3}{|c|}{8\,Bq/kg} &  {45\,Bq/kg}\\

\hline

Nuclide & \multicolumn{2}{|c|}{Tl-208}  &   \multicolumn{2}{|c }{Bi-214} &   {Pb-214} & {K-40}\\

$E_\gamma$,\,keV & 2614 &583 & 2204 & 609 &351&1460\\

\hline

Calculated flux, $[m^2\,s]^{-1}$ &  450 & 190&
100& 470& 300& 1000\\

Measured flux, (Tab.~\ref{TabFlux})&  362 $\pm$12& 108$\pm$8&
120$\pm$8& 546$\pm$40& 349$\pm$16&921$\pm$50\\
\hline
\end{tabular}
\caption{Calculated and measured in 2007 fluxes of photons
from natural radioactivity in Hall A.}\label{FluxFromRock}
\end{table}
The calculated fluxes are similar to those obtained from
direct measurements, but the accuracy of the calculated flux
is limited by the high variability and uncertainty of the
natural contamination in the rock.\\

\section{Contribution of scattered photons to the total flux}
The analysis of the count rate of full energy peaks has
yielded so far the flux of just the unscattered photons. The
total $\gamma$-flux, which includes also scattered photons
continuum was determined with a Monte-Carlo simulation. The
photons were simulated uniformly in the 0.5 m concrete slab,
and the flux was determined at 5 m from the slab in the air.
The EGSnrc simulation code \cite{EGS} was used. The spectra of
the flux for photons with an initial energy of 609, 1460 and
2614 keV are shown in Fig. \ref{fig:EnergyDistribFlux}. The
fraction of unscattered photons in the total flux is about
30\%. The scattered photons with energies above 2039 keV
contribute to the continuum background around the \qbb\ value
of neutrinoless double beta decay of \Ge. Their fraction is
about 5\% of the total primary flux. The fraction of the
scattered photons will change inside the GERDA shield as was
shown with the detailed MC
calculations of the full GERDA setup \cite{Luciano, Barabanov}.\\

\begin{figure}[h]
  \centering
  \includegraphics[width=14 cm]{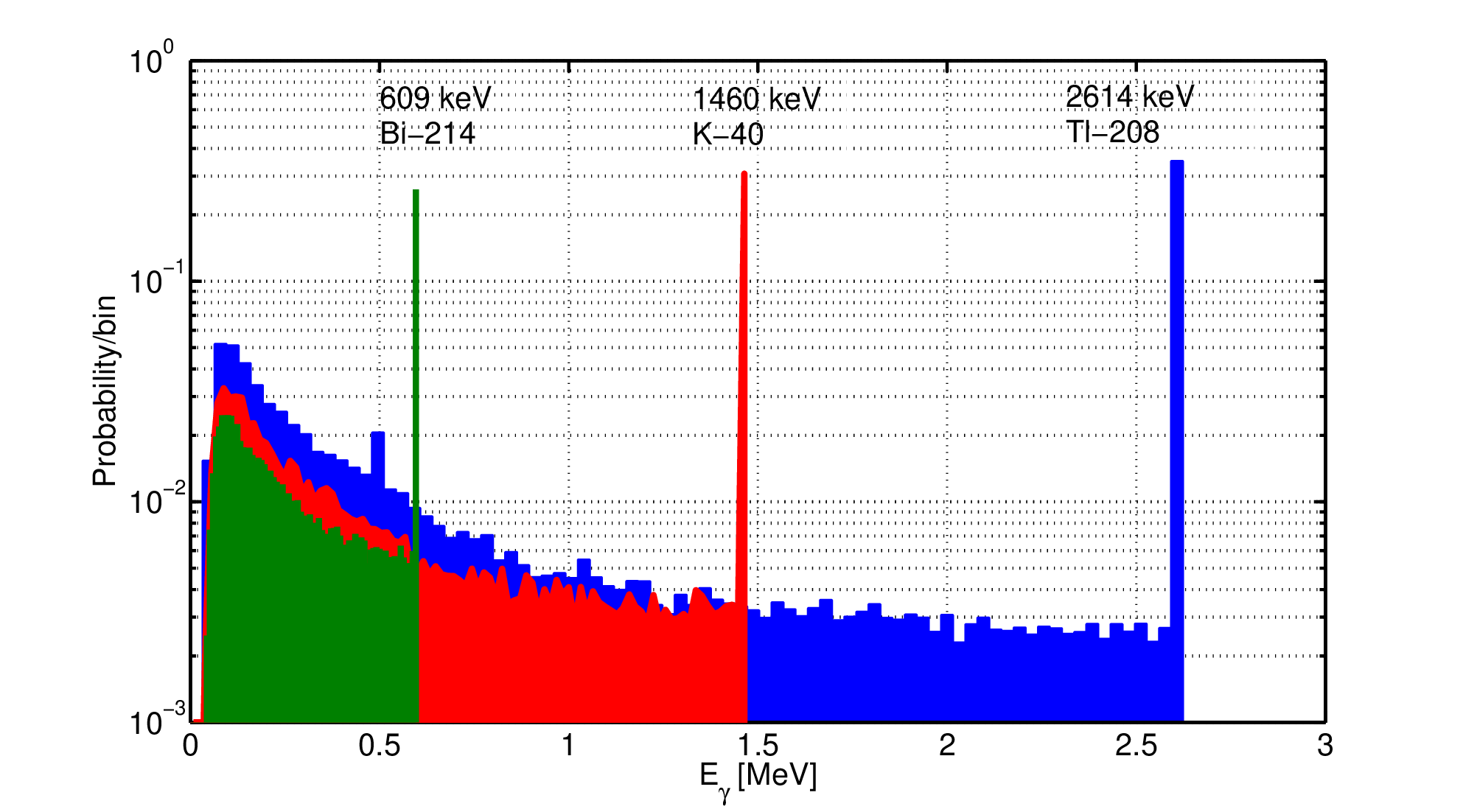}\\
  \caption{Monte Carlo probability distribution of the photon flux
  energy in air at 5 m from a radiative concrete slab
  of 0.5 m thickness for the 609, 1460 and 2614 keV
 primary photons. Each spectrum contains one hundred bins.
  }\label{fig:EnergyDistribFlux}
\end{figure}

\section{Summary}

The shielding needed for the GERDA experiment depends on the
intensity of the external radiation. The $\gamma$-ray flux at
the GERDA experimental site has been determined for the first
time in 2004 with a collimated spectrometer. The flux had to
be remeasured in 2007 after the reconstruction of the Hall A.
An HPGe detector was used to perform these measurements with
and without a collimator. The collimator was used in 2004 for
angular flux distribution measurements. The detector response
was determined with Monte-Carlo simulations. The detector
sensitivity was found to be almost isotropic, which allowed
simple flux calculations.\\
The spectra in Hall A were measured and the fluxes were
calculated. The 2614 keV photon flux is
(362$\pm$12)\,m$^{-2}$s$^{-1}$. The flux of the photons from
the ceiling is two times higher than the flux from the walls
and the floor. It was found that the intensity of the
$\gamma$-flux from thorium was 20\% higher in 2007 than in
2004. The difference could be explained by the absorption in
the massive iron parts
of the OPERA experiment stored there in 2004.\\
The flux was also determined with a simple model of a uniform
source distribution in the rock, and using measured
concentrations of \U,\Th\ and \K. The calculated flux is
similar to the numbers given by the direct measurements, but
the accuracy of the calculated flux is limited by the high
uncertainty on the measured radioactivity concentrations.\\
Based on the flux measured with the collimated and
uncollimated detector, the contribution of the GERDA
background index for the external $\gamma$-radiation has been
calculated as $1.1\cdot10^{-5}$ [keV\,kg\,y]$^{-1}$
\cite{Barabanov}. Details of the external GERDA shielding were
optimized based on these measurements. A lead shield, which
was originally conceived to reduce the radiation from the
concrete floor and ceiling, could for example be omitted.

\newpage
\chapter{Characterization of the HdM and IGEX detectors for GERDA Phase~I} 
\label{ch:detchar}
The enriched diodes from the past HdM and IGEX experiments
were produced in the period from 1990 to 1995. Because the
detector parameters might change with time, all the enriched
detectors have been characterized prior to the refurbishment
for GERDA. Leakage current, counting characteristic and energy
resolution have been measured as a function of the applied
bias voltage. The active mass of the GERDA Phase-I enriched
detectors, which defines the sensitivity of the experiment,
has been determined. This chapter summarizes the operations
and measurements carried out with the enriched detectors in
the GERDA underground detector laboratory (GDL) in preparation
for GERDA Phase~I.\\

\section{Overview}
The five HdM and the three IGEX detectors have been kept
underground at LNGS and at Canfranc laboratory since the end
of the experiments. The detectors were then moved to GDL for
maintenance and characterization. Subsequently the cryostats
were opened and the diodes were taken out. The dimensions and
the masses of the diodes as well as the dimensions of the
cryostats have been measured. The diodes were stored under
vacuum in special containers and were transported to the High
Activity Disposal Experimental Site (HADES), Geel, Belgium, in
which the Institute for Reference Materials and Measurements
(IRMM) operates a laboratory 233 m below ground. In
preparation for \gerda\ Phase~I, the enriched diodes are being
reprocessed to the Canberra type, which have more reliable
contacts than the ORTEC ones. The detectors were reprocessed
at Canberra Semiconductor NV, Olen, near HADES. During the
refurbishment process, the diodes were
stored underground in HADES in between the refurbishment steps.\\
The active mass of a germanium detector is the mass of the
volume within which energy depositions result in detector
signals.  A p-type HPGe detector has the n$^+$-contact on its
outer side, which is a thick ($\sim$1 mm) conductive layer,
insensitive to radiation, also known as the dead layer. This
contact is created by diffusion of lithium on the surface of
the germanium crystal. During extended storage at room
temperature, the lithium diffuses into the germanium,
increasing the dead layer thickness and reducing the active
mass of the detector. Detection efficiency was measured with
sources and the results were compared to Monte-Carlo
simulations. Based on these results, the active masses have
been determined and compared
to the previous values.\\

\section{Operations and measurements}
\subsection{Spectrometry parameters} The leakage current,
the count rate in the full energy peaks and the energy
resolution were measured as a function of the applied bias
voltage (HV). The HdM and IGEX detectors were equipped with
resistive feedback charge sensitive preamplifiers CANBERRA
Model 2002C and Princeton GammaTech Model RG11, respectively.
A simplified scheme of the first stage of the preamplifier
(Fig.~\ref{TPVPreamp}) shows how the leakage current $I_d$ can
be derived from the test point
voltage (TPV) measurements as follows,\\
\begin{equation}
I_d = \frac{TPV(HV)-TPV(0)}{R_f},\\
\label{TPVformula}
\end{equation}
where $R_f$ is the resistivity of the feedback resistor (2 G$\Omega$).\\
\begin{figure}[!h]
\centering
  \includegraphics[width=100 mm]{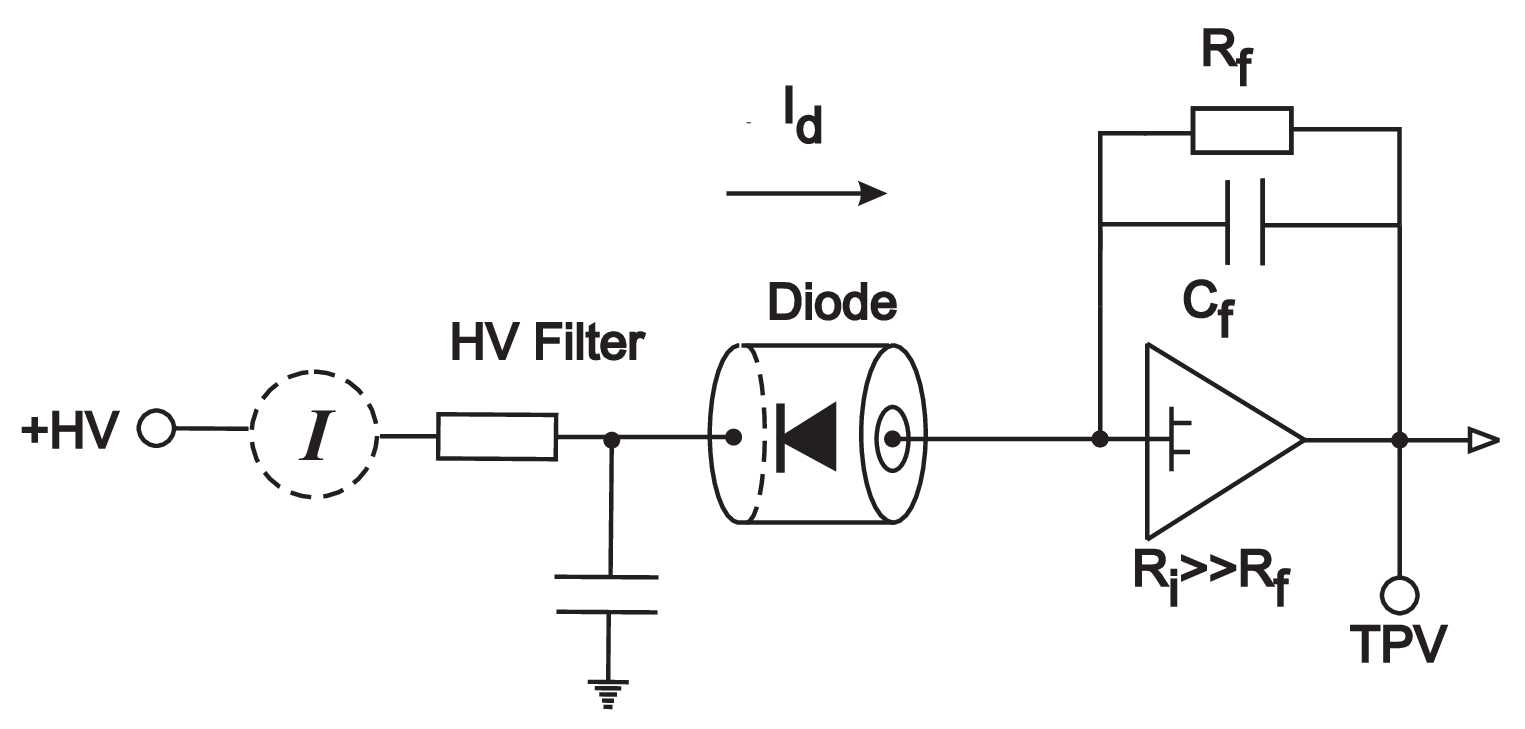}\\
  \caption{A simplified scheme of the diode connection to the first stage of resistive feedback charge sensitive preamplifier.
   The TPV depends linearly on the feedback resistor $R_f$ and the detector current $I_d$. An amperemeter was used optionally.}\label{TPVPreamp}
\end{figure}
\\
The leakage current was also measured directly. This was done
with a precision of $\pm$10 pA using an amperemeter, connected
to the HV side of the diode (see Fig. \ref{TPVPreamp}). The
picoamperemeter was made of a commercial voltmeter placed in a
case with HV isolation from the ground, which was reading the
voltage drop on a 100\,M$\Omega$ resistor. The leakage
currents measured with both methods provided
comparable results, but the TPV measurements are more accurate ($\pm$2 pA).\\
The operational voltage of the detectors has been determined.
The operational voltage is the voltage to be applied to obtain
saturated peak count rate with the best energy resolution. In
order to obtain good energy resolution, the leakage current
must be as low as possible
($\lesssim$100 pA) to minimize the shot noise of the detector.\\
ORTEC model 659 high voltage power supplies, model 672
amplifiers, and model 919 Multichannel Buffer ADC modules were
used for the
measurements. Data were recorded and analyzed with the GAMMA VISION program.\\
\Co\ and \Ba\ sources were used for efficiency and dead layer
measurements. The sources were enclosed in plastic cases of 1
mm thickness. The parameters of the sources are presented in
Table~\ref{CalibSources}.

\begin{table}
  \centering
  \begin{tabular}{|c|c|c|c|}
    \hline
    Calibration & Certified Activity & Activity & Uncertainty \\
    Source &     kBq       &  kBq         &3$\sigma$ \\
           &  (01.06.1990) & (08.02.2005) & \\
\hline
    \Co\ (1173.2, 1332.5\,keV) & 419.9 & 61.44 & 3\% \\
    \Ba\ (80, 356\,keV) & 454 & 172.2 & 5\% \\
    \hline
  \end{tabular}
  \caption{Activities of the \Co\ and \Ba\ calibration sources used for detector characterization.}\label{CalibSources}
\end{table}

\subsection{Using heating and pumping cycles for cryostat
vacuum restoration}

Sometimes an HPGe detector shows high leakage current and
consequently deteriorated energy resolution. The reason is
often poor vacuum in the detector cryostat. In order to
improve the spectroscopic performance of such detectors,
thorough heating and simultaneous pumping of the cryostat is
recommended as a common practice for HPGe detectors. The
pumping station, including a PFEIFFER oil free forepump and a
turbo pump, was assembled and tested. The heating was
performed using boiling water. This procedure allows to keep
the temperature of the cryostat constant, while avoiding its
overheating. Several heating-pumping cycles have been
performed with all IGEX detectors and the ANG1 detector. The
dipsticks of the IGEX cryostats were put into the boiling
water. The cryostat of the ANG1 detector was heated with
boiling water inside the dewar. The heating process was
continued until the vacuum stabilized at a pressure of
typically 10$^{-6}$ mbar. After each pumping-heating cycle,
the detectors were cooled with liquid nitrogen for at least
two days before measuring the detector parameters.

\section{HdM detectors}

The original parameters of the HdM enriched detectors are presented
in Table \ref{tab:HdMhist}. The detectors
were produced by EG\&G ORTEC, Tennessee, USA.
They are named historically as ANG -- "angereichert" 
-- and numbered from 1~to~5. From 1990 to 1995, the five
detectors were installed in the HdM experiment setup at LNGS.
Between 1995 and 2006 all HdM detectors stayed underground.

\begin{table}[!ht]
\centering
\small
\begin{tabular}{|l|c|c|c|c|c|}
\hline
  & ANG1 & ANG2 & ANG3 & ANG4 & ANG5\\
 \hline
 Crystal grown at & 11.04.90 & 07.02.91 & 21.03.91 & 11.11.93 & 06.10.93\\
 EG$\&$G ORTEC & & & & &\\
 Diode completed & 12.07.90 $^*$ & 13.02.91 & 19.07.91& 12.01.94& 10.11.93\\
 Serial Number & b8902 & P40239A & P40270A & P40368A & P40496A \\
 Diameter, mm & 58 & 80 & 78.5& 75&78.8\\
 Length, mm &  68 &  108 &  93.5& 100.5& 105.7\\
 Hole Diam., mm & 8 & 8 & 9& 8&8\\
 Hole Length, mm & 43 &  94 &  81.5& 88.9& 93.5\\
 Total Mass, g & 980 & 2905 & 2447 & 2400&2781\\
 Enrichment \Ge & 85.9$\pm$1.3\% $\!\!$& 86.6$\pm$2.5\% $\!\!$& 88.3$\pm$2.6\% $\!\!$& 86.3$\pm$1.3\% $\!\!$&85.6$\pm$1.3\% $\!\!$\\
 FWHM (1332 keV)$\!\!$& 1.99 keV & 1.99 keV & 1.99 keV &1.99 keV &2.29 keV\\
 Operating Bias, V & +4000 & +4000 & +3500& +3000& +2500 \\
 Leakage Current & no info &  138 pA & 7 pA &301 pA &8 pA\\
 Impurities/cm$^3$ & no info & no info & 0.5$\div1.3\cdot10^{10}$ &4.1$\div9.8\cdot10^{9}$&1.4$\cdot10^{9}$\\
 Installed in LNGS&07.90; 01.91& 09.91 & 08.92 & 01.95 & 12.94\\
 Moved to GDL & 29.09.04 & 29.09.04 & 29.09.04&29.09.04&29.09.04\\
  \hline
 \end{tabular}
\caption{History and original parameters of the HdM detectors
prior to the handover to the GERDA collaboration
\cite{Dipl-Echt,DissMuller,Diss-Dipl-Maier,Diss-Dipl-Petry,Diss-Dipl-Helmig}.
(*) Detector ANG1 was completed at Canberra Semiconductor NV.}
\label{tab:HdMhist}
\end{table}
All diodes were transported from USA to Germany by ship to
minimize the cosmic ray exposure. ANG1 crystal ingot was
pulled in April, 1990. After two trials, the ANG1 diode was
completed at Canberra Semiconductor NV, in July, 1990. In both
trials the detector was transported by plane. Total
irradiation time in flight was 24 hours \cite{DissZuber}. The
measured ANG1 diode mass in GDL is 968.7~g, which is 11.3~g
less then was reported by ORTEC (Table \ref{tab:HdMhist}). The
difference is due to the additional machining of the diode at
Canberra Semiconductor NV in 1991 after the damage of the
detector cryostat \cite{DissMuller}. The ANG2 diode mass
measured in GDL is 2878.3~g, which is 26.7~g less then was
reported by ORTEC (Table \ref{tab:HdMhist}). The reason for
this difference is unknown. All the other detectors
have masses within one gram of the ones reported by ORTEC.\\
On September 29, 2004, the enriched detectors were handed over
from HdM to GERDA and transferred to the LUNA1 underground
experimental site at LNGS, where their spectrometric
parameters and leakage currents have been measured. Unlike the
other HdM detectors, ANG1 was handed over warm, without liquid
nitrogen in its cryostat dewar. After cooling down, ANG1
showed a high leakage current ($\sim$0.8 nA) and deteriorated
energy resolution ($\sim$4 keV FWHM at 1332 keV). Normal
resolution and leakage current were measured with
ANG2-5($\sim$2--3 keV). The ANG2 detector showed sporadic
jumps of the full energy peak position of $\pm$10 keV. During
the period from December 2004 to November 2005, the detectors
were moved three times between the LUNA1 and the LENS sites
because of the transformation of the LENS site into the GDL.
After each relocation of the detectors, measurements of their
spectrometric parameters were repeated. As of February 2005,
the HdM detectors were permanently located in the GDL. The
instability of the peak position of ANG2 was not observable
anymore.
\begin{figure}[!ht] \centering
  \includegraphics[width=80mm]{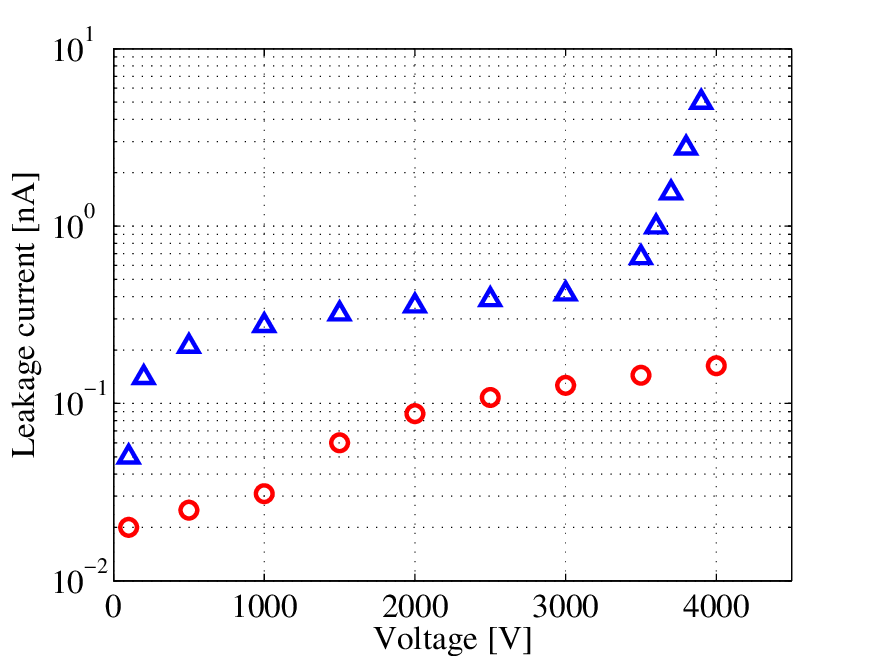}
  \caption{Leakage current for ANG1 detector before ($\vartriangle$) and
   after ($\circ$) heating and pumping procedure.}\label{ANG1rest}
\end{figure}
To reduce the leakage current of ANG1, a heating-pumping cycle
was performed. After 10 hours of heating and 24 hours of
pumping, the leakage current decreased as shown in
Fig.~\ref{ANG1rest} and an energy resolution of 2.5 keV FWHM
at 1332 keV has been obtained. In April 2006, ANG1 diode was
dismounted from its cryostat, its dimensions and mass were
measured and it was stored in a special container.
\begin{table}[!h]
\centering
\begin{tabular}{|c|c|c|c|c|c|}
\hline
Date & 1995  & 30.09.04 & 12.02.05 & 18.12.05 \\ 
Location & HdM setup \cite{Diss-Dipl-Helmig} & LUNA1 & LENS(GDL) & GDL\\
Detector, FWHM & [keV] & [keV] & [keV] & [keV] \\ 
\hline
ANG1 & \bf 2.2   & 4.0 (3000 V) & 2.7 (3000 V) & \bf 2.5 \\ 
ANG2 & \bf 2.4   & 2.4  & 2.3  & \bf 2.3 \\ 
ANG3 & \bf 2.7   & 3.1  & 2.5  & \bf 2.9 \\ 
ANG4 & \bf 2.2   & 2.4  & 2.5  & \bf 2.5 \\ 
ANG5 & \bf 2.6   & 2.9  & 2.8  & \bf 2.6 \\ 
\hline
\end{tabular}
\caption{The chronology of the energy resolution (FWHM) at
1332 keV of the HdM detectors measured prior to their
refurbishment at Canberra Semiconductor NV. The measurements
were performed at the operating voltage
(Tab.~\ref{tab:HdMhist}) at different locations at LNGS. The
standard error of FWHM is 5\%.}\label{hdmores}
\end{table}
Then, on August 18, ANG1 was tested in the radon free test
bench of GDL. It was mounted in the new low mass holder and
inserted in liquid argon. The leakage current, $\sim$1 nA at
3000 V, increased up to 3 nA within two days. An energy
resolution (FWHM) at 1332 keV of 3.76 keV was measured at
reduced (3000 V) bias. The operational voltage (4000 V) was
not reached because of a steep increase of the leakage
current. Finally, on August 21, ANG1 was transported in 26
hours by a courier car to HADES underground laboratory for
the refurbishment at Canberra Semiconductor NV.\\
In November 2006, the spectroscopic parameters of ANG2,3,4 and
5 detectors were measured for the last time before the opening
of the cryostats. Table~\ref{hdmores} summarizes the energy
resolution FWHM history of the HdM detectors.
Figure~\ref{fig:HdMcharact} shows the detector leakage
currents, the 1332 keV peak count rates and the energy
resolutions as a function of the applied bias for ANG1-5
detectors. The peak count rate is normalized to one. The
detectors leakage currents at the operational bias ranged from
0.2 to 1.2 nA and FWHM ranged from 2.3 to 2.9 keV. As
expected, the high leakage current ($\sim$1 nA) of the ANG3
detector causes an increase of FWHM due to the higher shot
noise contribution.

\begin{figure}[!ht]
\centering
  \includegraphics[width=69 mm]{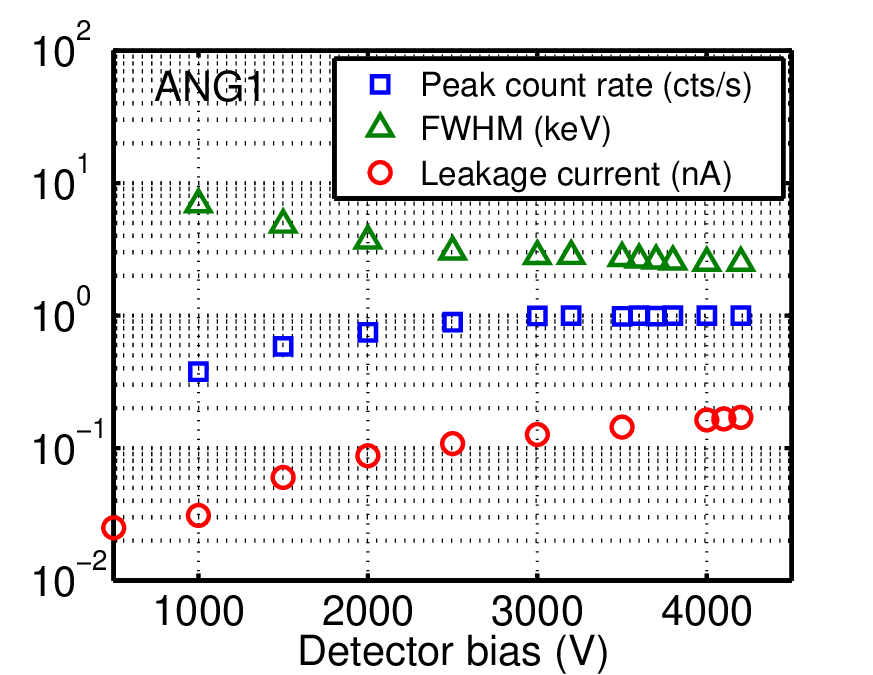}
  \includegraphics[width=69 mm]{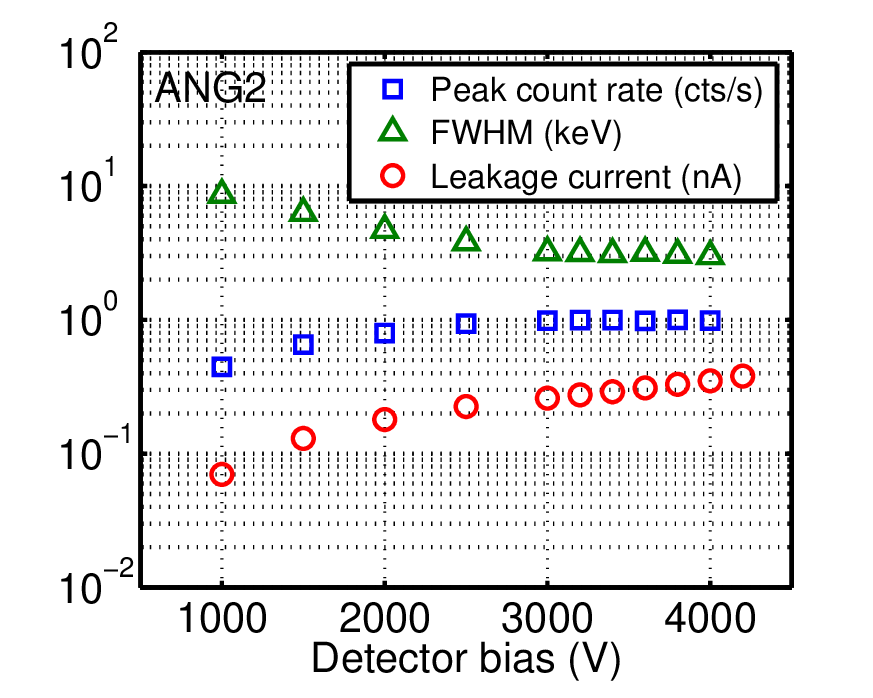}
  \includegraphics[width=69 mm]{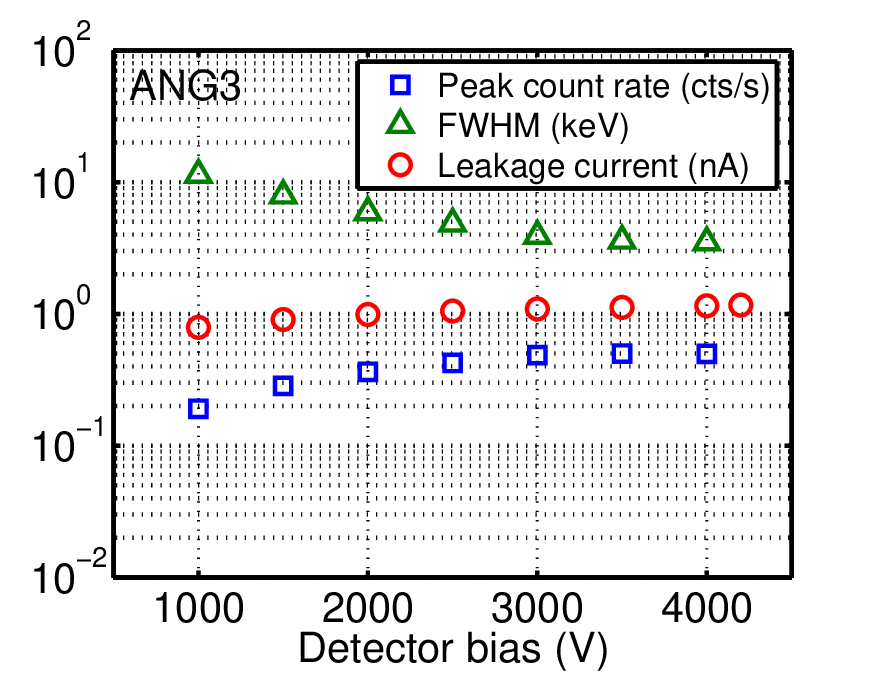}
  \includegraphics[width=69 mm]{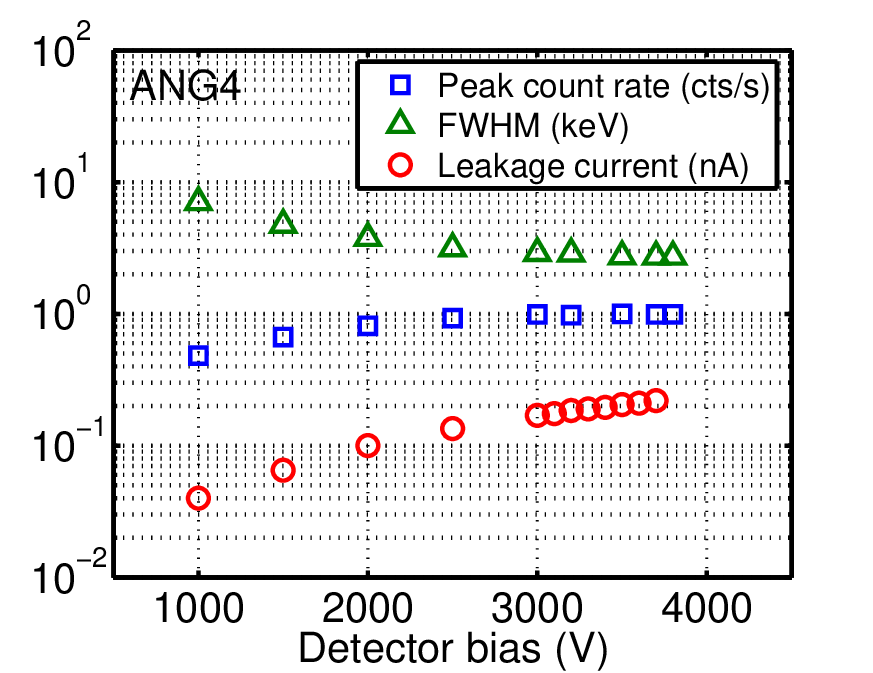}
  \includegraphics[width=70 mm]{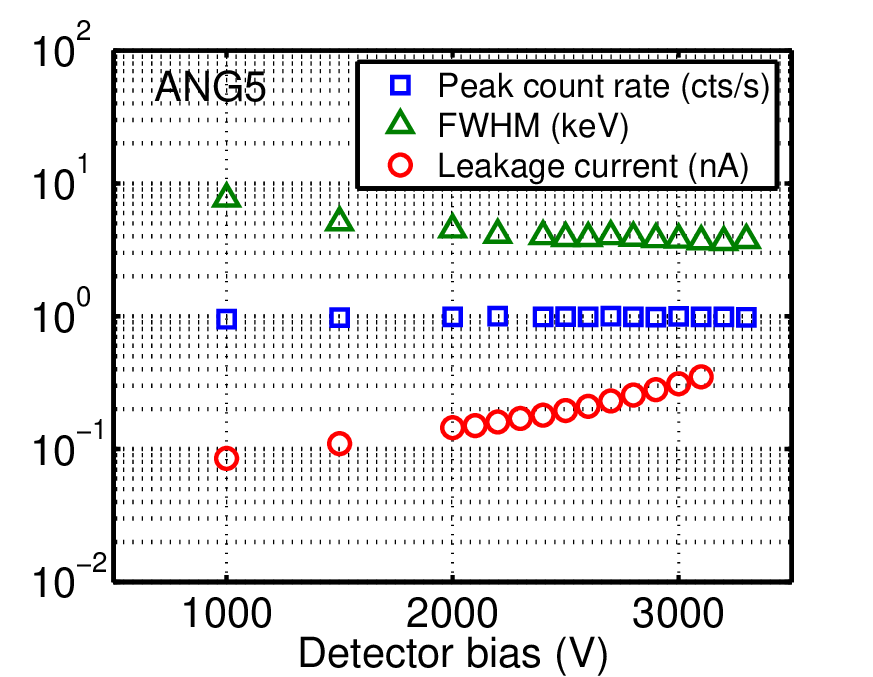}
  \caption{The leakage current, the energy resolution (FWHM) and the count rate
   of 1332 keV peak of the HdM detectors vs. HV bias. Measurements were
  performed in GDL at LNGS.}\label{fig:HdMcharact}
\end{figure}

\clearpage

\section{IGEX detectors}
Three enriched IGEX detectors were transported in 19 hours by
a courier car from Canfranc underground laboratory, Spain, to
the LNGS on November 18, 2005. The activation of the germanium
crystals by cosmic rays was negligible (the cosmic ray
activation rate of Ge at sea level is $\sim$1 nucleus/day/kg
\cite{MileyAct,AvignoneAct}). The detectors were stored
underground in the GDL, immediately after arriving at the
LNGS. The history and the specification of the IGEX detectors
are presented in Table~\ref{tab:IGEXhist}.
\begin{table}[!ht]
\centering
\begin{tabular}{|l|c|c|c|}
\hline
 & RG1 & RG2 & RG3\\
 \hline
\small Crystal grown at Oxford Inc., Oak Ridge, USA & 25.09.93 & 16.02.94 & 01.12.94\\
Detector completed & 02.11.93 & 10.05.94 & 15.04.95\\
Serial Number & 28005-S & 28006-S & 28007-S \\
Diameter and Length, mm & 77.6; 84.3 & 78.6; 84.0 & 79.2; 82.5\\
Total Mass, g & 2149.9 & 2194.0 & 2121.0\\
Dead Layer, $\mu$m & $\sim$800 & $\sim$800 & $\sim$500\\
FWHM at 1332 keV, keV & 2.16 & 2.37 & 2.13\\
Operating Voltage, V & +5000 & +4000 & +3800 \\
Installed in Homestake & 09.11.93 & 22.05.94 & N.A.\\
Removed from Homestake & 15.06.97 & 27.12.96 & N.A.\\
Installed in Canfranc & 15.07.97 & 25.01.97 & 10.05.95\\
Moved from Canfranc to GDL at LNGS & 18.11.05 & 18.11.05 & 18.11.05\\
  \hline
 \end{tabular}
\caption{History and specifications of the three IGEX detectors
\cite{AvignoneDet}.}\label{tab:IGEXhist}
\end{table}

The IGEX detectors were stored without cooling for a prolonged
period of time ($\sim$2 years) in the Canfranc underground
laboratory. The vacuum levels in the cryostats were unknown.
Therefore, before cooling the detectors at LNGS, a thorough
heating and simultaneous pumping of the cryostats was needed.
After the first pumping-heating cycle, the vacuum in the
cryostats of RG1 and RG3 was stable at 4$\cdot$10$^{-6}$ mbar
and the detectors showed good energy resolution. On the other
hand, RG2 detector showed degraded energy resolution ($\sim$10
keV FWHM at 1332 keV). Using a PFEIFFER helium leak detector,
a leak through the cryostat's cap flange was found. After
tightening the copper bolts on the flange and performing a
heating-pumping cycle, the vacuum became stable at
5$\cdot$10$^{-6}$ mbar. Table \ref{igexres} summarizes the
chronology of the energy resolution at 1332 keV for the IGEX
detectors. An improvement of the energy resolutions after the
heating-pumping cycles was obtained. The leakage current of
RG2 was significantly higher than in RG1 and RG3, however the
resolution is practically the same within $\pm$0.1\,keV.
Finally, the energy resolutions of all IGEX detectors were
restored close to their original values.
\begin{table}[!ht]
\centering
\begin{tabular}{|c|c|c|c|c|c|c|}
\hline
    & Operating & 1993-94 & 25.11.05 & 04.12.05 & 10.12.05 & 24.12.05\\
Detector, FWHM& bias\,[V] & [keV] & [keV] & [keV] & [keV] & [keV] \\
\hline
RG1 & 4800 & \bf 2.16 &  2.6 & 2.34 & 2.38 & \bf 2.21 \\
RG2 & 4000 & \bf 2.37 &  10  & 7.94 & 3.01 & \bf 2.31 \\
RG3 & 3800 & \bf 2.13 &  4.7 & 2.74 & 3.9  & \bf 2.26 \\
\hline
\end{tabular}
\caption{The chronology of the energy resolution (FWHM) at 1332 keV
of IGEX detectors prior to their dismounting from
cryostats.}\label{igexres}
\end{table}
Figure \ref{fig:RGcurv1} presents the leakage current, the energy
resolution and the count rate of IGEX detectors as a function of the
applied bias voltage.

\begin{figure}[!h]
\centering
  \includegraphics[width=69mm]{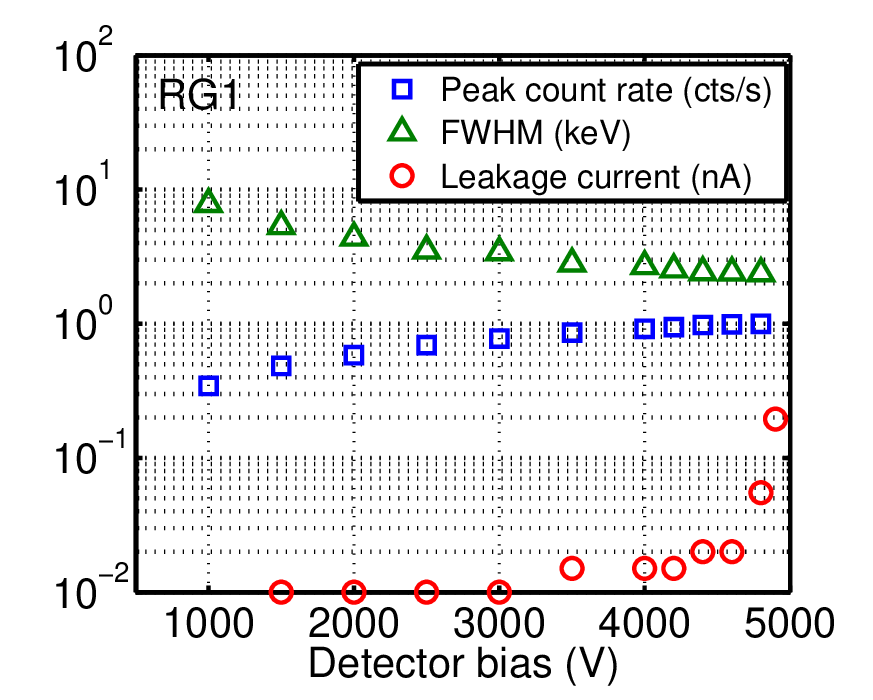}
  \includegraphics[width=69mm]{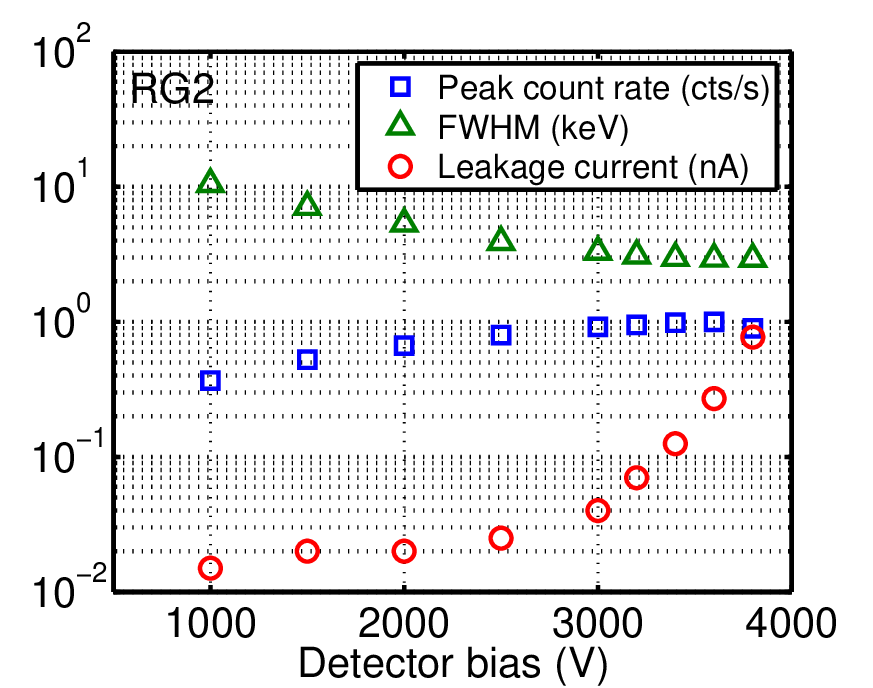}
  \includegraphics[width=69mm]{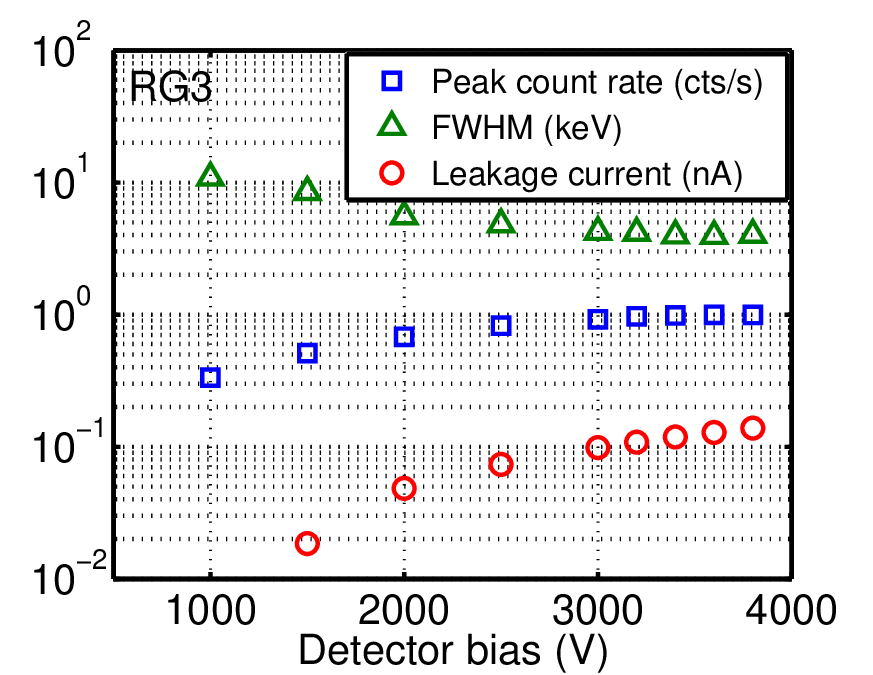}
  \caption{The leakage current, the energy resolution (FWHM) and the count rate
   of 1332 keV peak for the IGEX detectors vs. HV bias measured in GDL, LNGS, 10.12.05.}\label{fig:RGcurv1}
\end{figure}

\section{Dismounting of diodes and dimension measurements} From
April 2006 to November 2006, the eight enriched diodes were
dismounted from their cryostats in the clean room environment of the
GDL. The dimensions of the diodes were measured (see
Fig.~\ref{DetMeas}) using a sliding gauge made of plastic to avoid
scratches of the crystal surface during measurements. The accuracy
of the dimension measurements is $\pm$0.1 mm. The borehole is
covered with a thin boron implantation layer. Because of the
sensitivity of this layer, the diameter and the length of the well
were measured approximately, without touching the surface.
Therefore, the accuracy of the hole measurement is within $\pm$1 mm.\\
\begin{figure}[!h]
\centering
  \includegraphics[width=140 mm]{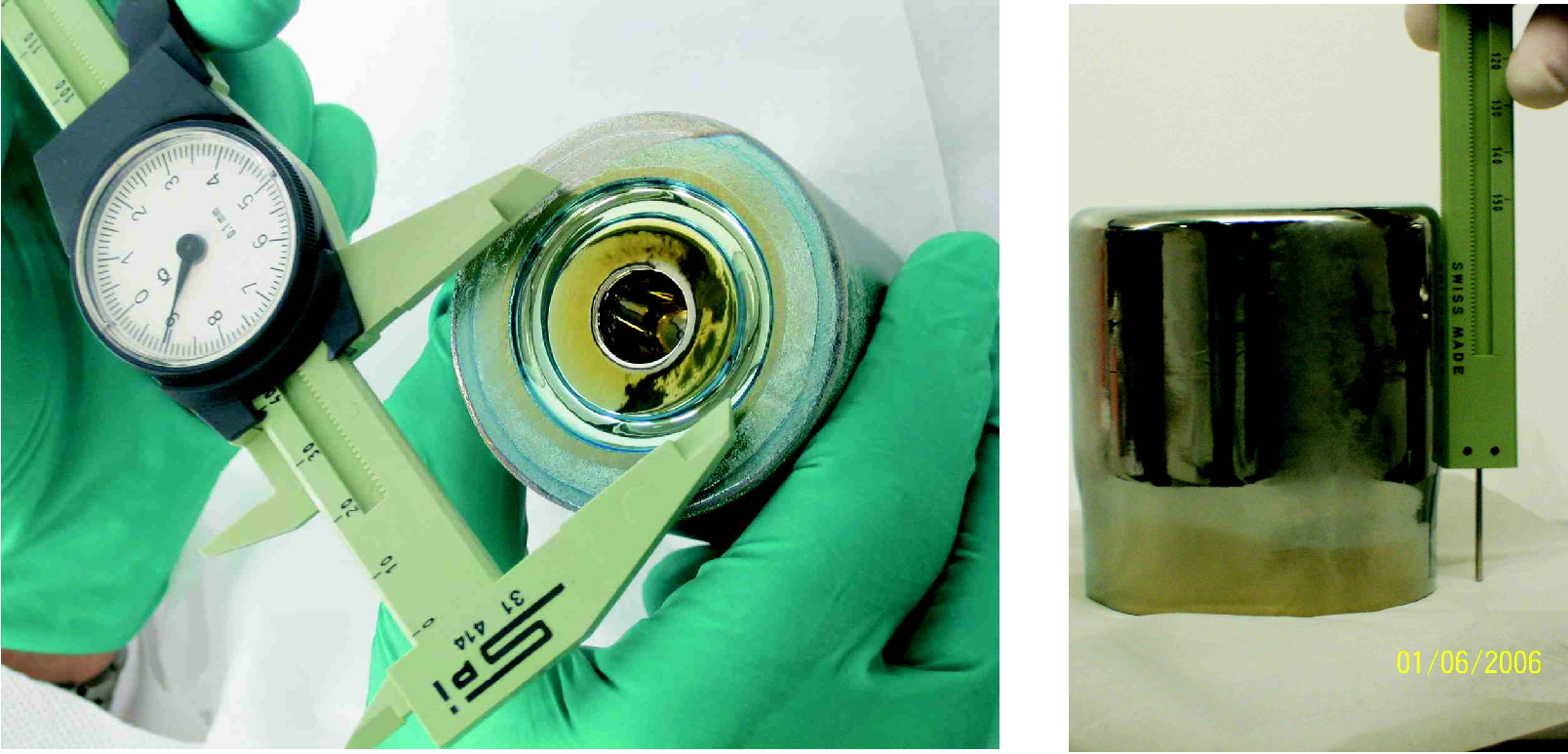}
  \caption{The enriched detectors ANG1\,(left) and RG3\,(right) after opening of the cryostats and detectors holders.
  Dimensions measurements were performed on the Class 10 clean bench in GDL at LNGS.}\label{DetMeas}
\end{figure}
Table~\ref{DimDetPar} gives the dimensions and the masses of
the detectors. Each HdM and IGEX crystal has a specific shape
which is slightly different from a cylinder. In addition, the
cryostats  have been measured (Fig.~\ref{cryostats}). The
dimensions of the cryostats were verified with previous
drawings and then used for a Monte Carlo simulation.
\begin{figure}[!h]
\centering
  \includegraphics[width=140 mm]{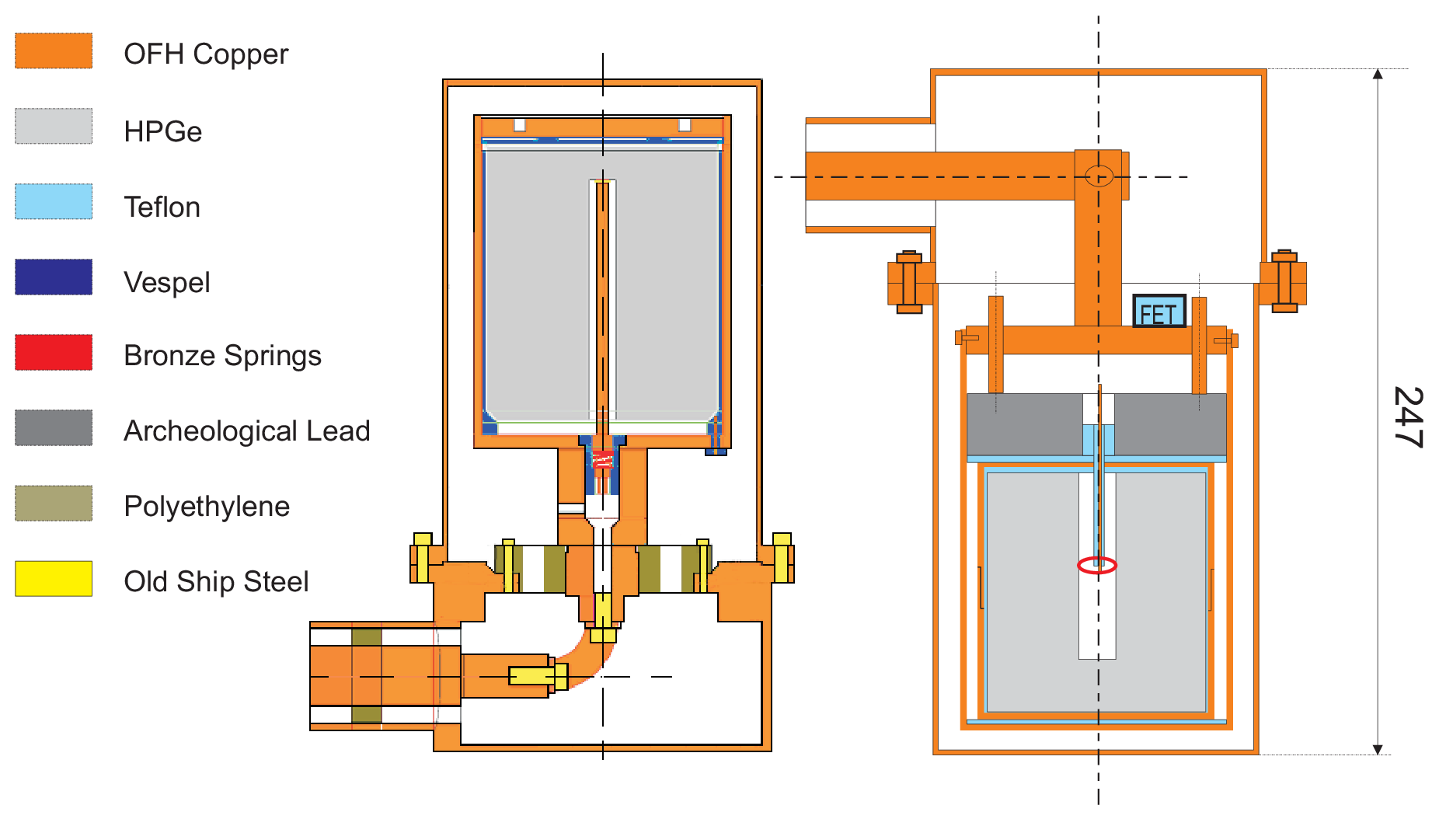}
  \caption{Scaled schematic diagram of the HdM(left) and IGEX(right) cryostat heads
   and detector holders.}
  \label{cryostats}
\end{figure}

\begin{table}[!h]
\centering
\small
\begin{tabular}{|c| c |c| c| c| c| c| c| c| c|}
  \hline
  Detector & Mass & Mass & Crystal & Crystal &  Hole &  Hole \\
            & Measured & Manufacturer & Diameter & Length & Diameter & Depth\\
             & [g] & [g] &[mm]    &   [mm] &  [mm] & [mm]\\
  \hline
   ANG1 &968.7  & 980  & 58.3 & 68.2 & 11 & 51 \\
   ANG2 &2878.3 & 2905 & 79.3 &107.4 & 10 & 93 \\
   ANG3 &2446.5 & 2447 & 78.1 & 93.3 & 11 & 81 \\
   ANG4 &2401.2 & 2400 & 75.0 &100.3 & 11 & 88 \\
   ANG5 &2782.1 & 2781 & 78.3 & 105.5& 10 & 93 \\
    RG1 &2152.3 & 2149.9 &78.5 & 84.0& 12 & 32 \\
    RG2 &2194.2 & 2194.0 &78.5 &84.5 & 10 & 74 \\
    RG3 &2120.9 & 2121.0 &79.8 &82.6 & 12 & 77 \\
 \hline
\end{tabular}
\caption{Masses and dimensions of the HdM and IGEX crystals
measured after the dismounting from their cryostats in 2006.
The error in the masses is 0.1 g. Diameter and length of
crystals are measured with an accuracy 0.1 mm. The holes were
measured with an uncertainty 1 mm.} \label{DimDetPar}
\end{table}
Finally, crystals were stored in stainless steel containers under
vacuum and transported to HADES for their refurbishment at Canberra
Semiconductor NV.

\section{Active mass determination}

\subsection{Motivation and method}
The active mass of the HPGe detector is affected by the dead
layer and incomplete charge collection. The latter is assumed
to be negligible in a fully depleted germanium detector. The
HdM and IGEX enriched detectors are fully depleted at their
operational biases. Because the IGEX and some of HdM detectors
have been stored for a prolonged period without cooling, their
dead layers have likely increased. Therefore, measurements of
detection efficiency and dead layer thickness were performed.
The results were compared with the results of Monte Carlo
simulations and the active masses and dead layers were
determined. Another motivation for active mass study was the
discrepancies between the measured and previously reported
detector masses for ANG1 and ANG2 diodes (Table
\ref{DimDetPar}).

\subsection{Experimental setup and measurements}
The active masses of the HPGe detectors were determined using
measurements with \Co\ and \Ba\ sources. One detector (RG1)
was scanned with a collimated \Ba\ source to check variation
of the dead layer. The dimensions of the cryostat parts and
their relative positions were measured with an accuracy of
0.05 mm and 0.5 mm respectively, giving a precision of
$\pm$0.1 mm for the dead layer thickness determination. To
measure their performance parameters and the background, the
HdM detectors were placed in a lead shield as shown in
Fig.~\ref{fig:SetupWithSources}. The measurements with the
IGEX detectors were performed without a shield. The \Co\
source was placed $\sim$20--25 cm away from the detector
copper cap. The time normalized background spectra were
subtracted from the spectra measured with the source for each
detector. The intensity of the summation peak at 2505 keV was
0.07\% of the 1173 and the 1332 keV peak intensities. The
statistical uncertainty of the measurements was better than
0.5\%. The systematic uncertainty of the source activity was
3\% (3$\sigma$). The activity of the \Co\ source was also
verified using a \Ra\ source with 3\% (3$\sigma$) uncertainty.
\begin{figure}
  \includegraphics[width=140 mm]{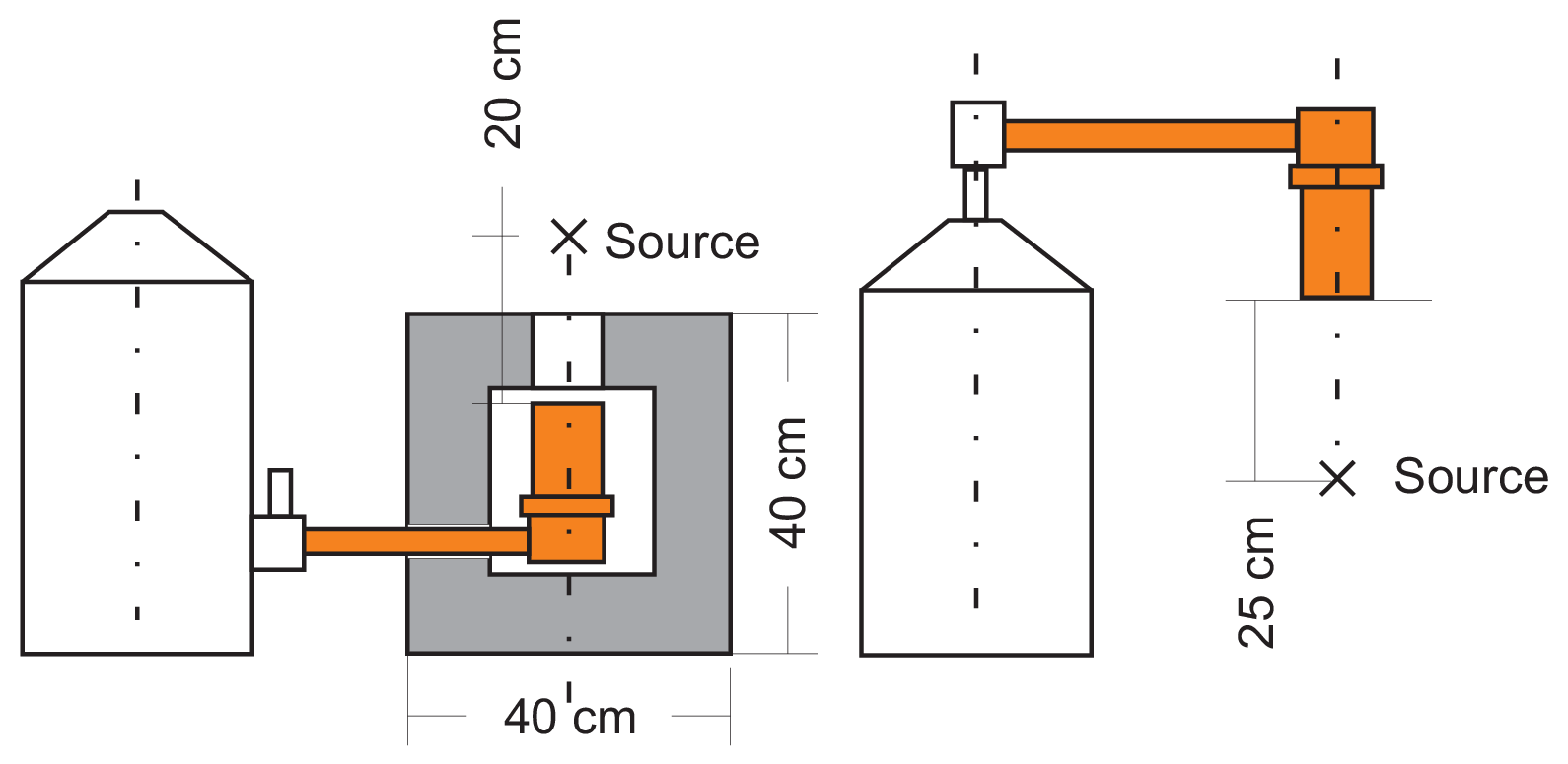}\\
  \caption{The HdM (left) and IGEX (right) detectors and the source setups.
  The HdM detectors were shielded with 10 cm of lead with a window for the
  source.}\label{fig:SetupWithSources}
\end{figure}
The variations of the dead layers thickness were measured with
a collimated \Ba\ source. The 81 keV gamma-rays were
collimated by a 30 mm thick copper collimator with a 2 mm
diameter hole. The collimator was moved along the RG1 cap
diameter in 5 mm steps. The setup and the count rate profile
of the 81~keV \Ba\ line are shown in Fig.~\ref{ScanBaFig}. To
determine the average dead layer thickness the measurements
with the \Ba\ source were carried out without a collimator.
The \Ba\ source was placed $\sim$20 cm from the center of the
cryostat copper cap (Fig.~\ref{fig:SetupWithSources}). The
\Ba\ 81\,keV peak was measured with an accuracy of 10\%. A
simulation was performed for several values of the dead layer
thickness with EGSnrc code. The measured and simulated \Ba\
spectra are presented in Fig.~\ref{SpectraCoBa}.

\clearpage

\subsection{Monte Carlo simulation}

Monte Carlo simulations were performed using two independent
MC codes:  MaGe framework \cite{MaGe} based on Geant4
(ver.~4.6.2) and EGSnrc simulation code \cite{EGS}. Detailed
geometry models were developed for the HdM and IGEX detectors
based on the measurement of the crystals and cryostats
dimensions. The cryostat schematic drawings are presented in
Fig.~\ref{cryostats}. The dead layer thickness was a parameter
of the simulation geometry. The Monte Carlo simulation results
were verified by comparing the results of two codes for two
detectors, ANG3 and ANG5 (Tab.~\ref{ActMassTab1}). The angular
correlation of the \Co\ photons was not implemented in the
Geant4 and EGSnrc codes. The angular correlation can
contribute up to $10\%$ of the summation effect, which itself
was measured to be as small as 0.07\% of the 1332 keV peak.
The results of the simulation and the measurement for the
summation peak agree within statistical errors. The error in
the summation peak is too big ($>10\%$) to observe the angular
correlation effect. The simulations with Geant4 were performed
without including the lead shield in the geometry model.
Therefore, the backscattering peak in the simulation spectra
was absent. The lack of backscattering in the shield also
affects the low energy part of the spectrum ($<300$ keV), as
seen in Fig.~\ref{SpectraCoBa}.

\subsection{Results and comparisons with MC simulation}

Table~\ref{ActMassTab1} presents the simulated and measured
efficiencies for the \Co\ 1332 keV peak and for the integral
count rate in energy range 400-1100 keV. The simulations were
performed using the dead layer thicknesses provided by the
manufacturer. The results of the simulation and the
measurement agree within 3-6\% for ANG1-4 and RG2. The results
of simulations for ANG5, RG1,
and RG3 are $\sim$10\% higher than the measurements.\\
Figure \ref{SpectraCoBa} shows the \Co\ and \Ba\ spectra
measured with ANG3 detector in the lead shield. As expected,
the backscattering peaks at 216 keV and 149 keV from \Co\ and
\Ba\ respectively are not observed in the simulations. The
scan of RG1 with the collimated \Ba\ source
(Fig.~\ref{ScanBaFig}) shows that variations of the dead layer
thickness are within 10\%.\\

\begin{figure}
\centering
  \includegraphics[width=14 cm]{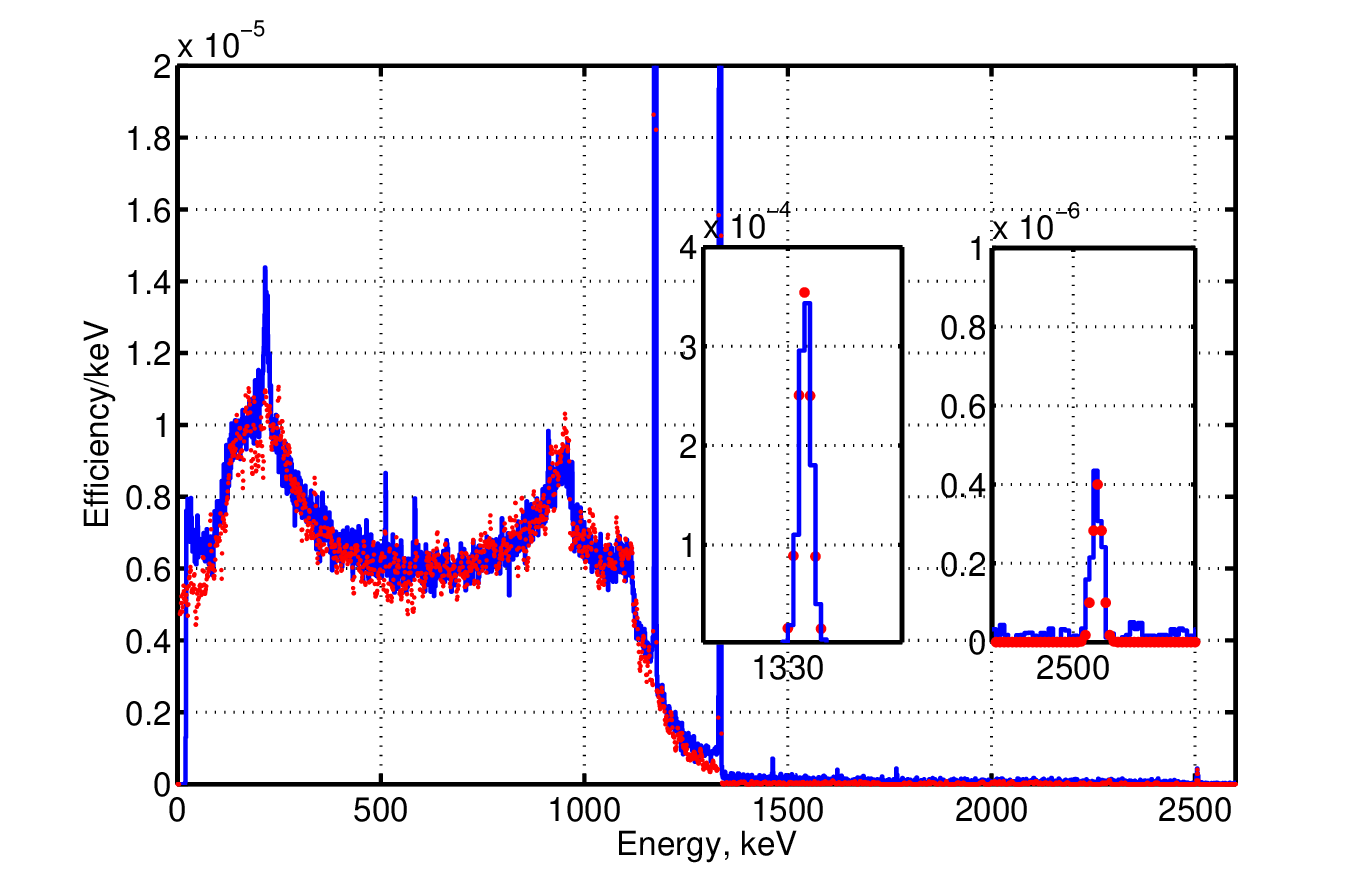}
  \includegraphics[width=14 cm]{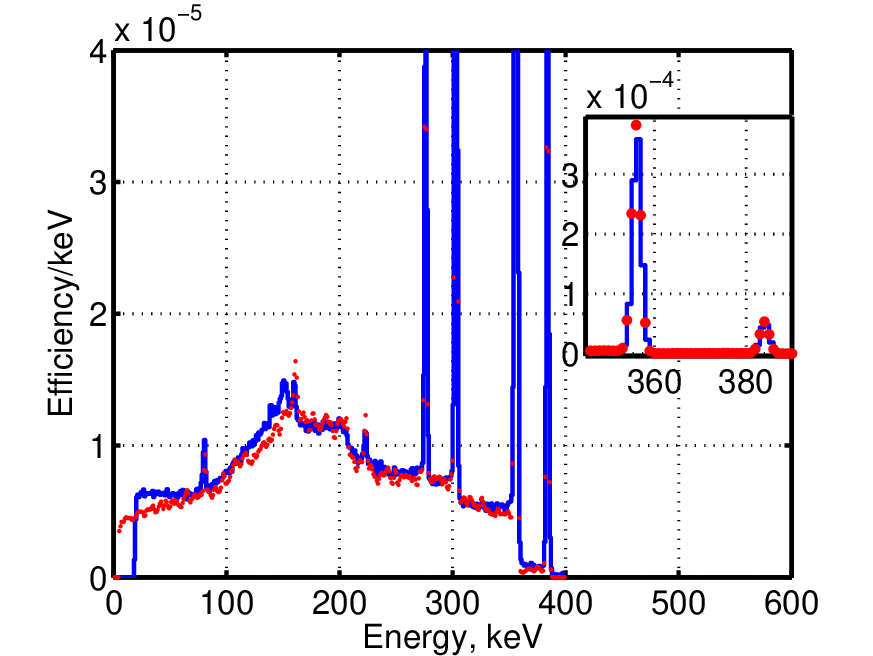}\\
  \caption{\Co\ (top) and \Ba\ (bottom) spectra measured with the ANG3
detector in the lead shield (solid line) and Monte Carlo
simulation (red dots). The MC simulations were performed
without a lead shield, therefore backscattering peaks at 216
and 149\,keV from \Co\ and \Ba\, respectively are not observed
in the simulation spectra.}\label{SpectraCoBa}
\end{figure}

\begin{figure}
\centering
  \includegraphics[width=140mm]{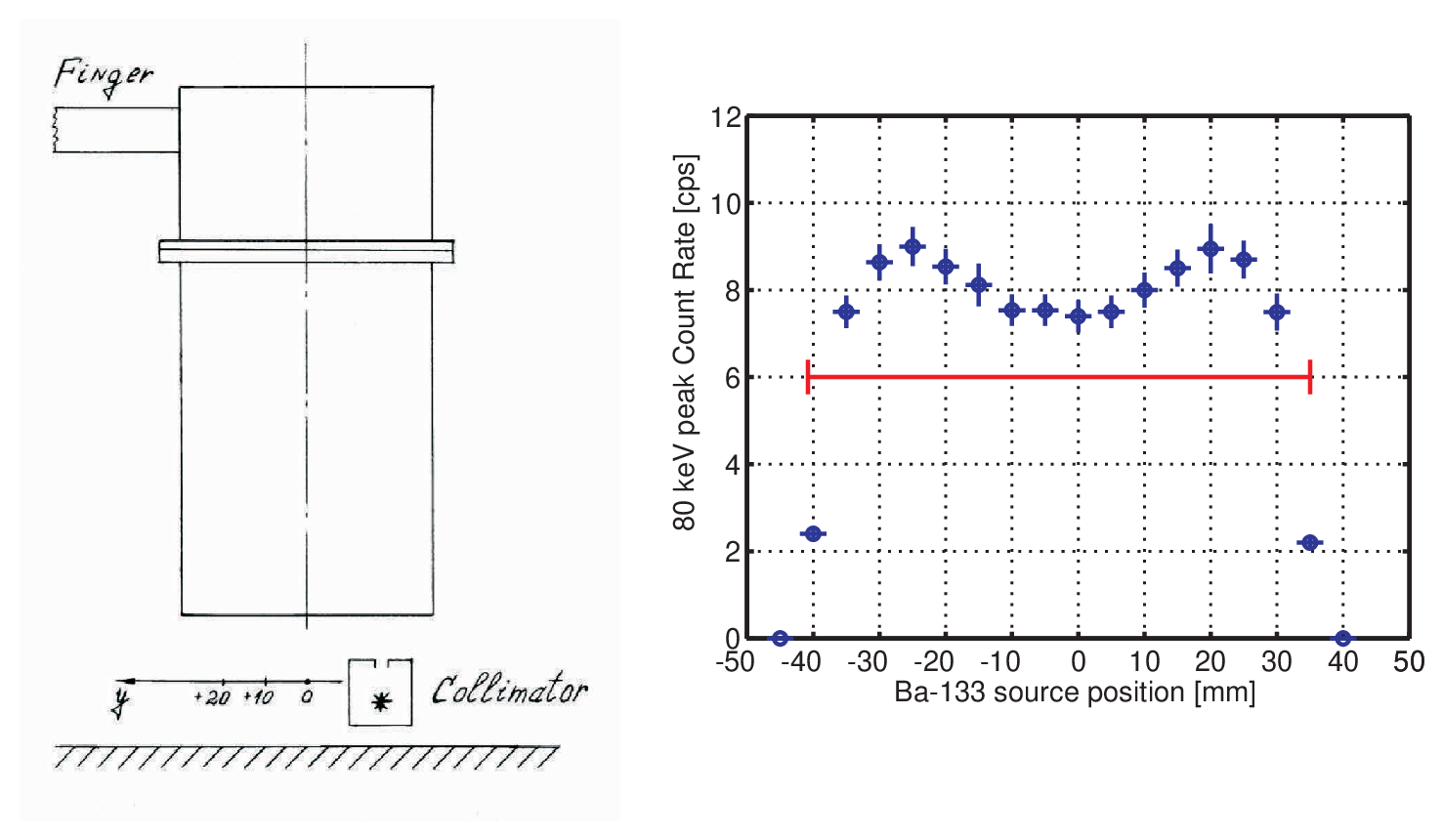}\\
  \caption{Left: schematic diagram of the setup for the dead layer
measurements with a collimated source. Right: 81\,keV peak
count rate measured along the RG1 detector diameter with the
collimated \Ba\ source. The crystal diameter is shown below
the data points.}\label{ScanBaFig}
\end{figure}

\begin{sidewaystable}
\centering \small
\begin{tabular}{|c|c|c|c|c|c|c|c|c|c|c|}
  \hline
  Detector & Dead layer,& Distance,  &  Meas. Eff. & Simul. Eff. & $\delta$,  & Meas. Eff.& Simul. Eff. & $\delta$, & Simulation & Active\\
           & manufac. & to source & \Co\ peak &\Co\ peak &  &   Contin.  & Contin. &  & code & mass,\\
           & [mm]      & [cm] & E=1332 keV  & E=1332 keV  &\large{$\frac{\epsilon_m-\epsilon_s}{\epsilon_m}$} &400-1100 keV&400-1100 keV  & $\frac{\epsilon_m-\epsilon_s}{\epsilon_m}$& & calcul.\\
           &           &      &$\epsilon_m$, [10$^{-3}$]&$\epsilon_s$,[10$^{-3}$] & [\%] &$\epsilon_m$, [10$^{-3}$]&$\epsilon_s$,[10$^{-3}$]& [\%] &  & [g]\\
  \hline
  ANG1     & 0.7       & 19.5 & 0.473      & 0.461& 2    & 2.85 & 2.80 & 2& Geant4 & 921\\
  &&&&&&&&&&\\
  ANG2     & 0.7       & 19.7 & 1.16       & 1.18 & -2    & 5.14 & 5.24 & 2& Geant4 & 2765\\
  &&&&&&&&&&\\
  ANG3     & 0.7       & 21.4 & 0.97       & 1.03 & -6    & 4.50 & 4.59 & -2& Geant4 & 2338\\
  ANG3     & 0.7       &         &               & 0.99 & -2    &         & 4.36 & 3 & EGS4 & \\
  &&&&&&&&&&\\
  ANG4     & 0.7       & 19.9 & 0.94       & 0.97 & -3    & 4.42 & 4.31 & 2& EGS4 & 2292\\
  &&&&&&&&&&\\
  ANG5     & 0.7       & 19.9 & 1.08       & 1.18 & -9     & 4.88 & 5.30 & -8& Geant4 & 2662\\
  ANG5     & 0.7       &         &         & 1.20 & -10    &         & 4.74 & 3& EGS4 &\\
  &&&&&&&&&&\\
  RG1      & 0.8       & 25.0 & 1.00       & 1.11 & -11   & 3.82 & 3.58 & 7& EGS4 & 2043\\
  RG2      & 0.8       & 25.0 & 1.06       & 1.08 & -2    & 3.94 & 3.74 & 5& EGS4 & 2082\\
  RG3      & 0.5       & 25.0 & 1.02       & 1.11 & -9    & 3.81 & 3.62 & 6& EGS4 & 2054\\
  \hline
\end{tabular}
\caption{Results of measurement and simulation of the
detection efficiency for HdM and IGEX detectors. The \Co\
source was positioned at distance R from the detector's copper
cap. Efficiency for the 1332 keV peak and the Compton
continuum sum in the energy range from 400 keV to 1100 keV are
presented. The simulation for ANG3 and ANG5 was performed by
two codes for simulation verification. Active masses and
efficiencies have a $\sim1\%$ error.} \label{ActMassTab1}
\end{sidewaystable}
\clearpage

\begin{table}[!h]
\centering 
\begin{tabular}{|c| c |c| c| c| c| c| c| c| c|}
  \hline
  Detector & $^{76}$Ge & $^{74}$Ge & $^{73}$Ge & $^{72}$Ge & $^{70}$Ge & $\rho$  \\
             & [\%] & [\%] &[\%]    &   [\%] &  [\%] & [g/cm$^3$]\\
  \hline
   ANG1 & 85.92  & 13.06  & 0.25 & 0.46 & 0.31 &  \\
   ANG2 & 86.44 & 13.32 & 0.12 & 0.08 & 0.04 &  \\
   ANG3 & 88.15 & 11.58 & 0.10 & 0.10 & 0.06 & 5.5 \\
   ANG4 & 86.3 & na & na & na & na &  \\
   ANG5 & 85.6 & na & na & na & na &  \\
\hline
    RG1 &  &  & & &  &  \\
    RG2 & 87.43 & 12.51 & 0.05 & 0.005 & 0.005 &  5.5 \\
    RG3 &  &  & & &  &  \\
\hline
  Natural Ge & 7.76 & 36.54 & 7.76 & 27.43 & 20.52 & 5.3 \\
 \hline
\end{tabular}
\caption{Isotopic content of the HdM and IGEX diodes. The
uncertainty is $\sim1\%$ for \Ge\ \cite{Dipl-Echt,
Diss-Dipl-Maier}.} \label{EnrichmentDet}
\end{table}

\clearpage

\subsection{Discussion of results}
The 10\% deficit of efficiency for ANG5, RG1 and RG3 detectors
compared to the simulations can be explained by an increase in
the detector dead layer by a factor of two. The difference can
also result from incorrect dimensions in the detector model.
Using the current measurements, one centimeter uncertainty in
the source position results in 10\% efficiency error. The
dimensions of the detectors were verified to exclude
systematic errors in the simulation. The cryostats were opened
again and the dimensions were compared with previous
measurements. No discrepancies in the dimensions were found.

\begin{figure}

\centering
  \includegraphics[width=69mm]{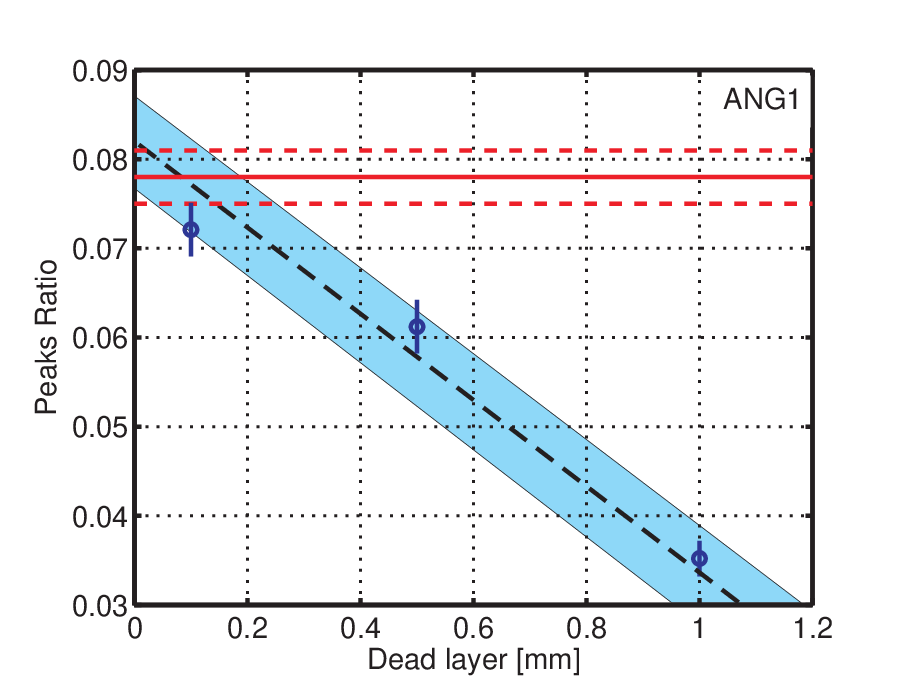}
  \includegraphics[width=69mm, height=52mm]{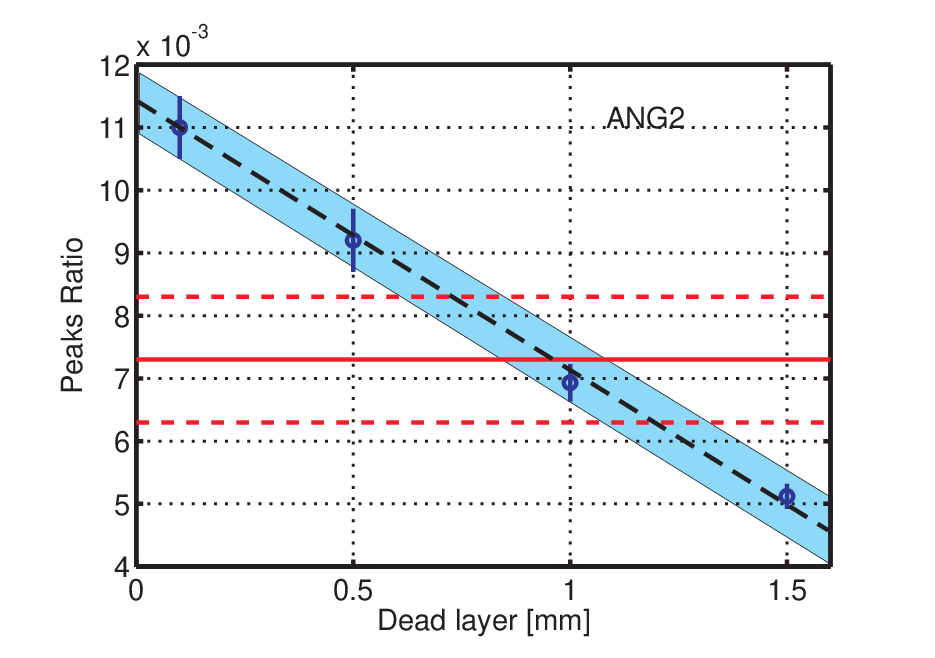}
  \includegraphics[width=69mm]{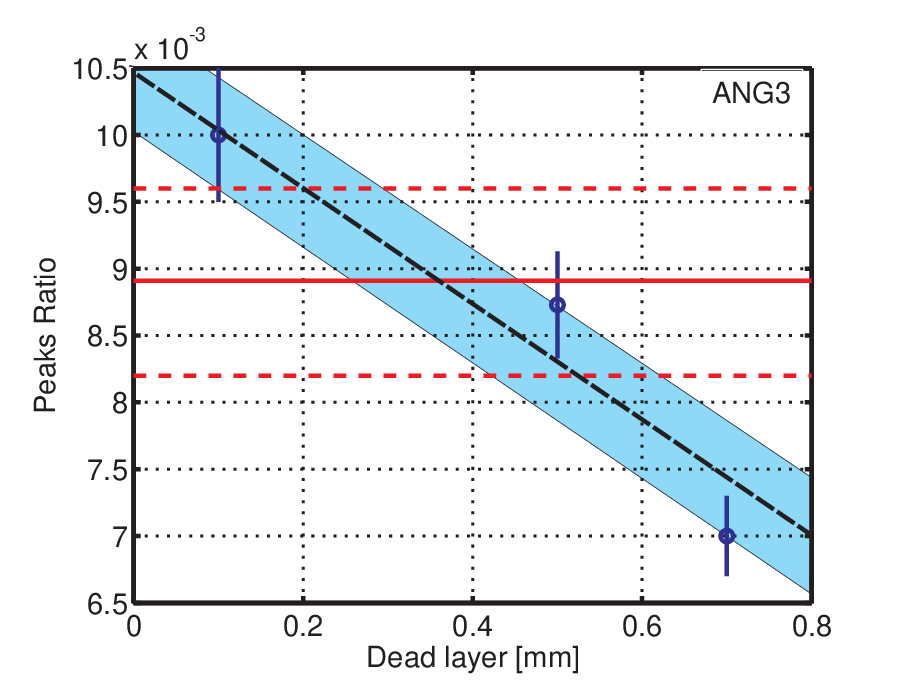}
  \includegraphics[width=69mm]{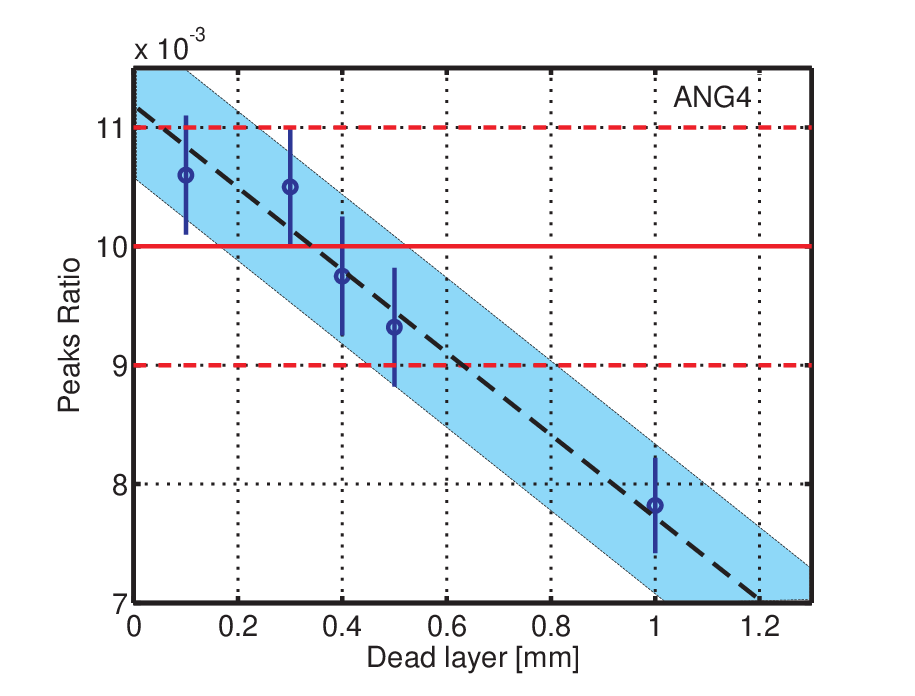}
  \includegraphics[width=69mm]{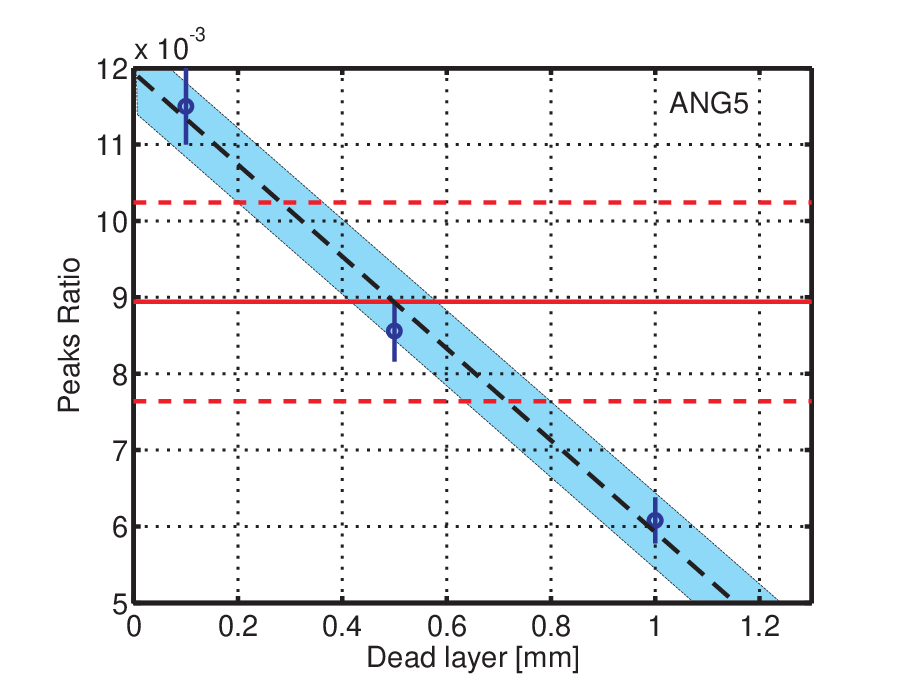}
  \includegraphics[width=69mm]{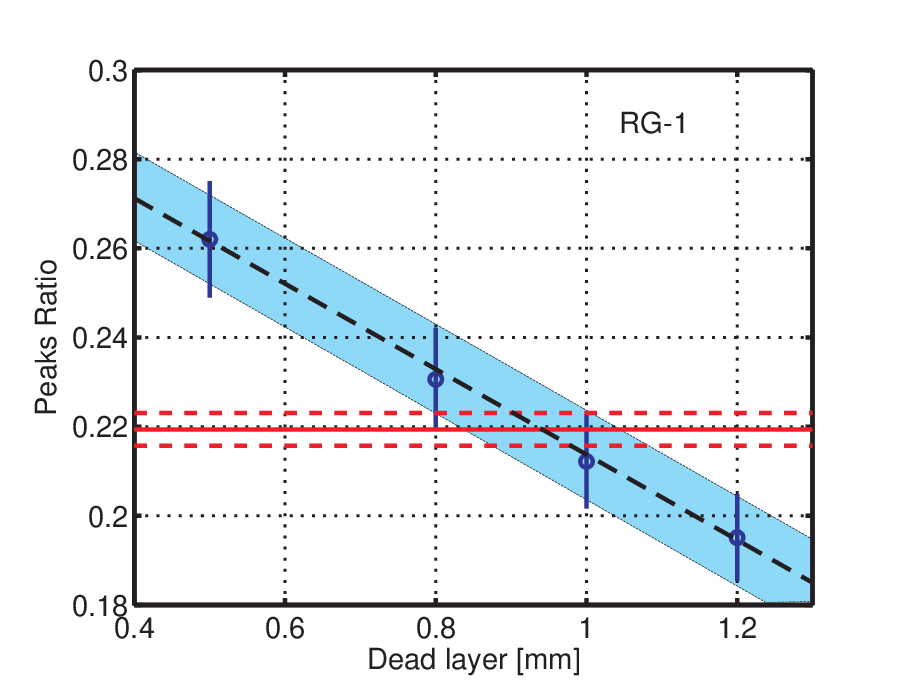}\\
  \caption{Ratio of the 81 keV and 356 keV \Ba\ peaks simulated as a function of the dead
   layer thickness for the ANG1-5 and RG1 detectors.
   The measured ratio is shown as the horizontal band with 1$\sigma$ error.}\label{DeadLayerFig}
\end{figure}

\begin{table}[!h]
\centering
\begin{tabular}{|c|c|c|c|c|c|c|}
  \hline
  Detector & Dead layer, & Dead layer \\
           & manufac.   &  \Ba\ meas.\\
           & [mm]        & [mm] \\
  \hline
  ANG1     & 0.7         & $\leq${0.2} \\
  ANG2     & 0.7         & 0.9$\pm$0.3 \\
  ANG3     & 0.7         & 0.4$\pm$0.3 \\
  ANG4     & 0.7         & $\leq${0.8} \\
  ANG5     & 0.7         & 0.5$\pm$0.3 \\
  RG1      & 0.8         & 0.9$\pm$0.2 \\
  \hline
\end{tabular}
\caption{The dead layer thicknesses provided by the
manufacturers and determined with the \Ba\ source and MC
simulations (Fig. \ref{DeadLayerFig}).} \label{DeadLayerTab}
\end{table}

The average dead layer thickness was determined using
measurements with a \Ba\ source. To exclude the systematic
uncertainty of the \Ba\ source activity the ratio of peak
intensities was used instead of the absolute intensity of the
peak. The peak ratio is also less sensitive to the uncertainty
of the source position. The measured ratio of 80.9 keV and 356
keV \Ba\ peaks was used for comparison with the simulation
results. The dependence of the peak ratio on the dead layer
thickness is presented in Fig.~\ref{DeadLayerFig}. The dead
layer thicknesses for the closed end surface of the detectors
were determined and are presented in Table \ref{DeadLayerTab}.
The obtained values are in agreement (68\% c.l.) with those
stated by the manufacturer, except for ANG1, for which the
dead layer thickness upper limit is 0.2~mm~(68\%~c.l.). An
investigation of the ANG1 detector history revealed an
additional grinding of the surface of the detector in 1991,
when the detector was broken \cite{DissMuller,StreckerPrivat}.
The thin dead layer and the reduced mass of the ANG1 detector
both can be explained by the $\sim0.7$\,mm layer of germanium,
which had been removed from the closed end part of the diode.\\
The dead layers of the ANG5 and RG1 detectors are found in
agreement with the values provided by the manufacturer
(0.7\,mm and 0.8\,mm, respectively), the observed
10\%~efficiency reduction for these two detectors cannot be
explained. The dead layer on the closed end of the detector is
not necessarily the same as the rest of the surface, therefore
additional measurements of the dead layer on the side surface
are recommended.\\
Kirpichnikov et al.\cite{KirpichActivMass} used a different
approach to determine the active masses. The distributed
sources of natural radioactivity in the surrounding rock are
utilized instead of artificial point sources. One of the
detectors has to be assigned as reference detector for the
relative measurements. The measurements have to be carried out
at the same place and at the same time to be comparable. The
active masses of RG1-3 and ANG2 were determined using
background measurements in GDL. The RG3 was selected as the
reference detector. The active masses of RG1 and RG2 were
found in agreement with the source measurements, but ANG2 and
RG3 are 10\% off (Fig.~\ref{ActMassRelative}). The uncertainty
of $\pm5\%$ for the active mass is a reasonable estimate.\\
\begin{figure}
  \includegraphics[width=14 cm]{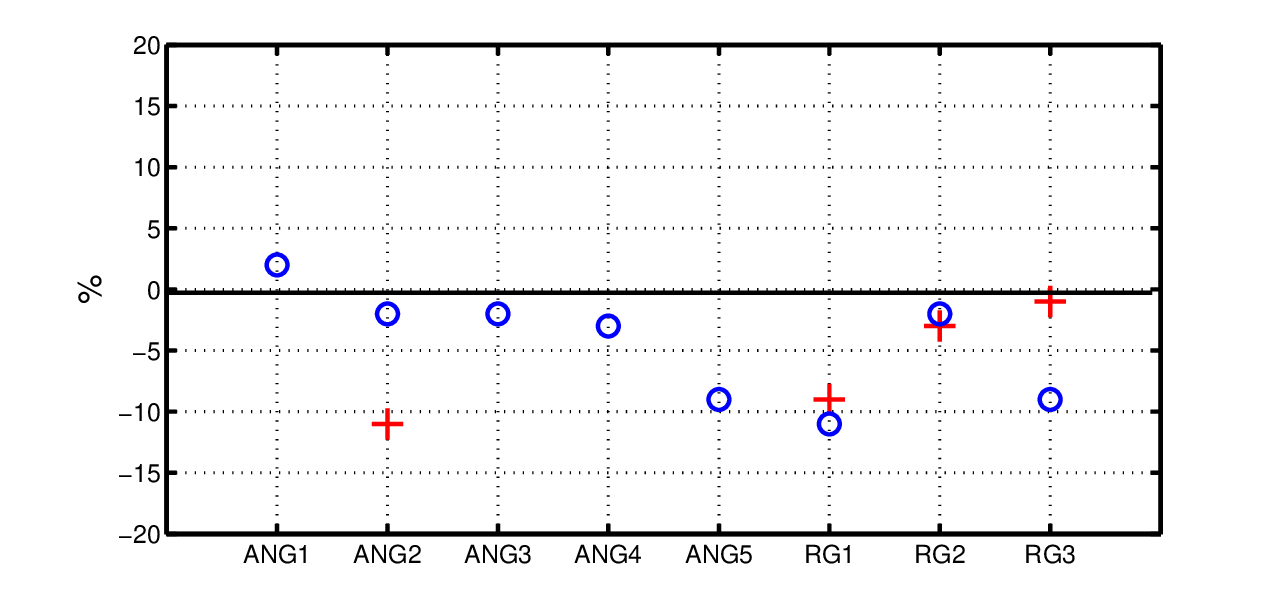}\\
  \caption{The relative deviations of measured and calculated active masses.
  ($\circ$) - measured with \Co\ source, ($+$) - determined with background measurements \cite{KirpichActivMass}.
  The errors are within the size of the symbols.}\label{ActMassRelative}
\end{figure}
The active mass is decreasing if the detector is not fully
depleted. However, the measurements were performed at
operating bias, when count rate is saturated, as shown in
previous sections, and therefore this effect can be excluded.
Another possibility is the presence of high concentration of
the charge traps in some parts of the detector volume. The
trapping effect is possible to diagnose using
detailed detector scanning with a collimated source.\\
 Table~\ref{ActMassTab2} gives the active masses of the detectors.
\begin{table}[!h]
\centering
\begin{tabular}{|c | c |c| c | c|}
  \hline
  Detector & Measured   & Calculated          & Previously Reported        &  Active mass, \\
           &   mass       & active mass, [g]     & active mass, [g]     &   \%  \\
           &    [g]         & $\rho^*$ = 5.5 g/cm$^3$ & (Tables \ref{tab:HdMhist},\,\ref{tab:IGEXhist}) & of measured  \\
  \hline
  ANG1     &\bf 968.7$\pm$0.1          &\bf 921$\pm$9          & 920      & 95$\pm$1 \\
  ANG2     &\bf 2878.3$\pm$0.1         &\bf 2756$\pm$28        & 2758     & 96$\pm$1 \\
  ANG3     &\bf 2446.5$\pm$0.1         &\bf 2338$\pm$23        & 2324     & 96$\pm$1 \\
  ANG4     &\bf 2401.2$\pm$0.1         &\bf 2292$\pm$23        & 2295     & 96$\pm$1 \\
  ANG5     &\bf 2782.1$\pm$0.1         &\bf 2662$\pm$27        & 2666     & 96$\pm$1 \\
  RG1      &\bf 2152.3$\pm$0.1         &\bf 2043$\pm$20        & $\sim$2000    &95$\pm$1 \\
  RG2      &\bf 2194.2$\pm$0.1         &\bf 2082$\pm$21        & $\sim$2000    & 95$\pm$1\\
  RG3      &\bf 2120.9$\pm$0.1         &\bf 2054$\pm$21        & $\sim$2000    & 97$\pm$1\\
  \hline
  Total    &\bf 17944.2$\pm$0.3        &\bf 17148$\pm$56       &   ---         & 95.5$\pm$0.3\\
  \hline
\end{tabular}
\caption{The active mass of HdM and IGEX detectors calculated
using the measured dimensions from Tab.~\ref{DimDetPar}. (*)
The specific density of enriched germanium (86\% \Ge) is
slightly higher than the natural germanium density,
$\rho_{^{76}Ge}$\,=\,5.5~g/cm$^3$,
$\rho_{natGe}$\,=\,5.3~g/cm$^3$.} \label{ActMassTab2}
\end{table}

 The active masses
were calculated using the dead layer thickness given by the
manufacturers with 10\% uncertainty, which was determined
using the collimated
\Ba\ source measurements.\\

\section{Summary}
The \gerda\ Phase~I detectors from HdM and \igex\ experiments
were maintained and their performance parameters characterized
between September 2004 and November 2006. The main
spectrometric and working parameters such as energy
resolution, operation voltage and leakage current have been
measured. The ANG1 and RG2 detectors initially showed
deteriorated energy resolution and high leakage current. To
restore their parameters, heating and pumping operations have
been performed. Finally, all the detectors demonstrated
parameters which correspond to the specifications
at the time of production.\\
Measurements of detector efficiencies with a \Co\ source were
performed for all the enriched detectors. The measured
efficiencies were compared with Monte Carlo simulations. For
the verification of the simulations, two independent codes
were used: Geant4 and EGSnrc. The two codes produce results
within statistical errors. Using the efficiency and the dead
layer values the active masses of the enriched detectors for
the \gerda\ Phase~I were determined. The obtained active
masses were compared with those provided by the manufacturer
and were found to agree. The deterioration of the dead layers
were found to vary by less than 0.2 mm from the values
provided by the manufacturers. The total and active masses of
the \gerda\ Phase~I detectors before their refurbishment are
17944.2$\pm$0.3~g and
17.2$^{+0.1}_{-0.8}$~kg respectively.\\
Afterwards, the enriched diodes were dismounted from their
cryostats and sent to Canberra Semiconductor NV, Olen,
Belgium, in preparation for \gerda\ Phase~I.

\newpage
\chapter{Searching for Neutrinoless Double Electron Capture
of $^{36}$Ar} \label{ch:ar36}
Double electron capture (2$EC$) is a process inverse to double
beta decay. This chapter presents a search for the
neutrinoless mode of the double electron capture (\onECEC) in
\Ar. The decay of \Ar\ has not been investigated yet and no
measured limits of the half-life exist in the literature.
Measurements with a bare HPGe detector in liquid argon were
performed in the \gerda\ underground detector laboratory (GDL)
at LNGS during a long term stability test of the detector
parameters. The measured background spectra were analyzed and
the first limit for the radiative mode of the \onECEC\ decay
in \Ar\ is reported.

\section{Introduction to radiative \onECEC\ decay in $^{36}$Ar}
The study of the \onECEC\ decays could give additional
information about mechanisms of the lepton number violation
(see e.g. \cite{Hirsh94}). In the process of 2$EC$, the two
atomic electrons are absorbed by the nucleus:
\begin{equation}\label{ECEC}
    e^- + e^- + A(Z)  \Rightarrow A(Z-2) + (2\nu) + Q.
\end{equation}
In the two neutrino 2$EC$ decay, the energy is carried away by
two neutrinos (neglecting atom's recoil) and X-rays. In the
neutrinoless case, the momentum-energy conservation requires
that the energy has to be released somehow. There are various
\onECEC\ decay modes with emission of $e^+e^-$ pairs, internal
conversion electrons or photons. The latter mode is a
radiative \onECEC\ decay. The bremsstrahlung photon is emitted
by one of the captured electrons. The capture of the two
K-shell electrons in the $0^+\rightarrow0^+$ transition with
emission of one photon is forbidden by the conservation of the
angular momentum, therefore, one of the electrons has to be
captured from the L shell. The detailed discussion of the
$2EC$ processes with emission of photons can be found e.g. in
\cite{Vergados83, DoiKotani1993}.
\\
Natural argon contains the isotope \Ar\ with an abundance of
0.336\%, which is expected to be unstable undergoing the
double electron capture \cite{TretyakTables}. The energy
levels of the $A=36$ isobar triplet are shown in
Fig.~\ref{Ar36Levels} \cite{TOI99}. The energy of the \Ar\
decay is 433.5 keV. Therefore, the radiative and the internal
conversion modes of the \onECEC\ decay are energetically
allowed. The search of the radiative \onECEC\
decay can be performed with a gamma-ray detector.\\
\begin{figure}[!h]
\begin{center}
  \includegraphics[width=100 mm]{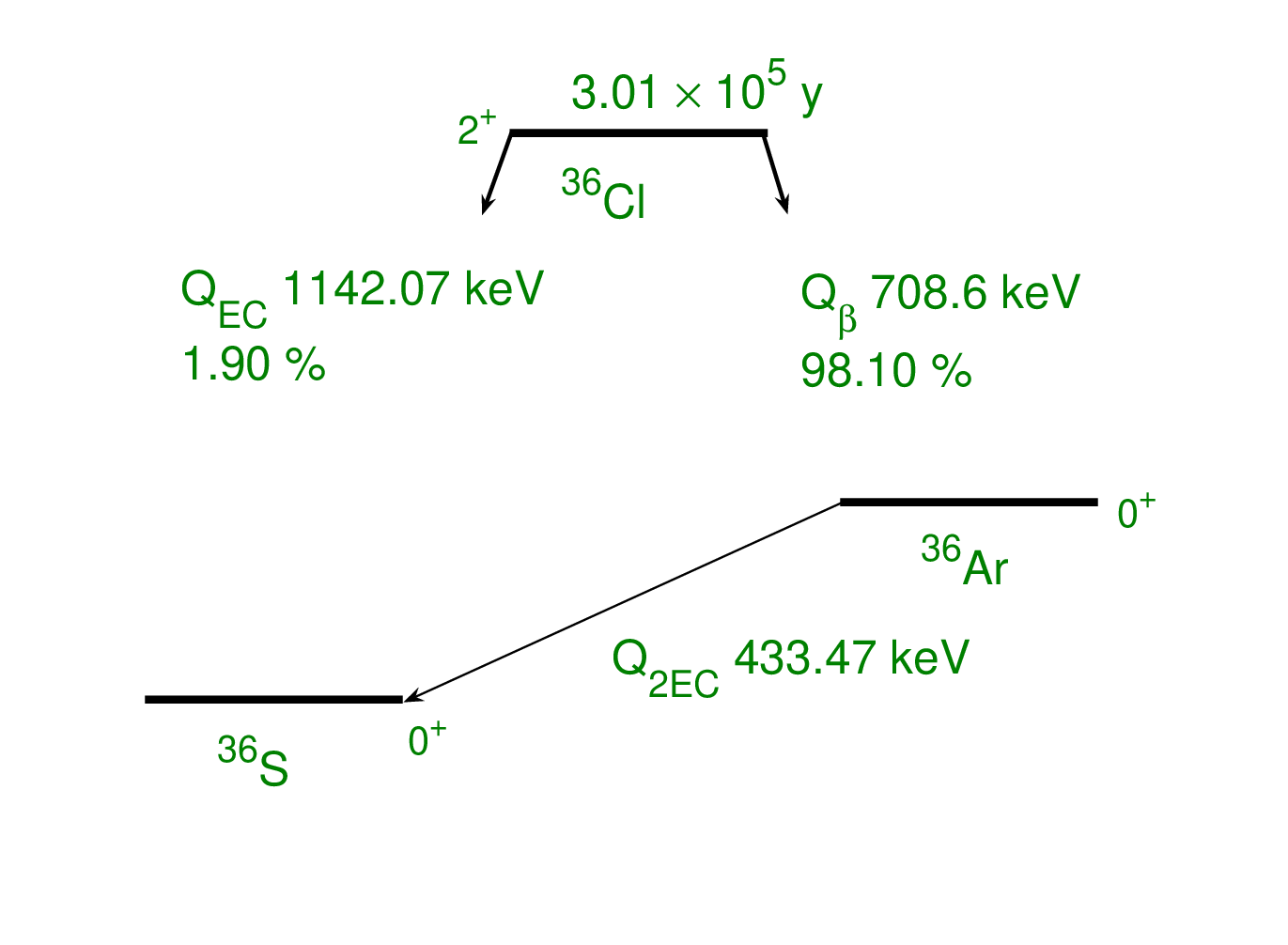}
  \caption{Lowest energy levels of isobar triplet A=36 with double
  electron capture decay of $^{36}$Ar \cite{TOI99}.}\label{Ar36Levels}
\end{center}
\end{figure}
The amplitude of the 2$EC$ transition with the energy release
$Q$ close to the 2$P$--1$S$ atomic level difference could be
resonance enhanced, as was shown in \cite{Polish2005}. In the
case of \Ar, the daughter $^{36}$S has no excited states
available for the 2$EC$ transition. Therefore, the 2$EC$ decay
must be a ground-state-to-ground-state transition ($0^+_{g.s.}
\rightarrow 0^+_{g.s.}$). One electron is captured from the
$K,L$... shells and the other from the $L,M$... shells.
Theoretical calculations have been carried out for the two
neutrino 2$EC$ decay mode with the half-live in the order of
$10^{29}$ years \cite{TheorAr36}. A preliminary estimate of
the neutrinoless mode gives a half-live in the order of
$10^{35}$ years \cite{Theor0nuAr36}, which is far beyond the
current
experimental sensitivity.\\
The expected decay of \Ar\ will produce three photons with
energies: $E_K = 2.47$~keV, $E_L = 0.23$~keV and $E_\gamma =
430.8$~keV given the $Q_{2EC}$ = 433.5~keV. The experimental
signature is the monochromatic photon with the energy of
$E\gamma = Q - E_K - E_L$ = 430.8~keV, which is detected with
a high-purity germanium (HPGe) detector submerged in liquid
argon. In addition, the two X-rays of the daughter atom could
give a coincidence trigger, if the liquid
argon scintillation light is detected.\\

\section{Experimental setup in the {\sc Gerda} detector laboratory}
The measurements were performed in the {\sc Gerda} underground
detector laboratory (GDL) located at LNGS. The laboratory is
equipped with a radon-reduced clean bench mounted to a dewar
system. It is designed to operate bare germanium detectors
submerged in liquid argon or nitrogen. For the past two years,
the leakage current, the energy resolution and the detector
stability have been investigated in this setup. Table
\ref{Ar36param} gives the experimental parameters and the
values used for the present measurement.
\begin{table}
\centering
\begin{tabular}{|l|c|}
  \hline  
  Experimental parameter & Value  \\
  \hline
  Volume of liquid argon & $70\pm7$ liters \\
  Mass of liquid argon & $100\pm10$ kg\\
  \hline
  Detector diameter and length & 75 mm, 69 mm\\
  Measurement live-time & 237 hours\\
  Energy resolution (FWHM) at 430.8 keV & 4.1$\pm$0.5 keV\\
  Full energy peak efficiency at 430.8 keV & 0.26$\pm$0.03 \%\\
  \hline
  \Ar\ abundance & 0.336 \%\\
  \Ar\ 2EC reaction Q value & 433.5 keV\\
  $^{36}$S Binding energy $E_K$, $E_L$ \cite{TOI99}& 2.47 keV, 0.23 keV\\
  \hline
\end{tabular}
\caption{Experimental parameters and values used for
analysis.}\label{Ar36param}
\end{table}
The experimental setup is shown in Fig. \ref{Ar36setup}.
\begin{figure}
\centering
  \includegraphics[width=8.8 cm]{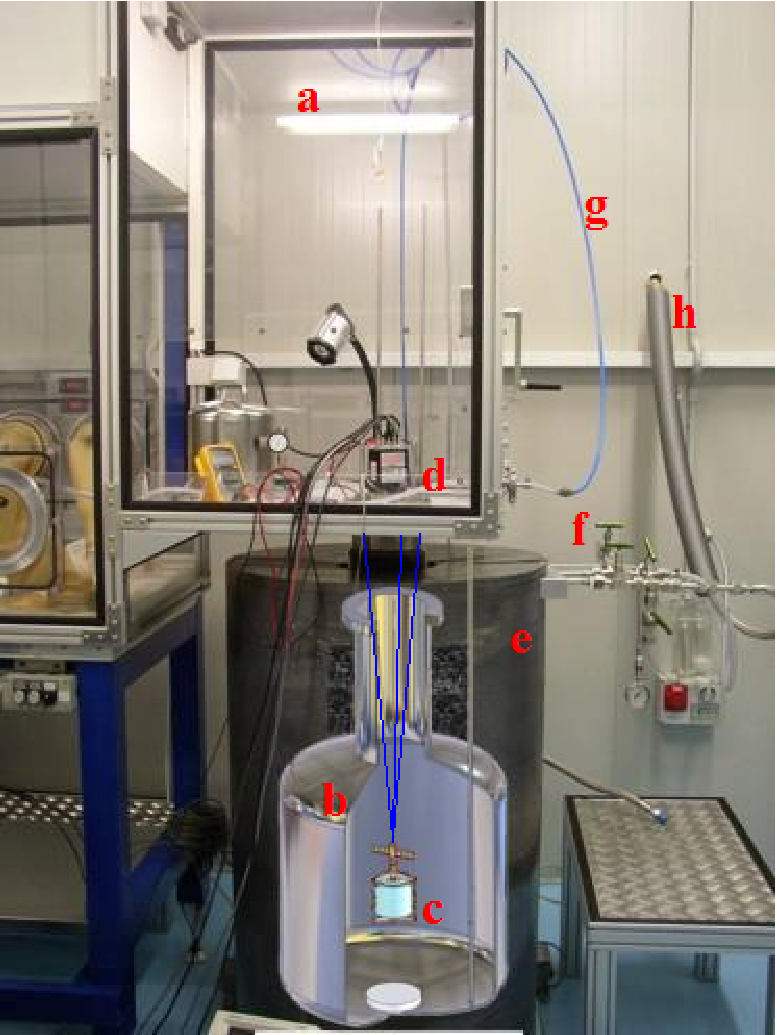}
  \includegraphics[width=5 cm]{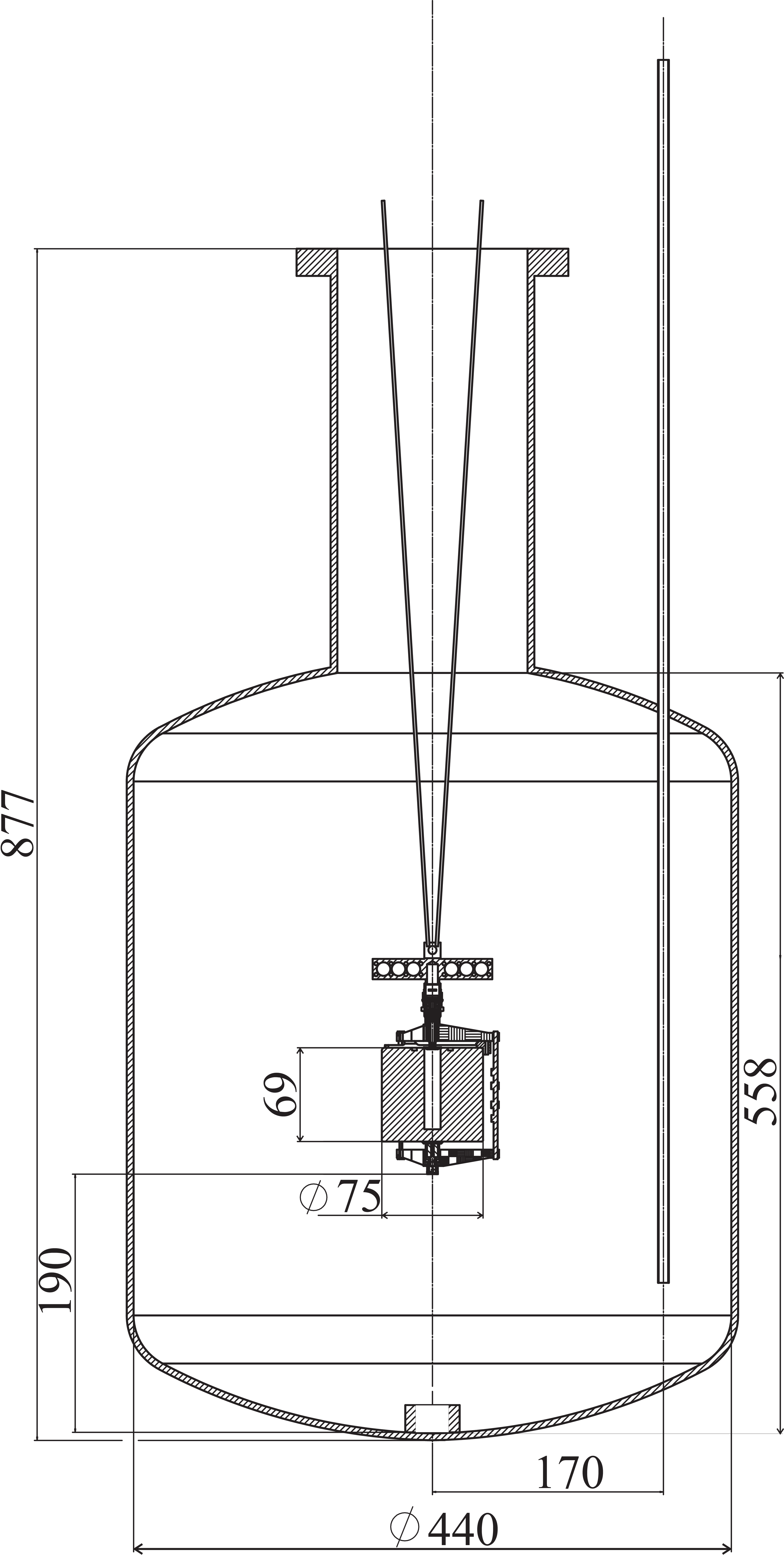}\\
  \caption{The experimental setup for measurements with a bare
  HPGe detector in liquid argon. On the left, the detector test
   bench of GDL
   in which a bare HPGe detector is operated in liquid argon.
   On the right, a detailed view of the detector assembly in the dewar.
   The figure shows:
   a) the radon free detector test bench,
   b) the 70 liter dewar,
   c) the HPGe detector in LAr suspended on Kevlar strings,
  d) the warm FET preamplifier, CANBERRA 2002,
   e) the lead shield barrel,
  f) the liquid argon filling/venting lines,
   g) the nitrogen gas flushing line,
  h) the line to the external refilling dewar.}\label{Ar36setup}
\end{figure}
A 300 $cm^3$ high-purity germanium diode was mounted in the
GERDA Phase~I low mass holder to provide suspension, high
voltage and signal contacts. The holder made of copper,
silicon and PTFE was suspended on 80 cm long Kevlar strings.
The strings were attached to a dewar flange on which a warm
FET preamplifier (CANBERRA 2002) was installed. The detector
was positioned in the center of a 70 l dewar. The moderate
shield of the test bench consisted of 2.5 cm of lead
surrounding the dewar. It slightly suppresses the external
radiation by a factor around 10. The dewar was refilled with
liquid argon every four days. Between fillings 14\,l of argon
evaporates.\\
The ORTEC spectrometry amplifier and Maestro multichannel
analyzer (MCA) were used to collect spectra continuously
between fillings. A ten day background spectra were acquired
during a long term stability test ordinarily performed with a
\Co\ source. The resolution of the $^{40}$K 1460.8 keV
background line was 4.56 keV (FWHM). The energy calibration
was performed using a \Co\ source and the background peaks.
The energy resolution dependance on energy was fitted by the
function: $FWHM(keV)=\sqrt{15+0.0039\cdot E_\gamma}$, where
$E_\gamma$ is the energy in keV. For example, the energy
resolution of the expected peak at 430.8 keV from the
radiative \onECEC\ decay is 4.1 keV. The detector parameters,
energy resolution and leakage current, were stable during
the measurement.\\


\section{Results and analysis}
Three measured energy spectra with respective live times of
70.4, 96.6 and 70.0 hours were individually calibrated in
energy and then summed together. Fig.~\ref{SpecAr} displays
the resulting spectrum and the region of interest around 430
keV. In the region of interest the background index amounts to
440 counts/(keV$\cdot$day). The spectrum shape is dominated by
the $\gamma$-rays from the external natural radioactivity. The
intensity and the resolution of the major peaks are shown in
Table \ref{LArPeaksTab}.
\begin{figure}
\centering
  \includegraphics[width=100mm]{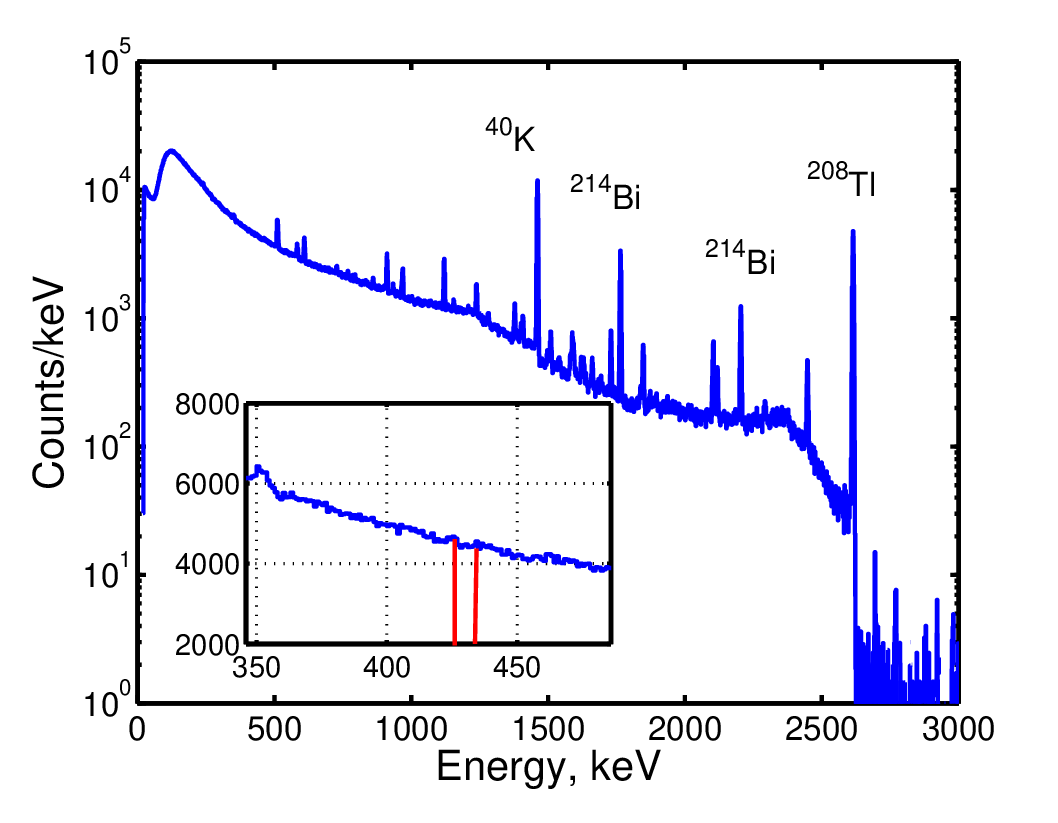}\\
  \caption{The measured background spectrum of a bare HPGe
  detector operated in liquid argon over a 10.0 days. {\it In the inset}:
   Low energy part of the spectrum together with the region of interest
   around Q-value of the \onECEC\ decay of \Ar.}\label{SpecAr}
\end{figure}
\begin{table}
\centering
\begin{tabular}{|c|c|c|}
  \hline
  &&\\
  Measured energy [keV] & FWHM [keV] & Peak area \\
  &&\\
  \hline
  609.4 & 4.32 & 6656$\pm$260 \\
  1120.5 & 4.33 & 7906$\pm$187 \\
  1461.1 & 4.56 & 55850$\pm$261 \\
  1764.7 & 4.68 & 15589$\pm$146 \\
  2204.2 & 4.98 & 5850$\pm$100 \\
  2614.5 & 5.05 & 25664$\pm$161 \\
  \hline
\end{tabular}
\caption{The intensity and the resolution of the major peaks
in the background spectrum measured with the HPGe detector in
liquid argon for 237 hours.}\label{LArPeaksTab}
\end{table}
The simulation codes EGSnrc \cite{EGS} and TEFF \cite{TEFF}
were used to determine the full energy peak (FEP) efficiency
$\varepsilon$ of detection for the photons emitted in the
liquid argon volume. The energy dependence of the FEP
detection efficiency is shown in Fig.~\ref{EffLAr}.\\
\begin{figure}
\centering
  \includegraphics[width=8 cm]{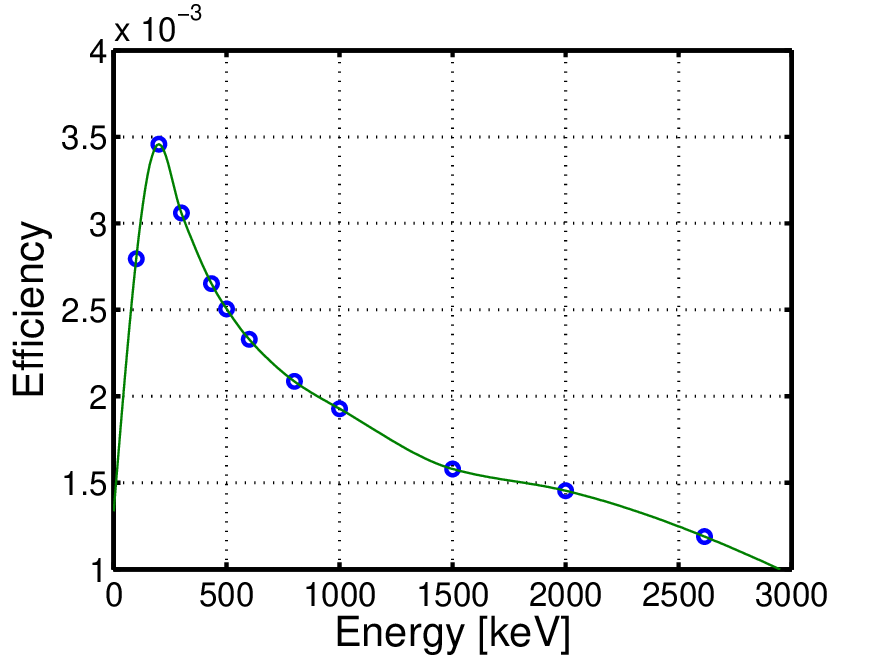}\\
  \caption{Calculated full energy peak efficiency of the
  300 $cm^3$ HPGe detector in the 70 liter dewar filled with argon.
  Simulated $\gamma$-rays are uniformly distributed in the argon volume.}\label{EffLAr}
\end{figure}
The spectrum in the energy range of 350--500 keV is
essentially featureless, as seen on the inset of
Fig.~\ref{SpecAr}. The lower half-life limit is expressed as a
function of the upper limit for counts in the peak ($\lim S$)
at a given confidence level:
\begin{equation}\label{ArLimit}
\lim T_{1/2}=\ln 2 \cdot N \cdot \varepsilon \cdot
\frac{\Delta T}{\lim S},\quad N = N_A\cdot\frac{{\it a}\cdot
M}{A},
\end{equation}
where $N$ is the number of the $2EC$ source nuclei,
$\varepsilon$ is the FEP efficiency, $\Delta T$ is the
measurement time, $\it{a}$ is the isotopic abundance of the
isotope, $M$ is the total mass of the argon source material,
$N_A$ is the Avogadro number, and $A$ is the molecular weight
of the source material.
\begin{figure}
\centering
  \includegraphics[width=100mm]{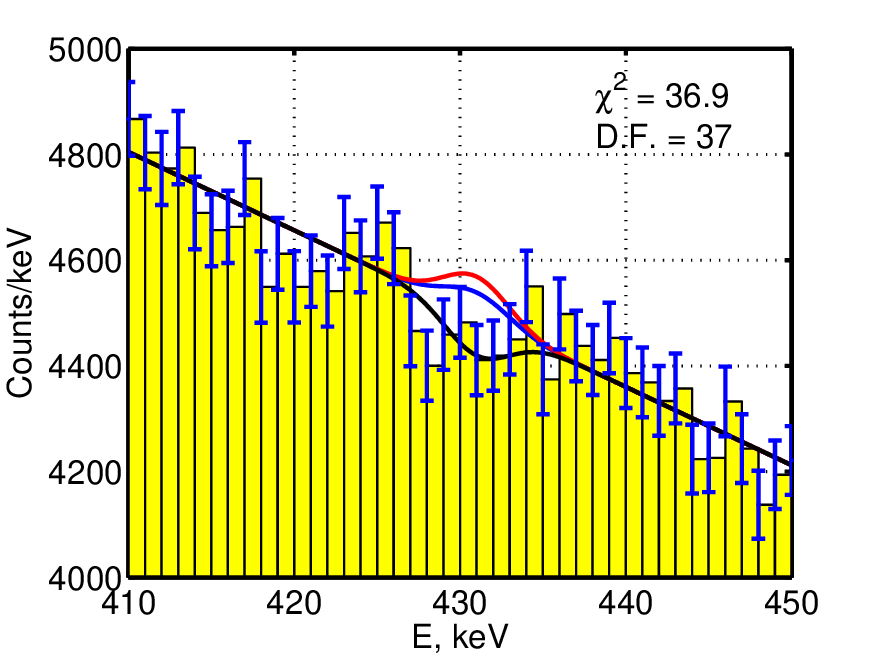}\\
  \caption{The fit of the measured energy spectrum
   of the HPGe detector in the region of interest
   around the Q-value of the \onECEC\ decay of \Ar.
   The sum of a Gaussian and a linear slope (black curve)
   fitted to the measured energy spectrum.
   The blue and red lines are the 68\% and 90\% C.L. bounds
   on the peak counts, respectively.}\label{SpecArFit}
\end{figure}
The upper limit on the number of counts in the expected peak
was obtained by fitting the experimental energy distribution
with the sum of a Gaussian and a linear background slope. The
Gaussian mean and standard deviation are fixed to 430.8 keV
and 1.84 keV, respectively. The amplitude of the gaussian and
the parameters of the linear continuum were free parameters of
the least square fitting in the energy interval of 410--450
keV (Fig. \ref{SpecArFit}).\\
In the region between 410 and 450\,keV, assuming a linear
background, the best fit yields a negative number of counts
under the peak $-376\pm230(382)$ counts at 68\%(90\%) C.L. The
resulting upper bound on the number of candidate events in the
\onECEC\ decay peak is a 1$\sigma$ statistical fluctuation $S$
of background counts in the energy window $\Delta E$=5\,keV,
$S$ = $\sqrt{B\cdot\Delta E}$ = $\sqrt{4500\cdot5}$=150 at
68\% C.L. The resulting lower limit on the half-life is
computed according to (Eq.~\ref{ArLimit}):
T$_{1/2}^{0\nu}$(\Ar) = $\geq$ (1.85 $\pm$ 0.25(syst) $\pm$
0.01(stat)) $\cdot10^{18}$\,y (68\% C.L.). The systematic
uncertainties reside in the source mass, efficiency
calibration and background spectral shape. The main factors
influencing the limit are the uncertainty in the source mass
($\pm$10\%) and the efficiency of detection ($\pm$10\%). The
quoted 68\% C.L. lower bound was computed using the best
value, 430\,keV. There is a small dip in the data centered at
430\,keV as shown in Figure~\ref{SpecArFit}. This has been
treated as a statistical fluctuation.\\
With Eq.~\ref{ArLimit} and the values given in
Tab.~\ref{Ar36param}, the half-life limit for the radiative
\onECEC\ decay of \Ar\ with emission of single photon is:
\begin{equation}\label{ArResult} \nonumber
T_{1/2}(0^+ \rightarrow g.s.~with~emission~of~single~\gamma)~
\geq ~ 1.85\cdot10^{18}~y~(68\%~C.L.).
\end{equation}
The obtained half-live limit is comparable to recent results
from dedicated experiments, which are summarized in
Tab.~\ref{ArResultsTab}. The half-life limits for most of
\onECEC\ decay experiments are in the range of
10$^{16}$--10$^{19}$ years. The best limit ($\sim$10$^{21}$ y)
was achieved with an enriched $^{78}$Kr \cite{78-Kr}.
\begin{table}[ht]
\small
\begin{tabular}{|c|c|c|c|l|}
  \hline
  &&&&\\
  Isotope & Abundance [\%] & Mode  & $T_{1/2}$ limit [y] (C.L.) & Reference  \\
  &&& &\\
  \hline
  $^{36}$Ar & 0.336 &0$\nu$2EC  & $1.85 \cdot 10^{18}$ (68\%) & this work \cite{GSTR-Ar36} \\
  $^{50}$Cr & 4.345 &(0$\nu$+2$\nu$)EC$\beta^+$  & $1.3 \cdot 10^{18}$ (95\%) & Bikit et al. (2003) \cite{50-Cr}\\
  $^{64}$Zn & 48.63 &0$\nu$2EC  & $4.1 \cdot 10^{18}$ (68\%) & Belli et al. (2008) \cite{64-Zn}\\
            & &0$\nu$EC$\beta^+$  & $6.1 \cdot 10^{20}$ (68\%) & --"--\\
  $^{74}$Se & 0.89 &0$\nu$2EC  & $6.4 \cdot 10^{18}$ (90\%) & Barabash et al. (2006) \cite{74-Se}\\
             & &(0$\nu$+2$\nu$)EC$\beta^+$  & $1.9 \cdot 10^{18}$ (90\%) & --"--\\ 
  $^{78}$Kr & 0.35 &2$\nu$2EC  & $1.5 \cdot 10^{21}$ (90\%) & Gavriljuk et al. (2006) \cite{78-Kr}\\
  $^{106}$Cd & 1.25 &2$\nu$2EC  & $4.8 \cdot 10^{19}$ (90\%) & Stekl et al. (2006) \cite{106-Cd}\\
  $^{108}$Cd & 0.89 &0$\nu$2EC  & $2.5 \cdot 10^{17}$ (68\%) & Danevich et al. (2003) \cite{108-Cd}\\
  $^{112}$Sn & 0.97 &(0$\nu$+2$\nu$)EC$\beta^+$  & $1.5 \cdot 10^{18}$ (68\%) & Kim et al. (2003) \cite{112-Sn}\\
  $^{120}$Te & 0.09 &2$\nu$2EC  & $9.4 \cdot 10^{15}$ (90\%) & Kiel et al. (2003) \cite{120-Te}\\
  $^{130}$Ba & 0.106 &0$\nu$EC$\beta^+$  & $2.0 \cdot 10^{17}$ (90\%) & Cerulli et al. (2004) \cite{130-Ba}\\
  $^{136}$Ce & 0.185 &2$\nu$2EC  & $4.5 \cdot 10^{16}$ (68\%) & Belli et al. (2003) \cite{136-Ce}\\
  $^{138}$Ce & 0.251 &2$\nu$2EC  & $6.1 \cdot 10^{16}$ (68\%) & --"-- \\ 
  $^{180}$W  & 0.12 &0$\nu$2EC  & $1.3 \cdot 10^{17}$ (68\%) & Danevich et al. (2003) \cite{180-W}\\
  \hline
\end{tabular}
\caption{Recent results of half-life measurements for $2EC$
and $EC\beta^+$ processes with transition to ground
state.}\label{ArResultsTab}
\end{table}


\clearpage

\section{Outlook} 
The sensitivity of the experiment presented here is limited by
the radiation from the outside of the setup, which is not
designed for ultra low-background measurements. The {\sc
Gerda-LArGe} facility, which will be used to test the
backgrounds of the Phase~I detectors prior to their operation
in {\sc Gerda}, will provide improved limits. External
radiation will be suppressed by a massive passive shield, the
mass of argon will increase to approximately one ton and up to
nine detectors could in principle be operated simultaneously.
The sensitivity will then be limited by the bremsstrahlung of
$^{39}$Ar beta decays (Q = 565 keV, T$_{1/2}$ = 269 y). The
Monte-Carlo simulation gives a count rate which is on the
order of 3 counts/(keV$\cdot$y$\cdot$kg) in the HPGe detectors
at 430 keV \cite{SimgenAr39}. An additional source of
background is the 2$\nu\beta\beta$\ decay of $^{76}$Ge. For
enriched detectors this background is $\sim$1.5
counts/(keV$\cdot$y$\cdot$kg) in the region of interest, which
is comparable to the $^{39}$Ar background. The background can
be reduced by an order of magnitude using detectors made of
natural germanium. For one year of measurements, the expected
sensitivity is in the range of 10$^{22}$--10$^{23}$ years. If
the X-ray scintillations in liquid argon can be detected with
a reasonable efficiency, an X-ray -- gamma coincidence could
be exploited as an additional signature to reduce the
$^{39}$Ar bremsstrahlung background. The ultimate sensitivity
will be achieved in {\sc Gerda} with the operation of 40~kg of
HPGe detectors in Phase~I and Phase~II.

\section{Summary}
An experiment to search for double electron capture with the
emission of one photon in \Ar\ was proposed and carried out in
the {\sc Gerda} underground detector laboratory at LNGS using
a bare HPGe detector submerged in liquid argon. For the first
time a limit on the neutrinoless double electron capture decay
of \Ar\ has been determined from the experimental data with
$T_{1/2}\geq1.85\cdot10^{18}$ yr at 68\% C.L.. The given limit
is comparable to those for different isotopes obtained in
dedicated experiments. The sensitivity in the {\sc LArGe}
setup and the limiting backgrounds of $^{39}$Ar bremsstrahlung
and \tnbb\ have been estimated.

\newpage
\chapter{Conclusions}
The thesis focuses on the search for neutrinoless double beta
(\onbb) decay of \Ge\ with HPGe detectors. The germanium
detectors are proven to be an excellent instrument for double
beta decay searches because they have superior energy
resolution and very high purity of the detector material.
Because \Ge\ acts both as the source and the detector, high
detection efficiency is reached. Presently the best limits on
\onbb\ decay come from germanium experiments:
Heidelberg-Moscow and IGEX. Their sensitivities reach about
0.3~eV of the effective neutrino mass $m^\nu_{\beta\beta}$.
Both collaborations have reported almost the same lower limit
on the \onbb\ decay half-life of $\sim$1.6$\cdot$10$^{25}$~y
(68\% C.L), corresponding to an effective neutrino mass upper
limit range of 0.33 to 1.3~eV (68\% C.L).\\
Two different experiments located at LNGS - HdM and \gerda,
both using enriched HPGe diodes but with different operating
techniques, were presented. The first one, HdM experiment, has
operated from 1990 to 2003 and had provided new results on
double beta decay search. The second one, \gerda, is now under
construction in Hall A of LNGS, and will search for \onbb\
decay with higher sensitivity, using bare germanium detectors
submerged in liquid argon. The liquid argon will serve as a
shield against external radioactivity
as well as a cooling medium.\\
This thesis presents improved analysis of the HdM raw data
collected in the period from 1990 to 2003. More accurate
energy calibration was performed resulting in better energy
resolution of the summed spectrum. The fitted background
spectrum gives accurate peak positions and widths. The energy
resolution determines the sensitivity of the experiment. 20\%
improvement in the energy resolution was obtained which led to
10\% increase in sensitivity. This improvement led to
publications \cite{KK-NewAn-NIM04,KK-NewAn-PL04}, which claim
4.2$\sigma$ evidence of \onbb\ decay.\\
A new background model of the HdM spectrum was developed using
measurements with sources and MC simulations. The model
accounts for the \Ra\, \Th\ and \Co\ contaminations, the muon
induced neutrons and the \tnbb\ decay. The model deviation
from the HdM spectrum is within 1\% in the energy interval
250-2800\,keV. The background continuum in the region of
interest around \qbb\ value was determined using the model.
The obtained value of the background is
(11.8$\pm$0.5)~counts/keV, which is higher than the background
used in publication \cite{KK-NewAn-NIM04} --
(10.0$\pm$0.3)~counts/keV. The model is still not accounting
for all data in the full range of energies (e.g. $^{207}$Bi,
$^{125}$Sb, $^{134}$Cs). However, it accounts for
contributions from \Th, \U, \Ra, \Co\ and muon induced
neutrons to the region of interest around \qbb\ value. The fit
of the HdM spectrum with the model background gives better
agreement of the \Bi\ lines intensities. The intensity of the
$\sim$2039\,keV peak, (15$\pm$12)\,counts for 71.7\,kg\,y, is
less significant than the published value,
(28.8$\pm$6.9)\,counts \cite{KK-NewAn-NIM04} which was
obtained before the model was developed in 2006. The
corresponding half-life of the \onbb\ decay is $2.2\cdot
10^{25}$\,y and the 68\% C.L. interval is $(0.4 - 4.0)\cdot
10^{25}$\,y, the sensitivity of the HdM experiment for \onbb\
decay is $4.6\cdot 10^{25}$ y (68\%, C.L.). The effective
neutrino mass is 0.32\,eV within (0.19-0.45)\,eV 68\% C.L.
interval using the nuclear matrix elements reported in \cite{Sta90}.\\
The \genius-TF setup with four HPGe detectors immersed in
liquid nitrogen was presented. The author's work covers the
first year of operation, including the detector assembly, at
the Gran Sasso underground laboratory. The measurements
performed during the first year shown that the detectors and
the electronics were highly sensitive to environmental
interferences (vibrations in liquid nitrogen, electromagnetic
pickups and infrared radiation). Techniques were applied to
reduce these interferences and an acceptable energy resolution
was obtained ($\sim$4\,keV). An analysis of the \genius-TF
background after the completion of the shield was presented.
The contribution to the background coming from the
$^{222}$\rm{Rn} decay chain was identified.\\
During one year, the four bare HPGe detectors operated in
liquid nitrogen were biased at their nominal voltage and their
parameters were stable. No deterioration of their energy
resolution and leakage current was observed. It shows the
principal feasibility of the GERDA experiment, which will use
bare germanium detectors in liquid argon.\\
The shielding needed for the GERDA experiment depends on the
intensity of the external radiation. The $\gamma$-ray flux at
the GERDA experimental site has been determined for the first
time in 2004 with a collimated $\gamma$-spectrometer. The flux
had to be remeasured in 2007 after the reconstruction of Hall
A. An HPGe detector was used to perform these measurements
with and without a collimator. The collimator was used in 2004
for angular flux distribution measurements. The detector
response was determined with Monte-Carlo simulations. The 2614
keV photon flux is (362$\pm$12)\,m$^{-2}$s$^{-1}$. The
contribution to the GERDA background index from external
$\gamma$-radiation has been calculated as $1.1\cdot10^{-5}$
[keV\,kg\,y]$^{-1}$. Details of the external GERDA shielding
are optimized based on
these measurements.\\
The GERDA Phase~I enriched detectors from HdM and IGEX
experiments were maintained and their performance parameters
characterized in the period from September 2004 to November
2006. The main spectrometric and working parameters such as
energy resolution, operation voltage and leakage current have
been measured. Detector efficiency measurements with a \Co\
source were performed for all the enriched detectors. The dead
layers of most detectors were determined and compared to
previous values. After more than ten years, the increase in
dead layer is negligible. The measured efficiencies were
compared with Monte Carlo simulations. With the efficiency and
the dead layer values, the total active mass of all enriched
detectors were determined to be 17.2$^{+0.1}_{-0.8}$~kg.\\
An experiment to search for double electron capture with the
emission of one photon in \Ar\ was proposed and carried out in
the {\sc Gerda} underground detector laboratory at LNGS using
a bare HPGe detector submerged in liquid argon. For the first
time a limit on the neutrinoless double electron capture decay
of \Ar\ has been determined from the experimental data with
$T_{1/2}\geq 1.85\cdot10^{18}$ yr at 68\% C.L. The given limit
is comparable to those for different isotopes obtained in
dedicated experiments. The sensitivity in the {\sc LArGe}
setup and the limiting backgrounds of $^{39}Ar$ bremsstrahlung
and \tnbb\ have been estimated.



\newpage
\section*{Acknowledgments}

It was a pleasure for me to work with all the wonderful people
here in Heidelberg. First of all, I would like to thank
Karl-Tasso Kn\"opfle for being a great advisor. His tremendous
support had a major influence on this thesis. He spend a lot
of time helping me as well as all the other people in our lab.
I would like to thank him for his "Scientific Writing"
introduction. I would like to thank Stefan Sch\"onert for
reviewing my thesis. I am happy to have such a supportive
supervisor. I enjoyed his interest in my research as well as
the fruitful discussions.\\
My thanks to my friends and colleagues for the great time I
had in our group. I enjoyed the atmosphere, their friendship,
and their support. My thanks to: Marik Barnabe-Heider, Claudia
Tomei, Tina Pollmann, Andrey Vasenko, Sergey Vasiliev, Dushan
Budjas, Grzegorz Zuzel, Mark Heisel, Werner Maneschg, Peter
Peiffer, Alexander Merle, and Alexander Dietz for the great
collaboration over the years. It was a pleasure to work with
all these people and to benefit from their knowledge.
Especially, I would like to thank Marik Barnabe-Heider and
Konstantin Gusev for the interesting and fruitful discussions,
for the great collaboration on GERDA, and for the
cool time we had in Gran Sasso.\\
Special thanks to Herbert Strecker and Ute Schwan for their
help on different technical problems.\\
My thanks to Marik Barnabe-Heider for proof-reading of thesis
and for providing helpful suggestions for improving this
manuscript. Any ``linguistic crimes'' are mine alone.\\
My thanks to Askhat Gazizov for his helpful insights on the subject.\\
My thanks to Vladimir Tretyak for providing me his code on
spectra relocation and delightful discussions on double beta
decay subjects.\\
My thanks to Prof. H.\,V.\,Klapdor-Kleingrothaus for being
open-minded in discussions.\\
My thanks to Frau Anja Berneiser, Frau Brigitte Villaumie and
Frau Ruth H\"afner for help on administrative matters.\\
Last but not least, I wish to thank my family who have always
supported me, Maryna and Maxim, and most of all Natalia for
enjoying life together with me.\\
This thesis has been supported by Max-Planck Gesellschaft
(MPG) and Max-Planck-Institut f\"ur Kernphysik. Their support
is gratefully acknowledged.


\newpage
\addtocontents{toc}{\bf Bibliography}



\end{document}